\documentclass[12pt]{report}

\usepackage{silence}
\usepackage[margin=2.5cm]{geometry}

\usepackage{dcolumn}
\usepackage{bm}

\usepackage{amsmath}
\usepackage{amssymb}
\usepackage{float}
\usepackage{graphicx}
\usepackage{breqn}
\usepackage{mathtools}
\usepackage{listings}
\usepackage{comment}
\usepackage{fancyhdr}
\usepackage[numbers,square,sort&compress]{natbib}
\usepackage{subcaption}
\usepackage{titlesec}
\usepackage{notoccite}
\usepackage[bottom]{footmisc} 

\titlespacing*{\section}{0pt}{5.5ex plus 1ex minus .2ex}{4.3ex plus .2ex}
\titlespacing*{\subsection}{0pt}{5.5ex plus 1ex minus .2ex}{4.3ex plus .2ex}

\usepackage{xcolor}
\usepackage[normalem]{ulem}

\usepackage[hidelinks]{hyperref}
\usepackage{cleveref}

\usepackage{placeins}

\newcommand{\infinity}{\infty}

\makeatletter
\renewcommand\tagform@[1]{\maketag@@@{\ignorespaces#1\unskip\@@italiccorr}}
\makeatother

\crefname{figure}{Fig.}{Fig.}
\Crefname{figure}{Figure}{Figures}
\crefname{equation}{Eq.}{Eq.}
\Crefname{equation}{Equation}{Equations}
\crefname{section}{Sec.}{Sec.}
\crefname{appendix}{App.}{App.}
\Crefname{appendix}{Appendix}{App.}
\Crefname{section}{Section}{Sections}
\crefname{chapter}{Ch.}{Ch.}
\Crefname{chapter}{Chapter}{Chapters}
\creflabelformat{equation}{#2\textup{#1}#3}

\begin{document}

\newcommand{\reftexts}{Ref.~}
\newcommand{\lips}[1]{\tilde{d} #1 \;}
\newcommand{\vecIV}[1]{#1} %
\newcommand{\vecIII}[1]{\vec{#1}} %
\newcommand{\vecII}[1]{\boldsymbol{#1}} %

\newcommand{\comm}[1]{%
{}%
}

\newcommand{\cole}[1]{%
{#1}%
}

\newcommand{\coleTwo}[1]{%
{#1}%
}

\newcommand{\wah}[1]{%
{#1}%
}

\newcommand{\rev}[1]{{#1}}

\makeatletter

\renewcommand\subsubsection{\@startsection{subsubsection}{3}{\z@}%
  {-3.25ex\@plus -1ex \@minus -.2ex}%
  {1.5ex \@plus .2ex}%
  {\normalfont\itshape}}

\newcommand{\coll}[2]{%
	$\coll@process{#1} + \coll@process{#2}$%
}

\newcommand{\collThree}[3]{%
	$\coll@process{#1} / \coll@process{#2} + \coll@process{#3}$%
}

\newcommand{\collFour}[4]{%
  $\coll@process{#1} / \coll@process{#2} / \coll@process{#3} + \coll@process{#4}$%
}

\renewcommand{\bibsection}{\chapter*{References}}

\newcommand{\coll@process}[1]{%
  \ifx#1A%
    #1%
  \else\ifx#1B%
    #1%
  \else\ifx#1x%
    #1%
  \else\ifx#1p%
    #1%
  \else\ifx#1h%
    #1%
  \else\ifx#1d%
    #1%
  \else
    \coll@checkHeThree{#1}%
  \fi\fi\fi\fi\fi\fi%
}

\newcommand{\coll@checkHeThree}[1]{%
  \ifnum\pdfstrcmp{#1}{He3}=0 %
    {}^3\mathrm{He}%
  \else
    \mathrm{#1}%
  \fi
}

\makeatother

\pagestyle{fancy}

\fancyhf{}

\newcommand{\fullthechapter}{Preamble}
\fancyhead[L]{\fullthechapter}
\fancyhead[R]{\thepage}

\renewcommand{\headrulewidth}{0.5pt}

\renewcommand{\footrulewidth}{0pt}

\newlength{\twosubht}
\newsavebox{\twosubbox}

\begin{titlepage}
    \centering

    {\scshape\Large University of Cape Town \par}
    \vspace{1.0cm}
    
    {\Huge\bfseries Energy Loss and Theoretical Uncertainties in Small Quark-Gluon Plasmas\par}
    \vspace{1.0cm}

    {\LARGE \textbf{Coleridge Faraday} \par}
    \vspace{1.5cm}

    \textbf{Supervisor:}\\
    {\Large Assoc.\ Prof.\ W.\ A.\ Horowitz\par}
    \vfill

    {\Large September 2024 \par}
		\vspace{1cm}

			\textit{ Thesis presented for the degree of Master of Science in the Department of Physics, University of Cape Town}

			\vspace{1cm}

    \includegraphics[width=7cm]{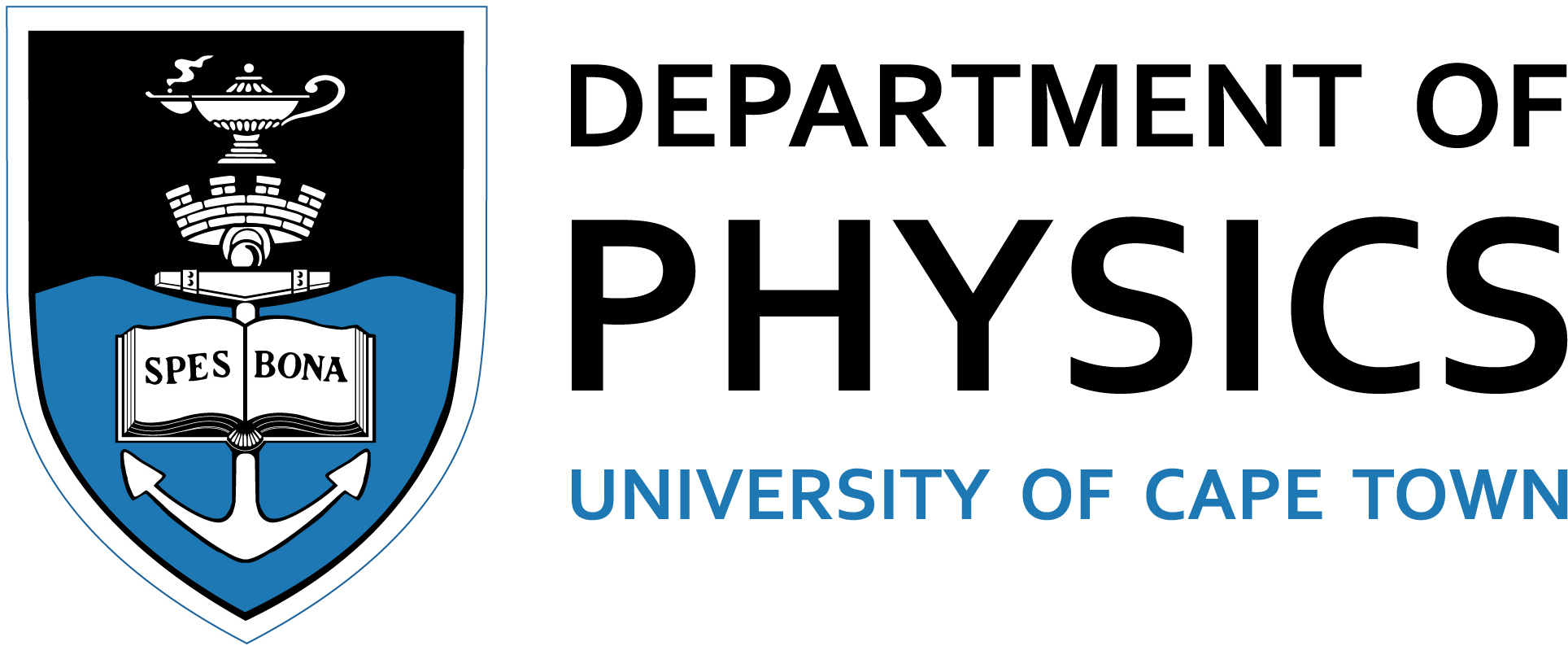}\par
\end{titlepage}

\renewcommand{\abstractname}{\LARGE\textbf{Abstract}}
\begin{abstract}

		There is a wealth of evidence that a Quark Gluon Plasma (QGP) is formed in heavy-ion collisions at RHIC and the LHC. Recently, there have been observations of QGP signatures in much smaller collision systems—including proton proton, and proton heavy-ion collisions---wherein QGP was not expected to form. Experimentally measuring suppression in small systems is more difficult than in large systems, motivating the need for theoretical guidance on the problem. The goal of this thesis is to systematically improve current energy loss models, particularly in how they pertain to small systems.

	We present a perturbative Quantum Chromodynamics (pQCD) based  energy loss model which receives small system size corrections to both the radiative and elastic energy loss, and which takes into account realistic collision geometry, production spectra, and fragmentation. We use the Djordjevic-Gyulassy-Levai-Vitev (DGLV) radiative energy loss model, and include a small system size correction which adds back in previously neglected terms that are suppressed according to the system size. We find that the correction is extremely large for pions at high momenta, which leads us to question the validity of various approximations in the model. We investigate the self-consistency of the various approximations used in the derivation of the Djordjevic-Gyulassy-Levai-Vitev (DGLV) radiative energy loss model, where we find that a particular approximation---the large formation time approximation---is \emph{not} satisfied self-consistently within the model. We explore a kinematic cutoff on the transverse radiated gluon momentum, which restores the self-consistency of this approximation, but at the cost of an increased sensitivity to the exact cutoff chosen. The exploration and quantitative treatment of theoretical uncertainties in the energy loss model is a central theme of this thesis.

In the same vein of uncertainty quantification, we investigate the common application of the central limit theorem to approximate the elastic energy loss as a Gaussian distribution. We find that all our results are remarkably insensitive to this approximation, not because we are in the regime of many scatters where the central limit theorem is applicable, but rather understood from an expansion of the $R_{AA}$ in terms of the moments of the underlying energy loss distributions. 
We also investigate the uncertainty in the elastic energy loss to the crossover between HTL and vacuum propagators.

Finally, we perform a one-parameter fit of the strong coupling $\alpha_s$ to available large system data from RHIC and LHC, keeping track of all of the aforementioned uncertainties. We find that the uncertainties may largely be absorbed into a different value of the strong coupling $\alpha_s$, but small uncertainty bands remain in any case. We explore differences in the energy loss models that remain even after the fit of the model to data, and find that the different elastic energy losses lead to different $p_T$ and system size dependencies. We also plot large-system constrained model results for small \collFour{p}{d}{He3}{A} collisions, where we find that our results are quantitatively consistent with small system data from RHIC and inconsistent with small system data from LHC.

\end{abstract}

\pagenumbering{roman}

\tableofcontents

\clearpage

\listoffigures

\clearpage

\section*{\centering Acknowledgments}
\vspace{3em}

First and foremost I would like to thank my supervisor, Will Horowitz, for all of the time and energy that he has invested into my studies. He has shown me what research is, made it possible to present at international conferences and visit CERN, and made me believe that a career in physics is possible.

I gratefully acknowledge funding from the SA-CERN collaboration for this M.Sc.\ degree as well as for funding trips to present at international conferences and to visit CERN. I also wish to thank the University of Cape Town and the National Research Foundation for their generous support related to travel. Computations in this thesis were performed using facilities provided by the University of Cape Town’s ICTS High Performance Computing team: \href{http://hpc.uct.ac.za}{hpc.uct.ac.za}.

I am grateful to all the people at CERN who have made my trips there possible and my stays productive and enjoyable, with special thanks to Urs Wiedemann who was essential in coordinating these trips. Thank you to everyone that I met at CERN for being extremely welcoming to me and for all of the interesting conversations, especially Govert Nijs, Jasmine Brewer, Wilke van der Schee, Sanyong Jeon, Krishna Rajagopal, João Barata, Enrico Speranza, and Shahin Iqbal. Thank you to Claire Lee for giving me a tour of CERN and the CMS experiment when I first visited CERN.
I am also appreciative for all the incredible people I have met in the field at conferences and other places of which there are too many to name. Some who stand out are Ivan Vitev, Magdelena Djordjevic, Isobel Kolbé, Borris Hippolyte, Jörg Aichelin, Micheal Murray, Matthew Sievert, and Xin-Nian Wang.

Thank you to all the other UCT masters students who have made this masters such an enjoyable experience. I am especially appreciative of all of the lunchtime philosophy and crosswords, campus work sessions till late, and rock climbing and hikes which have made the last few years memorable. Thank you to everyone in the physics department at the University of Cape Town for all the help that I have been provided with during this degree.

Thank you to my friends and family, but especially my parents for helping me through this degree and my undergraduate degree, and for the uncountable other things that they have done for me. Thank you to Ash for keeping me calm and sane, and for always being there for me. I could not have done this masters without your support.

\clearpage
\pagenumbering{arabic}
\renewcommand{\fullthechapter}{Chapter \thechapter}

 \chapter{Introduction}
 \label{sec:introduction}

 \section{Atoms and Beyond}
\label{sec:atoms_and_beyond}

There is a long history in physics, and indeed generally in science, of gaining insight through a careful understanding of objects in terms of more simple, and fundamental, constituents. 
This reductionist approach has led to the development of the Standard Model of particle physics (SM) \cite{Glashow:1961tr, Gross:1973id, Higgs:1964pj, Weinberg:1967tq} and General Relativity (GR) \cite{Einstein:1916vd}, two of the crowning achievements of 20th-century science. This \emph{Core Theory} \cite{Wilczek:2015,Carroll:2022} of $\text{SM} + \text{GR}$ consists of a few simple components which explain almost all phenomena that we see around us.

One of the earliest and most enduring ideas in this reductionist approach was the concept of atoms---indivisible pieces of matter that make up everything in the universe. Richard Feynman starts his famous lecture notes with the following \cite{Feynman:1963uxa}:
\begin{quote}
If, in some cataclysm, all of scientific knowledge were to be destroyed, and only one sentence passed on to the next generations of creatures, what statement would contain the most information in the fewest words? I believe it is the \emph{atomic hypothesis} (or the atomic fact, or whatever you wish to call it) that \emph{all things are made of atoms---little particles that move around in perpetual motion, attracting each other when they are a little distance apart, but repelling upon being squeezed into one another}. In that one sentence, you will see, there is an enormous amount of information about the world, if just a little imagination and thinking are applied.
\end{quote}
The idea of atoms is often attributed to the ancient Greek philosopher Democritus, who proposed that all matter is composed of tiny, indivisible units called ``atomos", meaning ``uncuttable" in Greek. These ideas were purely philosophical, and would remain untestable for centuries.
In the early 19th century, Dalton solidified the concept of atoms as a scientific theory by postulating that each chemical element consists of unique atoms with specific masses \cite{Dalton:1808}. 
This postulate explained why elements, which combine in fixed ratios to form compounds, have the masses that they do.

The next set of insights into atoms would arrive in the late 19th and early 20th centuries.
Initially conceived of as indivisible units by Dalton, atoms were discovered to be composite particles with smaller, and more fundamental constituents. We will see that this decomposition of objects into more fundamental constituents is a common theme in physics.
Thomson discovered negatively charged electrons in 1897 \cite{Thomson:1897}. This led him to propose the ``plum pudding" model, where the negative, point-like electrons (plums) were embedded in a positively-charged medium (pudding), leading to an overall neutral atom. Rutherford's now famous 1911 gold foil experiment involved firing alpha particles at a thin sheet of gold foil. The observation that the beam was mostly undeflected, but with rare, high-momentum-transfer scatters, suggested a small, dense, positively charged core--the \emph{nucleus}.

Rutherford's notion of the nucleus, in combination with orbiting electrons, was classically not stable as the accelerating electrons should radiate, losing energy and falling into the nucleus. Niels Bohr addressed this instability in 1913 by introducing quantized energy levels \cite{Bohr:1913}. This concept explained discrete spectral lines and provided the building blocks for quantum mechanics. In the 1920s, Heisenberg, Schrödinger, Planck, Einstein, de Broglie,  and others, further refined atomic models with the introduction of Quantum Mechanics (QM) \cite{Heisenberg:1925,Schrodinger:1926,Planck:1901,Einstein:1905,deBroglie:1924}.
\rev{The electrons in an atom were now described with fixed energy levels (as well as other fixed quantum numbers), which protected them from falling into the nucleus. Quantum mechanics offered an explanation of why atoms were stable as well as other famous experiments, including the double slit experiment.}

James Chadwick's 1932 discovery of the neutron \cite{Chadwick:1932}  was the final component for the basic structure of the atom. It was observed to be neutral and of a similar mass to the proton.

The latter half of the 20th century saw further refinements, notably the development of quantum electrodynamics (QED) by Schwinger, Tomonaga, and Feynman, which unified special relativity and classical electrodynamics with quantum mechanics \cite{Feynman:1948,Schwinger:1948,Tomonaga:1948}. The Standard Model (SM) of particle physics would later be completed by the description of the weak \cite{Glashow:1961tr} and strong \cite{Politzer:1973fx} forces.

\section{The Discovery of Quarks}
\label{sec:the_discovery_of_quarks}

\begin{figure}[!htbp]
	\centering
	\includegraphics[width=.6\linewidth]{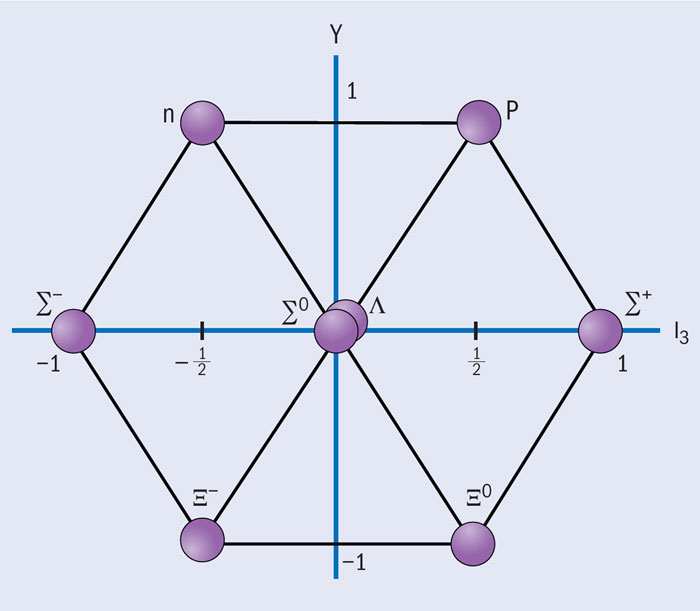}
	\caption{The \emph{Eightfold Way}: SU(3) octet of the ground state baryons. Figure taken from \cite{cerncourier_history_qcd}.}
	\label{fig:eightfold-way}
\end{figure}

Around sixty years ago, experiments at the Radiation Laboratory in Berkeley \cite{Rochester:1947mi} and cosmic ray studies \cite{Lattes:1947mw} led to the discovery of several new particles, including $\Delta$ baryons, $K$ mesons, and hyperons. These discoveries prompted Murray Gell-Mann \cite{Gell-Mann:1961omu} and Yuval Ne'eman \cite{Neeman:1961jhl} to propose that these particles could be classified within the framework of SU(3) symmetry, in a scheme known as the \emph{Eightfold Way}. In this scheme, baryons were organized into octets and decuplets, while mesons were grouped into octets and singlets. \Cref{fig:eightfold-way} illustrates the baryon octet, akin to a periodic table for particle physics. The interested reader may refer to \cite{cerncourier_history_qcd} for Harald Fritzsch's recount of the history of the development of QCD, which this section is largely based on.

At that time, however, it was not clear why the simplest representation of SU(3)—the fundamental (triplet) representation—was not observed in experiments. Gell-Mann \cite{Gell-Mann:1964ewy} and George Zweig \cite{Zweig:1964jf} later proposed the quark model independently, in which baryons and mesons are composed of more fundamental constituents called quarks.  Again, the indivisible units---protons and neutrons this time around---had been divided. The fact that the triplet representation is not observed, would later be explained by the presence of color charge and the fact that all hadrons must be color-neutral.

The community was initially skeptical of the quark model, as quarks had never been directly observed. The definitive evidence for quarks came from deep inelastic scattering experiments at the Stanford Linear Accelerator Center (SLAC) in the 1960s. These experiments showed that protons were composite particles containing point-like particles---the quarks.

\section{Color charge and Quantum Chromodynamics}
\label{sec:the_development_of_quantum_chromodynamics_qcd}

\begin{figure}[!htbp]
	\centering
	\includegraphics[width=0.75\linewidth]{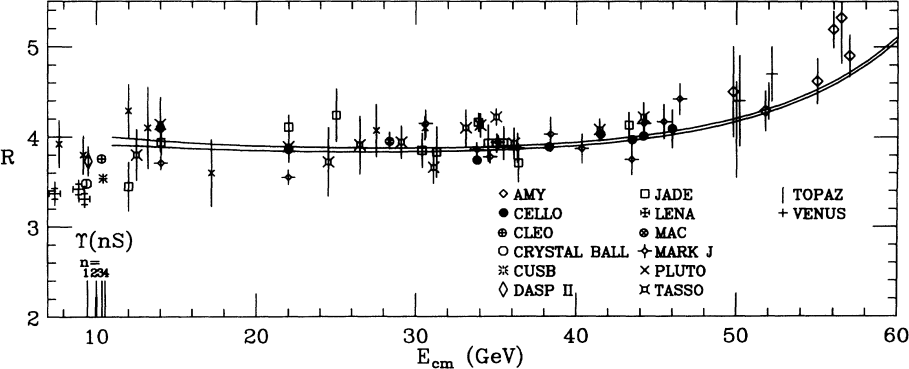}
	\caption{Plot of the ratio $R = \sigma_{\text{hadron}} / \sigma_{\text{muon}}$ as a function of center of mass energy $E_{\text{CM}}$. Data is shown as indicated in the legend. Solid line is the quark model prediction for 3 colors and 5 quark flavors (energies are not large enough to produce $t \bar{t}$ pairs). Figure taken from III.80 of \cite{ParticleDataGroup:1992tph}.}
	\label{fig:five_quarks_pdg}
\end{figure}

A spin statistics problem related to the wave function of the $\Omega^-$ meson, composed of three strange quarks, was proving to be a problem for the quark model in the 1970s \cite{cerncourier_history_qcd}. Gell-Mann and Fritzsch \cite{Fritzsch:1973pi} solved this issue by introducing a new charge, dubbed ``color charge", and requiring that all hadrons are color neutral. A compelling and simple calculation which helped bolster this picture of color-charged quarks, was the ratio of the hadron to muon production cross section
\begin{equation}
	R \equiv \frac{\sigma_{e^+ e^- \to \text {hadrons }}}{e^+ e^- \to \sigma_{\text {muons}}}=\frac{\sum \sigma_{q \bar{q}}}{\sigma_{\text {muons }}}=\sum\left(\frac{q_q}{e}\right)^2.
	\label{eqn:ratio_of_hadron_to_muon_production_cross_section}
\end{equation}
This calculation depends on the center of mass energy $\sqrt{s}$ of the collision in terms of how many quark flavors enter the calculation. If one performs this simple estimate for $\sqrt{s} \sim 10 \text{--}100 ~\mathrm{GeV}$, all quarks but the top quark contribute, and we obtain $R \approx 3 \times [ 4/9 +1/9 + 1/9 + 4/9 + 1/9 ] = 33/9 \simeq 3.7$ for three colors while for one color we obtain $R \approx 1 \times [ 4/9+1/9+1/9+4/9+1/9 ]=11/9 \simeq 1.2$. \Cref{fig:five_quarks_pdg} shows experimental measurements of the ratio $R$, alongside a more detailed QCD calculation (solid curve). We see that for most energies our simple estimate of $R$ provides compelling evidence that there are three quark colors and five active quark flavors. A detailed scan over a range of $\sqrt{s}$ also showed the rough masses we should expect the quarks to have, as their contributions turn off for $\sqrt{s}$ less than their mass.

\begin{figure}[!htbp]
	\centering
	\includegraphics[width=0.5\linewidth]{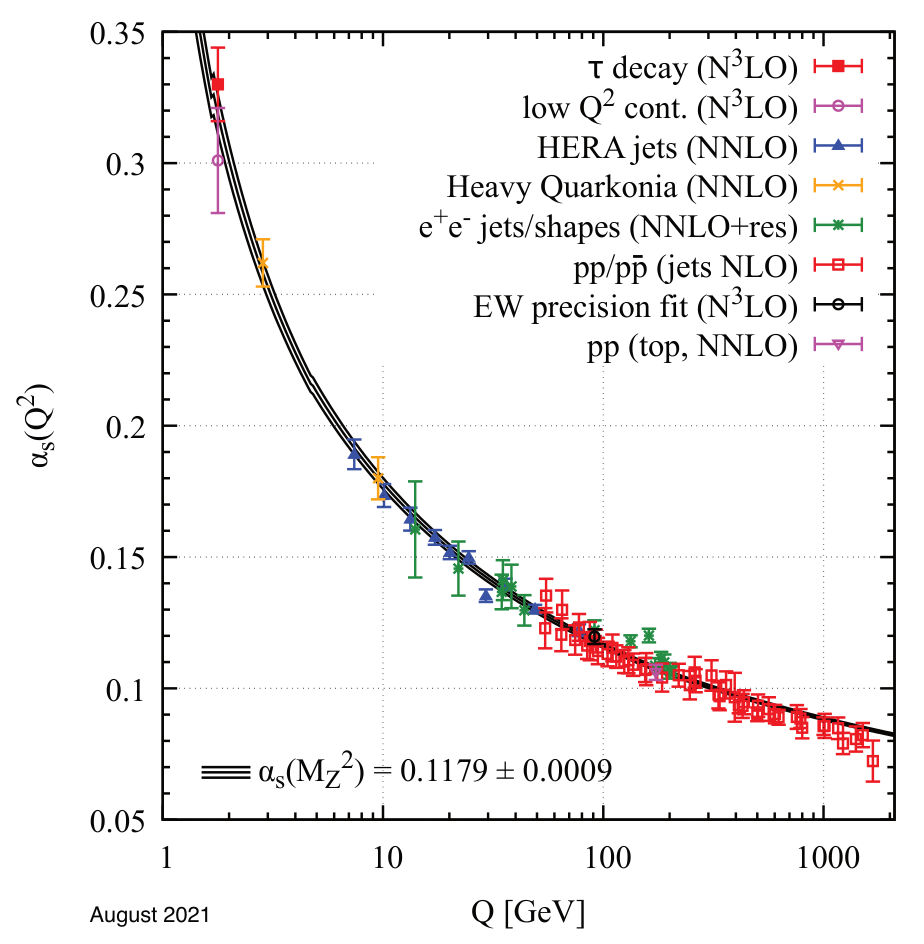}
	\caption{Summary plot of the experimental determination of the running of the strong coupling $\alpha_s$ as a function of the energy scale $Q$. Figure taken from \cite{ParticleDataGroup:2022pth}.}
	\label{fig:running_coupling_pdg_2022}
\end{figure}

``Confinement" is one of the most unique aspects of QCD \cite{Gross:1973id,Gross:1973ju}. Confinement forces quarks to form color-neutral hadrons which explains why quarks are never found in isolation. One may think of this as arising from the length dependence of the strong force: as color-charged particles move apart, the force between them increases. If one were to fully separate the particles, so much energy would have been invested that extra particles are created from the vacuum to neutralize the color-charged particles. QCD also explains the concept of ``asymptotic freedom," discovered by Gross, Wilczek, and Politzer in 1973. Asymptotic freedom refers to the property that quarks behave almost as free particles when they are extremely close together, which is because the strong force becomes weaker as distances become smaller or equivalently as energies become larger  \cite{Gross:1973id, Politzer:1973fx,Gross:1974cs,Gross:1973ju}. This discovery was crucial in establishing QCD as the correct theory of the strong interaction. \Cref{fig:running_coupling_pdg_2022} plots the running of the strong coupling---which captures the strength of the interaction between two color charged objects---as a function of energy scale. The fact that $\alpha_s(Q \to \infty) \to 0$ is a manifestation of asymptotic freedom. In contrast, the QED coupling increases as a function of energy scale $Q$ \cite{ParticleDataGroup:2022pth}. 

\section{Emergent strongly interacting matter}
\label{sec:phase_diagram_of_qcd}

Having introduced the fundamental properties of QCD, we now turn our attention to one of the most interesting aspects of strongly interacting matter: the emergence of collective behavior from the simple rules of QCD. This emergent behavior provides the background context for the field of heavy-ion physics, with one of the primary goals being to understand how strongly interacting matter behaves at various temperatures. Heavy-ion physics collides heavy ions (for instance lead or gold) and aims to study the resulting emergent medium of strongly interacting particles. The Relativistic Heavy-Ion Collider (RHIC) was built for the purpose of studying this emergent matter \cite{PHENIX:2004vcz, PHOBOS:2004zne, STAR:2005gfr,BRAHMS:2004adc} and the Large Hadron Collider (LHC) also studies this matter \cite{ALICE:2022wpn}. The emergent strongly interacting matter which is formed in heavy-ion collisions will be discussed in detail in \cref{sec:background}.

 \chapter{Background}
 \label{sec:background}
 \section{Why study the Quark Gluon Plasma?}

\begin{figure}[!b]
	\centering
	\includegraphics[width=0.75\linewidth]{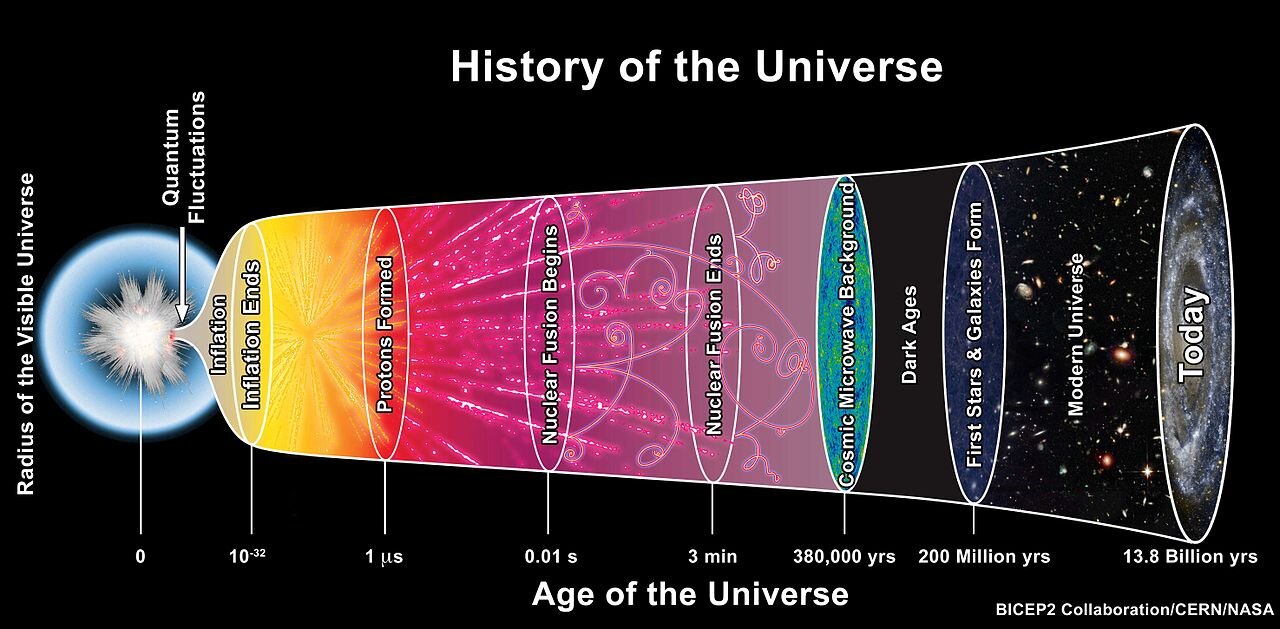}
	\caption{Illustration of the different phases of matter that the universe was in as a function of time. (Figure taken from \cite{recreating_big_earth} and originally produced by BICEP2 Collaboration/CERN/NASA)}
	\label{fig:qgp_universe_history}
\end{figure}

Extremely high temperatures and densities are expected to have been present in the early universe, according to the Big Bang model \cite{Kolb:1990vq}. Specifically, between $10^{-10}$ and $10^{-6}$ seconds after the Big Bang, it is expected that a quark-gluon plasma (QGP) was present \cite{Collins:1974ky}. The QGP is a novel state of matter which is formed at extremely high temperatures, where hadrons melt down, and the fundamental degrees of freedom of QCD---the quarks and gluons---move around freely. 
Insight into this period of the universe cannot be directly gained through measurements by telescopes, as is the standard approach for cosmological observations, because the universe was opaque to electromagnetic radiation until about three hundred thousand years after the Big Bang. There was hope in the early days of the field that a first order phase transition, which was assumed to occur between the QGP and regular hadronic matter, might leave an imprint on the matter distribution in the universe \cite{Busza:2018rrf}. However, lattice QCD \cite{Aoki:2006we} calculations showed that the crossover was continuous, removing the possibility of such an imprint. Therefore, understanding the properties of the QGP through experimental observation in heavy-ion collisions is likely the best---if not the only---way to probe the conditions of the primordial universe. \Cref{fig:qgp_universe_history} illustrates the expansion of the universe and the state of matter present in the universe at different times.

\begin{figure}[!htbp]
	\centering
	\includegraphics[width=0.65\linewidth]{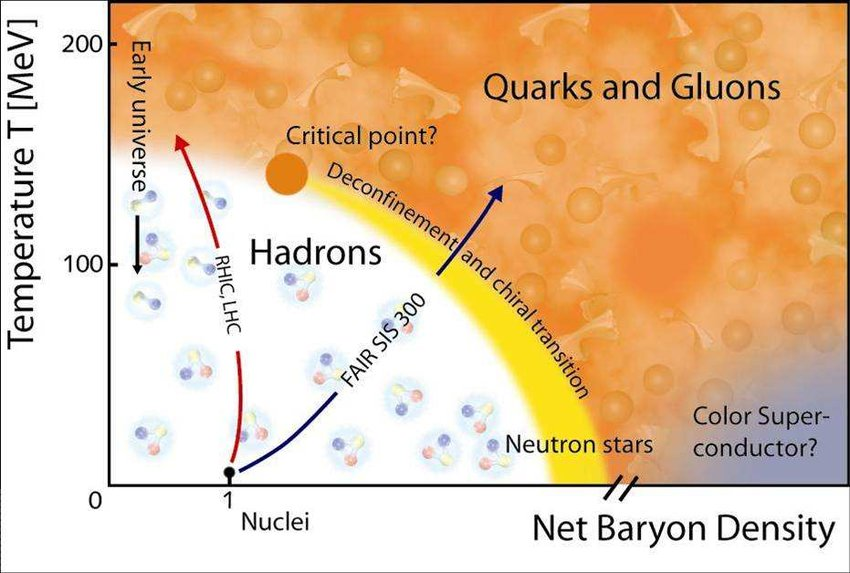}
	\caption{Possible sketch of the phase diagram of QCD. The vertical axis is the temperature $T$ and the horizontal axis is the net baryon density. Figure taken from \cite{Aarts:2014fsa}.}
	\label{fig:qcd_phase_diagram}
\end{figure}

The phase diagram of QCD has been a driving goal in the field of heavy-ion physics since its inception \cite{Busza:2018rrf}. As previously mentioned, there was a strong motivation to understand whether a first order phase transition existed between the QGP and regular hadronic matter, which amounts to a partial mapping of the phase diagram. \Cref{fig:qcd_phase_diagram} shows a schematic representation of the QCD phase diagram. Collider experiments explore the high temperature, low net baryon density region of the phase diagram, which is similar the conditions of the early universe. Other features indicated on the phase diagram include a potential critical point, as well as a first order chiral phase transition, and the unknown make up of neutron stars. The presence and behavior of such features on the phase diagram is an important topic for the field. The beam energy scan (BES) \cite{Bzdak:2019pkr} at RHIC aims to locate a potential critical point, by lowering the center of mass energy $\sqrt{s}$, although currently the results from this study and similar ones are inconclusive.

Another reason for studying the QGP is to perform precision tests of the Standard Model. While Quantum Electrodynamics (QED) has been tested to a few parts in a billion and the weak force to a few parts in a thousand, perturbative forms of Quantum Chromodynamics (QCD) have been tested to only a few percent \cite{ParticleDataGroup:2022pth}. The quark-gluon plasma, containing the fundamental degrees of freedom of QCD, may provide a valuable system for precision tests of QCD. 
Other theoretical tools may also be improved by studying the QGP including thermal field theory, perturbation theory, as well as more phenomenological pieces of QCD such as parton distribution functions and fragmentation functions. 
Research on the QGP may also help measure properties of QCD and nuclear matter, including nuclear modifications to fragmentation and parton distribution functions \cite{Eskola:2012rg}, and precision nuclear structure measurements \cite{Giacalone:2024ixe,Giacalone:2024luz}.

The QGP is a relatively simple system in which we can study the onset of collectivity \cite{Busza:2018rrf}. Hydrodynamic simulations of the QGP have shown that surprisingly few particles are needed for collectivity to emerge in a system. Understanding this onset of collectivity is crucial for the broader study of complex systems. Understanding whether a QGP is formed in even the smallest of collision systems---such as proton-proton (\coll{p}{p}) and proton-heavy-ion (\coll{p}{A})---and the extent to which such small QGPs are described by hydrodynamics, will be invaluable for understanding the onset of collectivity.

At present, there is a wealth of evidence supporting the formation of a novel, dense state of matter in heavy-ion collisions at the LHC and RHIC \cite{Pasechnik:2016wkt, Qin:2015srf, Busza:2018rrf, ALICE:2022wpn, ALICE:2022wpn}. This state of matter is identified as a strongly-coupled fluid made of quarks and gluons, known as a quark-gluon plasma (QGP). Since the QGP formed in these collisions is short-lived, with a lifetime of approximately $10~\mathrm{fm}/c$ or $10^{-23}~\mathrm{s}$\footnote{From here on, we will use natural units with $\hbar = c = 1$, which implies $\hbar c \simeq 0.197 \,\mathrm{fm} \; \mathrm{GeV}$.}, its existence must be inferred from the imprint it leaves on the particles emitted from the plasma. Moreover, the quarks and gluons that make up the plasma are not directly detected; instead, we observe color-neutral hadrons. This necessitates a detailed understanding of the various components of the collision, the formation of the QGP, its subsequent evolution, and the fragmentation into hadrons.

\section{Bulk observables}
\label{sec:bulk_observables}

\emph{Bulk observables} in the context of heavy-ion physics refers to observables of the low-$p_T$ particles which emerge from a heavy-ion collision. This includes particle spectra, mean $p_T$, and flow coefficients, all as a function of transverse momentum $p_T$, pseudorapidity $\eta$, centrality, and particle species. For reviews, read \cite{Gale:2013da, Muller:2011tu,Herrmann:1999wu}. For a detailed discussion of the experimental evidence for QGP formation, including bulk observables but also high-$p_T$ observables, refer to the white papers on QGP formation from the four experiments at RHIC \cite{PHENIX:2004vcz, PHOBOS:2004zne, STAR:2005gfr,BRAHMS:2004adc} and the comprehensive ALICE review \cite{ALICE:2022wpn}.
Since this thesis is focused on QGP formation in small systems, we emphasize the results from small systems in this section.

\subsection{Strangeness enhancement}
\label{sec:strangeness_enhancement}

Strangeness enhancement was first suggested as a potential signature of QGP formation by Rafelski and Hagedorn \cite{Rafelski:1980rk} in 1980. This was later refined \cite{Rafelski:1982pu} to produce a set of predictions for heavy-ion collisions at the SPS at CERN. The quark content of the collided heavy-ions accelerators is simply that of protons and neutrons, i.e.\ up and down quarks. The idea behind strangeness enhancement is then, because the critical temperature $T_c \simeq 170~\mathrm{MeV}$%
 is similar to the strange quark mass $m_s \simeq 100~\mathrm{MeV}$, strange quarks should be hard to produce for temperatures below the QCD critical temperature \cite{Muller:2011tu}. For temperatures greater than the critical temperature $T_C$, strange quarks should be produced thermally, leading to an enhancement of strange hadrons when a QGP is produced.

\begin{figure}[!htbp]
	\centering
	\includegraphics[width=0.5\linewidth]{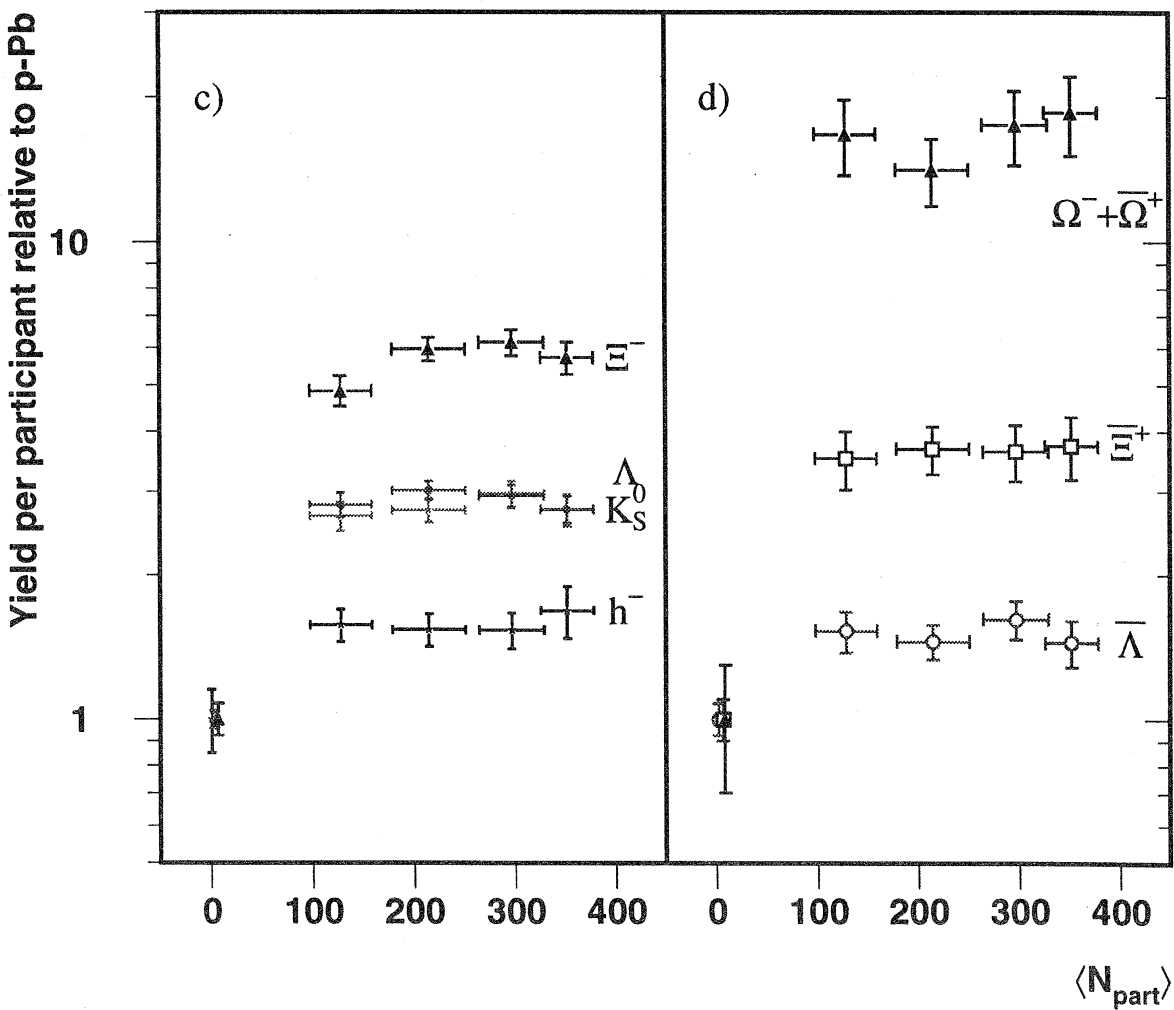}
	\caption{Yields per participant, expressed in units of yields observed in \coll{p}{Pb} collisions, as a function of the number of participants for (c) $\mathrm{h}^{-}$, $\mathrm{K}_{\mathrm{S}}^0, \Lambda$ and $\Xi^{-}$; (d) $\bar{\Lambda}$, $\bar{\Xi}^{+}$and $\Omega^{-}+\bar{\Omega}^{+}$. The proton points are juxtaposed on the horizontal scale. Figure is taken from Fig.\ 3 in \cite{WA97:1999uwz}.}
	\label{fig:wa97_strangeness_enhancement}
\end{figure}

\begin{figure}[!htbp]
	\centering
	\includegraphics[width=0.55\linewidth]{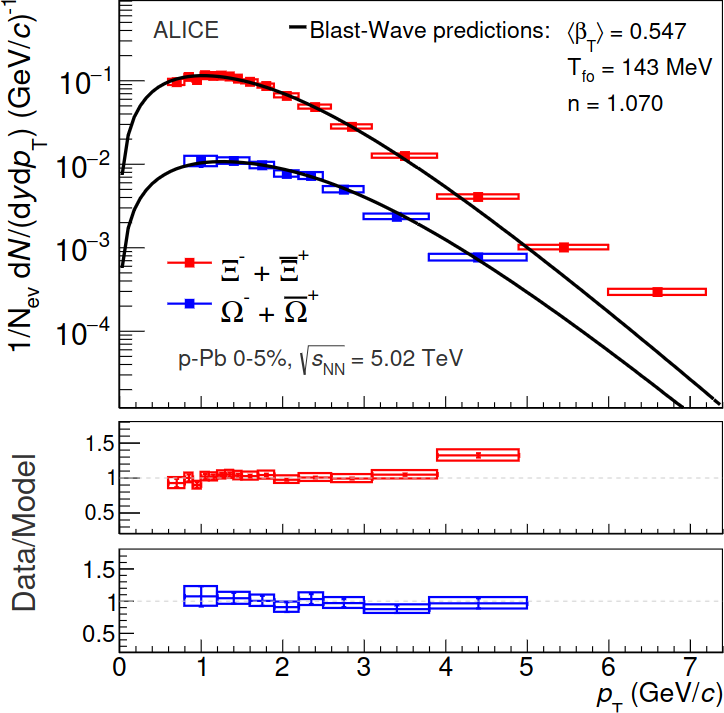}\hfill
	\includegraphics[width=0.44\linewidth]{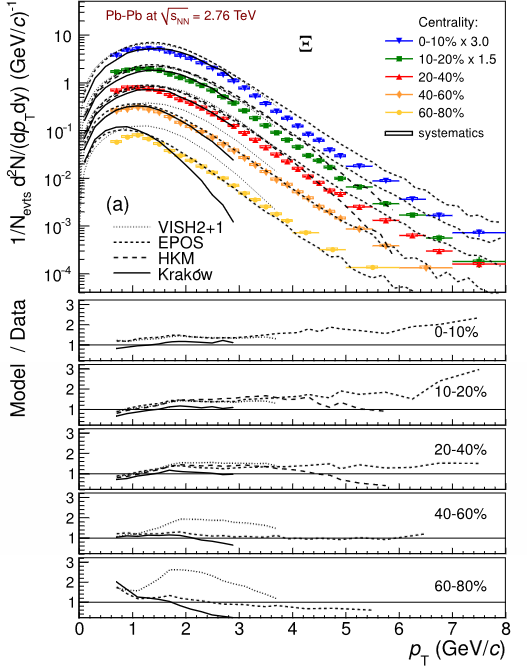}
	\caption{(left) $\left(\Xi^{-}+\bar{\Xi}^{+}\right)$and $\left(\Omega^{-}+\bar{\Omega}^{+}\right) p_{\mathrm{T}}$ spectra in the $0-5 \%$ multiplicity classes compared to predictions from the BG-BW model (upper panels) with the ratios on a linear scale (lower panels). Figure is taken from Fig.\ 3 in \cite{ALICE:2015mpp}. (right) $\Xi$ $p_T$ spectra for various centrality classes compared to hydrodynamic models. Figure taken from Fig. 4 of \cite{ALICE:2013xmt}.}
	\label{fig:alice_strangeness_enhancement_ppb}
\end{figure}

 This was first observed by the WA97 experiment at the SPS at CERN in 1999 \cite{WA97:1999uwz}, with a statement from CERN in 2000 that ``we now have evidence of a new state of matter where quarks and gluons are not confined" \cite{cernPress2000}. \Cref{fig:wa97_strangeness_enhancement} plots the measured yield of strange and multistrange hadrons normalized according to the number of participants. We observe that all strange hadrons are significantly enhanced above the expected number from proton-proton collisions and that this enhancement increases as a function of strangeness content. That is, the normalized yield per participant of $\Omega = s s s$, where $s$ is the strange quark, is significantly higher than that of $\Xi = (u \text{ or } d) ss $, which is in turn significantly higher than that of $\Lambda = u d s$. This is consistent with the thermal production of strange quarks according to the QGP scenario, but inconsistent with other scenarios such as hadronic rescattering models (see \cite{WA97:1999uwz} and references within).

 More recently, the ALICE collaboration has observed strangeness enhancement in high-multiplicity collisions \coll{p}{p} \cite{ALICE:2016fzo} and \coll{p}{Pb} collisions \cite{ALICE:2015mpp, ALICE:2013wgn}.
 \Cref{fig:alice_strangeness_enhancement_ppb} shows the measured $p_T$ spectra of strange hadrons as a function of $p_T$ in central \coll{p}{Pb} (left) and multiple centrality \coll{Pb}{Pb} (right) collisions. Various hydrodynamic model predictions are also shown; see \cite{ALICE:2013xmt} and the citations within. Amazingly, hydrodynamic models can quantitatively predict the strange hadron yield in \emph{both} \coll{p}{Pb} and \coll{Pb}{Pb} collisions. This is indicative of QGP formation in high-multiplicity \coll{p}{Pb} collisions.

\subsection{Anisotropic flow}
\label{sec:collective_phenomena}

Before RHIC turned on in 2000, it was widely anticipated that because of the high center of mass energy, $\sqrt{s_{NN}} = 200~\mathrm{GeV}$, and the asymptotic freedom of QCD, the medium which was formed in heavy-ion collisions would be a weakly-coupled\footnote{Here we use the term ``weak coupling" not in the sense of the weak or strong force, but in the sense of the QCD coupling being much less than one.} gas of quarks and gluons (``partons" collectively) \cite{Gale:2013da}. This gas of weakly-coupled partons expands isotropically, leaving the initial momentum distribution of particles---which before the evolution of the medium is isotropic---unchanged. In actuality, the first results from the STAR experiment at RHIC \cite{STAR:2000ekf} indicated a large azimuthal anisotropy which had been predicted by ideal hydrodynamics models \cite{Jacobs:2000wy,Huovinen:2001cy}. The success of hydrodynamics indicated that the medium formed in heavy-ion collisions at RHIC was not a weakly-coupled gas of partons, but rather a strongly-coupled fluid.

Subsequent measurements of $v_2$ for various particle species revealed a mass-dependent splitting, consistent with hydrodynamic predictions and demonstrating that the particles emerging from the QGP originated from a common velocity field \cite{Gale:2013da}. Refinement of hydrodynamic models since the first predictions of $v_2$ included viscous corrections \cite{Luzum:2008cw, Romatschke:2007mq,Chaudhuri:2007qp}, event-by-event simulation and fluctuation of the initial state \cite{Gale:2012rq}, and more realistic initial state models \cite{Schenke:2012wb}. 

Fluctuations were found to be important in \coll{Cu}{Cu} collisions at RHIC \cite{PHOBOS:2006dbo}, where the measured $v_2$ in central collisions was nonzero. This was not expected from hydrodynamic models which used smooth initial conditions arising from the Woods-Saxon distributions. The resolution of this apparent inconsistency was to include fluctuating initial conditions and to calculate the $v_2$ not relative to the impact parameter but rather to an axis determined on an event-by-event basis \cite{Schenke:2012wb}.

\subsection{Bayesian analysis for all bulk observables}
\label{sec:era_of_bayesian_analysis}

\begin{figure}[H]
    \centering
    \begin{subfigure}[t]{0.45\textwidth}
        \centering
        \includegraphics[width=\linewidth]{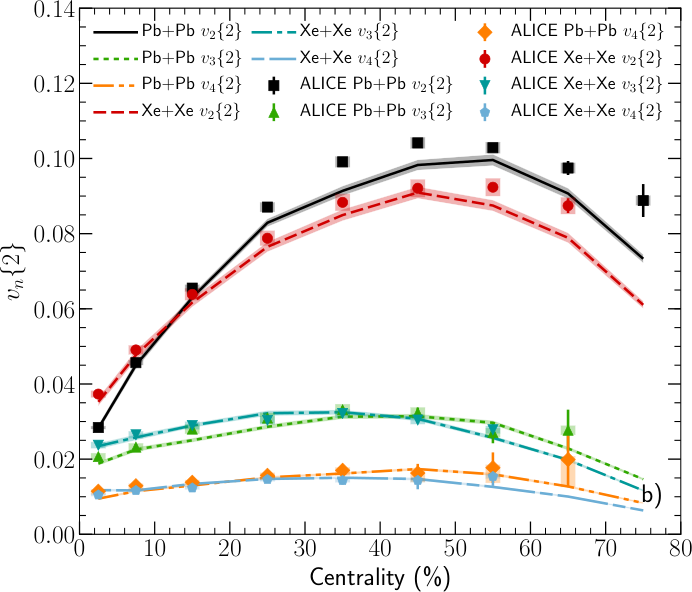}
        \caption{The elliptic flow $v_2$ as a function of centrality (\%) for different particle species in heavy-ion collisions, comparing experimental data from ALICE with theoretical predictions. Figure taken from \cite{Schenke:2020mbo}.}
    \end{subfigure}\hfill
    \begin{subfigure}[t]{0.53\textwidth}
        \centering
        \includegraphics[width=\linewidth]{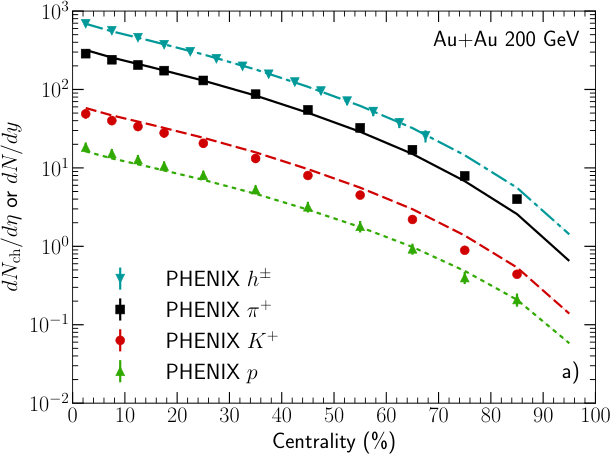}
        \caption{Transverse momentum ($p_T$) spectra of identified particles ($\pi^\pm$, $K^\pm$, $p$, and $\bar{p}$) measured by the PHENIX experiment in Au+Au collisions at 200 GeV as a function of centrality. Figure taken from \cite{Schenke:2020mbo}.}
    \end{subfigure}
		\vspace{1em}
    \begin{subfigure}[t]{0.53\textwidth}
        \centering
        \includegraphics[width=\linewidth]{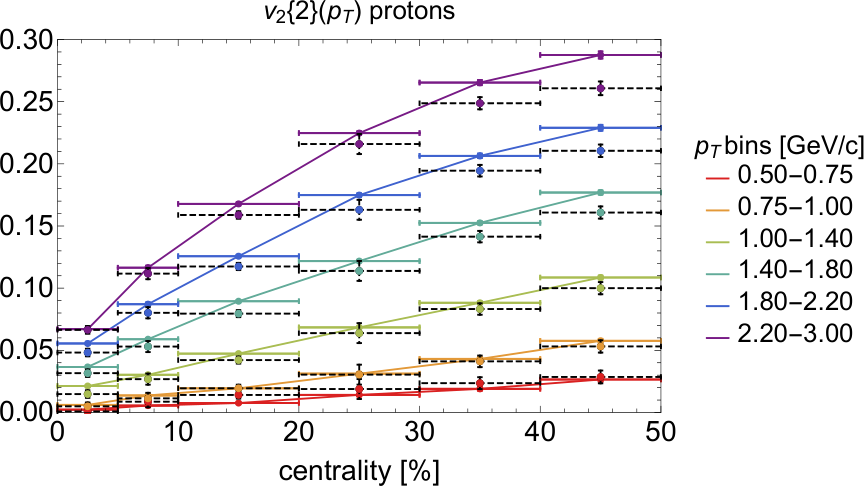}
        \caption{$v_2(p_T)/n$ protons as a function of centrality (\%) for different transverse momentum ($p_T$) bins in heavy-ion collisions. Figure taken from \cite{Nijs:2020roc}.}
    \end{subfigure}\hfill
    \begin{subfigure}[t]{0.45\textwidth}
        \centering
        \includegraphics[width=\linewidth]{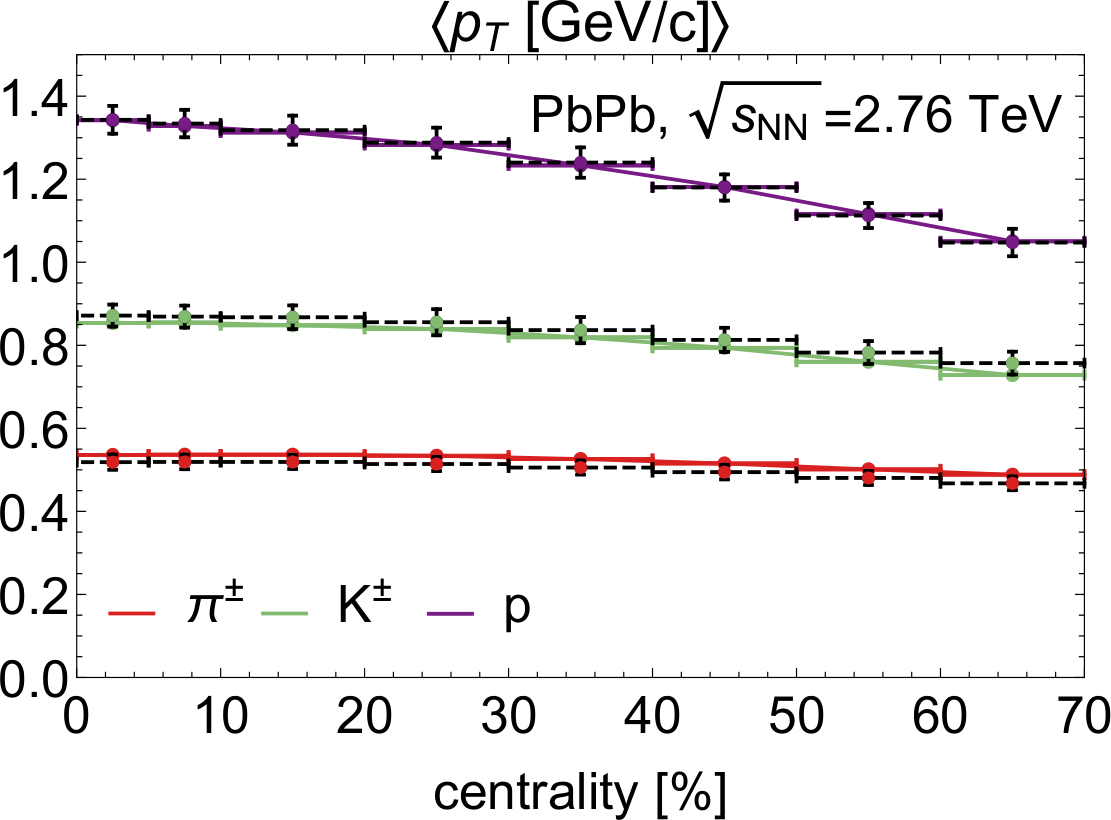}
        \caption{Mean transverse momentum $\langle p_T \rangle$ of $\pi^\pm$, $K^\pm$, and $p$ as a function of centrality (\%) in \coll{Pb}{Pb} collisions at $\sqrt{s_{NN}} = 2.76$ TeV. Figure taken from \cite{Nijs:2020roc}.}
    \end{subfigure}
    \caption{Plot of a small subset of posterior hydrodynamic model results for a variety of low-$p_T$ data from two different hydrodynamic models, \cite{Schenke:2020mbo} (top) and \cite{Nijs:2020roc} (bottom). }
    \label{fig:hydro_model_comparison}
\end{figure}

Many current hydrodynamics models \cite{Schenke:2020mbo, Nijs:2020roc, Bernhard:2019bmu} show remarkably good agreement with the vast majority of low-$p_T$ observables. Typically, current hydrodynamic models attempt to fit a variety of data using Bayesian analysis in order to constrain the properties of the hydrodynamic model. The observables used include mean $p_T$, $v_n$, various particle spectra, all as a function of $p_T$, pseudorapidity $\eta$, and particle species. \Cref{fig:hydro_model_comparison} shows a subset of results from two hydrodynamic model \cite{Schenke:2020mbo, Nijs:2020roc} compared with experimental data. The fit is extremely good, and this is representative of the fit to all low-$p_T$ experimental observables.

\begin{figure}[!htbp]
	\centering
			\includegraphics[width=\linewidth]{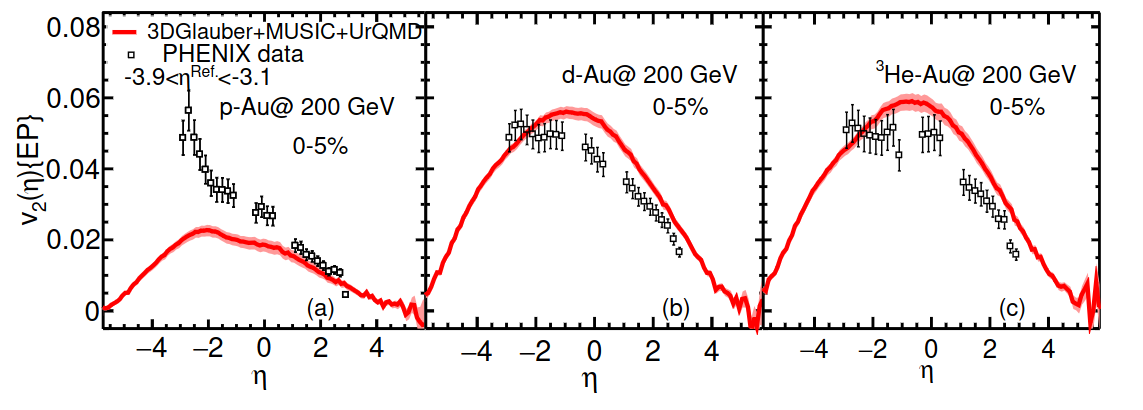}
			\caption{Elliptic flow $v_2\{2\}$ as a function of pseudorapidity for \coll{p}{Au} (left), \coll{d}{Au} (middle) and \coll{He3}{Au} (right) collisions at RHIC at $\sqrt{s_{NN}} = 200 ~\mathrm{GeV}$. Data from PHENIX experiment is also shown. Figure taken from \cite{Zhao:2022ugy}.}
	\label{fig:music_v2_small_systems}
\end{figure}

It was not theoretically expected that azimuthal flow would be produced in \coll{p}{p} and \coll{p}{Pb}, as it was not expected that QGP forms in these systems. However, in 2010 and subsequent years many experimental measurements from experiments at the LHC and RHIC found azimuthal anisotropy in high-multiplicity \coll{p}{p}, \coll{p}{Pb}, and \coll{d}{Au} collisions \cite{ATLAS:2014qaj, CMS:2010ifv, CMS:2012qk, CMS:2013jlh, CMS:2015yux, PHENIX:2013ktj}. The same hydrodynamic models which successfully described heavy-ion data \cite{Schenke:2020mbo, Weller:2017tsr}, also can be used to make predictions for the observed results in small systems \cite{Zhao:2022ugy}. \Cref{fig:music_v2_small_systems} shows the $v_2$ as a function of pseudorapidity $\eta$ for central \coll{p}{Au} (left), \coll{d}{Au} (middle) and \coll{He3}{Au} (right) collisions at RHIC at $\sqrt{s_{NN}} = 200 ~\mathrm{GeV}$. The $v_2$ predicted by the hydrodynamics model \cite{Schenke:2020mbo} is qualitatively consistent with the observed data, although there are some deviations. This qualitative consistency is suggestive of QGP formation in these small systems, but is far from definitive proof.

\section{The Glauber Model}
\label{sec:back_the_glauber_model}

The Glauber model is one of the most successful models in relativistic heavy-ion physics. It is a simple framework which calculates various geometrical quantities from the collision of composite particles, with remarkably accurate predictions. For a review read \cite{Miller:2007ri}.

Roy Glauber developed the ``Glauber model" to make theoretical predictions for high energy scattering involving composite particles. This was presented in his 1958 lectures \cite{Glauber:1959,Miller:2007ri}. The Glauber model views the collision of two colliding nuclei in terms of the interactions of the individual nucleons. The model works in the high-energy or Eikonal limit, and therefore assumes that the nucleons are \rev{not} deflected during the collision. One also assumes that the nucleons move independently inside the nucleus, with a probability distribution given by the density of nuclear charge. This means that the only necessary inputs to the model are the nuclear charge density $\rho_N(\vec{r})$ and the nucleon-nucleon cross section $\sigma_{NN}$. The nuclear charge densities are typically characterized by Fermi distributions with parameters measured in cold nuclear matter experiments. From this simple set of assumptions, one may calculate two important geometrical quantities: the number of \emph{binary collisions} and the number of \emph{participants}. The number of binary collisions is the number of nucleon-nucleon collisions which occur during the full \coll{A}{B} heavy-ion collision. The number of participants (or wounded nucleons) refers to the number of nucleons which undergo at least one binary collision. 

An important theoretical quantity which determines the number of participants and binary collisions, is the \emph{impact parameter} $b$, which is defined as the transverse distance between the centers of the two nuclei. Collisions with a smaller impact parameter (\emph{central} collisions) lead to a larger number of participants and binary collisions, as there is a greater overlap region; collisions with a larger impact parameter (\emph{peripheral} collisions) have a smaller overlap region and therefore have correspondingly fewer participants and binary collisions.

\begin{figure}[!htbp]
	\centering
	\includegraphics[width=0.75\linewidth]{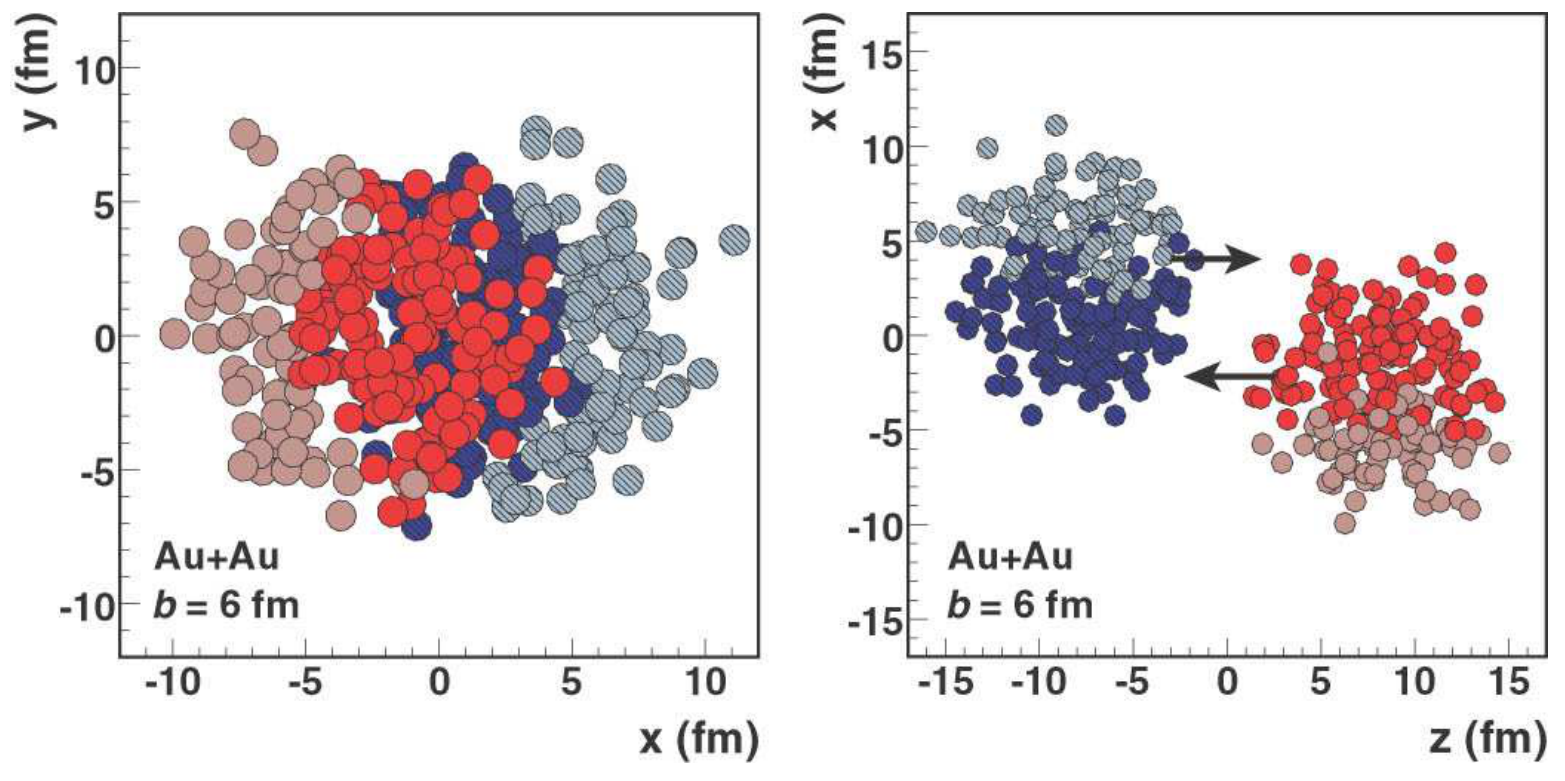}
	\caption{Illustration of the (Monte-Carlo) Glauber model from two different perspectives for an impact parameter of $b = 6 ~\mathrm{fm}$ (semi-central). The left pane shows the view down the beam axis, and the right pane shows the view perpendicular to the beam axis. The more saturated red and blue circles correspond to participating nucleons, while the less saturated corresponding to nucleons which do not participate. Figure taken from \cite{Miller:2007ri}.}
	\label{fig:glauber_monte_carlo}
\end{figure}

\Cref{fig:glauber_monte_carlo} shows a visualization of a Monte-Carlo implementation of the Glauber model from two different perspectives for an impact parameter of $b = 6 ~\mathrm{fm}$ (semi-central). Here, the more saturated red and blue circles correspond to participating nucleons, while the less saturated corresponding to nucleons which do not participate (also called \emph{spectator} nucleons). 

There are two important experimental phenomena which the Glauber model successfully describes. The first is the phenomenon of \emph{participant scaling}. In the 1970s Fermilab measured the total multiplicity for the collision \coll{h}{A} for various hadrons $h$ \cite{Elias:1978ft}. They saw that the total multiplicity was proportional to the number of participants. Even more interesting were the subsequent measurements at RHIC in the 2000s, which revealed the same proportionality over a wide span of $\sqrt{s}_{NN}$ in \coll{A}{A} collisions \cite{PHOBOS:2005zhy}, and similar measurements from the LHC in the 2010s further confirming this empirical law \cite{ALICE:2013jfw}. 

\begin{figure}[!htbp]
	\centering
	\begin{subfigure}[t]{0.47\textwidth}
			\centering
			\includegraphics[width=\linewidth]{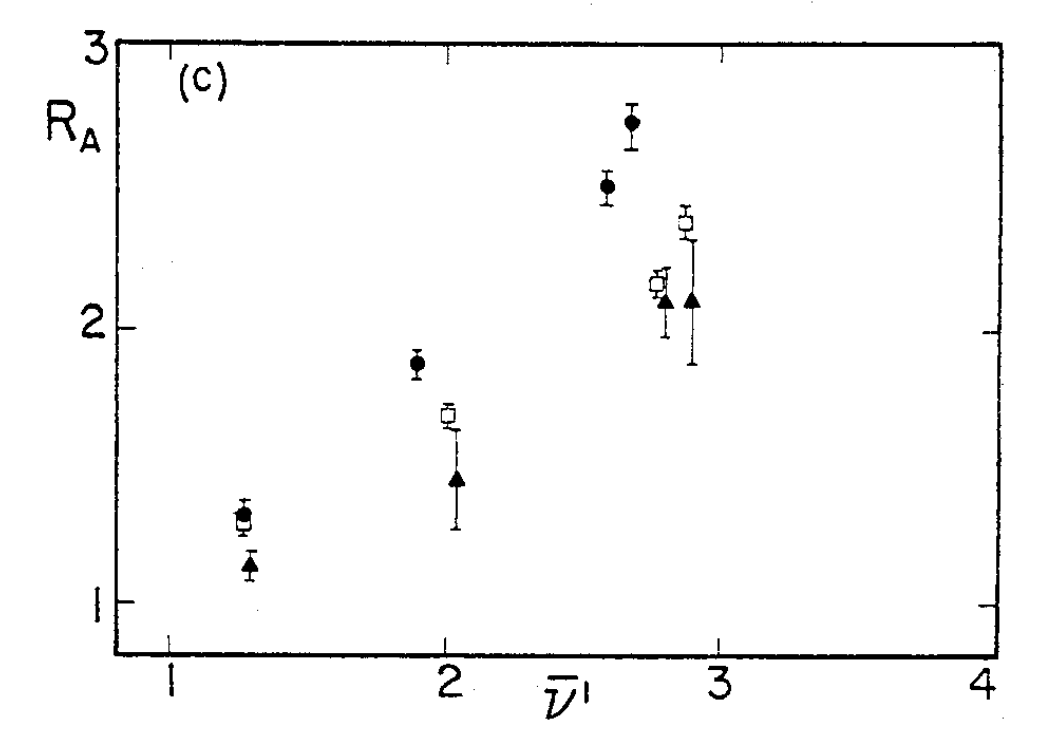}
			\caption{$R_A$ (proportional to total charged particle multiplicity) as a function of $\bar{\nu}$, where $\bar{\nu}$ represents the average number of inelastic collisions a hadron undergoes while passing through the nucleus. Figure taken from \cite{Elias:1978ft}.}
	\end{subfigure}\hfill
	\begin{subfigure}[t]{0.51\textwidth}
			\centering
			\includegraphics[width=\linewidth]{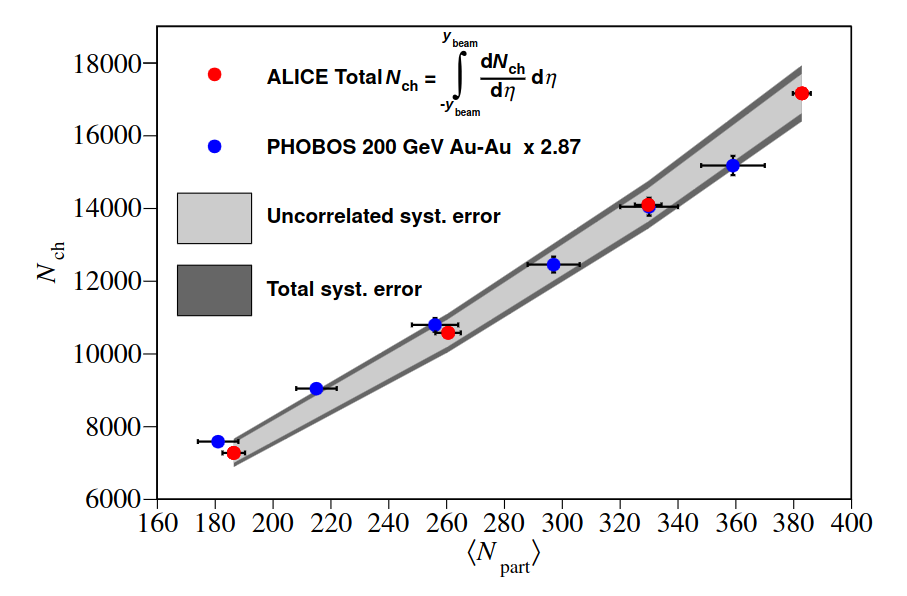}
			\caption{Total charged particle multiplicity $\langle N_{\text{ch}} \rangle$ as a function of $\langle N_{\text{part}} \rangle$, where $\langle N_{\text{part}} \rangle$ represents the average number of participants in the collision. Data from ALICE (red circles) and PHOBOS (blue circles) experiments are shown. Figure taken from \cite{ALICE:2013jfw}.}
	\end{subfigure}
	\caption{Results from Fermilab \cite{Elias:1978ft} (left) and LHC \cite{ALICE:2013jfw} and RHIC \cite{PHOBOS:2005zhy} (right) experiments demonstrating the linear dependence of the measured total charged multiplicity on the number of participants.}
	\label{fig:participant_scaling}
\end{figure}

The left pane of \cref{fig:participant_scaling} plots the total charged particle multiplicity as a function of the average number of inelastic collisions that the hadron undergoes as it passes through the nucleus, if interactions are completely governed by the hadron-nucleon cross section. The right pane of \cref{fig:participant_scaling} plots the total charged particle multiplicity as a function of the average number of participants. The linear relationship between the total charged multiplicity and the number of participants is clear, over the wide range of center of mass energies shown in the figures.

The second important experimental phenomena which the Glauber model successfully describes is \emph{binary scaling}. Factorization theorems in QCD \cite{Collins:1989gx,Collins:2011zzd} separate the soft (low-momentum and long-range) and hard (high-momentum and short-range) physics. Such theorems have been proven for only a subset of processes for which experimental data exists, including Deep Inelastic Scattering (DIS), Drell-Yan, and electron-positron annihilation to name a few. Commonly, in heavy-ion phenomenology a similar factorization is assumed, although currently there is no rigorous proof for such a factorization. This assumption allows one to express the number of hard collisions in \coll{A}{B} in terms of the number of hard collisions in \coll{p}{p} as $N_{\text {hard }}^{A+B, \text { enc }}(b) \propto N_{\text{coll}}(b) \sigma_{\text {hard }}^{\mathrm{pp}}$. Experimental evidence for binary scaling was found as early as the 1970s \cite{May:1975ju}, with overwhelming evidence in RHIC and LHC eras. Typically, binary scaling is captured by the \emph{nuclear modification factor} or $R_{AB}$, which we will discuss in detail in \cref{sec:jet_quenching}.

\begin{figure}[!htb]
	\begin{subfigure}[t]{0.47\textwidth}
		\centering
		\includegraphics[width=\linewidth]{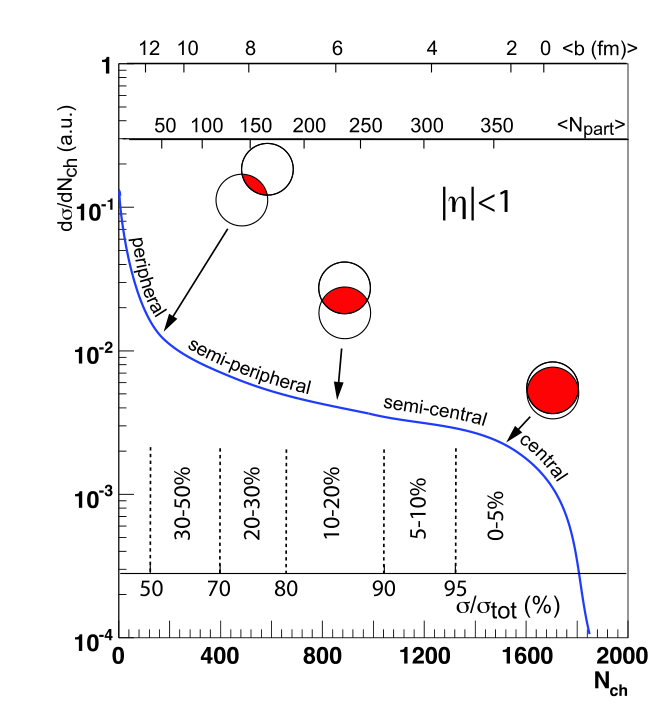}
		\caption{Figure showing experimental centrality determination. Figure taken from \cite{Miller:2007ri}.}
	\end{subfigure}\hfill
	\begin{subfigure}[t]{0.42\textwidth}
		\centering
		\includegraphics[width=\linewidth]{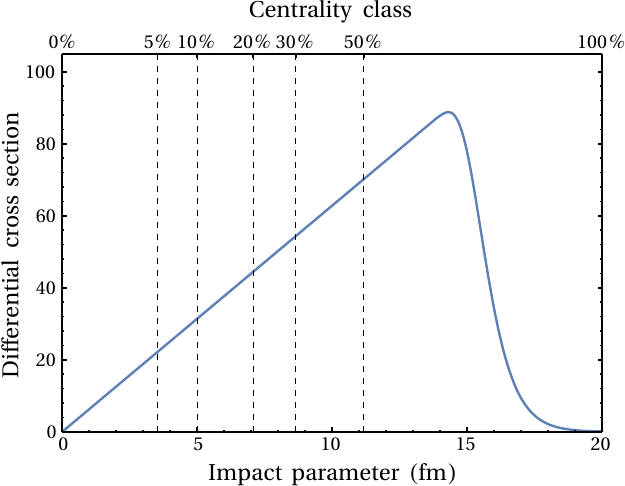}
		\caption{Theoretical centrality determination. The vertical axis is the inelastic cross section which is proportional to the number of participants.}
	\end{subfigure}
	\caption{Illustration of how experiments determine centrality, and how theorists use this to calculate impact parameters.}
	\label{fig:centrality_determination}
\end{figure}

One of the most important uses of the Glauber model is for centrality determination in heavy-ion colliders. The size of the QGP which is formed in heavy-ion collisions is strongly correlated with the impact parameter, however the impact parameter is not something which can be set by the experiment. The nuclei collide with a random impact parameter, and one must infer from the observed particles what the corresponding impact parameter was. This is practically done by experiments through a series of correlations. Participant scaling implies that $N_{\text{ch}} \propto N_{\text{part}}$, and it is true in the \rev{optical} Glauber model that $N_{\text{part}}$ and the impact parameter $b$ are in a one-to-one correspondence. \rev{For the more realistic Monte Carlo Glauber model, $b$ and $N_{\text{part}}$ are not in precise one-to-one correspondence due to fluctuations in the nucleon positions; however, this is a relatively small effect in central heavy-ion collisions \cite{Miller:2007ri}. Note that in small \collFour{p}{d}{He3}{A} collisions and peripheral \coll{A}{A} collisions, fluctuations cause a significant decorrelation in the correspondence between $b$ and $N_{\text{part}}$ \cite{Miller:2007ri}.}
The \rev{approximate one-to-one correspondence between $N_{\text{ch}}$ and $N_{\text{part}}$} allows experiments to use the measured number of charged particles to divide the collisions into ``centrality classes". This process is illustrated in \cref{fig:centrality_determination}.

\begin{figure}[!htbp]
\begin{subfigure}[t]{0.49\textwidth}
\centering
\includegraphics[width=\linewidth]{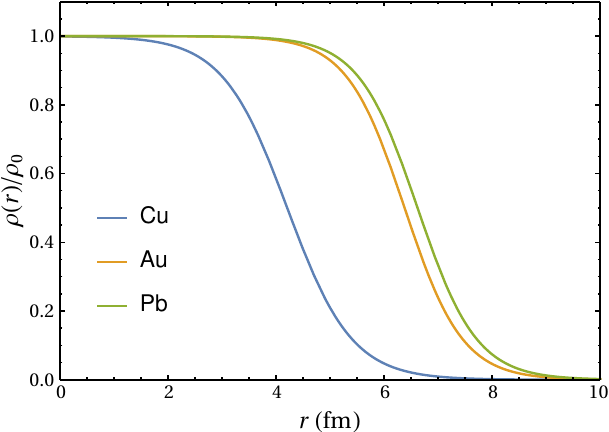}
\caption{Nucleon density $\rho(r)/\rho_0$ as a function of radius $r$ for Cu, Au, and Pb nuclei. $\rho_0$ represents the central density.}
\end{subfigure}\hfill
\begin{subfigure}[t]{0.49\textwidth}
\centering
\includegraphics[width=\linewidth]{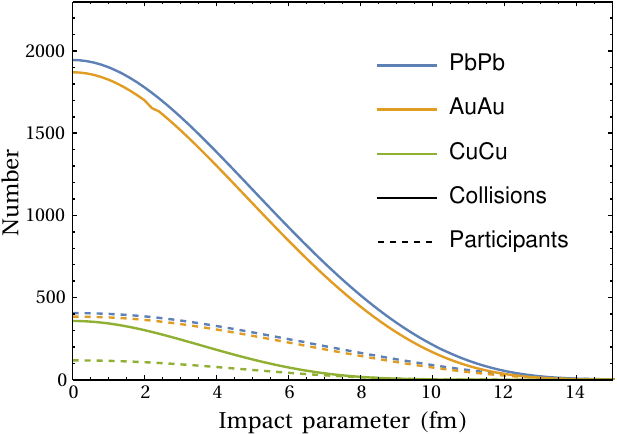}
\caption{Number of participants and binary collisions as a function of impact parameter $b$ for \coll{Pb}{Pb}, \coll{Au}{Au}, and \coll{Cu}{Cu} collisions.}
\end{subfigure}
\caption{Nuclear density profiles and collision geometry from the optical Glauber model. (a) shows the radial distribution of nucleons in different heavy nuclei, while (b) illustrates how the number of participating nucleons and binary collisions vary with collision centrality for different collision systems.}
\label{fig:nuclear_density_and_collision_geometry}
\end{figure}

From the experimental side, one orders the collisions according to the number of charged particles, and then divides these into centrality classes. For instance the collisions with the top $10\%$ of charged particle multiplicity would fall into the $0\text{--}10\%$ centrality class. From the theoretical side, one undergoes the same procedure but for $N_{\text{part}}$. By equating the centrality classes determined on the theoretical and experimental side, one establishes the appropriate mapping between the impact parameter and total multiplicity (or event activity). \rev{The left panel of \cref{fig:nuclear_density_and_collision_geometry} shows the nucleon density as a function of radius $r$ for various phenomenologically relevant heavy-ions. The right panel shows the number of binary collisions and participants as a function of impact parameter. This illustrates the theoretical calculation of the binary collisions and participants in the Glauber model, which can then be used for centrality determination and understanding various geometrical properties of heavy-ion collisions.}

\FloatBarrier

\section{Hard probes}
\label{sec:jet_quenching}

One critical area of investigation in heavy-ion collisions is the energy loss of high-momentum partons—quarks or gluons—that traverse the QGP. These high-transverse momentum $p_T$ partons are produced in the initial stages of the collision from hard scattering processes and, as they propagate through the QGP, they lose energy via interactions with the medium. This process, known as \emph{jet quenching}, is a key signature of QGP formation \cite{Busza:2018rrf}, as the high-$p_T$ partons are produced in the early stages of the collision and carry a memory of the evolution of the plasma. One of the goals of studying jet quenching is to extract the properties of the plasma formed in heavy-ion collisions---such as the density, temperature, and the interaction strength. Other goals are to understand how partons lose energy in the medium, to what extent fragmentation is modified by the presence of the medium, and how jet substructure is modified by the medium. For a review of jet quenching, refer to \cite{Connors:2017ptx, Qin:2015srf, Cao:2020wlm}.

Jet quenching manifests itself in several observable phenomena, including the suppression of high-$p_T$ hadrons, the modification of jet shapes, and the redistribution of energy within jets. These observables are collectively referred to as \textit{hard probes}. The degree of jet quenching is typically quantified using the nuclear modification factor, denoted as $R_{AB}$ for the collision \coll{A}{B}, which compares the yield of high-$p_T$ particles in heavy-ion collisions to that in proton-proton collisions. The $R_{AB}$ is defined as
\begin{equation}
	R^h_{A B}\left(p_T, y, \phi \right) \equiv \frac{1}{N_{\text {coll }}} \frac{d N^h_{A B} / d p_T d y d \phi}{d N^h_{pp} / d p_T d y d \phi},
	\label{eqn:back_rab}
\end{equation}
where $dN_{AB} / d p_T dy d \phi~(dN_{pp} / d p_T dy d \phi)$ is the spectra of hadrons $h$  detected in the collision \coll{A}{B} (\coll{p}{p}), and $N_{\text{coll}}$ is the number of binary collisions. The number of binary collisions is typically calculated with the Glauber model; see \cref{sec:back_the_glauber_model} for details. 
If $R^h_{AB} = 1$, it indicates that the \coll{A}{B} collision is simply a transparent superposition of the nucleon-nucleon collisions (of which there are $N_{\text{coll}}$ such collisions) that occur in the \coll{A}{B} collision. A measured $R^h_{AB}$ significantly less than one at high momentum is indicative of final state energy loss of high momentum particles due to interactions with the medium. 

\begin{figure}[!htbp]
\centering
\begin{subfigure}[t]{0.49\textwidth}
    \centering
    \includegraphics[width=\linewidth]{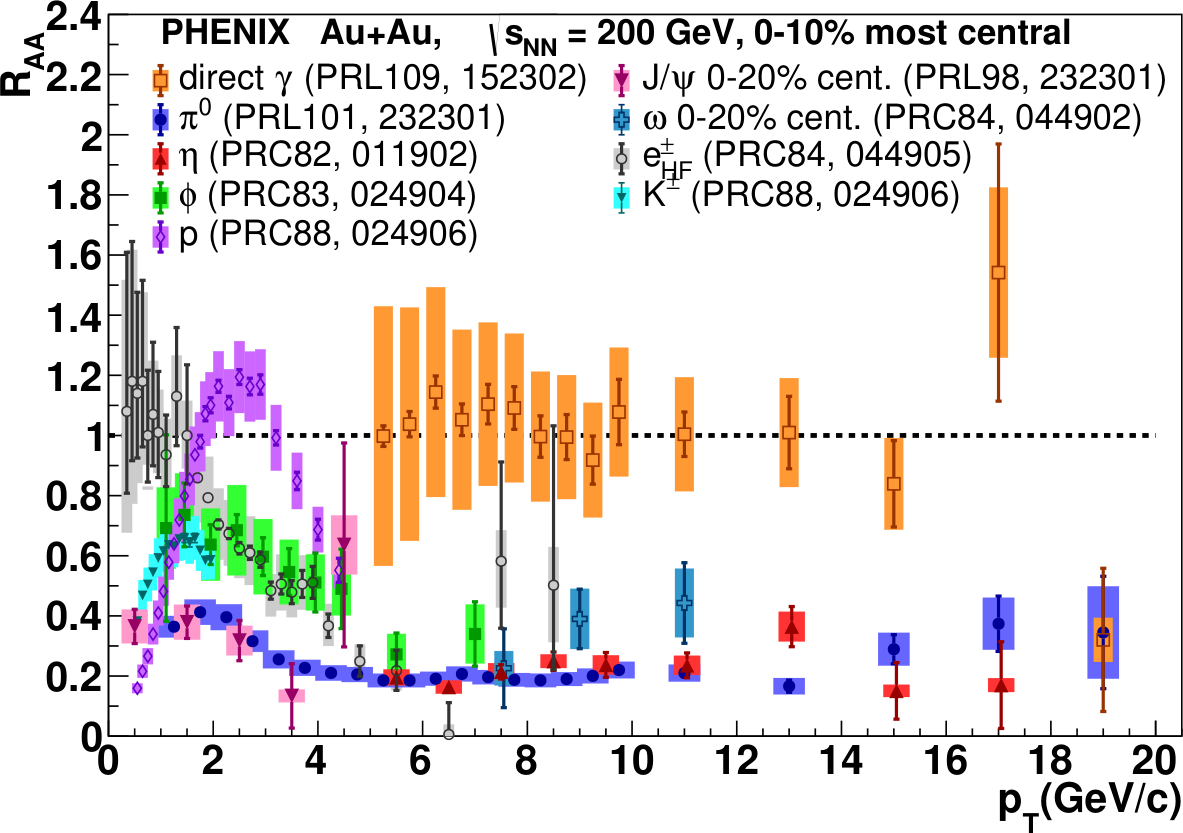}
    \caption{RHIC}
    \label{fig:back_RHIC_raa_summary}
\end{subfigure}\hfill
\begin{subfigure}[t]{0.49\textwidth}
    \centering
    \includegraphics[width=\linewidth]{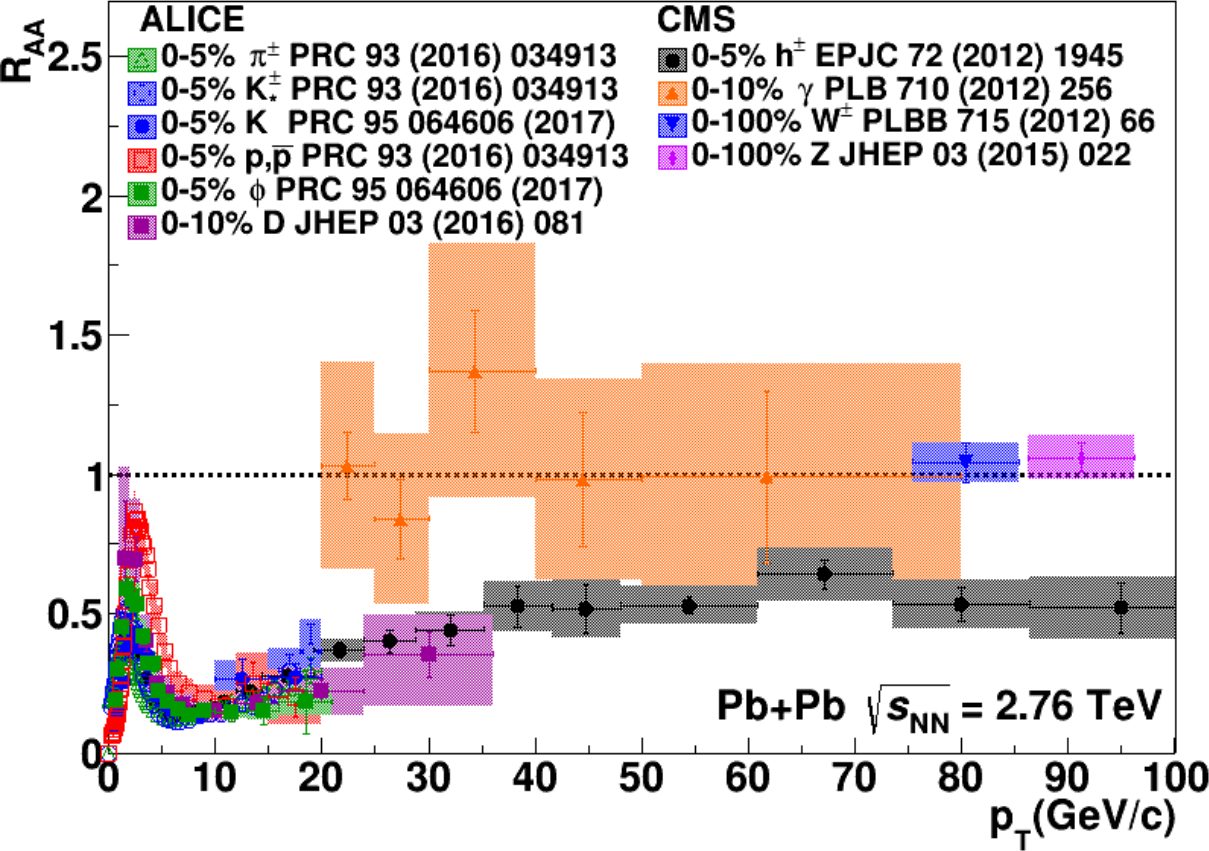}
    \caption{LHC}
    \label{fig:back_LHC_raa_summary}
\end{subfigure}
\caption{Summary plot of $R_{AA}$ as a function of $p_T$ for various particle species as measured by experiments at LHC and RHIC. Figure is taken from \cite{Connors:2017ptx}.}
\label{fig:raa_summary}
\end{figure}

The nuclear modification factor $R_{AA}$ has been extensively studied at various collision energies and for different particle species, providing crucial insights into the properties of the QGP created in heavy-ion collisions. \Cref{fig:raa_summary} presents a summary of $R_{AA}$ measurements from RHIC (left) and LHC (right).
Key observations from \cref{fig:raa_summary} include:
\begin{enumerate}
	\item \textit{$R_{AA}$ consistent with unity for photons, $W$ bosons, and $Z$ bosons.} This is predicted by the Glauber model \cite{Glauber:1970jm,Miller:2007ri}, which predicts that the high-$p_T$ particles are produced by binary nucleon-nucleon collisions. Weakly interacting particles---including photons and $W$ and $Z$ bosons--- do not interact strongly with the QGP, and so the number of particles produced is simply the transparent superposition of the number of particles produced in proton-proton collisions, leading to $R_{AA} = 1$.
	\item \textit{Factor five suppression of high-$p_T$ hadrons in heavy-ion collisions compared to proton-proton collisions.} An $R_{AA} \sim 0.2$ is reflective of the energy loss suffered by partons as they traverse the QGP. This is qualitatively well predicted by various energy loss models, which we will detail in \cref{sec:back_elastic_energy_loss,sec:back_radiative_energy_loss}. 
	\item \textit{Independence of the light-flavor suppression on the light-flavor hadron mass.} If the medium which was formed in heavy-ion collisions was a hadron gas, one would expect that the degree of energy loss depends on the light flavor hadron which is detected. However, if the medium is made from quarks and gluons and fragmentation happens similarly to how it does in vacuum, then one expects that suppression depends only on the partons which compose the hadron. The independence of suppression on the light-flavor hadron mass shows that the scenario where the medium is comprised of quarks and gluons is favored by data.
\end{enumerate}

In the future, hard probes may be used to quantitatively extract properties of the medium such as the temperature, strong coupling, thermalization time, as well as help in allowing data to guide theory on the preferred initial conditions, and energy loss mechanism. 
Such extractions \cite{JETSCAPE:2024cqe,JET:2013cls,JETSCAPE:2021ehl} have broadly focused on extracting the jet transport coefficient $\hat{q}$.
This approach is explicitly model dependent, although sometimes various different models are considered \cite{JET:2013cls}, and is not nearly as broad as similar analyses conduct in the soft sector \cite{Schenke:2020mbo, Nijs:2020roc, Bernhard:2019bmu}. Numerous other model dependent extractions with various energy loss models, which will be discussed in \cref{sec:back_radiative_energy_loss,sec:back_elastic_energy_loss}, have been performed \cite{Bass:2008rv,Armesto:2009zi,Chen:2010te,Qin:2009bk,Kang:2014xsa,Karmakar:2024jak,Karmakar:2023ity}. Future work may perform a Bayesian analysis in a less model dependent manor. The gold standard would be an extraction of plasma properties and energy loss mechanism in a model independent manor using observables in both the soft and hard sector.

Such an extraction of medium properties may be currently out of reach due to the significant uncertainties associated with various choices in contemporary energy loss models, mostly because of the approximations in such models. Some examples of these theoretical uncertainties include the angular cutoff on radiated gluon emission which stems from the collinear approximation \cite{Horowitz:2009eb}, the choice of potential to model interactions with the plasma, the use of vacuum vs HTL propagators \cite{Wicks:2008zz,Romatschke:2004au}, and the treatment of multiple gluon emission. We will also discuss throughout this thesis an uncertainty in the Djordjevic-Gyulassy-Levai-Vitev (DGLV) \cite{Djordjevic:2003zk} radiative energy loss model used in this work, which stems from a particular approximation---the large formation time approximation---which results in an even larger theoretical uncertainty on the angular cutoff on radiated gluon emission than the collinear approximation does.

There are two forms of energy loss which partons undergo as the traverse the plasma---radiative (or inelastic) and collisional (or elastic). 
Historically, radiative energy loss has been considered the dominant energy loss mechanism for light flavors \cite{Qin:2015srf}. However, for realistic calculations, elastic energy loss has been shown to be important at lower momenta (particularly pertinent at RHIC energy scales) \cite{Wicks:2005gt}, and therefore, elastic energy loss should not be neglected.
For heavy flavor (charm and bottom) partons, it is generally considered that elastic energy loss is the dominant energy loss mechanism, due to the deadcone effect \cite{Dokshitzer:2001zm,Djordjevic:2003zk} which suppresses radiation in a relatively large cone around the parent parton. However for larger momenta, particularly important at LHC energies, radiative energy loss becomes an important contribution to the heavy-flavor energy loss. We will see in \cref{sec:EL_relative_elastic_radiative} that the ratio of the contribution of elastic over radiative energy loss ranges from $0.25\text{--}4$, indicating the importance of including both forms of energy loss in phenomenological models. 

\subsection{Radiative energy loss}
\label{sec:back_radiative_energy_loss}

High-$p_T$ partons with a large virtuality (offshellness) are produced from initial hard collisions between partons in the colliding nuclei. These high-$p_T$ partons will undergo subsequent radiation, even without the presence of a medium, which results in them losing virtuality. The amount of radiation is modified by the presence of the medium. 
An important effect is the Landau-Pomeranchuk-Migdal (LPM) effect \cite{Landau:1953um, Migdal:1956tc}, which leads to a finite formation time before gluon emission can occur. Detailed comparisons and reviews of radiative energy loss are included in \cite{Qin:2015srf, Armesto:2011ht, Cao:2020wlm}, which this section is based off of.

\subsubsection{BDMPS-Z and ASW-MS}
\label{sec:bdmpsz}

The Baier-Dokshitzer-Mueller-Peigne-Schiff-Zakharov (BDMPS-Z) formalism \cite{Baier:1996kr, Baier:1996sk, Zakharov:1997uu, Baier:1996vi, Zakharov:1996fv} gluon radiation is written in terms of a path integral which resums all these scatterings on static color-charged scattering centers. In-medium scatters are modeled as interactions with the Gyulassy-Wang (GW) model of Debye screened scattering centers \cite{Gyulassy:1993hr}, represented by the potential
\begin{align}
  V_n=& V(q_n) e^{i q_n . x_n} \nonumber\\
  =& 2 \pi \delta\left(q^0\right) v\left(\vec{q}_n\right) e^{- i \vec{q_n} \cdot \vec{x_n}} T^R_{a_n} \otimes T^n_{a_n},
  \label{eqn:back_gyulassy_wang_potential}
\end{align}
where $T^R_{a_n}$ ($T^n_{a_n}$) are the SU(3) generators in the representation of the hard parton ($n^{\text{th}}$ scattering center), $x_n = (x_0, \mathbf{b}, \Delta z)$ is the location of the $n^{\text{th}}$ scattering center, and $q_n = (q^0_n, \vec{q}_n) = (q^0_n, \mathbf{q}_n, q^z_n)$ is the momentum exchanged with the $n^{th}$ scattering center. Here the $\delta(q^0)$ arises form the fact that the coordinate space representation of the scattering potential is independent of time, i.e.\ the potential is ``static". 

A common numerical implementation of the BDMPS-Z formalism is the Armesto-Salgado-Wiedemann-Multiple-Scattering (ASW-MS) formalism \cite{Wiedemann:2000za,Salgado:2003gb, Armesto:2003jh}. In the calculation of the BDMSP-Z path integral, one typically can only evaluate the path integral in saddle-point approximation, which assumes that the incident parton interacts with the medium via multiple soft scattering processes. \rev{The \textit{opacity} $L / \lambda$ is a measure of the number of expected scatterings in a medium and quantifies the thickness of the medium, where $L$ is the pathlength that the parton travels through the QGP and $\lambda$ is the mean free path between scatterings. The BDMPS-Z formalism is valid in the limit of thick media or equivalently large opacity; i.e., $L / \lambda \gg 1$.}

\subsubsection{(D)GLV and ASW-SH}
\label{sec:glv}

The Gyulassy-Levai-Vitev (GLV) formalism \cite{Gyulassy:2000er}, and the Djordjevic-Gyulassy-Levai-Vitev (DGLV) extension \cite{Djordjevic:2003zk} to include non-zero quark and thermal gluon masses, can be shown to be an expansion of the same BDMPS-Z path integral in the \textit{opacity} or expected number of scatterings. This was also independently developed by Wiedemann \cite{Wiedemann:2000za,Wiedemann:2000tf}, in what is referred to as the Armesto-Salgado-Wiedemann-Single-Hard (ASW-SH) approach. Both the GLV and ASW-SH approach include contributions from the same diagrams, however differ mostly in how they extend the calculation to the full gluon radiation phase space \cite{Armesto:2011ht}. 

The GLV and ASW-SH formalisms are based on the GW potential and consider an expansion in the number of hard (high momentum transfer) scatters with static scattering centers. Most common implementations of the opacity expansion includes only terms first order in the opacity, however higher order terms have been computed \cite{Wicks:2008ta}. The motivation for this is the oscillatory nature of the opacity expansion, which means that including the first term is similar to the sum of the first three terms \cite{Gyulassy:2001nm, Wicks:2008ta}.

Djordjevic \cite{Djordjevic:2007at,Djordjevic:2008iz,Djordjevic:2009cr} extended this formalism using a Hard Thermal Loop based potential (HTL) \cite{Braaten:1989mz, Klimov:1982bv, Pisarski:1988vd, Weldon:1982aq, Weldon:1982bn}, which allows for so-called ``dynamic" scattering centers. The Improved Opacity Expansion (IOE) \cite{Mehtar-Tani:2019ygg,Mehtar-Tani:2019tvy} aims to create a single framework which connects the GLV-based and BDMPS-Z-based formulations of energy loss. This utilizes an expansion around the harmonic oscillator solution, which allows for contributions from multiple soft scatters and single hard scatters.

\subsubsection{AMY}
\label{sec:amy}

The Arnold-Moore-Yaffe (AMY) approach \cite{Arnold:2001ms,Arnold:2002ja,Arnold:2001ba} is formulated in the framework of HTL, and treats the medium as weakly-coupled in the high temperature limit. While GLV and BDMPS-Z formalisms consider the case of a highly-offshell incident parton, the AMY framework originally considered a parton which is close to on shell \cite{Majumder:2010qh}. Modern formulations of AMY \cite{Caron-Huot:2010qjx} now correctly model the parton as being highly offshell.

\subsubsection{HT}
\label{sec:ht}

Guo and Wang \cite{Guo:2000nz,Wang:2001ifa} developed the Higher Twist (HT) formalism. In this approach the interacting of the incident parton with the medium is characterized by higher-twist contributions to the scattering. One assumes that the process $A+A \to h + X$ where $h$ is a hadron factorizes at next-to-leading twist \cite{Armesto:2011ht, Qin:2015srf}. Factorization has been proven for certain processes, including single hadrons produced in $e^+ e^-$ annihilation, Drell-Yan, Deep Inelastic Scattering (DIS), hadron-hadron collisions, etc.\  \cite{Collins:2011zzd, Collins:1989gx}. The higher-twist approach applies these techniques to calculate the medium modification of the fragmentation function of a quark produced in a hard collision in a QGP. In this approach the medium is completely characterized by a single parameter $\hat{q}$, the ``jet quenching parameter", which is the transverse momentum squared imparted by the medium per unit length. Typically $\hat{q}$ is a free parameter in the model. Subsequently, Majumder extended the HT formalism to include multiple gluon emission and multiple scattering \cite{Majumder:2010qh}.

\subsection{Multiple gluon emission}
\label{sec:back_multiple_gluon_emission}

Most current radiative energy loss models calculate the spectrum of radiated gluons for a \emph{single} hard scatter. Gyulassy, Levai, and Vitev \cite{Gyulassy:2001nm} first included the effects of multiple gluon emission by assuming that radiated gluons are emitted independently, leading to a Poisson distribution of gluon emissions. This model choice is motivated by an exact analytic calculation for vacuum radiation in QED (see \cite{Peskin:1995ev}, chapter 6). This does not effect the average energy loss of a parton, which corresponds to the first moment of the probability of energy loss distribution, but it does effect higher order moments. Gyulassy, Levai, and Vitev \cite{Gyulassy:2000fs} found that because of the steepness of the production spectrum, which leads to the $R_{AA}$ depending on higher order moments of the energy loss distribution, the inclusion of multi-gluon emission according to an assumed Poisson distribution decreased the predicted $R_{AA}$ significantly. We also provide a detailed discussion of the dependence of the $R_{AA}$ on the various moments of the energy loss probability distributions in \cref{sec:EL_distribution_dependence}.

Explicitly, if the emission spectrum for a single hard scatter is $dN^g / dx(x \mid E)$, then the total emission spectrum (for an arbitrary number of hard scatters) is \cite{Gyulassy:2001nm}
\begin{equation}
	P(\epsilon \mid E)=\sum_{n=0}^{\infty} \frac{e^{-N_g} \left(N_g\right)^n}{n!} p_n(\epsilon \mid E),
	\label{eqn:back_poisson}
\end{equation}
where
\begin{align}
	p_0(\epsilon \mid E) \equiv& \delta(\epsilon)\\
	p_1(\epsilon \mid E) \equiv& \frac{1}{N_g} \frac{dN^g}{d \epsilon} (\epsilon \mid E)\\
	p_{n+1}(\epsilon \mid E) \equiv& \int dx \; p_n(\epsilon - x) p_1(x).
	\label{eqn:back_poisson_more_detailed}
\end{align}
In the above we have altered the form that the Poisson convolved distribution is typically written, to emphasize the Poisson nature of \cref{eqn:back_poisson}. In this representation, the $p_n$ are the probability distributions for radiating a total fraction of energy $\epsilon$ \emph{given} that there are exactly $n$ scatters. If the fluctuations in radiated gluon number are independent, it then follows that \cref{eqn:back_poisson} is the total probability for losing a fraction $\epsilon$ of energy. It is also evident in the above that the probability distribution is normalized
\begin{equation}
	\int d \epsilon \; P(\epsilon \mid E) = \sum_{n=0}^{\infty} e^{-N_g} N_g^n / n! = 1.
	\label{eqn:back_poisson_normalized}
\end{equation}
Defining $g(\epsilon) \equiv d N^g / d\epsilon (\epsilon)$, we may also show that the fractional energy loss $\Delta E / E$ is unchanged
\rev{
\begin{align}
	\int d \epsilon \; \epsilon P(\epsilon \mid E) =& \int d \epsilon \; \epsilon \sum_{n=0}^{\infty} \frac{e^{-N_g} N_g^n}{n!} p_n(\epsilon)\nonumber\\
	=& \sum_{n=0}^{\infty} \frac{e^{-N_g} N_g^n}{n!} \int d \epsilon \; d x_1 \cdots d x_{n} \; \epsilon \frac{g(x_1)}{N_g} \cdots \frac{g(x_{n})}{N_g} \delta(\epsilon - x_1 - \cdots - x_{n-1})\nonumber\\
	=& \sum_{n=0}^{\infty} \frac{e^{-N_g}}{n!} \int d \epsilon d x_1 \cdots d x_{n-1} \; \epsilon \; g(x_1) \cdots g(x_{n-1}) g\left(\epsilon - (x_1 + \cdots + x_{n-1})\right)\nonumber\\
	=& \sum_{n=0}^{\infty} \frac{e^{-N_g}}{n!} \int d \epsilon'  \; (\epsilon' + x_1 + \cdots + x_{n-1}) g(x_1) \cdots g(x_{n-1}) g(\epsilon')\nonumber\\
	=& \sum_{n=1}^{\infty} \frac{e^{-N_g} }{n!} \times \left(n \frac{\Delta E}{E}\right) N_g^{n-1}\nonumber\\
	=& \sum_{n' = 0}^{\infty} \frac{e^{-N_g}  N_g^{n'}}{n'!} \times \left(\frac{\Delta E}{E}\right)\nonumber\\
	=& \frac{\Delta E}{E}.
	\label{eqn:back_poisson_energy_loss}
\end{align}
Note that in the above we have used $\int d \epsilon \; \epsilon dN^g/d \epsilon = \Delta E / E$.
}

We note that in the above formulation the energy of the parton is not updated after each scatter, which is a byproduct of the soft radiation approximation and the Eikonal approximation. The fact that the energy of the parton is not updated, leads to a finite probability weight for $\epsilon > 1$. This probability leakage is typically treated in two common ways: \emph{reweighting} or \emph{amputation} \cite{Armesto:2011ht,Horowitz:2010dm}.  The reweighting procedure involves normalizing the portion of the probability distribution between $\epsilon = 0$ and $\epsilon = 1$ (the physical region). The amputation procedure involves putting the integrated portion of the distribution past $\epsilon = 1$ as a coefficient of a Dirac delta function at at $\epsilon = 1$. The fact that the average fractional energy loss is left unchanged by the Poisson procedure is only strictly true when there are no kinematic constraints on $\epsilon$, which allows us to perform the change of variables $\epsilon^' = \epsilon - (x_1 + \cdots  + x_n)$ without changing the bounds on the integral $\epsilon \in (-\infty, \infty)$. In \cref{sec:mod_numerical_elastic_radiative} we will see that for phenomenologically relevant parameters, the average energy loss is left unchanged (to $< 1\%$).

\subsection{Elastic energy loss}
\label{sec:back_elastic_energy_loss}

Bjorken performed the earliest calculation of elastic energy loss of an incident quark moving through a QGP in 1982 \cite{Bjorken:1982tu}. This work utilized vacuum pQCD propagators, and shielded the infrared divergence in the radiated momentum with a soft scale $\mu$, of order the Debye mass. The result obtained by Bjorken was
\begin{equation}
	\frac{dE}{dz} = 2 \pi \alpha_s^2 T^2 C_R \log \frac{\left\langle k \right\rangle E}{m_g^2}\left(1+\frac{n_f}{6}\right),
	\label{eqn:back_bjorken}
\end{equation}
where $E$ is the incident energy, $\alpha_s$ is the strong coupling, $C_R$ is the Casimir of the incident parton (either $4 / 3$ for quarks or $3$ for gluons), $T$ is the temperature, $m_g$ is the effective gluon mass, $\left\langle k \right\rangle \approx 3 T$ is the average momentum of the medium particles, and $n_f$ is the number of active quark flavors in the plasma (typically taken to be $2\text{--}3$ [either $u, d$ or $u, d, s$]). This calculation involved an \emph{ad hoc} estimate of the gluon mass $m_g$, which is used as an infrared cutoff, and typical plasma momentum $\left\langle k \right\rangle$. Additionally the angular dependence was not properly accounted for \cite{Wicks:2008zz}, which means there are missing terms in \cref{eqn:back_bjorken}. Despite these approximations, \cref{eqn:back_bjorken} has the correct parametric dependence on $T$, and $\alpha_s$, and later developments to the elastic energy loss mostly changed the coefficient of \cref{eqn:back_bjorken}.

Braaten, Thoma, and Gyulassy \cite{Thoma:1990fm,Braaten:1991we, Braaten:1991jj} calculated the elastic energy loss from within Braaten and Pisarski's framework of Hard Thermal Loops (HTL) \cite{Braaten:1989mz, Klimov:1982bv, Pisarski:1988vd, Weldon:1982aq, Weldon:1982bn}, which naturally gives rise to a gluon mass of order the Debye mass $\mu$, and thereby shields the propagator. This removed the need to artificially regulate the infrared divergence in the momentum transfer with the gluon mass. Gyulassy and Thoma \cite{Thoma:1990fm} perform a similar calculation to Bjorken \cite{Bjorken:1982qr}, but with the HTL propagators. They find that the energy loss is
\begin{equation}
	\frac{\mathrm{d} E}{\mathrm{d} z}=\frac{4 \pi}{3} C_{R} \alpha_{s}^2 T^2 \log \left(\frac{k_{\text{max}}}{k_{D}}\right) \frac{1}{v^2}\left(v+\frac{v^2-1}{2} \log \frac{1+v}{1-v}\right)
	\label{eqn:back_gyulassy_thoma_elastic_energy_loss},
\end{equation}
where $v$ is the velocity of the hard parton, and $k_{\text{max}}$ and $k_{D}$ are the maximum and infrared cutoff momentum, respectively. This expression was evaluated with $m_g = 0$, and an artificial cutoff $k_D \sim m_g$, in order to obtain a simple analytic expression.

Braaten and Thoma \cite{Braaten:1991we, Braaten:1991jj} calculated the energy loss for both hard and soft momentum transfers. They argued that for hard momentum transfers $q \gtrsim T$ screening effects are less important, and one may calculate the energy loss using standard QCD vacuum propagators. For soft momentum transfers $q \lesssim gT \sim \mu$ screening effects are once again important and the resummed HTL propagator is used. For the soft result one has an upper bound of $q^*$, where $gT \lesssim q^* \lesssim T$ is an intermediary momentum scale, and for the hard result $q^*$ is the lower bound of the momentum transfer integral. Schematically one may then write
\begin{equation}
	\frac{dE}{dz} (q^*) = \left.\frac{dE}{dz}\right|_{\text{hard}}(q^*) + \left.\frac{dE}{dz}\right|_{\text{soft}} (q^*).
	\label{eqn:back_hard_plus_soft}
\end{equation}
Miraculously, one finds that the $q^*$ dependence of the hard and soft contributions is $\sim \log( 2 T E / q^* M)$ and $\log (q^* M / 2 TE)$, respectively, where $M$ is the mass of the incident parton. This dependence on the intermediary $q^*$ scale implies that the full $dE / dz$ result is \emph{independent} of $q^*$. To obtain a simple analytic result, Braaten and Thoma calculated the energy loss in two regimes: $E \gg T^2 / M$ and $E \ll T^2 / M$, where the intermediary dependence is typically inferred by requiring continuity. The result for $E \ll T^2 / M$ is 
\begin{equation}
    \frac{\mathrm{d} E}{\mathrm{d} z} = 2 C_R \pi \alpha_s^2 T^2 \left(1 + \frac{n_f}{6}\right) \log \left(2^{\frac{n_f}{2(6+n_f)}} \, 0.92 \frac{\sqrt{E T}}{m_g}\right),
    \label{eqn:back_elastic_energy_loss_high}
\end{equation}
and the result for $E \gg T^2 / M$ is
\begin{equation}
\frac{\mathrm{d} E}{\mathrm{d} z} = 2 C_R \pi \alpha_s^2 T^2 \left[\frac{1}{v}- \frac{1-v^2}{2 v^2}\log \frac{1+v}{1-v}\right] \log \left(2^{\frac{n_f}{6+n_f}} B(v) \frac{E T}{m_g M}\right)\left(1+\frac{n_f}{6}\right),
\label{eqn:back_elastic_energy_loss_low}
\end{equation}
where $B(v)$ is some smooth function with features described in \cite{Braaten:1991jj,Braaten:1991we}, and which are unimportant for the current discussion.

Comparing the canonical energy loss calculations from Bjorken \cite{Bjorken:1982qr}, Thoma and Gyulassy \cite{Thoma:1990fm}, and Braaten and Thoma \cite{Braaten:1991we, Braaten:1991jj} in \cref{eqn:back_bjorken,eqn:back_gyulassy_thoma_elastic_energy_loss,eqn:back_gyulassy_thoma_elastic_energy_loss} shows agreement, for the most part, on the parametric dependence of the energy loss on $\alpha_s$ and $T$. As noted by Wicks, Horowitz, Djordjevic, and Gyulassy (WHDG) \cite{Wicks:2005gt}, the different energy loss calculations differ only in the factors involved in the logarithm. Therefore, while their dependence on various quantities such as $E$, $T$, and $\alpha_s$ are the same asymptotically, the coefficient which multiplies these quantities may differ significantly. At non-asymptotic energies even the dependence will differ. We will see in \cref{sec:EL_paper} that these different coefficients and dependencies in the elastic energy loss calculation, can lead to dramatically different results for suppression---at both the qualitative and quantitative level.

Wicks \cite{Wicks:2008zz} moved past the approximation of the elastic energy loss distribution as Gaussian according to the central limit theorem, modeling the number of elastic scatters with a Poisson distribution, in an analogous equation to \cref{eqn:back_poisson}. Also pointed out by Wicks was the importance of including terms past leading order in the thermal distributions which occur in the various integrals. For this reason, many integrals must be performed numerically, and so there is no simple ``pocket formula" as there was for the previous elastic energy loss results. We discuss the elastic energy loss implementation of Wicks further in \cref{sec:back_elastic_energy_loss}.

Elastic energy loss cannot be naturally captured in the GLV, BDMPS-Z, and ASW formalisms, as the scattering centers are modeled as infinitely massive and, therefore, by construction, ignore energy loss due to recoil of the scattering center. Djordjevic \cite{Djordjevic:2006tw} introduced a dynamical scattering center reformulation of the GLV radiative energy loss, which allows for both radiative and elastic energy loss to be captured by the same formalism. The AMY and LBT models naturally allow for both radiative and elastic energy loss to be included on equal footing. The HT model can include elastic contributions through longitudinal drag coefficients \cite{Majumder:2010qh}.
Phenomenological models based on BDMPS-Z, GLV, and ASW typically treat the elastic and radiative energy loss as independent processes \cite{Wicks:2005gt,Armesto:2011ht,Qin:2015srf}.

\subsection{Strongly coupled approaches}
\label{sec:strongly_coupled_approach}

An open question in the field of heavy-ion physics is at which scale the strong coupling runs in the QGP. All previously discussed energy loss models have taken the viewpoint that the large momentum scale $p_T$ of the hard parton causes the coupling to be small, analogous to the case in \coll{p}{p} collisions. There is strong evidence from hydrodynamics calculations that the soft modes of the plasma are strongly coupled (see \cref{sec:bulk_observables} and \cite{Busza:2018rrf, Casalderrey-Solana:2011dxg}), which implies that there are at least partial contributions to the energy loss which are not perturbative. The procedure followed by pQCD-based models is to extrapolate the result down to nonperturbative scales. Another approach is to extrapolate up from the soft modes of the plasma, and assume that the high-$p_T$ physics may be described by a strongly coupled plasma. This is particularly applicable for smaller momentum exchanges, where the ordering of scales assumed in perturbative approaches, $E \gg x E \gg (T, \Lambda_{\text{QCD}})$, may not be well controlled \cite{Casalderrey-Solana:2011dxg}. 
Typically, the QGP is modeled using $\operatorname{SU}(N_C)$ $\mathcal{N} = 4$ supersymmetric Yang-Mills (SYM) where $\mathcal{N} = 4$ refers to the number of supersymmetric charges, which is the maximum allowed in 4 spacetime dimensions \cite{DHoker:2002nbb}.
While $\mathcal{N} = 4$ SYM and QCD have fundamentally different properties in the vacuum, these differences may become unimportant at temperatures close to the critical temperature $T_c$ \cite{Casalderrey-Solana:2011dxg}. The calculation is performed in the strong-coupling limit by utilizing the Anti de Sitter/Conformal Field Theory (AdS/CFT) correspondence, which is a conjectured relationship between a certain string theory in an Anti de Sitter spacetime geometry (AdS) and a conformal gauge theory in flat space (CFT) \cite{Klebanov:2000me}.

The AdS/CFT correspondence allows one to perform calculations in the strongly-coupled limit of $\mathcal{N} = 4$ SYM by performing the corresponding calculation in the gravity dual of the theory. The original treatment of the energy loss of a heavy quark in a strongly coupled plasma was developed in the limit where the heavy quark mass is much larger than the temperature scale \cite{Gubser:2006bz, Herzog:2006gh}. This allows one to treat the heavy quark as a Wilson line along the worldline of the quark, which on the gravity dual of the theory corresponds to a classical string hanging from the boundary of AdS into the bulk of AdS \cite{Casalderrey-Solana:2011dxg}. 
In this scenario, energy loss occurs due to a drag force on the heavy quark. This presents a fundamentally different picture from the radiative-dominated energy loss in perturbative QCD calculations. 
Other work expanded the calculation to include transverse and longitudinal momentum broadening \cite{Casalderrey-Solana:2007ahi}, accounted for radiative energy loss \cite{BitaghsirFadafan:2008adl, Mikhailov:2003er}, and extended the calculation to the case of light quarks and gluons \cite{Chesler:2008uy, Chesler:2008wd, Gubser:2008as}. A calculation of the jet quenching parameter $\hat{q}$ \cite{Liu:2006ug} was important for making more quantitative comparisons between pQCD and strongly coupled approaches, and comparing to experimental data. Results related to the disassociation of heavy quark bounds states (quarkonia) involved the treatment of a $q \bar{q}$ pair with a ``hot wind" of QGP blowing over it \cite{Casalderrey-Solana:2014bpa}. Calculations of the jet $R_{AA}$ \cite{Morad:2014xla,Casalderrey-Solana:2014bpa} and heavy quark $R_{AA}$ \cite{Horowitz:2015dta} have been performed which are qualitatively consistent with data for $p_T \lesssim 10 ~\mathrm{GeV}$.

This thesis adopts the perspective that the large momentum scale $p_T$ of the hard parton causes the coupling to be small.
However, we acknowledge that it is an open problem in the field as to the correct approach. For a review of strongly coupled techniques refer to \cite{Casalderrey-Solana:2011dxg} or for specific applications to energy loss refer to \cite{Mes:2020vgy}.

\subsection{Parton showers and jet substructure observables}
\label{sec:parton_showers_and_jet_substructure_observables}

Partons produced from binary collisions have a large offshellness (or virtuality), which they lose by successive splitting until they have $p^2 = m^2$ and are onshell. The vacuum radiation is preferentially emitted collinear to the parent parton, which results in a spray of final state hadrons known as a \emph{jet}. Monte Carlo implementations of the parton shower, included in \texttt{PYTHIA} \cite{Sjostrand:2007gs} for instance, may keep track of the various particles that arise from splittings in order to model the substructure of jets. This is a generally accepted procedure for \coll{p}{p} collisions.
Generally there are two approaches to modify the parton shower in the medium \cite{Cao:2020wlm}: (a) modify the splitting functions, which describe the probability for a parton to split in vacuum; or (b) keep the splitting functions the same, and apply medium modifications / energy loss to the partons in between scatters. 
Approaches which follow the (a) approach include \rev{\texttt{MARTINI} \cite{Schenke:2009gb,Park:2018acg}}, \texttt{MATTER} \cite{Cao:2017qpx, Majumder:2013re}, \texttt{Q-PYTHIA} \cite{Armesto:2009fj}, and \texttt{YAJEM-FMED} \cite{Renk:2010mf, Renk:2013pua}, and approaches which follow (b) include \texttt{JEWEL} \cite{Zapp:2011ya, Zapp:2012ak}, \texttt{HYBRID} \cite{Casalderrey-Solana:2014bpa, Casalderrey-Solana:2016jvj, Hulcher:2017cpt}, and \texttt{YAJEM-RAD/DRAG/DE} \cite{Renk:2008pp, Renk:2009nz}.
For a detailed review of such models refer to \cite{Cao:2020wlm}.

This thesis focuses on the leading (highest momentum) particle in the jet. This is a scenario which, while under better theoretical control than the full jet, misses out on the multitude of jet substructure observables that have been measured \cite{ALICE:2022wpn, Connors:2017ptx}. We take the stance that the large theoretical uncertainties which we will show are present in even this more simple theoretical situation, make the extension of the formalism to substructure observables---a situation under even less theoretical control---difficult to control.

 \chapter{ Energy loss framework}
 \label{sec:model}
 
\section*{Comment}

Chapter 3 of this thesis incorporates the model sections of the paper which was published in The European Physics Journal C (EPJ C) \cite{Faraday:2023mmx} and a preprint manuscript \cite{Faraday:2024gzx}. The EPJ C paper was written in collaboration with Antonia Grindrod, and Dr.\ W.\ A.\ Horowitz, and I have permission from my coauthors to reproduce this work in this thesis. The preprint manuscript was written in collaboration with Dr.\ W.\ A.\ Horowitz, who was similarly granted permission for its inclusion. I wrote the first draft of both papers, and WAH was responsible for minor editing and comments in subsequent drafts. Some of the numerics were produced independently by both me and AG, and some of the numerics was performed only by me. All presented results were numerically computed by me with codes that I wrote. The codes for fragmentation and production spectra was provided to me by WAH. Chun Shen provided me with hydrodynamic backgrounds \cite{Schenke:2020mbo, shen_private_communication}. Conversations with WAH and AG were invaluable for the ideas in the manuscript and the numerical implementation of these ideas. WAH conceived of the project.
\newpage

\section{Introduction}

The energy loss model used throughout this thesis is based on the Wicks-Horowitz-Djordjevic-Gyulassy (WHDG) energy loss model \cite{Wicks:2005gt}. There are a few key areas in which our model differs from the WHDG model. We incorporated a short pathlength correction to the radiative energy loss \cite{Kolbe:2015rvk, Kolbe:2015suq}, the effects of which are explored in \cref{sec:SPL_paper}. We included a new elastic energy loss calculation \cite{Wicks:2008zz} based on Hard Thermal Loops \cite{Braaten:1989mz, Klimov:1982bv, Pisarski:1988vd, Weldon:1982aq, Weldon:1982bn}, which is discussed in \cref{sec:EL_paper}. Additionally the model utilizes IP-Glasma, fluctuating initial conditions \cite{Schenke:2012wb, Schenke:2020mbo,shen_private_communication} instead of the Glauber model \cite{Miller:2007ri, Glauber:1970jm} in order to generate the collision geometry. In this section we review the components of the energy loss model, which are also presented in our previous work \cite{Faraday:2023mmx, Faraday:2024gzx}.

\section{Radiative energy loss}
\label{sec:mod_radiative_energy_loss}

\subsection{DGLV Radiative Energy Loss}
\label{sec:mod_dglv_radiative_energy_loss}

The Djordjevic-Gyulassy-Levai-Vitev (DGLV)
opacity expansion \cite{Djordjevic:2003zk, Gyulassy:1999zd}
gives the inclusive differential distribution of radiated gluons from a high-$p_T$ parent parton moving through a smooth brick of QGP. The expansion is in the expected number of scatterings or the \textit{opacity} $L / \lambda_g$, where $L$ is the length of the QGP brick and $\lambda_g$ is the mean free path of a gluon in the QGP.

The 4-momenta of the radiated gluon, the final hard parton, and the exchanged Debye medium quasiparticle with the $i^{\text{th}}$ scattering center are given respectively in lightfront coordinates (using the same conventions as in \cite{Kolbe:2015rvk}) by
\begin{subequations}
\begin{gather}
	k=\left[x P^{+}, \frac{m_g^2+\mathbf{k}^2}{x P^{+}}, \mathbf{k}\right] \label{eqn:mod_momenta_k}\\
 p=\left[(1-x) P^{+}, \frac{M^2+\mathbf{k}^2}{(1-x) P^{+}}, \mathbf{q}-\mathbf{k}\right] \label{eqn:mod_momenta_p}\\
 \rev{q_i=\left[q_i^{+}, q_i^{-}, \mathbf{q_i}\right],} \label{eqn:mod_momenta_q}
\end{gather}
\label{eqn:mod_four_momenta}
\end{subequations}
where $M$ is the mass of the hard parton, $m_g$ is the gluon mass, $P^+$ is the initial hard parton momentum in the $+$ direction, and $x$ is the radiated momentum fraction. 

The DGLV approach makes a number of assumptions related to the physical setup of the problem:
\begin{itemize}
	\item The large pathlength assumption, that $L \gg \mu^{-1}$\rev{, where $\mu$ is the Debye mass.}
    \item The well separated scattering centers assumption, that $\lambda_g \gg \mu^{-1}$. %
    \item The eikonal assumption, that $P^+ = E^+ \simeq 2E$ is the largest scale in the problem.
    \item The soft radiation assumption, that $x\ll1$.
    \item The collinear radiation assumption, that $k^+ \gg k^-$.
    \item The large formation time assumption, that $\mathbf{k}^2 /x E^+ \ll \mu$ and $(\mathbf{k}-\mathbf{q}_1)^2/x E^+ \ll \sqrt{\mu^2+\mathbf{q}_1)^2}$.
\end{itemize}
The DGLV single inclusive gluon radiation spectrum is then \cite{Gyulassy:2000er,Djordjevic:2003zk}
\begin{gather}
  \frac{\mathrm{d} N^g_{\text{DGLV}}}{\mathrm{d} x}=  \frac{C_R \alpha_s L}{\pi \lambda_g} \frac{1}{x} \int \frac{\mathrm{d}^2 \mathbf{q}_1}{\pi} \frac{\mu^2}{\left(\mu^2+\mathbf{q}_1^2\right)^2} \int \frac{\mathrm{d}^2 \mathbf{k}}{\pi} \int \mathrm{d} \Delta z \, \bar{\rho}(\Delta z) \nonumber\\
 -\frac{2\left\{1-\cos \left[\left(\omega_1+\tilde{\omega}_m\right) \Delta z\right]\right\}}{\left(\mathbf{k}-\mathbf{q}_1\right)^2+m_g^2+x^2 M^2}\left[\frac{\left(\mathbf{k}-\mathbf{q}_1\right) \cdot \mathbf{k}}{\mathbf{k}^2+m_g^2+x^2 M^2}-\frac{\left(\mathbf{k}-\mathbf{q}_1\right)^2}{\left(\mathbf{k}-\mathbf{q}_1\right)^2+m_g^2+x^2 M^2}\right].
 \label{eqn:mod_DGLV_dndx}
\end{gather}
In \cref{eqn:mod_DGLV_dndx} we have made use of the shorthand $\omega \equiv x E^+ / 2,~\omega_0 \equiv \mathbf{k}^2 / 2 \omega,~\omega_i \equiv (\mathbf{k} - \mathbf{q}_i)^2 / 2 \omega$, $\mu_i \equiv \sqrt{\mu^2 + \mathbf{q}_i^2}$, and $\tilde{\omega}_m \equiv (m_g^2 + M^2 x^2) / 2 \omega$ following \cite{Djordjevic:2003zk, Kolbe:2015rvk}. Additionally $\mathbf{q}_i$ is the transverse momentum of the $i^{\mathrm{th}}$ gluon exchanged with the medium; $\mathbf{k}$ is the transverse momentum of the radiated gluon; $\Delta z$ is the distance between production of the hard parton, and scattering; $C_R$ ($C_A$) is the quadratic Casimir of the hard parton (adjoint) representation ($C_F = 4 / 3$ [quarks], and $C_A = 3$ [gluons]); and $\alpha_s$ is the strong coupling.

 The quantity $\bar{\rho}(\Delta z)$ is the \emph{distribution of scattering centers} in $\Delta z$ and is defined in terms of the density of scattering centers $\rho(\Delta z)$ in a static brick,
\begin{equation}
  \rho(\Delta z) = \frac{N}{A_{\perp}} \bar{\rho}(\Delta z),
  \label{eqn:mod_density_scattering_centers}
\end{equation}
where $\Delta z$ is in the direction of propagation, $N$ is the number of scattering centers, $A_{\perp}$ is the perpendicular area of the brick, and $\int \mathrm{d}z \; \bar{\rho}(\Delta z) = 1$. The analysis of realistic collision geometries adds complexity to the scenario, as detailed in \cref{sec:mod_geometry}.

\subsection{Short pathlength correction to DGLV radiative energy loss}
\label{sec:mod_radiative_energy_loss_correction}

The derivation of the modification to the radiative energy loss in the DGLV \cite{Vitev:2002pf,Djordjevic:2003zk} opacity expansion approach with the relaxation of the large pathlength assumption $L \gg \mu^{-1}$ was considered in \cite{Kolbe:2015rvk,Kolbe:2015suq}.  In the derivation of the short pathlength correction, all assumptions and approximations made in the original GLV and DGLV derivations were kept, except that the short pathlength approximation $L\gg\mu^{-1}$ was relaxed.  The single inclusive radiative gluon distribution, including both the original DGLV contribution as well as the short pathlength correction, is
\begin{gather}
    \frac{\mathrm{d} N^g_{\text{DGLV+corr}}}{\mathrm{d} x}=  \frac{C_R \alpha_s L}{\pi \lambda_g} \frac{1}{x} \int \frac{\mathrm{d}^2 \mathbf{q}_1}{\pi} \frac{\mu^2}{\left(\mu^2+\mathbf{q}_1^2\right)^2} \int \frac{\mathrm{d}^2 \mathbf{k}}{\pi} \int \mathrm{d} \Delta z \, \bar{\rho}(\Delta z) \nonumber\\
   \times\left[-\frac{2\left\{1-\cos \left[\left(\omega_1+\tilde{\omega}_m\right) \Delta z\right]\right\}}{\left(\mathbf{k}-\mathbf{q}_1\right)^2+m_g^2+x^2 M^2}\left[\frac{\left(\mathbf{k}-\mathbf{q}_1\right) \cdot \mathbf{k}}{\mathbf{k}^2+m_g^2+x^2 M^2}-\frac{\left(\mathbf{k}-\mathbf{q}_1\right)^2}{\left(\mathbf{k}-\mathbf{q}_1\right)^2+m_g^2+x^2 M^2}\right] \right. \nonumber\\
   +\frac{1}{2} e^{-\mu_1 \Delta z}\left(\left(\frac{\mathbf{k}}{\mathbf{k}^2+m_g^2+x^2 M^2}\right)^2\left(1-\frac{2 C_R}{C_A}\right)\left\{1-\cos \left[\left(\omega_0+\tilde{\omega}_m\right) \Delta z\right]\right\}\right.\nonumber\\
   \left.\left.+\frac{\mathbf{k} \cdot\left(\mathbf{k}-\mathbf{q}_1\right)}{\left(\mathbf{k}^2+m_g^2+x^2 M^2\right)\left(\left(\mathbf{k}-\mathbf{q}_1\right)^2+m_g^2+x^2 M^2\right)}\left\{\cos \left[\left(\omega_0+\tilde{\omega}_m\right) \Delta z\right]-\cos \left[\left(\omega_0-\omega_1\right) \Delta z\right]\right\}\right)\right],
\label{eqn:mod_full_dndx}
\end{gather}
\noindent where the first two lines of the above equation are the original DGLV result \cite{Gyulassy:2000er,Djordjevic:2003zk}, \cref{eqn:mod_DGLV_dndx}, while the last two lines are the short pathlength correction.
We emphasize that contributions from all diagrams which are not suppressed under the relevant assumptions are included.  Of particular importance is the large formation time assumption, which allows one to systematically neglect a significant number of diagrams in both the original DGLV derivation \cite{Vitev:2002pf,Djordjevic:2003zk} and in the short pathlength correction \cite{Kolbe:2015rvk,Kolbe:2015suq}.

Since $\mathrm{d} N/\mathrm{d} x$ includes an integration over all $\Delta z$, the correction is present for the energy loss of a parton going through plasma of \emph{any} length; however, the relative contribution of the correction term does go to zero as the pathlength goes to infinity.

The finite pathlength correction originates from not neglecting the $q^z = i\mu_1$ pole in the Gyulassy-Wang potential, as was originally done \cite{Gyulassy:2000er,Djordjevic:2003zk}, which leads to the overall $\exp(-\mu_1\Delta z)$ scaling of the correction term in \cref{eqn:mod_full_dndx} \cite{Kolbe:2015rvk,Kolbe:2015suq}.  

There is a significant literature of energy loss derivations and corrections to earlier energy loss derivations.  Even though the focus of this work is the numerical implementation of \cref{eqn:mod_full_dndx} and the examination of its underlying assumptions, it is worth taking some time to contextualize the short pathlength correction in \cref{eqn:mod_full_dndx} within the literature.  In particular, there is currently no other derivation of short pathlength corrections to any energy loss formalism in the literature.  

The original BDMPS \cite{Baier:1996kr, Baier:1996sk, Baier:1996vi, Baier:1998kq} energy loss derivation explicitly neglects the $q^z=i\mu$ pole in the Gyulassy-Wang potential.  In principle, then, one could  derive a short pathlength correction in the original BDMPS-Z formalism analogous to the one derived in \cite{Kolbe:2015rvk,Kolbe:2015suq}.  Subsequent work within the BDMPS-Z formalism \cite{Baier:1996kr, Baier:1996sk, Baier:1996vi, Baier:1998kq,Zakharov:1996fv,Zakharov:1997uu} considered the saddle point approximation of the path integral, that in the limit of a large number of scatterings one could make a simple harmonic oscillator approximation (via the Central Limit Theorem).  This SHO approximation explicitly requires a large opacity $L / \lambda \gg 1$.  For a perturbative calculation, one requires the scattering centers are well-separated, $\lambda\gg1/\mu$, and so a large opacity implies a large pathlength; one therefore cannot determine a short pathlength correction to the SHO approximated BDMPS-Z approach.  %
If one assumes that the system is strongly coupled (see, e.g., \cite{Liu:2006ug,Casalderrey-Solana:2011dxg}) and $\lambda\ll\mu^{-1}$, then all paths are long and there is no short pathlength correction.  

In the Improved Opacity Expansion (IOE) \cite{Mehtar-Tani:2019ygg,Mehtar-Tani:2019tvy}, the starting point is already the $z$-integrated path integral; i.e.\ the IOE starts with the equation of motion for the propagator in transverse position space, and is completely insensitive to questions about the interplay between any longitudinal scales such as the pathlength, mean free path, and distance between scattering centers.  Within the IOE formalism the ``Gyulassy-Wang'' potential is taken to be $V\left(\mathbf{q}^2\right) \sim 1 / \left(\mathbf{q}^2+\mu^2\right)^2$ where, importantly, $\mathbf{q}$ is the transverse momentum exchanged with the scattering center; i.e.\ any potential pole from the $z$ component of the fully three dimensional Gyulassy-Wang potential is already neglected.  Thus the IOE formalism is unable to compute any short pathlength correction to the energy loss. %

Similar to the IOE approach, the finite size-improvement \cite{Caron-Huot:2010qjx} to the AMY formalism \cite{Arnold:2001ba,Arnold:2001ms,Arnold:2002ja} begins with Zakharov's path integral formalism and considers only the transverse momentum transfer $\mathbf{q}$ from the in-medium scattering cross section.  Thus, like in the IOE approach, unless the scattering center cross section is considered in full three dimensions, the finite size-improvement to AMY cannot capture the short pathlength corrections to energy loss.

There is a further extensive literature of work that utilizes only the 2D (rather than 3D) potential, and thus cannot capture the short pathlength corrections found in \cite{Kolbe:2015rvk,Kolbe:2015suq}.  Some of these works include the antenna problem \cite{Mehtar-Tani:2010ebp,Mehtar-Tani:2011hma,Mehtar-Tani:2011vlz,Armesto:2011ir,Mehtar-Tani:2011lic,Mehtar-Tani:2012mfa}, $\hat q$ resummation \cite{Blaizot:2014bha}, jet cascades \cite{Blaizot:2013vha,Iancu:2014kga}, running coupling effects in $\hat q$ \cite{Iancu:2014sha}, the radiative energy loss of neighboring subjets \cite{Mehtar-Tani:2017ypq}, and quark branching beyond the soft gluon limit \cite{Sievert:2018imd}.  

In \cite{Sadofyev:2021ohn} the authors couple the opacity expansion to the collective flow of the quark gluon plasma.  This work explicitly makes the assumption $\mu \Delta z \gg 1$; thus short pathlength corrections to this derivation are possible, but not included in the original derivation.

In \cite{Djordjevic:2007at, Djordjevic:2008iz, Djordjevic:2009cr} the Gyulassy-Wang static scattering center potential in canonical opacity expansion calculations was replaced by HTL propagators communicating between the high-energy parent parton and the in-medium thermal parton.  In the first derivation \cite{Djordjevic:2007at, Djordjevic:2008iz}, the authors work purely in momentum space to compute the interaction rate; since the authors do not Fourier transform into position space, their result is completely insensitive to the particulars of the path.  This work was improved in \cite{Djordjevic:2009cr} where a Fourier transform into position space was done and, as is done in the opacity expansion approach, the phases were kept.  However the limit $L\to\infinity$ is explicitly taken.  In principle one could derive a short pathlength correction to this derivation by relaxing the assumption of $L\to\infinity$.

The Higher Twist (HT) approach \cite{Guo:2000nz,Wang:2001ifa} in general only keeps the most length-enhanced contributions \cite{Majumder:2007hx}.  In principle, one may include less enhanced contributions from the assumed factorized nuclear expectation values of the various four point functions.

Gradient jet tomography \cite{He:2020iow,Fu:2022idl} couples a high momentum parton to an asymmetric medium.  The practical implementation of this procedure utilizes the dipole approximation in the path integral approach where the entire nuclear medium is treated as a sheet \cite{Casalderrey-Solana:2007knd}, and so any sensitivity to longitudinal physics is lost.

\section{Elastic energy loss}
\label{sec:mod_elastic_energy_loss}

We will consider three different elastic energy loss kernels in this work, all of which utilize the Hard Thermal Loops (HTL) effective field theory \cite{Braaten:1989mz, Klimov:1982bv, Pisarski:1988vd, Weldon:1982aq, Weldon:1982bn}. 

The 4-momenta of the radiated gluon, the final hard parton, and the exchanged Debye medium quasiparticle with the $i^{\text{th}}$ scattering center are given respectively in lightfront coordinates (using the same conventions as in \cite{Kolbe:2015rvk}) by
In \cref{sec:mod_braaten_thoma_elastic} we present a short review of the elastic energy loss derived by Braaten and Thoma (BT) \cite{Braaten:1991jj, Braaten:1991we}. The BT elastic energy loss kernel is computed with HTL propagators for soft momentum transfer and vacuum pQCD propagators for hard momentum transfer. The BT elastic energy loss kernel was used in our previous work \cite{Faraday:2023mmx}. %

\Cref{sec:mod_pure_htl_elastic_energy_loss} details the elastic energy loss kernel calculated purely with HTL propagators \cite{Wicks:2008zz}, which we will label \textit{Poisson HTL}. This corresponds to the ``HTL-X1" prescription in \cite{Wicks:2008zz}, which calculates all energy loss with HTL propagators but includes the full kinematics of the hard scatters. A brief summary of the derivation of the HTL elastic energy loss kernel is presented in \cref{sec:mod_pure_htl_elastic_energy_loss}, due to a few typographical errors in the original work \cite{Wicks:2008zz}. 
Finally we present results with the same average energy loss as the Poisson HTL results, but with a Gaussian distribution, which we will label \textit{Gaussian HTL}. The Poisson and Gaussian distribution implementations are discussed in \cref{sec:mod_probability_of_energy_loss_distributions}.

This set of elastic energy loss kernels, facilitates two important comparisons in this work. The comparison of the Gaussian BT and Gaussian HTL results shows the sensitivity of the $R_{AA}$ to the magnitude of the elastic energy loss, ignoring distributional differences. This is important in understanding the fundamental theoretical uncertainty in the transition region between the HTL and Vacuum propagators. Additionally we can compare the Gaussian HTL and Poisson HTL results to isolate the sensitivity of the $R_{AA}$ to the distribution used. This allows us to understand the importance of more precise modeling of the distribution, and the impact of the central limit theorem approximation used in our previous work \cite{Faraday:2023mmx} and the literature \cite{Wicks:2005gt}.

\subsection{HTL vs Vacuum propagators}
\label{sec:mod_htl_vs_vacuum_propagators}

In our previous work \cite{Faraday:2023mmx} we used the elastic energy loss calculated by Braaten and Thoma \cite{Braaten:1991jj, Braaten:1991we}. This formalism accounts for soft contributions to the elastic energy loss through HTL screened gluon propagators which shield the infrared divergence. The hard contribution is calculated with vacuum propagators. The hard region is defined for momentum transfers $q \gtrsim T$ while the soft region occurs at $q \lesssim g T$ where $T$ is the temperature and $g$ is the strong coupling \cite{Braaten:1991jj}. It is therefore natural to define an intermediary scale $q^*$ chosen such that $gT \ll q^* \ll T$ which partitions the hard and soft regions. %
In the small coupling and high temperature limit, the dependency on the transition scale $q^*$ falls out at the level of the fractional energy loss to leading logarithm accuracy \cite{Romatschke:2004au,Gossiaux:2008jv}. If one does not take the strict soft coupling and high temperature limit then one finds a significantly larger uncertainty to the cutoff momentum chosen \cite{Romatschke:2004au, Wicks:2008zz}. 

\rev{One} approach to treat this uncertainty \cite{Gossiaux:2008jv} is to use a ``semi-hard" gluon propagator, $\sim 1 / (q^2 - \kappa \mu^2)$ where $\kappa$ is a dimensionless parameter, is used for momentum transfers $|q| > |q^*|$ in conjunction with the standard HTL propagator for soft momentum transfers $|q| < |q^*|$. The parameter $\kappa$ is then chosen such that the dependence on the intermediary scale $q^*$ is minimized.

In this current work we will not make such a prescription, and rather treat the uncertainty as a fundamental theoretical uncertainty. We do this by examining two extreme cases: the Braaten and Thoma and the HTL elastic energy loss.
Future work could more rigorously capture this uncertainty by varying the intermediary scale $q^*$.

\subsection{Braaten and Thoma elastic energy loss}
\label{sec:mod_braaten_thoma_elastic}

The BT elastic energy loss \cite{Braaten:1991jj, Braaten:1991we} of a quark is calculated in the regions $E \ll M^2 / T$ and $E \gg M^2 / T$, where $M$ is the mass of the incident quark, and $T$ is the temperature of the medium. For $E \ll M^2 / T$ the differential energy loss per unit distance is
\begin{equation}
\frac{\mathrm{d} E}{\mathrm{d} z} = 2 C_R \pi \alpha_s^2 T^2 \left[\frac{1}{v}- \frac{1-v^2}{2 v^2}\log \frac{1+v}{1-v}\right] \log \left(2^{\frac{n_f}{6+n_f}} B(v) \frac{E T}{m_g M}\right)\left(1+\frac{n_f}{6}\right),
\label{eqn:mod_elastic_energy_loss_low}
\end{equation}
where $B(v)$ is a smooth function satisfying constraints listed in \cite{Braaten:1991we}, $v$ is the velocity of the hard parton, and $n_f$ is the number of active quark flavors in the plasma (taken to be $n_f = 2$ throughout). For $E \gg M^2/T$ the differential energy loss per unit distance is
\begin{equation}
    \frac{\mathrm{d} E}{\mathrm{d} z} = 2 C_R \pi \alpha_s^2 T^2 \left(1 + \frac{n_f}{6}\right) \log \left(2^{\frac{n_f}{2(6+n_f)}} \, 0.92 \frac{\sqrt{E T}}{m_g}\right).
    \label{eqn:mod_elastic_energy_loss_high}
\end{equation}
The energy loss at arbitrary incident energy is taken to be the connection of these two asymptotic results such that $\mathrm{d} E / \mathrm{d} z$ is continuous (determined numerically). 
The calculation was performed explicitly \cite{Braaten:1991jj, Braaten:1991we} for incident quarks with a Casimir of $C_R = C_F$, and incident gluons are taken into account by a Casimir change of $C_R = C_A$.

\subsection{Pure HTL elastic energy loss}
\label{sec:mod_pure_htl_elastic_energy_loss}

In this section we present an outline of the steps involved in the HTL calculation of the elastic energy loss, following \cite{Wicks:2008zz, Braaten:1991jj, Braaten:1991we}. We duplicate the key steps in the calculation in order to highlight the differences in approaches in the literature as well as to fix some minor typographical errors in the equations in \cite{Wicks:2008zz}. 
The HTL elastic energy loss calculation differs from the BT result in a few key ways. Firstly, the energy loss is computed for all momentum transfers (both hard and soft) with the HTL propagators (the ``HTL-X1" procedure in \cite{Wicks:2008zz}). Secondly, the full kinematics of the hard momentum transfers are kept, i.e.\ we do not make the following approximations which are made in the BT \cite{Braaten:1991jj, Braaten:1991we} calculation:
\begin{enumerate}
	\item The expansion $1 + n_B(\omega) \approx \frac{T}{\omega}$
	\item The assumption $M, p \gg T$ where $M$ is the incident mass, $p$ is the incident momentum, and $T$ is the temperature
\end{enumerate}
This procedure is similar to the procedure used in \cite{Djordjevic:2006tw}, while the BT procedure is more akin to that used in \cite{Schenke:2009gb}. %

We begin by considering the $t$-channel matrix element $\mathcal{M}$ for the collision of two partons in a QGP. This is the dominant contribution to the full elastic scattering process in the eikonal limit. We will also only do the calculation for the scattering of a quark off a quark and make the approximation that the gluon-quark and gluon-gluon scattering processes differ only by changing the relevant Casimirs. We can write the spin and color averaged matrix element for an incident parton $q$ scattering off a medium parton $m$ as
\begin{equation}
	\left\langle \left| \mathcal{M}_{q m} \right|^{2}  \right\rangle  = \frac{1}{2 N_q} \frac{1}{2 N_m} \sum_{\text{spins, colors}} \left| \mathcal{M}_{qm} \right|^{2},
	\label{eqn:mod_color-spin-average-matrix-squared}
\end{equation}
where $N_{q / m}$ is the number of colors of the parton $q / m$.

One may write down the interaction rate $dN / d z$, where $N$ is the number of elastic scatters and $z$ is the distance that the parton has traveled through the medium, for a hard parton $q$ as \cite{Wicks:2008zz} %
\begin{align}
\begin{split}
	\frac{dN}{dz} =& \frac{1}{2 E} \int \lips{p} \lips{k} \lips{k'} \times (2 \pi)^4 \delta^4(p + k - p' - k')\\
	&\times \sum_m n_m(k^0)(1 \pm_m n_m(k^{\prime 0}) \left\langle \left| \mathcal{M}_{q m} \right|^{2}  \right\rangle,
\end{split}
	\label{eqn:mod_scattering_rate_first_step}
\end{align}
where $E$ is the energy of the incident parton $q$, the index $m$ sums over all of the medium partons ($N_c^2 - 1$ gluons and $4 N_c N_f$  for the active quark flavors), $\lips{p} \equiv d^3 \vecIII{p} / (2 \pi)^3 2 \vecIV{p}^0$  is the Lorentz invariant phase space, $n_m$ is the statistical distribution (Bose distribution for gluons and Fermi distribution for quarks), and $\pm_m$ is $+$ for bosons and $-$ for fermions.

The HTL quark propagator in the Coulomb gauge is given by \cite{Wicks:2008zz, Blaizot:2001nr, Bellac:2011kqa} %
\begin{gather}
	D_{\mu \nu}(Q) = Q_{\mu \nu}(Q) \Delta_L (Q)  + P_{\mu \nu}(Q) \Delta_T(Q),\label{eqn:mod_HTL_propagator}\\
	\Delta_{L,T}(Q) = \frac{1}{Q^2 - \Pi_{L,T}(Q)}\label{eqn:mod_delta_L_T}\\
	Q_{00}=\left(1-x^2\right),\quad P_{i j}=-\left(\delta_{i j}-\hat{q_i} \cdot \hat{q_j}\right)\\
	\Pi_L= \mu^2 (1 - x^2) \left(1 - \frac{x}{2} \log \frac{x+1}{x-1}\right) \\
	\Pi_T = \frac{\mu^2}{2} \left( x^2 - \frac{x(1-x^2)}{2} \log \frac{x+1}{x-1}\right) + \mu_M^2
\end{gather}
where $\hat{q_i} \equiv \vecIII{\hat{q_i}} / |\vecIII{\hat{q_i}}|$, $x \equiv \omega / q$, and all other components of $Q_{\mu \nu}$ and $P_{\mu \nu}$ are zero. 
Following \cite{Wicks:2008zz}, we include an \textit{ad hoc} magnetic mass $\mu_M$ in the transverse propagator $\Delta_T$ which is not present in standard HTL, although lattice calculations \cite{Nakamura:2003pu, Hart:2000ha} indicate that there is a nonzero magnetic mass in phenomenologically relevant QGPs. In this work we will take $\mu_M = \mu$ in all results for simplicity.

One may then perform the integral in \cref{eqn:mod_scattering_rate_first_step}, collapsing the energy-momentum conserving delta functions in $\left\langle |\mathcal{M}_{qm}|^2 \right\rangle$, as described in \cite{Wicks:2008zz}. Doing this, we obtain
\begin{multline}
		\frac{dN}{d z} = \frac{4 C_R \alpha_s^2}{\pi} \int_q \frac{n_B(\omega)}{q} \left( 1 - \frac{\omega^2}{q^2} \right)^2 \sum_m\left[C_{L L}^{m}\left|\Delta_L\right|^2+2 C_{L T}^{m} \operatorname{Re}\left(\Delta_L \Delta_T\right)+C_{T T}^m\left|\Delta_T\right|^2\right],
	\label{eqn:mod_scattering_rate_final}
\end{multline}
where
\begin{equation}
	\int_q \equiv \int \frac{d^3 \vecIII{q} d \omega}{2 \pi} \frac{E}{E^{\prime}} \delta\left(\omega+E-E^{\prime}\right).
	\label{eqn:mod_integral_q}
\end{equation}
The transverse part of \cref{eqn:mod_integral_q} can be performed trivially under the assumption that the medium is isotropic \cite{Wicks:2008zz}. The integral over $q=| \vecIII{q}|$ is done numerically and the integral over $\omega$ will be changed to an integral over the fractional energy loss $\epsilon$ defined via $\omega = \epsilon E$, and left unevaluated since the quantity we want to calculate is $dN / d \epsilon$.

The coefficients $C_{L L}^{j m}$, $C_{L T}^{j m}$, and $C_{T T}^m$ are given in terms of various thermal integrals \cite{Wicks:2008zz}
\begin{equation}
	C^{m} \equiv \int_{\frac{1}{2}(\omega+q)}^{\infty} k^0 k d k\left(n_{m}\left(k^0-\omega\right)-n_{m}\left(k^0\right)\right) c^{\prime}
\end{equation}
where the coefficients $c^\prime$ are
\begin{align}
	c_{L L}^{\prime} \equiv  & \left(\left(1+\frac{\omega}{2 E}\right)^2-\frac{q^2}{4 E^2}\right)\left(\left(1-\frac{\omega}{2 k^0}\right)^2-\frac{q^2}{4\left(k^0\right)^2}\right) \\
	c_{L T}^{\prime} \equiv  & 0 \\
	c_{T T}^{\prime} \equiv  & \frac{1}{2}\left(\left(1+\frac{\omega}{2 E}\right)^2+\frac{q^2}{4 E^2}-\frac{1-v^2}{1-\frac{\omega^2}{q^2}}\right) \left(\left(1+\frac{\omega}{2 k^0}\right)^2+\frac{q^2}{4\left(k^0\right)^2}-\frac{1-v_k^2}{1-\frac{\omega^2}{q^2}}\right).
\end{align}
In the above $v_k (v)$ is the velocity of the medium parton (incident parton), and the $c^\prime$ are independent of whether the medium parton is a quark or a gluon since we have assumed that the quarks and gluons differ only by the statistical distribution and the Casimir. The $C$ coefficients may then be calculated in terms of various thermal integrals which may be performed analytically
\begin{equation}
	I_n^m \equiv \int_{\frac{1}{2} (\omega + q)}^\infty dk \; (n_m (k-\omega) - n_m(k)) k^n.
	\label{eqn:mod_general_thermal_integral}
\end{equation}
We provide the results of these integrals for bosons ($+$) and fermions ($-$)
\begin{align*}
I_0^\pm &= \pm T \log\left(\frac{1 \mp e^{-\kappa_+}}{1 \mp e^{-\kappa_-}}\right) \\
I_1^\pm &= \pm T^2 \left[\operatorname{Li}_2\left(\pm e^{-\kappa-}\right) - \operatorname{Li}_2\left(\pm e^{-\kappa+}\right)\right] + \kappa_+ T I_0^\pm \\
I_2^\pm &= \pm 2 T^3 \left[\operatorname{Li}_3\left(\pm e^{-\kappa-}\right) - \operatorname{Li}_3\left(\pm e^{-\kappa+}\right)\right] \\
&\quad+ 2 \kappa_+ T I_1^\pm - \kappa_+^2 T^2 I_0^\pm,
\end{align*}
where $\kappa_{\pm} = (\omega \pm q)/2 T$ and $\operatorname{Li}_n$ is the polylogarithm function.

Finally one may perform a change of variables from $\omega$ to $\epsilon$ where $\omega = \epsilon E$ to convert from the scattering rate $d N / d z$ to the single elastic scattering kernel $d N / d \epsilon$. Schematically this gives
\begin{align}
	\frac{dN}{d \epsilon}  &= \frac{d\omega}{d \epsilon} \frac{d}{d \omega} \int d z \frac{dN}{d z} \\
	&\simeq L E \frac{dN}{d z \; d\omega},
	\label{eqn:mod_dndx_pure_htl}
\end{align}
where practically $dN /d z d \omega$ is simply $dN / d z$ but with the integral over $\omega$ removed from $\int_q$ in \cref{eqn:mod_integral_q}. In the last step we have approximated that $d N /dz$ is independent of $z$ which is true in a static brick; more realistic models for the medium would include the $z$ dependence in the temperature $T(z)$. The single elastic scattering kernel $dN / d \epsilon$ is the analog of the single emission kernel in the radiative energy loss \cite{Djordjevic:2003zk, Gyulassy:2000er, Gyulassy:2001nm} in \cref{sec:mod_radiative_energy_loss}.

\section{Energy loss probability distributions}
\label{sec:mod_probability_of_energy_loss_distributions}

\subsection{Gaussian approximation for elastic energy loss}
\label{sec:mod_gaussian_distribution_approximation}

For the BT elastic energy loss and the Gaussian HTL elastic energy loss we proceed by assuming that there are enough elastic scatters such that the elastic energy loss follows a Gaussian distribution according to the central limit theorem. This assumption implies that the distribution of elastic energy loss is Gaussian with mean provided by average elastic energy loss $\Delta E$, and width by the fluctuation dissipation theorem \cite{Moore:2004tg, Xu:2014ica}
\begin{equation}
    \sigma = \frac{2}{E} \int \mathrm{d}z \; \frac{\mathrm{d} E}{\mathrm{d}z} T(z),
    \label{eqn:mod_sigma_elastic}
\end{equation}
where $z$ integrates along the path of the parton, and $T(z)$ is the temperature along the path. Therefore
\begin{equation}
  P_{\text{el}}(E_f | E_i,\,L,\,T) \equiv \frac{1}{\sqrt{2\pi}\sigma}\exp\left[{-}\left( \frac{E_f-(E_i+\Delta E)}{\sqrt{2} \sigma} \right)^2 \right],
\end{equation}
where $\Delta E = \int dz \; \frac{dE}{dz}$. We can perform a change of variables from $E_f$ to $\epsilon$ where $E_f = (1-\epsilon) E_i$ resulting in the probability for a particle of initial energy $E_i$ losing a fraction $\epsilon$ of its energy

\begin{align}
	P_{\text{el}}(\epsilon | E_i,\,L,\,T) &\equiv \frac{E_i}{\sqrt{2\pi}\sigma}\exp\left[{-}\left( \frac{(1-\epsilon)E_i-(E_i+\Delta E)}{\sqrt{2} \sigma} \right)^2 \right].\\
	&= \frac{1}{\sqrt{2\pi}\sigma'}\exp\left[{-}\left(\frac{\epsilon - \Delta E/E_i}{\sqrt{2} \sigma'} \right)^2 \right],
\end{align}
where $\sigma' = \sigma / E_i$.

\subsection{Poisson distribution for radiative and elastic energy loss}
\label{sec:mod_multi_scatters}

The DGLV energy loss kernel \cref{eqn:mod_full_dndx} gives the inclusive spectrum of emitted gluons.  Thus the expected number of radiated gluons may be different from one.  In fact, one sees that for hard partons emerging from the center of a central heavy ion collision, the expected number of emitted gluons is $\sim 3$ \cite{Gyulassy:2001nm}. In a similar way \cref{eqn:mod_dndx_pure_htl} gives the differential number of scatters for the elastic energy loss, which is generically also different from one \cite{Wicks:2008zz}. We may treat these two situations in an identical way following the procedure outlined in \cite{Gyulassy:2001nm}. This requires assuming that the gluon emissions and elastic scatters are both independent in order to apply this procedure to the radiative and elastic energy loss kernels, respectively.

Explicitly we can write 
\begin{equation}
  P(\epsilon | E)=\sum_{n=0}^{\infty} P_n(\epsilon | E),
  \label{eqn:mod_poisson}
\end{equation}
where the $P_n$ are found via the convolution
\begin{align}
  P_{n+1}(\epsilon) & =\frac{1}{n+1} \int \mathrm{d} x_n \; \frac{\mathrm{d} N^{g}}{\mathrm{d} x} \; P_n(\epsilon-x_n)
    \label{eqn:mod_pn}
\end{align}
and we have $P_0(\epsilon) \equiv e^{- \langle N^g \rangle} \delta(\epsilon)$. Here, and for the rest of the paper, we define $1-\epsilon$ as the fraction of initial momentum kept by the parton, such that in the eikonal limit the final energy of the parton in terms of the initial energy of the parton is $E_f\equiv(1-\epsilon)E_i$.  The Poissonian form of \cref{eqn:mod_poisson} guarantees the distribution is normalized to one, and the expected number of emitted gluons (number of elastic scatters) is $\sum_n\int \mathrm{d} \epsilon \: n \, P_n(\epsilon, E) = \langle N^g \rangle \; \left(\left\langle N_{\text{scatters}} \right\rangle\right)$. The support of $P(\epsilon)$ past $\epsilon = 1$ is unphysical, and we interpret this as the probability for the parton to lose all of its energy before exiting the plasma. The fact that the support of $P(\epsilon)$ is not naturally constrained to be less than one, stems from the fact that we do not update the partons energy after each collision. Under this interpretation we include the excess weight $\int_1^\infty \mathrm{d} \epsilon \; P_{\text{rad}}(\epsilon)$ as the coefficient of a Dirac delta function at $\epsilon=1$.

We note that the random variable $N^g$ should rigorously be thought of as $N^g = (N^g_{\text{vac}} + N^g_{\text{med}}) - N^g_{\text{vac}}$ where $N^g_{\text{med}}$ is the number of radiated gluons occurring due to medium interactions, and $N^g_{\text{vac}}$ is the number of DGLAP vacuum radiation gluons. With this understanding, the independent gluon emission assumption means that $(N^g_{\text{vac}} + N^g_{\text{med}})$ and $N^g_{\text{vac}}$ should each be modeled by a Poisson distribution.  Then $N^g$ is actually given by a Skellam distribution \cite{skellam:1946}, the difference between two Poisson distributions.
In the current model, energy gain relative to the vacuum at some $x^*$ corresponds to $\mathrm{d} N^g (x^*) / \mathrm{d} x < 0$; whereas it should rigorously correspond to $\mathrm{d} N^g (-x^*) / \mathrm{d} x > 0$. Following previous work \cite{Gyulassy:2001nm, Wicks:2005gt, Horowitz:2011gd}, we will simply model $P(\epsilon)$ as a Poisson distribution; the effect of this simplification is not obvious and requires future work. This disucssion only applies to the radiative energy loss, as there is no elastic energy loss in vacuum.

\subsection{Total energy loss distribution}%
\label{sec:mod_total_energy_loss}

As done in \cite{Wicks:2005gt}, we convolve the radiative and elastic energy loss probabilities to yield a total probability of energy loss,
\begin{align}
  P_{\text{tot}}(\epsilon) \equiv \int \mathrm{d} x \, P_{\text{el}}(x)P_{\text{rad}}(\epsilon-x).
\end{align}

Note that both the radiative distribution $P_{\text{rad.}}(\epsilon)$ and $P_{\text{el.}}(\epsilon)$ may contain Dirac delta functions at both $\epsilon = 0$ and $\epsilon = 1$, corresponding to the probability of no energy loss occurring and the probability of all energy being lost before exiting the QGP respectively. This implies that the total energy loss has a delta function at $\epsilon = 0$ with a weight of the product of the radiative and elastic delta functions at $\epsilon =0$. When a Gaussian form of the elastic energy loss probability distribution is used then the elastic energy loss does not contain a delta function at $\epsilon =0$, which in turn implies that there is no delta function at $\epsilon = 0$ in the total energy loss distribution.

\section{Numerical implementation of the elastic and radiative energy loss}
\label{sec:mod_numerical_elastic_radiative}

For all numerical calculations we neglect the running of the strong coupling constant and use $\alpha_s = 0.3$, consistent with \cite{Kolbe:2015rvk, Wicks:2005gt, Djordjevic:2003zk}. Additionally we use charm and bottom quark masses of $m_c = 1.2~\mathrm{GeV}/c^2$, and $m_b = 4.75~\mathrm{GeV}/c^2$, respectively. The effective light quark and gluon masses are set to the asymptotic one-loop medium induced thermal masses, $m_{\text{light}} = \mu / 2$ and $m_g = \mu / \sqrt{2}$, respectively \cite{Djordjevic:2003be}. The upper bounds on the $|\mathbf{k}|$ and $|\mathbf{q}|$ integrals are given by $k_{\text{max}}=2x (1-x) E$ and $q_{\text{max}}=\sqrt{3 E \mu}$, following \cite{Wicks:2005gt}. This choice of $k_{ \text{max}}$ guarantees that the momentum of the radiated gluon and the initial and final momenta of the parent parton are all collinear. %
In our previous work \cite{Faraday:2023mmx} we motivated an alternative upper bound for the transverse radiated gluon momentum of $\operatorname{Min}\left(\sqrt{2 x E \mu_1}, 2 x(1-x) E\right)$ which guarantees that the energy loss kernel receives no contributions from regions of phase space where either the large formation time assumption or the collinear assumption are invalid. We showed \cite{Faraday:2023uay} that this resulted in a slight decrease of the standard DGLV radiative energy loss, and a dramatic reduction of the short pathlength correction, particularly for gluons. Future work will explore the impact of utilizing such a bound at the level of the $R_{AA}$; however in this work we utilize the standard collinear upper bound in order to be comparable with previous work \cite{Faraday:2023mmx, Wicks:2005gt, Wicks:2008zz}.

In order to maintain consistency with the WHDG model, we assume the distribution of scattering centers is exponential: $\rho_{\text{exp.}} (z) \equiv \frac{2}{L} \exp[ - 2 z / L]$. The exponential distribution serves to make the integral in $\Delta z$ analytically simple. The physical motivation for this is that an exponentially decaying distribution of scattering centers captures the decreasing density of the QGP due to Bjorken expansion \cite{Bjorken:1982qr}. It is likely that using an exponential rather than, say, a power law decay distribution is an overestimate of the effect of the expansion, since Bjorken expansion obeys a power law decay along the incident partons path, not exponential. It turns out that in DGLV, the distribution of scattering centers $\bar{\rho}(\Delta z)$ affects the characteristic shape of $\mathrm{d} N^g / \mathrm{d} x$; however the radiative energy loss $\int dx x \frac{dN^g}{dx}$ is largely insensitive to the distribution of scattering centers \cite{Armesto:2011ht, Faraday:2023mmx}. Once the short pathlength correction is included, however, the energy loss becomes far more sensitive to the distribution of scattering centers, particularly at small $\Delta z$ \cite{Kolbe:2015rvk, Faraday:2023mmx}. The average energy loss can be calculated from the single emission kernel as
\begin{equation}
	\frac{\Delta E}{E} = \int dx \; x\frac{dN^g}{dx}
	\label{eqn:mod_fractional_energy_loss}
\end{equation}
or after the Poisson convolution as
\begin{equation}
	\frac{\Delta E}{E} = \int dx \; x P(x | E).
	\label{eqn:mod_fractional_energy_loss_poisson}
\end{equation}
These two results will differ only in the amount of the distribution ($dN^g / dx$ or $P(x | E)$) that extends past $x=1$.

\comm{make sure this isn't repeated later}

\begin{figure}[H]
\centering
\begin{subfigure}[t]{0.49\textwidth}
    \centering
    \includegraphics[width=\linewidth]{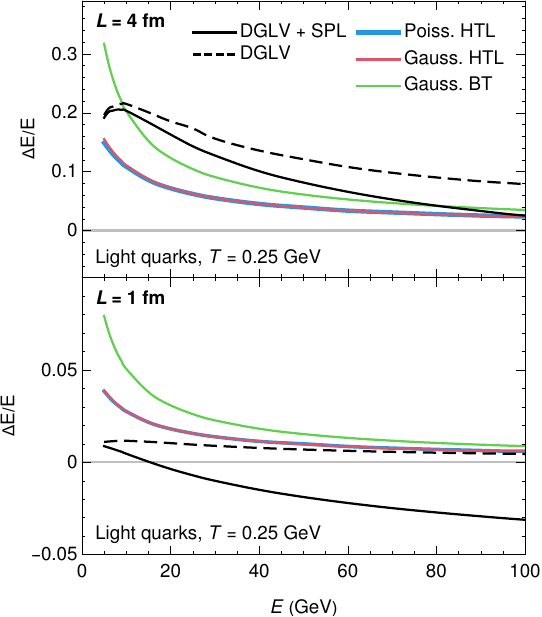}
    \caption{Light quarks}
    \label{fig:mod_deltaEoverE_small_large}
\end{subfigure}\hfill
\begin{subfigure}[t]{0.49\textwidth}
    \centering
    \includegraphics[width=\linewidth]{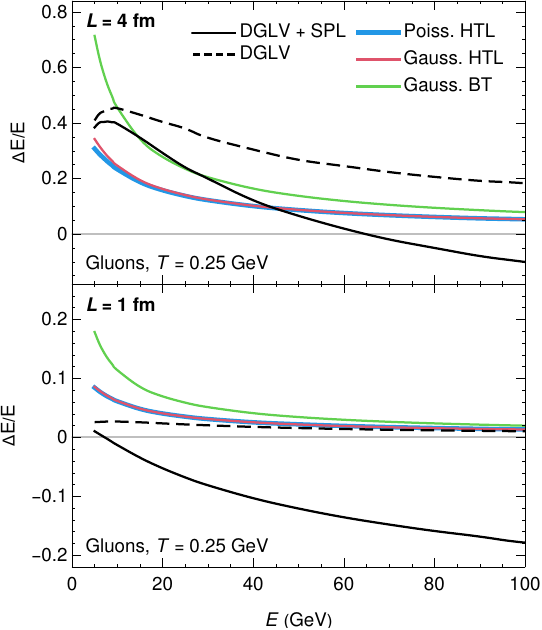}
    \caption{Gluons}
    \label{fig:mod_deltaEoverE_small_large_gluon}
\end{subfigure}
\caption{Plot of the fractional energy loss $\Delta E / E$ for elastic energy loss kernels of Gaussian Braaten and Thoma \cite{Braaten:1991jj, Braaten:1991we}, Gaussian HTL, and Poisson HTL \cite{Wicks:2008zz} and radiative energy loss kernels of DGLV \cite{Djordjevic:2003zk} and DGLV with the short pathlength correction \cite{Kolbe:2015rvk}. Top pane shows results for a large pathlength $L = 4$~fm and bottom pane for a small pathlength $L = 1$~fm. The temperature is kept constant at $T = 0.25$~GeV.}
\label{fig:mod_deltaEoverE_combined}
\end{figure}

\Cref{fig:mod_deltaEoverE_small_large} shows the fractional energy loss $\Delta E / E$ of light quarks for the various radiative and elastic energy loss kernels discussed in \cref{sec:mod_radiative_energy_loss,sec:mod_elastic_energy_loss}, for both a large pathlength of $L = 4$~fm (top pane) and a small pathlength of $L = 1$~fm (bottom pane). The fractional energy loss of gluons is similarly plotted in \cref{fig:mod_deltaEoverE_small_large_gluon}. The radiative energy loss curves shown are for the DGLV radiative energy loss \cite{Djordjevic:2003zk} and the DGLV radiative energy loss which receives a short pathlength correction \cite{Kolbe:2015rvk, Kolbe:2015suq}, both of which are modeled with a Poisson distribution \cite{Gyulassy:2001nm}. The elastic energy loss curves shown are for the HTL elastic energy \cite{Wicks:2008zz} loss with both a Poisson distribution and Gaussian distribution, and the Braaten and Thoma (BT) elastic energy loss \cite{Braaten:1991jj, Braaten:1991we} with a Gaussian distribution. 

From \cref{fig:mod_deltaEoverE_small_large,fig:mod_deltaEoverE_small_large_gluon} we see the various effects of the short pathlength correction to the radiative energy loss which were discussed in \cref{sec:mod_radiative_energy_loss_correction}. The short pathlength correction is much larger for gluons in comparison to light quarks, which is because of the breaking of color triviality in \cref{eqn:mod_full_dndx}; the short pathlength correction can lead to negative energy loss; the short pathlength correction grows faster in $p_T$ than standard DGLV; and the short pathlength correction is proportionally much larger at small lengths. \rev{We will see the downstream impacts of these effects at the level of the $R_{AA}$ and $R_{pA}$ in \cref{sec:SPL_paper,sec:EL_paper,sec:lft}. In particular, the much larger size of the short pathlength corrections for gluons, compared to light- and heavy-flavor quarks, will lead to a correspondingly larger impact of the correction for pion $R_{p / A A}$ compared to $D$ and $B$ meson $R_{p / A A}$; the negative energy loss of the short pathlength correction can lead to $R_{p / AA} > 1$; and the $p_T$ dependence causes a fast rise in the pion $R_{AA}$ as a function of $p_T$.}

Comparing the different elastic fractional energy loss kernels in \cref{fig:mod_deltaEoverE_small_large,fig:mod_deltaEoverE_small_large_gluon} we see that the BT elastic energy loss is significantly larger than the HTL elastic energy loss at low--moderate $p_T \lesssim \mathcal{O}(50)~\text{GeV}$. At asymptotically high momenta we see that the HTL and BT fractional energy loss results approach each other; however the convergence is slow. We see also that the effect of using a Gaussian compared to a Poisson distribution is negligible at the level of the fractional energy loss. 
This is expected, as both the Poisson and Gaussian elastic energy loss distributions are constrained to have an identical average energy loss if the kinematic bounds on the $\epsilon$ integral are ignored. Any minor observed differences in the Gaussian and Poisson average energy loss plots, therefore, arise from the portion of the distribution beyond the kinematic bound at $\epsilon = 1$.

Comparing the relative size of the elastic and radiative energy loss results, we see that for large pathlengths elastic energy loss is of similar importance to radiative energy loss at low momenta, while at large momenta the radiative energy loss is $\sim\!\! 2\text{--}4$ times larger than the elastic. For small $L=1$~fm pathlengths the elastic energy loss is $\sim\!\! 2\text{--}8$ times as large as the radiative energy loss for $5~\text{GeV} \leq p_T  \leq 25~\text{GeV}$, and becomes relatively less important at large momenta. The variation in the magnitude of elastic versus radiative energy loss, as a function of pathlength and energy, highlights the importance of including both forms of energy loss in order to make phenomenological predictions over a wide range of system sizes and momenta.

\begin{figure}[!htbp]
	\centering
	\includegraphics[width=0.5\linewidth]{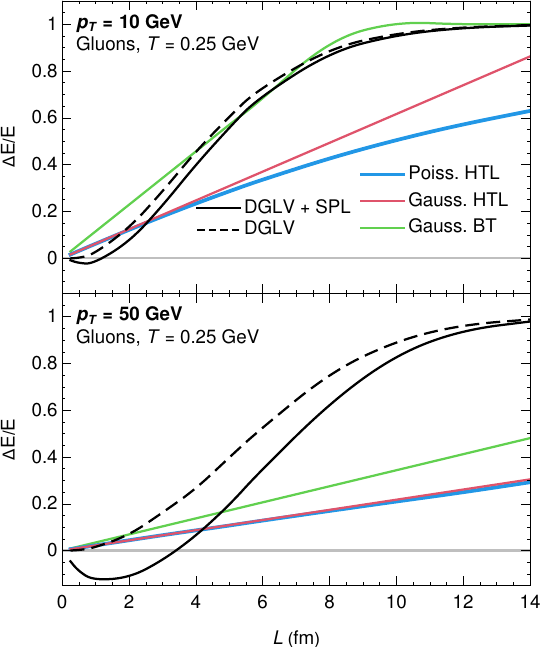}
	\caption{Plot of the fractional energy loss $\Delta E / E$ as a function of length $L$ for gluons with various elastic and radiative energy loss kernels at $p_T = 10$~GeV (top pane) and $p_T = 50$~GeV (bottom pane). The temperature is kept constant at $T = 0.25$~GeV.}
	\label{fig:mod_deltaEoverE_vs_L_gluon}
\end{figure}

\Cref{fig:mod_deltaEoverE_vs_L_gluon} shows the fractional energy loss $\Delta E / E$ as a function of the pathlength $L$ for set $p_T = 10$~GeV (top pane) and $p_T = 50$~GeV bottom pane, at constant temperature $T = 0.25$~GeV. In this figure we confirm that the short pathlength correction goes to zero at asymptotically large lengths, and that the short pathlength corrected DGLV radiative energy loss decreases linearly in $L$ at small lengths before the standard DGLV result overtakes the correction and the growth becomes quadratic in $L$.
At small lengths the DGLV radiative energy loss grows like $L^2$ due to the LPM effect, while the elastic energy loss is linear in $L$. These different $L$ dependencies lead to the elastic energy loss dominating at small pathlengths compared to the radiative energy loss.

\begin{figure}[!htbp]
    \centering
    \includegraphics[width=0.5\linewidth]{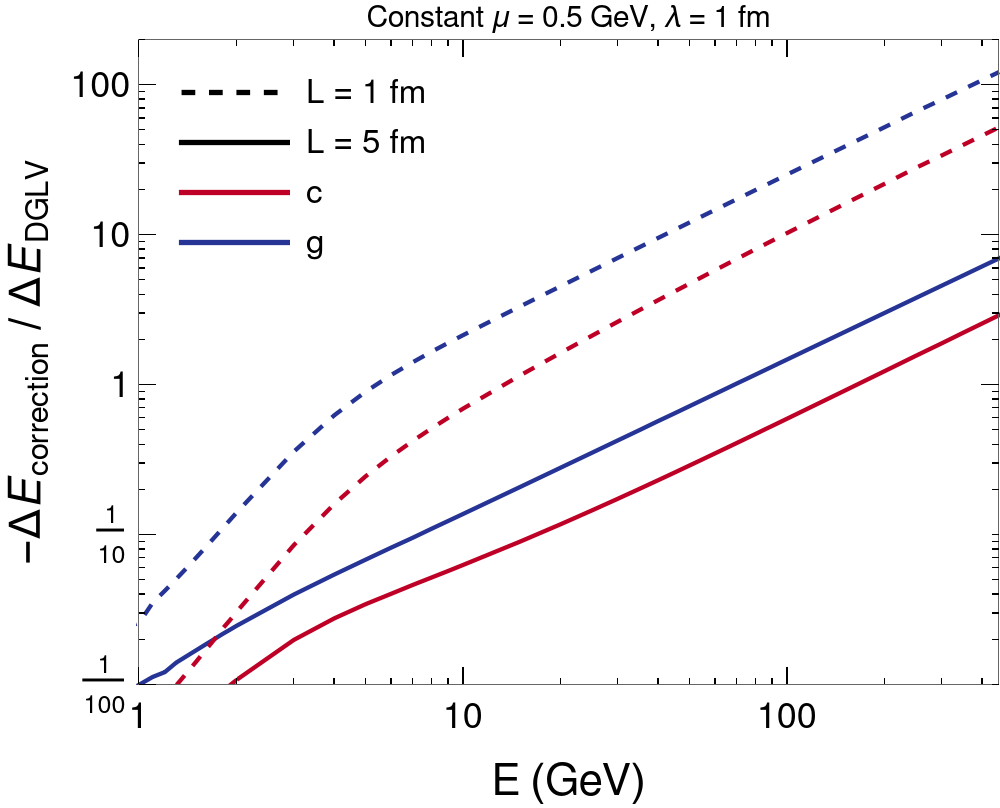}
    \caption{The ratio of the magnitude of the correction to the DGLV radiative energy loss $-\Delta E _{\text{corrected}}$ and the uncorrected DGLV radiative energy loss $\Delta E_{\text{DGLV}}$ is plotted as a function of incident energy $E$. This ratio is plotted for charm quarks (c), and gluons (g); and at $L = 1~\mathrm{fm}$ (dashed) and $L= 5~\mathrm{fm}$ (solid). Calculations were done with constant $\mu = 0.5 ~\mathrm{GeV}$ and $\lambda_g = 1~\mathrm{fm}$.
    }
    \label{fig:mod_deltaE_vs_energy}
\end{figure}

One may also consider the relative size of the short pathlength correction to the radiative energy loss in comparison to the uncorrected radiative energy loss, by calculating the ratio $\Delta E_{\text{corrected}} / \Delta E_{\text{DGLV}}$, shown in \cref{fig:mod_deltaE_vs_energy}. 

  To understand the energy $E$, length $L$, and incident Casimir $C_R$ dependence of the relative size of the short pathlength correction to the radiative energy loss, we examine the asymptotic dependence of both the DGLV and short pathlength corrected DGLV fractional radiative energy loss. For asymptotic energies the short pathlength corrected fractional energy loss is given by \cite{Kolbe:2015rvk}
\begin{subequations}
\begin{align}
  \frac{\Delta E_{\text{corrected}}}{E} =& \frac{C_R \alpha_s}{2 \pi} \frac{L}{\lambda_g}\left(-\frac{2 C_R}{C_A}\right) \frac{12}{2+\mu L}\nonumber\\
  &\times \int_0^1 \mathrm{d} x \log \left(\frac{L k_{\max }}{2+\mu L}\right)
  \label{eqn:mod_asymptotics_correction_xint}\\
  =& \frac{C_R \alpha_s}{2 \pi} \frac{L}{\lambda_g}\left(-\frac{2 C_R}{C_A}\right) \frac{\log \left(\frac{2 E L}{2+\mu L}\right)}{2+\mu L},
  \label{eqn:mod_asymptotics_correction}
\end{align}
\end{subequations}
which was calculated in this asymptotic analysis for simplicity with $k_{\text{max}}=2x E$ and, also for simplicity, an exponential distribution of scattering centers. The equivalent asymptotic result without the short pathlength correction is given by \cite{Gyulassy:2000er}
\begin{equation}
  \frac{\Delta E_{\text{DGLV}}}{E} = \frac{C_R \alpha_s}{4} \frac{L^2 \mu^2}{\lambda_g} \frac{1}{E}\log \frac{E}{\mu}.
    \label{eqn:mod_asymptotics_DGLV}
\end{equation}
The relative size of the correction is then given by
\begin{align}
  \frac{-\Delta E_{\text{correction}}}{\Delta E_{\text{DGLV}}} &= \frac{4}{\pi} \frac{C_R}{C_A} \frac{E}{\mu^2 L(2 + \mu L)},
  \label{eqn:mod_asymptotic_relative_correction}
\end{align}
keeping only leading terms in $E$. From \cref{eqn:mod_asymptotic_relative_correction}, the relative size of the short pathlength correction: increases linearly in energy, is $C_A / C_F = 9 /4$ times larger for gluons in comparison to quarks, and is $\sim 15$ times larger for a system with $\mu = 0.5$ GeV and $L = 1~\mathrm{fm}$ as opposed to a system with $L = 5~\mathrm{fm}$.  One may see in \cref{fig:mod_deltaE_vs_energy} that the detailed numerics display these three behaviors.

\section{Nuclear modification factor}
\label{sec:mod_nuclear_modification_factor}

The observable which we will be computing is the nuclear modification factor $R^h_{AB}(p_T)$ for hadrons $h$ produced in a collision system \coll{A}{B}, defined by 
\begin{equation}
    R^h_{AB}(p_T) \equiv \frac{\mathrm{d} N^{AB \to h} / \mathrm{d} p_T}{\langle N_{\text{coll}} \rangle \mathrm{d} N^{pp \to h} / \mathrm{d} p_T},
    \label{eqn:mod_nuclear_modification_factor}
\end{equation}
where $\mathrm{d} N^{AB / pp \to h} / \mathrm{d} p_T$ is the differential number of measured $h$ hadrons in \coll{A}{B} / \coll{p}{p} collisions, and $\langle N_{\text{coll}} \rangle$ is the expected number of binary collisions, usually calculated according to the Glauber model \cite{Glauber:1970jm,Miller:2007ri}.

To access the $R_{AB}$ \emph{theoretically} we make several assumptions about the underlying collision and partons.  In the following we will only refer to quarks, but all assumptions and formulae apply equally well to gluons.  We first assume, following \cite{Wicks:2005gt,Horowitz:2010dm}, that the spectrum of produced quarks $q$ in the initial state of the plasma formed by the collisions \coll{A}{B} (before energy loss) is $\mathrm{d} N^{q}_{AB\text{; init.}} / \mathrm{d} p_i = N_{\text{coll}} \times \mathrm{d}N^q_{pp} / \mathrm{d}p_i$ where $\mathrm{d} N^q_{pp} / \mathrm{d} p_i$ is the quark production spectrum in \coll{p}{p} collisions. This assumption is equivalent to neglecting initial state effects, which is justified in heavy-ion collisions by the measured $R_{AA}$ consistent with unity for probes which do not interact strongly with the QGP \cite{CMS:2012oiv, CMS:2011zfr}.
In addition we assume that the modification of the \coll{A}{B} production spectrum $\mathrm{d} N^{q}_{AB \text{; init.}}$ is due to energy loss,
\begin{equation}
d N^q_{AB; \text{ final}}\left(p_T\right)=\int  d \epsilon \; d N^q_{AB; \text{ init.}}\left(p_i\right) P_{\text{tot}}\left(\epsilon \mid p_i\right),
\end{equation}
where $P_{\text{tot}}(\epsilon | p_i)$ is the  probability of losing a fraction of transverse momentum $\epsilon$ given an initial transverse momentum $p_i$. Given these assumptions the expression for the $R_{AA}$ in \cref{eqn:mod_nuclear_modification_factor} simplifies to
\begin{equation}
    R_{AA} = \frac{1}{f\left(p_T\right)} \int \frac{d \epsilon}{1-\epsilon} f\left(\frac{p_T}{1-\epsilon}\right) P_{\text {tot}}\left(\epsilon \left| \frac{p_T}{1-\epsilon}\right.\right),
    \label{eqn:mod_full_raa_spectrum_ratio}
\end{equation}
where the $1 / 1 - \epsilon$ factor is a Jacobian and for brevity we have introduced the notation $f(p_T) \equiv dN^q / d p_T (p_T)$. In the literature \cite{Wicks:2005gt, Horowitz:2010dm} %
and in our previous work \cite{Faraday:2023mmx} an assumption was made that the production spectra followed a slowly-varying power law $f(p_T) \simeq A p_T^{-n(p_T)}$, which resulted in further simplifications. In this work all presented $R_{AA}$ results are calculated according to \cref{eqn:mod_full_raa_spectrum_ratio}, and the validity of the slowly-varying power law assumption is discussed in \cref{sec:EL_validity_of_power_law_approximation_to_r_aa}.

	Heavy quark production spectra $f(p_T)$ are calculated using FONLL at next-to-leading order \cite{Cacciari:2001td}; and gluons and light quark production spectra are computed at leading order \cite{wang_private_communication} as in \cite{Vitev:2002pf, Horowitz:2011gd}. 
	We note that for presented theoretical results on $\pi^0$ and charged hadron suppression, we used gluon and light quark production spectra $f(p_T)$ at $\sqrt{s_{NN}} = 5.5$ TeV and not $\sqrt{s_{NN}} = 5.02$ TeV, as this was conveniently available to us. This will lead to our results being slightly oversuppressed at $\sqrt{s_{NN}} = 5.02$ TeV.

There are uncertainties in both the fragmentation function fits and the generation of initial parton spectra's; however these uncertainties are very small compared to others in this problem (running coupling, first order in opacity, etc.), and so we will not take the fragmentation function uncertainties into account.
The spectrum $\mathrm{d} N^h / \mathrm{d} p_h$ for a hadron $h$ is related to the spectrum $\mathrm{d} N^q / \mathrm{d} p_q$ for a parton $q$ via \cite{Horowitz:2010dm}
\begin{equation}
        \frac{\mathrm{d} N^h}{\mathrm{d} p_h}\left(p_h\right) =\int \frac{\mathrm{d} N^q}{\mathrm{d} p_q}\left(\frac{p_h}{z}\right) \frac{1}{z} D^h_q(z, Q) \mathrm{d} z,
    \label{eqn:mod_parton_to_hadron_spectrum}
\end{equation}
where $z \equiv p_h / p_q \in (0,1]$, $p_h$ is the observed hadron momentum, $D^h_q(z,Q)$ is the fragmentation function for the process $q \mapsto h$, and $Q$ is the hard scale of the problem taken to be $Q = p_q =  p_h / z$ \cite{Horowitz:2010dm}. The hadronic $R^h_{AB}$ is then found in terms of the partonic $R^q_{AB}$ (\cref{eqn:mod_geometry_averaged_raa}) as \cite{Horowitz:2010dm}
\begin{align}
    R_{AB}^h\left(p_T\right)&=\frac{\sum_q \int d z \frac{1}{z} D_q^h(z) f\left(\frac{p_T}{z}\right) R_{AB}^q\left(\frac{p_T}{z}\right)}{\sum_q \int d z \frac{1}{z} D_q^h(z) f\left(\frac{p_T}{z}\right)}.
    \label{eqn:mod_parton_to_hadron_raa}
\end{align}
For details of the derivation needed for \cref{eqn:mod_parton_to_hadron_raa}, refer to Appendix B of \cite{Horowitz:2010dm}.

The fragmentation functions for $D$ and $B$ mesons are taken from \cite{Cacciari:2005uk}, and the fragmentation functions for $\pi$ mesons are taken from \cite{deFlorian:2007aj}. Note that the fragmentation functions for $\pi$ mesons were extrapolated outside their domain in $Q^2$, as we found that the extrapolation was smooth. %

	Additionally for fragmentation, all theoretical curves produced for charged hadrons and neutral pions are treated as neutral pions.

\section{Geometry}%
\label{sec:mod_geometry}

For large systems, using the Glauber model for the collision geometry is standard, with Wood-Saxon distributions for the nucleon density inside the heavy ions \cite{Glauber:1970jm,Miller:2007ri}. For small \collFour{p}{d}{He3}{A} collisions, the Glauber model cannot be applied in its most simple form since one expects subnucleonic features of the proton to be important \cite{Schenke:2020mbo}. Additionally, one may treat the subsequent evolution of the medium in a more sophisticated way than simply assuming a Bjorken expansion of the initial Glauber model geometry as was done, e.g., in \cite{Wicks:2005gt}.
 In this work, we will use collision profiles generated with \cite{Schenke:2020mbo} and sourced from \cite{shen_private_communication}.
In these calculations, initial conditions are given by the IP-Glasma model \cite{Schenke:2012hg, Schenke:2012wb}, which are then evolved with the \texttt{MUSIC} \cite{Schenke:2010rr,Schenke:2011bn, Schenke:2010nt} viscous relativistic (2+1)D hydrodynamics code, followed by UrQMD microscopic hadronic transport \cite{Bass:1998ca, Bleicher:1999xi}.  

Consistent with our previous work \cite{Faraday:2023mmx} we use only the initial temperature profile $T(\tau = \tau_0 = 0.4~\mathrm{fm})$ for our collision geometry, where $\tau_0$ is the turn-on time for hydrodynamics. The collision geometry used for all results presented in this work is, therefore, effectively IP-Glasma initial conditions \cite{Schenke:2012hg, Schenke:2012wb} coupled with Bjorken expansion time dependence, unless otherwise stated. 

From HTL perturbation theory \cite{Blaizot:2001nr} one may derive the leading order expression for the Debye mass 
\begin{equation}
	\mu  = g T \sqrt{\frac{2 N_c  + n_f}{6}} \text{ where } g = \sqrt{4 \pi \alpha_s}.
	\label{eqn:mod_debye_mass}
\end{equation}

The QGP is treated as an ultrarelativistic mixture of a Fermi and Bose gas with zero chemical potential, following \cite{Wicks:2005gt, Horowitz:2010dm, Faraday:2023mmx, Wicks:2008zz}. Then
\begin{subequations}
    \label{eqn:mod_thermodynamic_quantities}
		\begin{alignat}{2}
			\sigma_{gg}  =& \frac{C_A^2 \pi \alpha_s^2}{2 \mu^2} \quad \text{and} \quad \sigma_{qg}  = \frac{C_F}{C_A} \sigma_{gg}&\\
			\rho_g =& 2 (N_c^2 - 1) \frac{\zeta(3)}{\pi^2} T^3 \label{eqn:mod_rho_thermal_g}&\\
			\rho_q =& 3 N_c n_f \frac{\zeta(3)}{\pi^2} T^3 \label{eqn:mod_rho_thermal_q}&\\
			\rho =& \rho_g + \frac{\sigma_{q g}}{\sigma_{gg}} \rho_{q}& \nonumber\\
			=&\frac{\zeta(3) (N_c^2 - 1)}{\pi^2} T^3 \left( 2 + \frac{3n_f}{2N_c}\right)&\\
			\lambda_g^{-1}  =& \rho_g \sigma_{g g}+\rho_q \sigma_{q g} = \sigma_{gg} \rho,& \label{eqn:mod_mean_free_path}
    \end{alignat}
\end{subequations}
where $\zeta$ is the Riemann zeta function, $T$ is the temperature, $\rho_q$ ($\rho_g$) is the density of quarks (gluons), $\sigma_{qg}$ ($\sigma_{gg}$) is the gluon-gluon (quark-gluon) elastic cross section, $n_f$ is the number of active quark flavors (taken to be $n_f = 2$ throughout), and $N_c=3$ is the number of colors. We denote the cross section weighted density as $\rho$, which we will subsequently refer to as the density for simplicity. Due to color triviality, one need only calculate these results for an incident gluon, and the change for an incident quark is simply a change of Casimir in the relevant energy loss kernels in \cref{eqn:mod_DGLV_dndx,eqn:mod_full_dndx,eqn:mod_elastic_energy_loss_low,eqn:mod_elastic_energy_loss_high,eqn:mod_dndx_pure_htl}.

The radiative (\cref{eqn:mod_DGLV_dndx,eqn:mod_full_dndx}) and elastic (\cref{eqn:mod_elastic_energy_loss_low,eqn:mod_elastic_energy_loss_high,eqn:mod_dndx_pure_htl}) energy loss results were derived using a ``brick" model, which represents a medium with a fixed length $L$ and constant temperature $T$. In order to capture fluctuations in temperature and density, we need a mapping from the path that a parton takes through the plasma, to a brick with an effective length $L_{\text{eff}}$ and effective temperature $T_{\text{eff}}$.

We follow WHDG \cite{Wicks:2005gt} and define the effective pathlength as
\begin{equation}
	L_{\text{eff}} (\mathbf{x}_i, \boldsymbol{\hat{\phi}}) = \frac{1}{\rho_{\text{eff}}} \int_{0}^\infty \mathrm{d}z \; \rho(\mathbf{x}_i + z \boldsymbol{\hat{\phi}}, \tau_0),
    \label{eqn:mod_effective_length}
\end{equation}
and the effective density as
\begin{equation}
  \rho_{\text{eff}} \equiv \frac{\int \mathrm{d}^2 \mathbf{x} \; \rho^2(\mathbf{x}, \tau_0)}{\int \mathrm{d}^2 \mathbf{x} \; \rho(\mathbf{x}, \tau_0)} \iff T_{\text{eff}}^{3} \equiv \frac{\int \mathrm{d}^2 \mathbf{x} \; T^6(\mathbf{x}, \tau_0)}{\int \mathrm{d}^2 \mathbf{x} \; T^3(\mathbf{x}, \tau_0)}.
  \label{eqn:mod_effective_density}
\end{equation}
Here, the effective pathlength $L_{\text{eff}}$ includes all $(\mathbf{x}_i, \phi)$ dependence, and $\rho_{\text{eff}}$ is a constant for all paths that a parton takes through the plasma for a fixed centrality class. In principle, one can allow both the effective density and effective pathlength to depend on the specific path taken through the plasma. However, such a numerically intensive model is beyond the scope and objective of this work.

In WHDG \cite{Wicks:2005gt}, the prescription $\rho \equiv \rho_{\text{part}}$ was made, where $\rho_{\text{part}}$ is the participant density---the density of nucleons which participate in at least one binary collision. This prescription is not necessary in our case, since we have access to the temperature profile \cite{Schenke:2020mbo, shen_private_communication}. We extract the temperature profile $T(\mathbf{x}, \tau)$ from the hydrodynamics output \cite{Schenke:2020mbo,shen_private_communication}, and for $L_{\mathrm{eff}}$ one evaluates the temperature at the initial time set by the hydrodynamics simulation, $\tau_0=0.4~\mathrm{fm}$. 
There is no unique mapping from realistic collision geometries to simple brick geometries and more options are explored in \cite{Wicks:2008zz} and our previous work \cite{Faraday:2023mmx}. Future work will perform a careful analysis of the geometrical mapping procedure between realistic media and brick geometries wherein theoretical calculations take place \cite{Bert:2024}.

Bjorken expansion \cite{Bjorken:1982qr} is then taken into account by approximating
\begin{equation}
T_{\text{eff}}(\tau) \approx T_{\text{eff}}(\tau_0) \left( \frac{\tau_0}{\tau} \right)^{1/3} \approx T_{\text{eff}}(\tau_0) \left( \frac{2 \tau_0}{L_{\text{eff}}} \right)^{1/3}
  \label{eqn:mod_bjorken_expansion}
\end{equation}
where in the last step we have evaluated $T( \mathbf{x}, \tau)$ at the average time $\tau=L / 2$, following what was done in \cite{Wicks:2005gt, Djordjevic:2005db, Djordjevic:2004nq}. In \cite{Wicks:2005gt}, this average time approximation was found to be a good approximation to the full integration through the Bjorken expanding medium. For a given collision system, we can then calculate the distribution of effective pathlengths that a hard part will travel in the plasma. We assume, as is standard and consistent with WHDG \cite{Wicks:2005gt} and our previous work \cite{Faraday:2023mmx}, that the hard partons have starting positions weighted by the density of binary nucleon-nucleon collisions, provided by IP-Glasma \cite{shen_private_communication}. 

\begin{figure}[!htbp]
\centering
\begin{subfigure}[t]{0.49\textwidth}
    \centering
    \includegraphics[width=\linewidth]{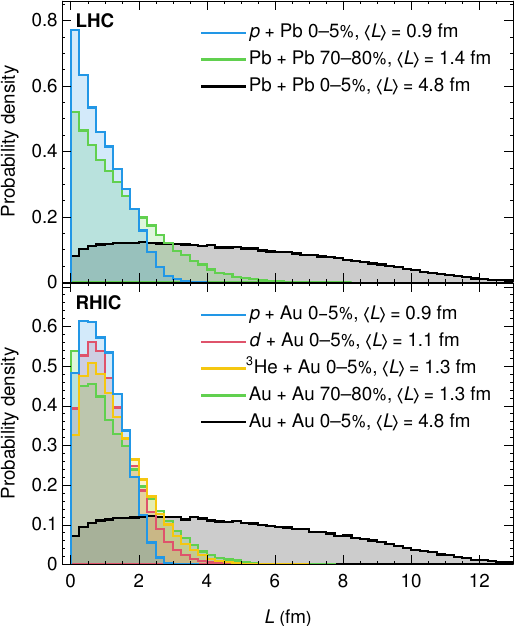}
    \caption{(Top) Distribution of the effective pathlengths in $0\text{--}5\%$ central \coll{p}{Pb}, $0\text{--}5\%$ central \coll{Pb}{Pb}, and $70\text{--}80\%$ peripheral \coll{Pb}{Pb} collision systems at $\sqrt{s_{NN}} = 5.02$ TeV. (Bottom) Distribution of the effective pathlengths in $0\text{--}5\%$ central \coll{p}{Au}, \coll{d}{Au}, \coll{He3}{Au}, and \coll{Au}{Au} as well as $70\text{--}80\%$ peripheral \coll{Au}{Au} collisions at $\sqrt{s_{NN}} = 200$ GeV. The average pathlength $\langle L \rangle$ is shown in the legend for each collision.}
    \label{fig:mod_path_length_distribution}
\end{subfigure}\hfill
\begin{subfigure}[t]{0.49\textwidth}
    \centering
    \includegraphics[width=\linewidth]{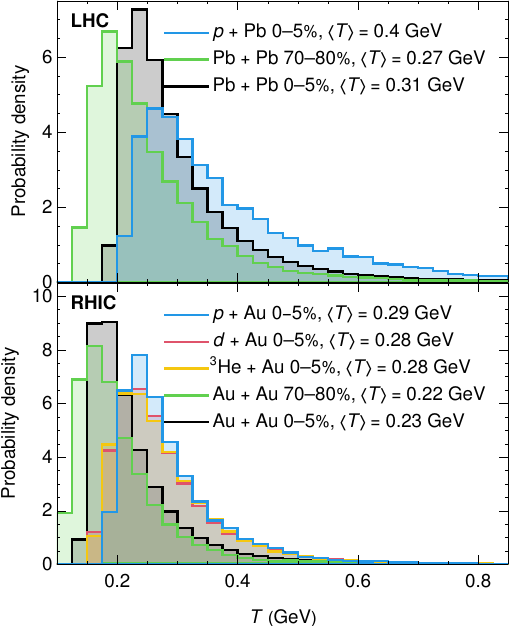}
    \caption{(Top) Distribution of the effective temperatures $T$ in $0\text{--}5\%$ central \coll{p}{Pb}, $0\text{--}5\%$ central \coll{Pb}{Pb}, and $70\text{--}80\%$ peripheral \coll{Pb}{Pb} collision systems at $\sqrt{s_{NN}} = 5.02$ TeV. (Bottom) Distribution of the effective temperatures $T$ in $0\text{--}5\%$ central \coll{p}{Au}, \coll{d}{Au}, \coll{He3}{Au}, and \coll{Au}{Au} as well as $70\text{--}80\%$ peripheral \coll{Au}{Au} collisions at $\sqrt{s_{NN}} = 200$ GeV. The average temperature $\langle T \rangle$ is shown in the legend for each collision.}
    \label{fig:mod_temperature_distribution}
\end{subfigure}
\caption{Distribution of effective pathlengths (left) and effective temperatures (right) for various collision systems at different energies.}
\label{fig:mod_combined}
\end{figure}

\Cref{fig:mod_path_length_distribution} shows the probability distribution of effective pathlengths for $0\text{--}5\%$ most central \coll{p}{Pb} and \coll{Pb}{Pb} collisions, as well as for $70\text{--}80\%$ peripheral \coll{Pb}{Pb} collisions in the top panel. The bottom panel shows the probability distribution for $0\text{--}5\%$ most central \coll{p}{Au}, \coll{d}{Au}, \coll{He3}{Au}, and \coll{Au}{Au} collisions, as well as for $70\text{--}80\%$ peripheral \coll{Au}{Au} collisions. 
In the legend, we indicate the average pathlength in each system. 
Note especially that the effective pathlength distribution for $0\text{--}5\%$ central \coll{p}{Pb} and \collFour{p}{d}{He3}{Au} are nearly indistinguishable from peripheral $70\text{--}80\%$ \coll{Pb}{Pb} and \coll{Au}{Au}. \Cref{fig:mod_temperature_distribution} plots the distribution of temperatures for the same set of collision systems as shown in \cref{fig:mod_path_length_distribution}, where the distributional shape of the figure is simply a mapping of the length distribution in \cref{fig:mod_path_length_distribution} according to \cref{eqn:mod_bjorken_expansion}.

The average pathlength in $0\text{--}5\%$ most central \collFour{p}{d}{He3}{A} collisions is $L \sim 1~\mathrm{fm}$, with an average temperature of $T \approx 0.3\text{--}0.4~\mathrm{GeV}$, corresponding to a mean free path $\lambda_g \approx 0.7\text{--}1~\mathrm{fm}$ and an inverse Debye mass $\mu^{-1} \approx 0.2\text{--}0.3~\text{fm}$.
We see from this simple analysis that in small \collFour{p}{d}{He3}{A} collision systems $L/\lambda_g\sim1$ for most of the distribution of effective pathlengths, which implies that approaches that assume many soft scatterings are inapplicable. Similarly, the small amount of expected scatters does not provide a good motivation for modeling the elastic energy loss as a Gaussian distribution according to the central limit theorem \cite{Moore:2004tg}.
Moreover, in small \collFour{p}{d}{He3}{A} collisions, the large pathlength assumption $\lambda \ll L \iff 0.2\text{--}0.3 ~\mathrm{fm} \ll 1 ~\mathrm{fm}$ is not particularly well founded, as discussed in our previous work \cite{Faraday:2023mmx}.

 The average pathlength of $\langle L \rangle \sim1$~fm in small systems, while substantially smaller than the $\langle L \rangle \sim5$~fm in large systems, is similar to the $\langle L \rangle \sim 1.4$~fm in $70\text{--}80\%$ peripheral \coll{Pb}{Pb} and \coll{Au}{Au} collisions. The average temperatures of these small collision systems are however notably larger: while $0\text{--}5\%$ \collFour{p}{d}{He3}{A} collisions and $70\text{--}80\%$ \coll{A}{A} collisions have comparable average lengths, the temperature in central \collFour{p}{d}{He3}{A} collisions is $\mathcal{O}(30\text{--}50\%)$ larger than that of peripheral \coll{A}{A} collisions. 

\Cref{fig:mod_temperature_distribution} shows how the small average pathlengths in small \collFour{p}{d}{He3}{A} collisions result in a temperature distribution with a larger average temperature since the parton spends a significantly larger fraction of its lifetime in the high-temperature region of the medium.

The temperature distributions between the various collision systems are more similar than the corresponding length distributions. This similarity is due to the $T \sim L^{- 1 /3}$ dependency from \cref{eqn:mod_bjorken_expansion}, which results in the temperature distributions having significantly less variation than the corresponding length distributions.

\comm{Incorporate this properly}

\begin{figure}[H]
    \centering
    \includegraphics[width=0.6\linewidth]{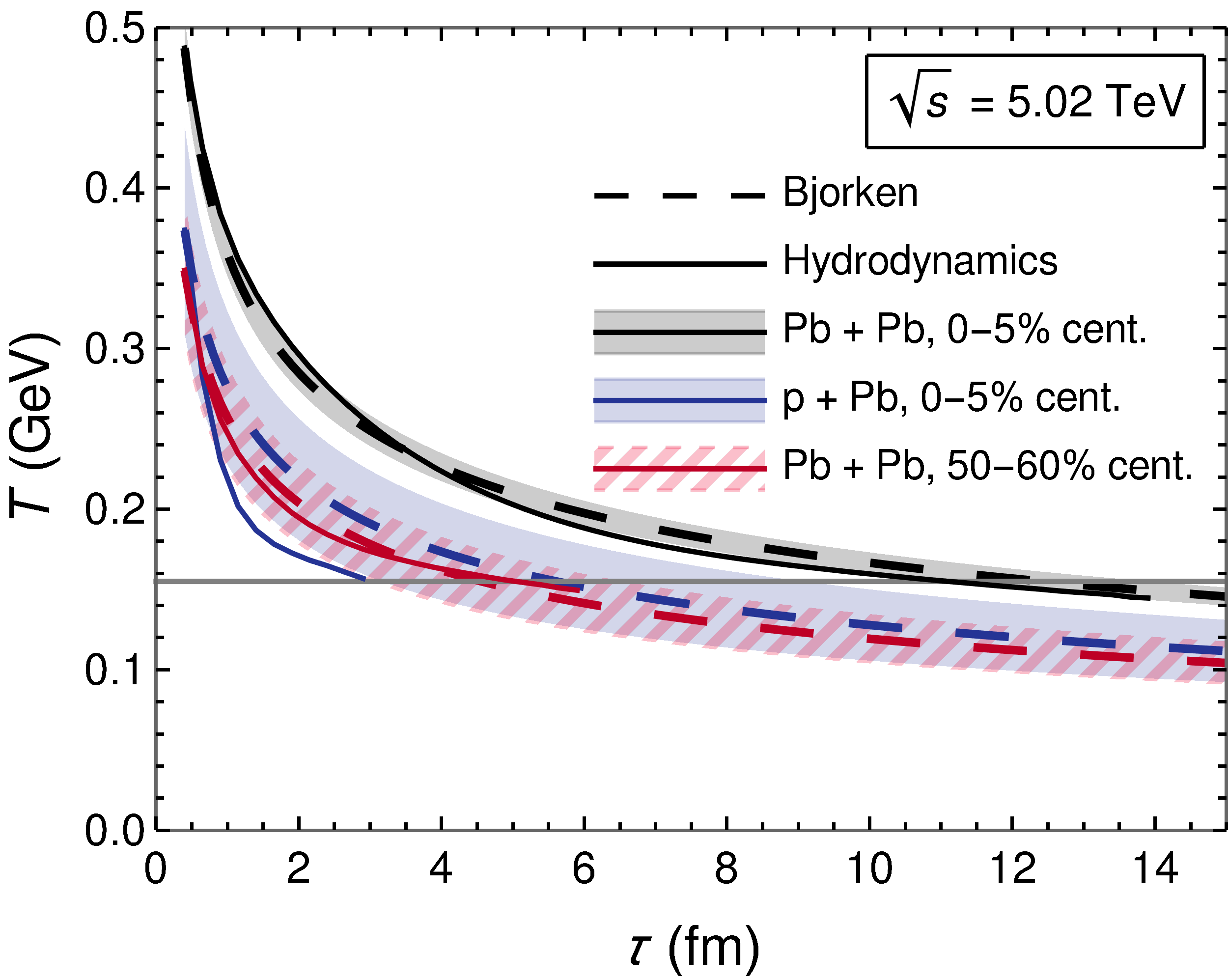}
    \caption{Plot of the temperature $T$ as a function of the proper time $\tau$ (in the plasma rest frame). The Bjorken expansion approximation (Bjorken) \cite{Bjorken:1982qr} to the $\tau$ dependence of temperature, $T(\tau) \approx T_{\text{eff}}(\tau_0) [\tau_0 / \tau]^{1 / 3}$, is plotted along with the effective temperature $T_{\text{eff}}(\tau)$ calculated as a function of time via hydrodynamics (Hydrodynamics) according to \cref{eqn:mod_effective_density}. Uncertainty bands represent a $2 \sigma$ (95\% CI), and are shown only for the Bjorken estimate only as the uncertainty on the Hydrodynamics result is negligible.
    Both curves are plotted for the collision systems $\mathrm{Pb}+\mathrm{Pb}$ at 0--5\%, and 50--60\% centrality; as well as $\mathrm{p} + \mathrm{Pb}$ at 0--5\% centrality. The freeze-out temperature $T_{\text{f.o.}} = 0.155~\mathrm{GeV}$ is shown as a horizontal gray line, which is the temperature at which it is assumed that hadronic degrees of freedom take over in the plasma. Hydrodynamic temperature profiles are taken from \cite{Schenke:2020mbo}. }
    \label{fig:mod_temperature_vs_t}
\end{figure}

\Cref{fig:mod_temperature_vs_t} compares the temperature of the plasma as a function of proper time in the rest frame of the plasma calculated via hydrodynamics (solid lines) versus the temperature from the Bjorken expansion formula (dashed lines). The effective temperature using hydrodynamics is calculated using \cref{eqn:mod_effective_density} and the Bjorken expansion approximation to the time dependence of the effective temperature is given by \cref{eqn:mod_bjorken_expansion}.
  Calculations are performed for the same three collision systems as in \cref{fig:mod_path_length_distribution}. Due to the fluctuations of the initial conditions of the plasma in our model---because of both nucleonic and subnucleonic fluctuations \cite{Schenke:2020mbo}---we obtain a distribution of effective temperatures at each point in proper time. We show the mean and the $2\sigma$ width of the Bjorken temperature estimates in \cref{fig:mod_temperature_vs_t} .  The width in the Bjorken result arises solely from the variation of the initial hydrodynamics temperature profile at $\tau = \tau_0$.  We computed the mean and standard deviation of the temperature distribution as a function of $\tau$ for the full hydrodynamics simulation; however the widths of these distributions are not plotted in \cref{fig:mod_temperature_vs_t} as for $\tau>\tau_0$ they are negligible compared to the width in the Bjorken expansion results. 

\subsection{Energy loss outside the QGP}
\label{sec:mod_prethermalization_time_energy_loss}

As an incident high-$p_T$ parton propagates through the QGP formed during a collision, it first traverses through a non-thermalized medium, then a thermalized QGP, and finally a hadronic medium of some sort. The first and last of these three phases are not rigorously modeled by pQCD energy loss, even though they must contribute to the energy loss, which means we must make a phenomenological choice on how to perform energy loss during this period of the collision.

There are a few common phenomenological choices on how to treat energy loss before the medium has thermalized, including \cite{Xu:2014ica}

\begin{enumerate}
	\item \textit{Free streaming}. The temperature distribution takes the form $T(\vecII{x}, \tau) \equiv \Theta(\tau - \tau_0) T(\vecII{x}) \left(\frac{\tau_0}{\tau}\right)^{1 / 3}$. This prescription is equivalent to assuming no energy loss occurs for the first $\tau_0$ of propagation distance through the plasma.
	\sloppy\item \textit{Linear thermalization}. The temperature distribution takes the form $T(\vecII{x}, \tau) \equiv T(\vecII{x})\left[ \Theta(\tau - \tau_0) \left(\frac{\tau_0}{\tau}\right)^{1 / 3}  + \Theta(\tau_0 - \tau) \left(\frac{\tau}{\tau_0}\right)^{1 / 3}\right] $.  This prescription assumes that the temperature starts at $0$ GeV and grows to the temperature at the thermalization time, after which the temperature decays according to Bjorken expansion.
	\item \textit{Instant thermalization}. The temperature distribution takes the form $T(\vecII{x}, \tau) \equiv T(\vecII{x}) \left(\frac{\tau_0}{\tau}\right)^{1 / 3}$. The pre-thermalization time behaviour is extrapolated from the post-thermalization time behavior of Bjorken expansion. 
\end{enumerate}

The prescriptions outlined above have implications for both the effective temperature in \cref{eqn:mod_bjorken_expansion} and the domain of the integrals in \cref{eqn:mod_effective_length,eqn:mod_effective_density}. In this work, all calculations presented are performed according to the instant thermalization prescription for simplicity, meaning that the integral in \cref{eqn:mod_effective_length} is evaluated from $\tau = 0$ to $\tau = \infty$. The instant thermalization prescription is the most theoretically consistent prescription for the exponential distribution of scattering centers, which is used for the radiative energy loss \cref{eqn:mod_density_scattering_centers}. It was found \cite{Xu:2014ica} that while $R_{AB}$ results were sensitive to the choice of pre-thermalization time energy loss, this sensitivity could largely be absorbed by a change in the strong coupling. Small systems likely have a greatly increased sensitivity to the choice of pre-thermalization time energy loss, and therefore the system size dependence of the $R_{AB}$ may not be able to be absorbed into changes in the strong coupling. Future work should examine the different pre-thermalization time energy loss scenarios, particularly in relation to the dependence on system size.

For energy loss at temperatures below the freezeout temperature of $T_{\text{f.o.}} \simeq 0.155$ GeV \cite{Borsanyi:2011sw}, one may choose to turn off energy loss, extrapolate the pQCD energy loss, or perform fragmentation and switch to a hadronic energy loss model. High-$p_T$ particles can interact with matter in the hadronic phase, and so a good compromise between simplicity and realism seems to be to extrapolate the pQCD energy loss calculation into the hadronic phase. We note that although the hydrodynamics temperature profiles have a turn-off at $T_{\text{f.o.}} = 0.155$ GeV \cite{Schenke:2020mbo}, the Bjorken expansion formula for the temperature has no such turn-off. Additionally, since all of our temperatures and lengths are calculated from the initial distribution, we are, to a good approximation, ignoring the freezeout temperature and extrapolating our energy loss into the hadronic phase.

\subsection{Geometry and event averaged nuclear modification factor}
\label{sec:mod_geometry_averaged_nuclear_modification_factor}

To incorporate our model for the collision geometry, we expand \cref{eqn:mod_full_raa_spectrum_ratio} to average over the effective pathlengths according to the length distribution described in \cref{sec:mod_geometry}. 

Additionally, one must perform an average over the various collision events due to statistical fluctuations in the IP-Glasma initial conditions \cite{Schenke:2020mbo}. Then

\begin{align}
  R&_{A B}^q\left(p_T\right) \nonumber\\
  & =  \langle R_{A B}^{q}(p_T, L, T) \rangle_{\text{geometry, events}}	\label{eqn:mod_geometry_averaged_raa}\\
	& = \left\langle \int \mathrm{d} L\; P_L(L) \; R_{AB}^q\left(p_T, L, T_{\text{eff.}} (L / 2)\right) \right\rangle_{\text{events}}\nonumber\\
	&\approx \int \mathrm{d} L\; \left\langle P_L \right\rangle_{\text{events}}(L) \; R_{AB}^q\left(p_T, L, \left\langle  T_{\text{eff.}}\right\rangle_{\text{events}} (L / 2)\right),\nonumber
\end{align}
where $P_L(L_{\mathrm{eff}})$ is the normalized distribution of effective pathlengths weighted by the binary collision density (\cref{fig:mod_path_length_distribution}), and $T(L_{\text{eff}} / 2)$ is the effective temperature (\cref{fig:mod_temperature_distribution}) determined according to \cref{eqn:mod_bjorken_expansion}. 

In the last line of the above, we have approximated the event average of the $R_{AB}$ by an event average of the distribution of effective lengths $\left\langle P_L \right\rangle_{\text{events}}$ and effective temperatures $\left\langle T_{\text{eff}} \right\rangle_{\text{events}}$.
This procedure is strictly correct in its application to $\left\langle P_L \right\rangle_{\text{events}}$, because the sum over the events and the integral over $L$ are linear operators and therefore commute. The approximation occurs in the fact that we average over the $T_{\text{eff.}}$ functions, which appear inside the non-linear $R_{AB}$ function. We make this approximation for computational simplicity, and because including the variation at the level of the initial temperatures would lead to an overestimate in the variation of the temperatures at later times \cite{Faraday:2023mmx}. This overestimation occurs because the Bjorken expansion \cite{Bjorken:1982qr} does not smooth out the initial temperature distribution, while a more realistic hydrodynamic evolution \cite{Schenke:2020mbo} would \cite{Faraday:2023mmx}.

 \chapter{Inconsistencies in, and short pathlength correction to, \texorpdfstring{$R_{AA}(p_T)$}{RAA( pT)} in \texorpdfstring{$A + A$}{A + A} and \texorpdfstring{$p + A$}{p + A} collisions}
 \label{sec:SPL_paper}
 \section*{Comment}

Chapter 4 of this thesis is an almost verbatim reproduction of the paper which was published in The European Physics Journal C \cite{Faraday:2023mmx}. The only difference between the paper and this chapter, is that the model section of the paper was combined with the model section of a more recent preprint manuscript \cite{Faraday:2024gzx} and was presented in \cref{sec:model}.

This paper was written in collaboration with Antonia Grindrod, and Dr.\ W.\ A.\ Horowitz, and I have permission from my coauthors to reproduce this work in this thesis. I was the first author of the paper, and wrote the first draft of the paper. AG and I both independently worked on the numerical implementation of the calculation of the $R_{AA}$. I implemented the utilization of the relativistic, viscous hydrodynamics background with backgrounds provided by Chun Shen. Conversations with Isobel Kolbé led to the suggestion that various approximations in the energy loss formalism may be breaking down. It was my idea to calculate the dimensionless ratios $\langle R \rangle$ from \cref{sec:SPL_assumptions} in order to interrogate the self-consistency of the various approximations. The cutoff on the transverse radiated gluon momentum in order to self-consistently satisfy the large formation time approximation was my idea (see \cref{sec:SPL_conclusions}).

All numerical results shown in the various plots in this chapter were produced by me, and the numerical calculations were performed by me with my own code. The codes for fragmentation and production spectra was provided to me by WAH. Conversations with WAH and AG were invaluable for the ideas in the manuscript and the numerical implementation of these ideas. WAH conceived of the project.
\newpage

\section{Introduction}

\comm{Make the introduction simpler?}

The modification of the spectrum of high transverse momentum (high-$p_T$) particles is one of the key observables used to understand the non-trivial, emergent, many-body dynamics of quantum chromodynamics (QCD) in high-energy collisions \cite{Gyulassy:2004zy,Wiedemann:2009sh,Majumder:2010qh,Busza:2018rrf}.  One of the most important findings of the Relativistic Heavy Ion Collider (RHIC) was a roughly factor of five suppression of leading light hadrons with $p_T\gtrsim 5$ GeV/c in central $\mathrm{Au}+\mathrm{Au}$ collisions \cite{PHENIX:2001hpc,STAR:2003pjh}.  This suppression, equal for pions and eta mesons \cite{PHENIX:2006ujp}, along with null controls of qualitatively no suppression of the weakly-coupled photons in $\mathrm{Au}+\mathrm{Au}$ collisions \cite{PHENIX:2005yls} as well as of leading hadrons in $\mathrm{d}+\mathrm{Au}$ collisions \cite{STAR:2003pjh,PHENIX:2006mhb}, clearly demonstrated that the suppression of leading hadrons in central collisions is due to final state energy loss of the high-$p_T$ partons interacting with the quark-gluon plasma (QGP) generated in the heavy ion collisions (HIC).  Models of leading hadron suppression based on final state energy loss derivations using perturbative QCD (pQCD) methods qualitatively describe a wealth of these high-$p_T$ RHIC data \cite{Dainese:2004te,Schenke:2009gb,Horowitz:2012cf}.

\sloppy One of the other major findings of RHIC was the near perfect fluidity of the strongly-coupled low momentum modes of the QGP formed in semi-central nucleus-nucleus collisions as inferred by the remarkable agreement of sophisticated, relativistic, viscous hydrodynamics models with the spectra of measured hadrons with $p_T\lesssim2$ GeV \cite{Romatschke:2007mq,Song:2007ux,Schenke:2010rr}.

The data from the Large Hadron Collider (LHC) has been no less impressive.  Of extraordinary importance have been the signs that the non-trivial, emergent, many-body QCD behavior associated with QGP formation in central and semi-central heavy ion collisions at RHIC and LHC are \emph{also} observed in small collision systems such as $\mathrm{p}+\mathrm{p}$ and $\mathrm{p}+\mathrm{A}$ for large final measured multiplicity.  For example, strangeness enhancement \cite{ALICE:2013wgn,ALICE:2015mpp} and quarkonium suppression \cite{ALICE:2016sdt} appear to depend only on multiplicity but not collision system.  And the same sophisticated, relativistic, viscous hydrodynamics models \cite{Weller:2017tsr} also describe the spectra of measured hadrons in high-multiplicity $\mathrm{p}+\mathrm{p}$ and $\mathrm{p}+\mathrm{A}$ collisions \cite{CMS:2015yux,ATLAS:2015hzw}.  

One may thus conclude that small droplets of QGP form even in these smallest of collision systems at the LHC.  If QGP is formed in high-multiplicity collisions of small systems, then high-$p_T$ partons should suffer some final state energy loss, as has been observed in large collision systems at RHIC and LHC.  (Models already demonstrate the importance of final state energy loss in forward hadron production in cold nuclear matter \cite{Arleo:2020eia,Arleo:2021bpv}.)  Experimentally, there are tantalizing signs of the non-trivial modification of high-$p_T$ particles in small collision systems \cite{ATLAS:2014cpa,PHENIX:2015fgy,ALICE:2017svf}.  However, there are likely non-trivial correlations between the multiplicity in small collision systems and the presence of high-$p_T$, high-multiplicity jets.  For example, these correlations likely impact the initial spectrum of high-$p_T$ partons \cite{ALICE:2018pal} that enters the numerator of the nuclear modification factor, while the minimum bias spectrum in the denominator is unchanged; thus one should be cautious in interpreting a standard $R_{AB}$ measurement.  

\sloppy From the experimental side, it is likely very interesting to consider the ratio of the spectrum of a strongly-interacting particle with a weakly-interacting particle, each from the same multiplicity class.

From the theoretical side, we may potentially make progress by considering the small collision system predictions of the energy loss models used to describe so well qualitatively the large collision system high-$p_T$ particle suppression.  One obvious challenge for directly comparing these energy loss models to small collision system data is the assumption made in the energy loss derivations that the high-$p_T$ particles travel a large pathlength in the QGP medium. For example, energy loss models built on BDMPS-Z-type energy loss \cite{Baier:1996kr, Baier:1996sk, Baier:1996vi, Baier:1998kq,Zakharov:1996fv,Zakharov:1997uu} utilize the central limit theorem \cite{Armesto:2011ht}, and so assume a very large number of collisions occur between the high-$p_T$ probe and the QGP medium.  Even in large collision systems of characteristic size $\sim5$ fm, since the mean free path for these high-$p_T$ particles given by weakly-coupled pQCD is $\sim1\text{--}2$ fm \cite{Armesto:2011ht}, the application of the central limit theorem is dubious.  

\sloppy Even for the thin plasma approach of DGLV that naively seems the best suited for modelling the radiative energy loss processes in systems of phenomenologically relevant size, there is an explicit assumption that the partonic pathlength in the QGP medium, $L$, is large compared to the natural scale set by the Debye screening mass $\mu$, $L\gg1/\mu$.  In the original derivation of the induced gluon radiation spectrum, contributions from the Gyulassy-Wang potential \cite{Gyulassy:1993hr} were dropped as they are $\mathcal O(e^{-L\mu})$.  For $\mu\sim gT\sim 0.5$ GeV \cite{Kapusta:2006pm}, the characteristic size $L\sim 1$ fm in high-multiplicity $\mathrm{p}+\mathrm{p}$ and $\mathrm{p}+\mathrm{A}$ collisions is not particularly large compared to $1/\mu\sim0.4$ fm.  Thus to create an energy loss model to compare to data in these small systems, one needed the small pathlength corrections to the DGLV opacity expansion.

This small pathlength correction to the first order in opacity DGLV radiative energy loss were derived for the first time in \cite{Kolbe:2015rvk}.  
For later noting, the derivation in \cite{Kolbe:2015rvk} benefited significantly from a simplification due to the large formation time assumption, $\tau_{\text{form}}\gg1/\mu$, an assumption made also in the original DGLV derivations.

The small pathlength correction to the usual DGLV energy loss contained four surprises: due to the LPM effect (interference between the induced radiation and the usual vacuum emissions due to the hard initial scattering), the small pathlength correction \emph{reduces} the energy lost; the reduction in energy loss is seen in \emph{all} pathlengths (although the relative importance of the correction decreases with pathlength, as expected); the correction \emph{grows linearly} with partonic energy (as opposed to the logarithmic growth of the usual DGLV energy loss); and the correction breaks color triviality, with the correction for gluons $\sim10$ times the size of the correction for quarks.

Having derived the correction to the radiative energy loss due to small pathlengths, it is of considerable interest to determine quantitatively the importance of the correction in phenomenological models of high-$p_T$ particle suppression.  It is the goal of this manuscript to provide just such predictions.  We are particularly interested in seeing the quantitative importance of the reduction in energy loss as a function of energy and of collision system: the reduction in energy loss from the short pathlength correction might provide a natural explanation for the surprisingly fast rise in the nuclear modification factor with $p_T$ for leading hadrons at LHC \cite{Horowitz:2011gd} and for the enhancement of the nuclear modification factor above unity seen in $\mathrm{p}+\mathrm{A}$ collisions \cite{Balek:2017man,ALICE:2018lyv}.

What we will see is that for light flavor final states at larger energies, $p_T\gtrsim30$ GeV/c, the small system ``correction'' becomes of order of the energy loss itself, which leads us to consider systematically the extent to which energy loss model energy losses are consistent with the various approximations used in the opacity expansion derivation.  We will consider in detail the approximations used in the opacity expansion derivation of radiative energy loss \cite{Vitev:2002pf,Djordjevic:2003zk,Kolbe:2015rvk}.  We will find that for the radiated gluons that dominate the energy weighted single inclusive emission distribution, the large formation time assumption is violated for $E\gtrsim30$ GeV/c in large collision systems and for $E\gtrsim10$ GeV/c in small systems, where $E$ is the energy of the radiating parton; which implies the need for yet another derivation of radiative energy loss in the thin plasma limit but with the large formation time assumption relaxed. We also see that the usual WHDG treatment of the average elastic energy loss, with fluctuations given by the fluctuation-dissipation theorem, is not appropriate for small collision systems where the number of scatterings is not large; thus in future work one must implement an elastic energy loss appropriate for small and large collision systems.

\section{Initial results}
\label{sec:SPL_results}

As a first exploration of the effect of the short pathlength correction to the DGLV radiative energy loss model, we consider $\mathrm{Pb}+\mathrm{Pb}$ and $\mathrm{p} + \mathrm{Pb}$ collision systems. We emphasize that in this work we are not trying to create the best possible fit to data (which could be done by tuning various parameters), but are instead focused on the impact of the short pathlength correction. For this reason all numerical values are used in consistency with the original WHDG predictions \cite{Wicks:2005gt,Horowitz:2011gd,Andronic:2015wma}, and agreement with data is not the primary focus of this work.

\subsection{Suppression of heavy flavor mesons}
\label{sub:SPL_results_heavy}

\begin{figure}[H]
    \centering
    \begin{subfigure}[t]{0.49\textwidth}
        \centering
        \includegraphics[width=\linewidth]{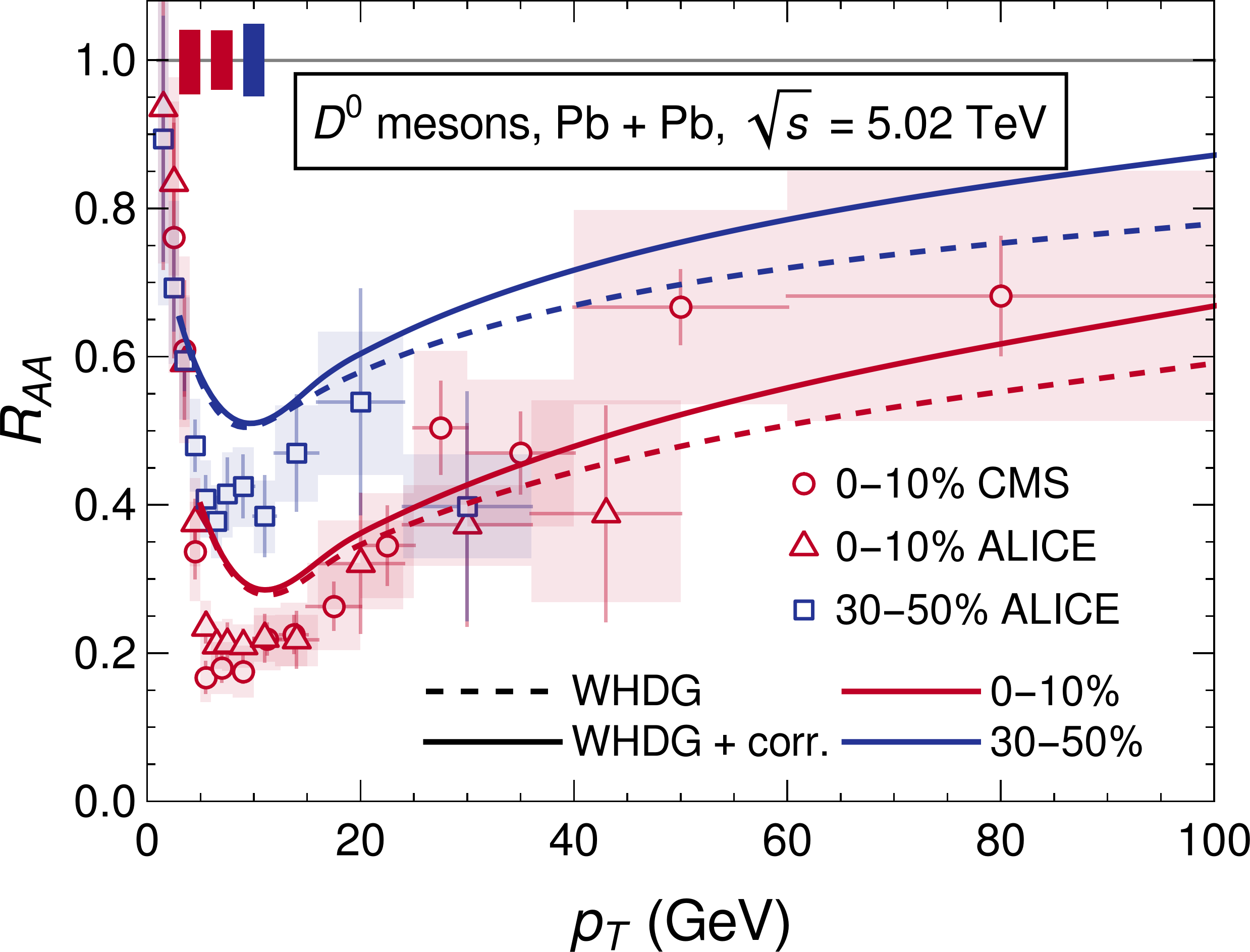}
        \caption{$R_{AA}$ for $D^0$ mesons in 0--10\% and 30--50\% central $\mathrm{Pb} + \mathrm{Pb}$ collisions.}
        \label{fig:SPL_D_raa_AA}
    \end{subfigure}\hfill
    \begin{subfigure}[t]{0.49\textwidth}
        \centering
        \includegraphics[width=\linewidth]{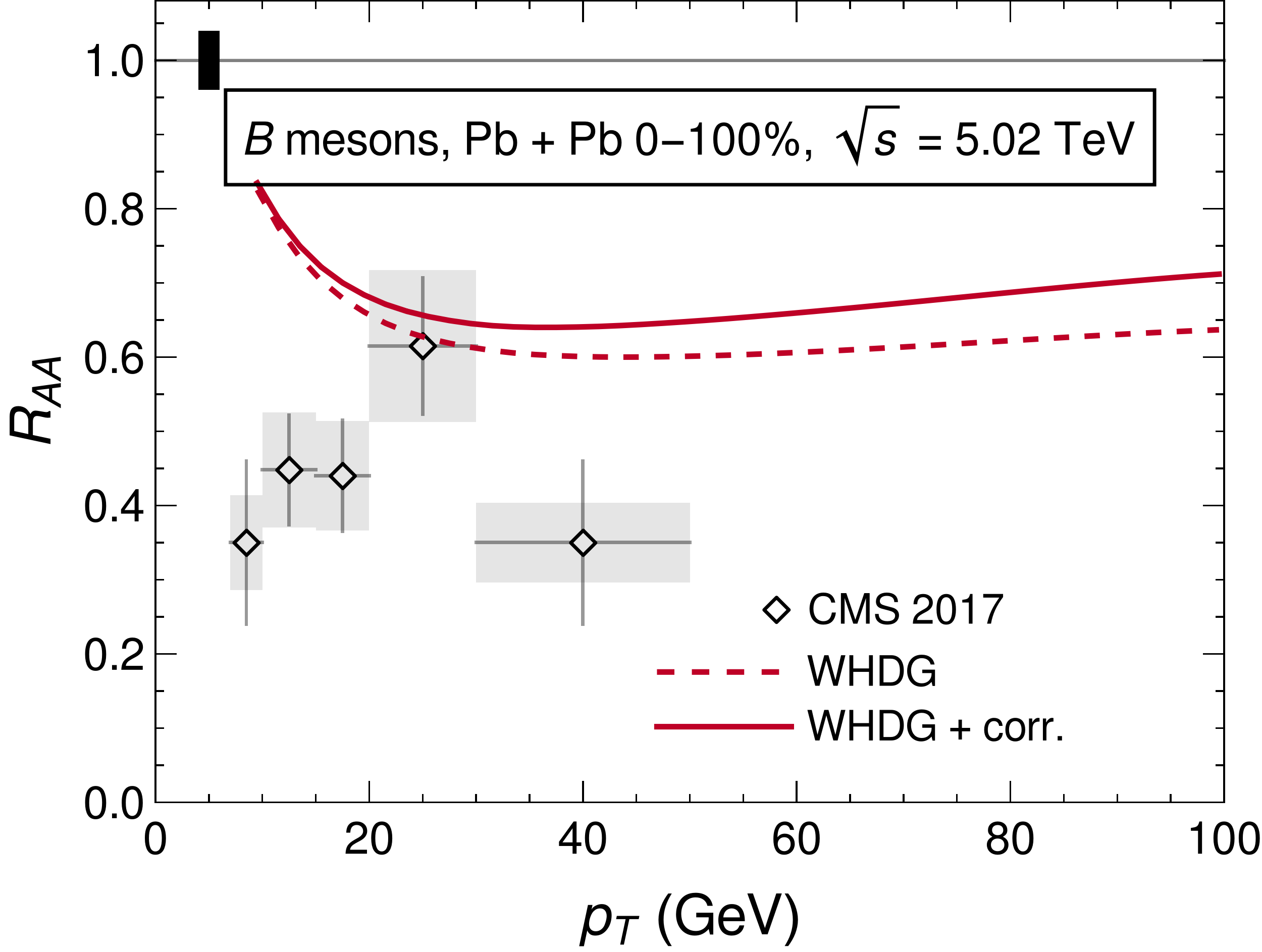}
        \caption{$R_{AA}$ for $B$ mesons in $0\text{--}100\%$ $\mathrm{Pb} + \mathrm{Pb}$ collisions.}
        \label{fig:SPL_B_raa_AA}
    \end{subfigure}
    \caption{The nuclear modification factor $R_{AA}$ as a function of final transverse momentum $p_T$ is calculated for $D^0$ and $B$ mesons with and without the short pathlength correction to the radiative energy loss. Calculations were done for $D^0$ mesons in 0--10\% and 30--50\% centrality classes and for $B$ mesons. Data from CMS \cite{CMS:2017qjw, CMS:2017uoy}, and ALICE \cite{ALICE:2018lyv} are shown for comparison, where error bars (boxes) indicate statistical (systematic) uncertainties. The global normalization uncertainty on the number of binary collisions is indicated by the solid boxes in the top left corner of the plots (left to right: 0--10\% CMS, 0--10\% ALICE, 30--50\% ALICE for $D^0$ mesons, and CMS for $B$ mesons).}
    \label{fig:SPL_DB_raa_AA_combined}
\end{figure}

In \cref{fig:SPL_D_raa_AA}, we show $R_{AA}(p_T)$ for $D$ mesons in $\sqrt{s} = 5.02$ TeV $\mathrm{Pb}+\mathrm{Pb}$ collisions at 0--10\% and 30--50\% centrality from our convolved radiative and collisional energy loss model, with and without the short pathlength correction to DGLV radiative energy loss, compared to data from CMS \cite{CMS:2017qjw}, and ALICE \cite{ALICE:2018lyv}.  In \cref{fig:SPL_B_raa_AA}, we show $R_{AA}(p_T)$ for $B$ mesons in $\sqrt{s} = 5.02$ TeV $\mathrm{Pb}+\mathrm{Pb}$ collisions at 0--10\% centrality from our convolved radiative and collisional energy loss model, with and without the short pathlength correction to DGLV radiative energy loss, compared to minimum bias data from CMS \cite{CMS:2017uoy}.  

From the figures we conclude that for heavy flavor final states, the short pathlength correction to the $R_{AA}$ is small up to $p_T \sim 100~\mathrm{GeV}$, and that the difference grows with $p_T$. This small difference was expected from the numerical energy loss calculations performed in \cite{Kolbe:2015rvk} and reproduced in \cref{fig:mod_deltaEoverE_small_large} that showed the small pathlength corrections are a relatively small correction for quarks. Agreement with data for $D^0$ mesons (\cref{fig:SPL_D_raa_AA}) is especially good for all $p_T \gtrsim 5~\mathrm{GeV}$, given that the calculation is leading order. The $D^0$ meson suppression prediction is underestimated for moderate $p_T \lesssim 20~\mathrm{GeV}$ in comparison to a previous prediction with the WHDG model in \cite{Andronic:2015wma}. We believe the main cause of this difference with past results is from not using the approximation $q_{\text{max}}\to \infty$ (see \cref{sec:mod_radiative_energy_loss_correction}), which overestimates the energy loss---especially at low $p_T$.

\begin{figure}[H]
    \centering
    \includegraphics[width=0.6\linewidth]{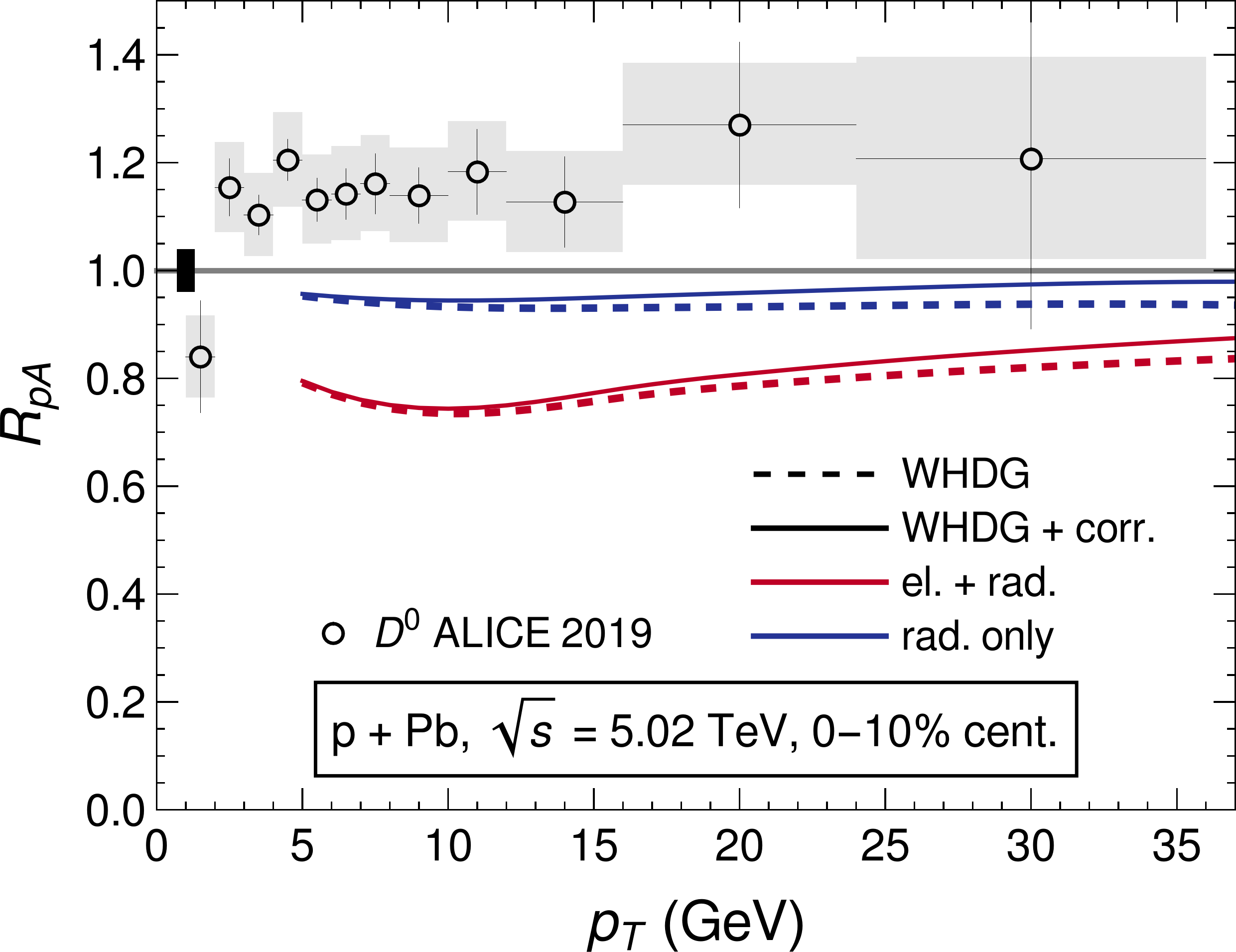}
    \caption{ The nuclear modification factor $R_{pA}$ for $D^0$ mesons as a function of final transverse momentum $p_T$ is calculated with and without the short pathlength correction. The $R_{pA}$ is calculated both with collisional and radiative energy loss (el.\ + rad.), and with radiative energy loss only (rad.\ only). Data from ALICE \cite{ALICE:2019fhe} for $D^0$ mesons is shown for comparison, where statistical (systematic) uncertainties are represented by error bars (boxes). The global normalisation uncertainty on the number of binary collisions is indicated by the solid box in the center left of the plot.}
    \label{fig:SPL_D_raa_pA}
\end{figure}

\Cref{fig:SPL_D_raa_pA} shows predictions for the $R_{pA}$ of $D$ mesons as a function of the final transverse momentum $p_T$ for central $\mathrm{p} + \mathrm{Pb}$ collisions at $\sqrt{s} = 5.02$ TeV from our convolved radiative and collisional energy loss model compared to ALICE data \cite{ALICE:2019fhe}.  In the same figure, we also show the prediction of $R_{pA}$ of $D$ mesons from our energy loss model with the collisional engery loss turned off.  One can see that the energy loss model that includes both collisional and radiative energy loss, both the small pathlength corrected and the uncorrected version, dramatically overpredicts the suppression in this small system.  At the same time, the predictions from the model with the collisional energy loss turned off are significantly less oversuppressed compared to the data.  The surprising sensitivity to the presence of the collisional energy loss process is due to our naive implementation of the collisional energy loss as discussed in \cref{sec:mod_gaussian_distribution_approximation}: the WHDG \cite{Wicks:2005gt} collisional energy loss model assumes an average collisional energy loss with Gaussian fluctuations, which is inappropriate for a small system with very few elastic scatterings.

\subsection{Suppression of light flavor mesons}
\label{sub:SPL_results_light}

\begin{figure}[H]
    \centering
    \begin{subfigure}[t]{0.49\textwidth}
        \centering
        \includegraphics[width=\linewidth]{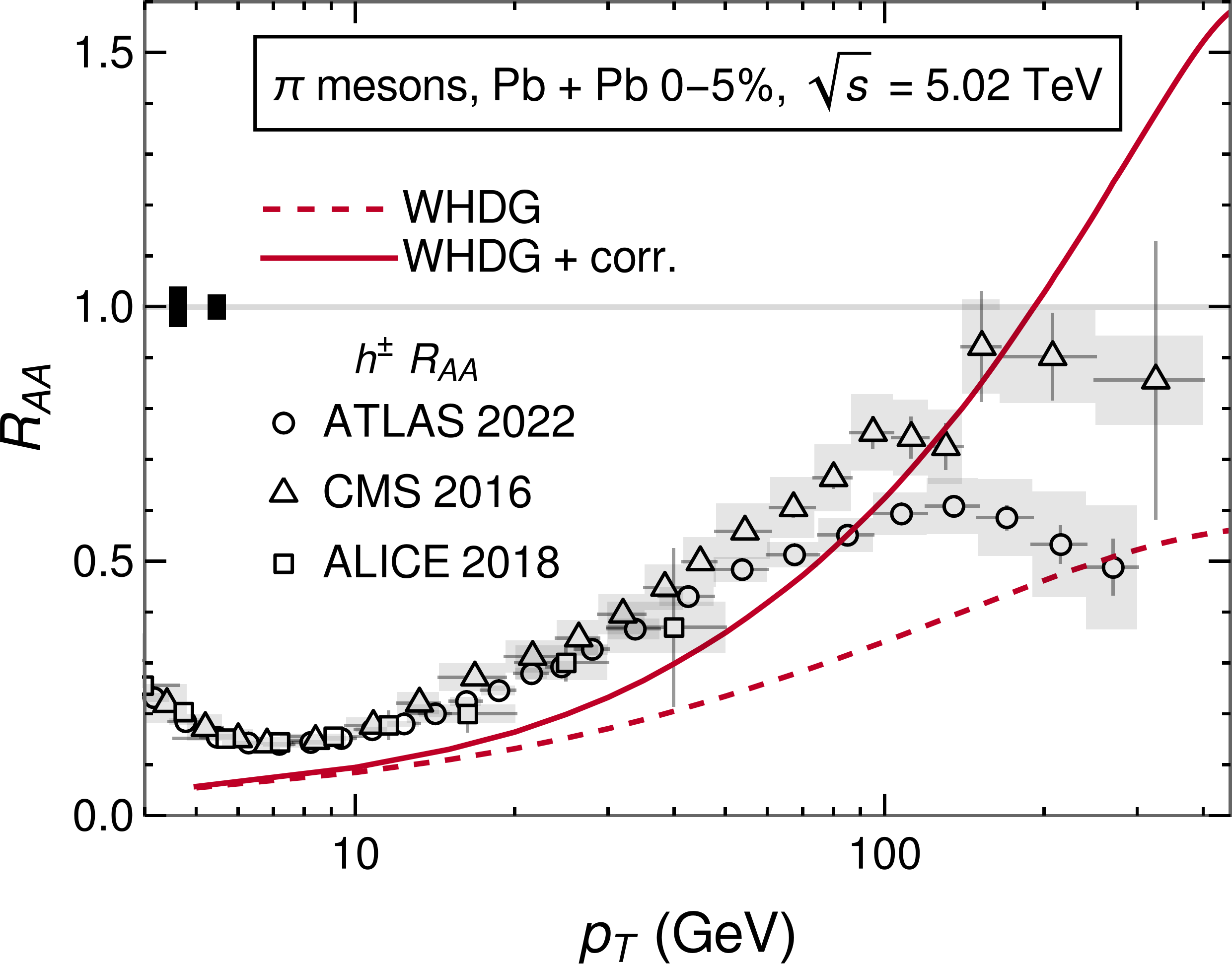}
        \caption{$R_{AA}$ for $\pi$ mesons produced in 0--5\% central $\mathrm{Pb} + \mathrm{Pb}$ collisions.}
        \label{fig:SPL_pion_raa_AA}
    \end{subfigure}\hfill
    \begin{subfigure}[t]{0.49\textwidth}
        \centering
        \includegraphics[width=\linewidth]{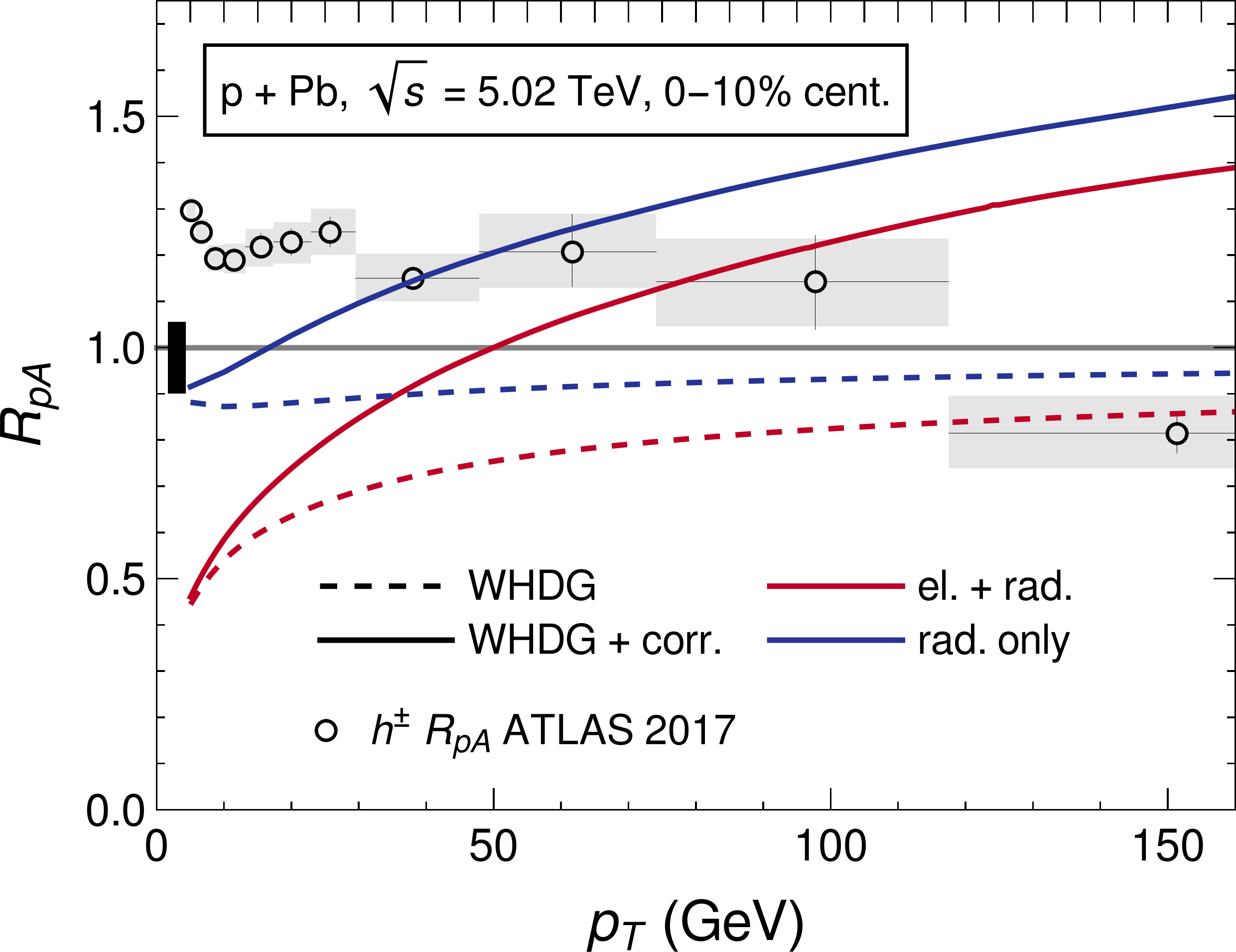}
        \caption{$R_{pA}$ for $\pi$ mesons produced in $0\text{--}5\%$ $\mathrm{p} + \mathrm{Pb}$ collisions.}
        \label{fig:SPL_pion_raa_pA}
    \end{subfigure}
    \caption{The nuclear modification factor $R_{AA}$ and $R_{pA}$ for $\pi$ mesons as a function of final transverse momentum $p_T$ in $\mathrm{Pb} + \mathrm{Pb}$ and $\mathrm{p} + \mathrm{Pb}$ collisions at $\sqrt{s}=5.02$ TeV. Predictions are calculated with and without the short pathlength correction to the radiative energy loss, using WHDG. Data from ATLAS \cite{ATLAS:2022kqu, Balek:2017man}, CMS \cite{CMS:2016xef}, and ALICE \cite{Sekihata:2018lwz} are shown for comparison, with statistical and systematic uncertainties indicated by error bars and boxes, respectively. The global normalization uncertainty on the number of binary collisions is represented by solid boxes.}
    \label{fig:SPL_pion_combined}
\end{figure}

\Cref{fig:SPL_pion_raa_AA} shows the $R_{AA}(p_T)$ for $\pi$ mesons as a function of $p_T$ in 0--5\% most central $\mathrm{Pb}+\mathrm{Pb}$ collisions at $\sqrt{s} = 5.02$ TeV from both our convolved radiative and collisional energy loss model compared to the $R_{AA}(p_T)$ for charged hadrons measured by ATLAS \cite{ATLAS:2022kqu}, CMS \cite{CMS:2016xef}, and ALICE \cite{Sekihata:2018lwz}. In our energy loss model, we used the fraction of $\pi$ mesons originating from light quarks versus gluons as in \cite{Horowitz:2011gd}.

\Cref{fig:SPL_pion_raa_AA} shows that the short pathlength correction does, in fact, lead to a stronger $p_T$ dependence in $R_{AA}(p_T)$ than the prediction without the correction.  In fact, the correction leads to a large change in predicted $R_{AA}(p_T)$, even at moderate momenta $20~\mathrm{GeV} \lesssim p_T \lesssim 100~\mathrm{GeV}$, when compared to the size of the correction for heavy flavor mesons.  This large change in $R_{AA}(p_T)$ is consistent with the large change in the average energy loss as calculated numerically in \cite{Kolbe:2015rvk} and reproduced in \cref{fig:mod_deltaEoverE_small_large_gluon,fig:mod_deltaEoverE_vs_L_gluon}; the correction is almost a factor of 10 times larger for gluons compared to quarks due to the specific way in which color triviality is broken by the short pathlength correction to the energy loss. For $p_T \lesssim 200~\mathrm{GeV}$ the corrected result is tantalizingly consistent with data, however for $p_T \gtrsim 200~\mathrm{GeV}$ the corrected result predicts anomalously large enhancement up to $R_{AA} = 1.5$ at $p_T \simeq 450~\mathrm{GeV}$, a shocking $\sim \!\! 200\%$ increase over the uncorrected result. 

\Cref{fig:SPL_pion_raa_pA} shows predictions for the $R_{pA}(p_T)$ of $\pi$ mesons in 0--10\% most central $\mathrm{p} + \mathrm{Pb}$ collisions from our convolved radiative and collisional energy loss model, with and without the short pathlength correction to the radiative energy loss, as well as predictions from our model with the collisional energy loss turned off.  \cref{fig:SPL_pion_raa_pA} also shows ATLAS charged hadron $R_{pA}(p_T)$ data \cite{Balek:2017man}. In this figure the breakdown of the elastic energy loss is even more obvious than for the heavy flavor mesons, since the elastic energy loss for gluons is enhanced by a factor $C_A / C_F \approx 2$. The uncorrected radiative-only result (no collisional energy loss) predicts mild suppression of $R_{AA} \approx 0.9$ for all $p_T$; and the corrected radiative-only result predicts enhancement that grows monotonically in $p_T$, reaching $R_{AA} \approx 1.5$ at $p_T = 150~\mathrm{GeV}$. The predicted enhancement for $p_T \gtrsim 100~\mathrm{GeV}$ is in excess of the measured enhancement \cite{Balek:2017man}, however for $p_T \lesssim 100~\mathrm{GeV}$ the presence of enhancement is qualitatively consistent with data.

The short pathlength ``correction'' leads to changes in $R_{AA}(p_T)$ of 100\% or greater in the light hadron sector, with predictions of $R_{AA}(p_T)$ well in excess of 1 for extremely high momenta, in both small and large collision systems leads us to question whether or not the energy loss calculation is breaking down in some fundamental way.  In particular, are we applying the energy loss formulae in our energy loss model in some regimes where the assumptions made in the derivation of those energy loss formulae no longer apply?

\section{Consistency of assumptions in DGLV}
\label{sec:SPL_assumptions}

The prediction of significant enhancement of high-$p_T$ light flavor mesons shown in \cref{fig:SPL_pion_raa_AA,fig:SPL_pion_raa_pA} stems from the asymptotic dependence of the short pathlength correction on energy \cite{Kolbe:2015rvk}.  
We see from \cref{eqn:mod_asymptotics_correction,eqn:mod_asymptotics_DGLV} that for asymptotically large values of energy, $\Delta E_{\text{corr.}} / E \sim E^0$ while $\Delta E_{\text{DGLV}} / E\sim \log E/E$.  Thus, inevitably, the correction becomes larger than the uncorrected result in the large $E$ limit.  Presumably, then, there's some intermediate value of the energy at which the assumptions that went into either the DGLV derivation, the derivation of the correction, or both are violated. As noted in the Introduction, the derivations of both DGLV energy loss and its small pathlength correction assumed: (1) the \textit{Eikonal approximation} which assumes that the energy of the hard parton is the largest scale in the problem; (2) the \textit{soft radiation approximation} which assumes $x \ll 1$; (3) \textit{collinearity} which assumes $k^+  \gg k^-$; (4) the impact parameter varies over a large transverse area; and (5) the \textit{large formation time assumption} which assumes $\omega_{0} \ll \mu \Leftrightarrow \mathbf{k}^2 / 2 x E \ll \mu$ and $\omega_1 \ll \mu_1 \Leftrightarrow (\mathbf{k} - \mathbf{q_1})^2 / 2 x E \ll \sqrt{\mu^2 + q_1^2}$. For the large formation time assumption we found that in the original calculation (details in \cite{Kolbe:2015suq}), $\omega_0 \ll \mu$ was only used in the weaker form $\omega_0 \ll \mu_1$; and so we will be considering this weaker assumption instead.

In this section we numerically check the consistency of these assumptions with the final radiative energy loss result.  In particular, the analytic properties of the matrix element mean that it may have non-zero support for momenta that are unphysical (even complex).  The relevant question for us is: does the matrix element (modulus squared) give a significant contribution to the energy loss in kinematic regions that are integrated over but for which the derivation of the matrix element is not under control? %

In an attempt to (partially) answer this question, we are motivated to calculate expectation values of ratios assumed small under the various assumptions, weighted by the absolute value of the mean radiative energy loss distribution, \cref{eqn:mod_fractional_energy_loss}\footnote{Just to be clear, we are weighting by the mean radiative energy loss as determined by the single inclusive gluon emission distribution, $\sim x \, \mathrm{d} N^g/\mathrm{d}x$; we are not weighting by the Poisson convolved distribution.}. Explicitly the procedure to calculate the expectation value of a function $R(\{ X_i \})$, depending on the set of variables $\{X _i\}$, is
\begin{equation}
  \langle R \rangle \equiv \frac{\int \mathrm{d} \{X_i\} ~ R(\{X_i\}) \; \left | \frac{\mathrm{d} E}{\mathrm{d} \{ X_i \}} \right |}{\int \mathrm{d} \{X_i\}~ \left | \frac{\mathrm{d} E}{\mathrm{d} \{X_i\}} \right |},
    \label{eqn:SPL_expected_value}
\end{equation}
where $\{X_i\}$ can be any of $\{\mathbf{k}, \mathbf{q}, x, \Delta z\}$ and $\mathrm{d} \{ X_i \} \equiv \prod_i \mathrm{d} X_i$. Also note that $R$ can depend on quantities that are not integrated over, such as $\{L, E, \mu\}$. It is important to note that this expectation value is not an expectation value in the usual sense, where the distribution is the distribution of radiated gluons, because we are weighting by the radiative energy loss and not radiated gluon number.  It is also important to note that even if a particular assumption is violated in the sense of this weighted average, that violation does not necessarily mean that the correction computed by relaxing the assumption is large; rather, we only know that the correction is not necessarily small.

\begin{figure}[H]
    \centering
    \includegraphics[width=0.5\linewidth]{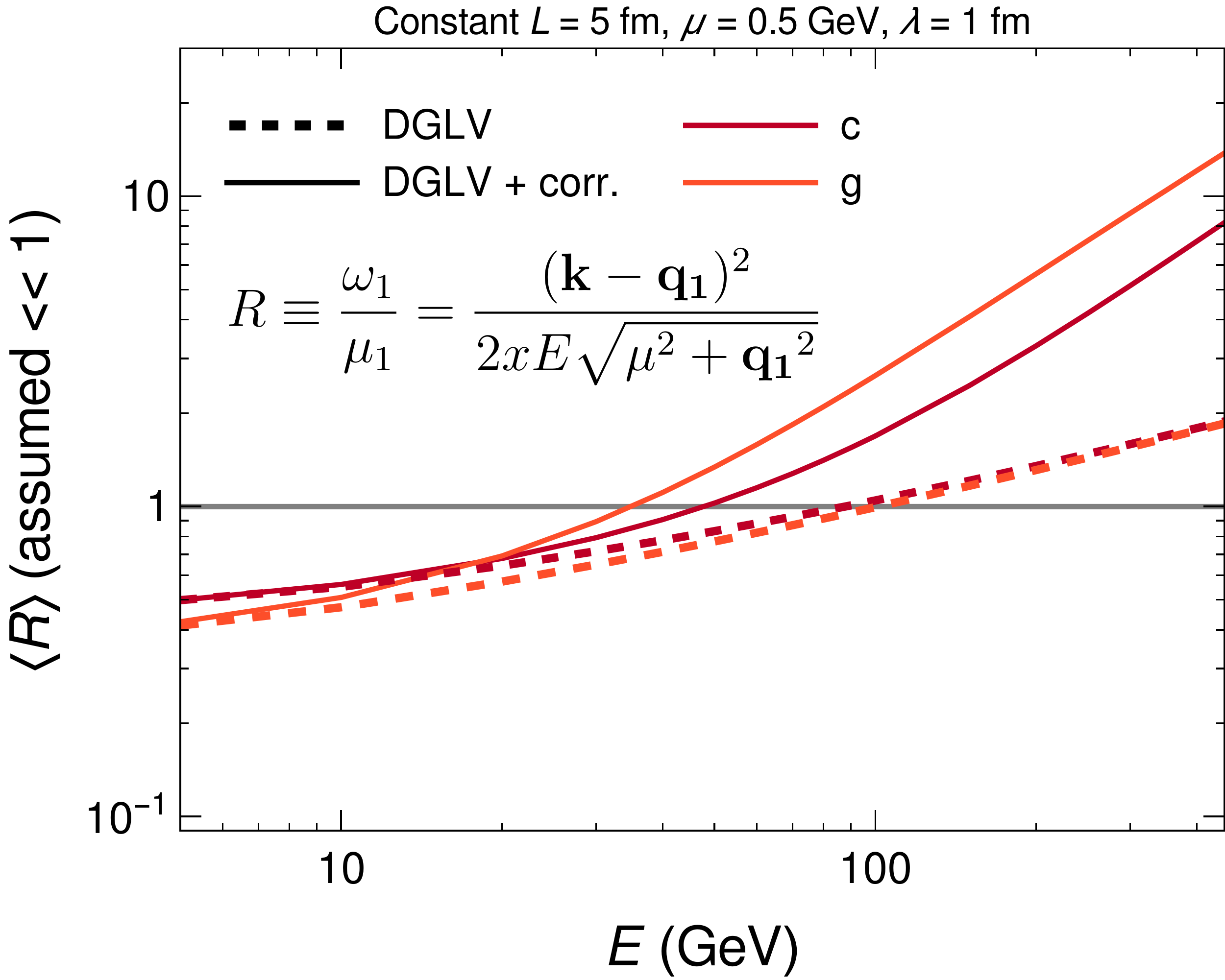}
    \caption{
    Plot of $\langle R\rangle \equiv \langle \omega_1/\mu_1 \rangle$ as a function of parent parton energy $E$.  $\langle R \rangle\ll1$ implies consistency with the large formation time assumption.  $\langle R \rangle$ is computed without (dashed) and with (solid) the short pathlength correction for charm quarks (dark red) and gluons (orange).  
    All curves use constant $L = 5~\mathrm{fm}$, $\lambda_g=1~\mathrm{fm}$, and $\mu=0.5~\mathrm{GeV}$.
  }
    \label{fig:SPL_largeformationtime2NoStep}
\end{figure}

\Cref{fig:SPL_largeformationtime2NoStep} investigates the large formation time assumption.  The figure shows the expectation value of $\omega_1 / \mu_1 = (\mathbf{k}-\mathbf{q}_1)^2/2xE\sqrt{\mu^2+\mathbf{q}_1^2}$, where the DGLV and correction derivations assume that $\omega_1 / \mu_1 \ll 1$, using $L = 5~\mathrm{fm}$, $\lambda_g=1~\mathrm{fm}$, and $\mu=0.5~\mathrm{GeV}$. 
The large formation time assumption is explicitly violated for the energy loss with and without the short pathlength correction. For the energy loss without the correction, the large formation time assumption is violated for $E \gtrsim 100$ GeV for both charm quarks and gluons, while for the energy loss with the correction the large formation time assumption is violated for $E \gtrsim 35~\mathrm{GeV}$ for gluons and for $E \gtrsim 50~\mathrm{GeV}$ for charm quarks. 
This breakdown calls into question the validity of the large formation time assumption in DGLV radiative energy loss for $p_T \gtrsim 100$ GeV, regardless of whether the energy loss receives a short pathlength correction. The increased rate and magnitude of large formation time assumption violation once the short pathlength correction is included in the energy loss indicates that the short pathlength corrected energy loss model predictions may be breaking down at moderate to high $p_T$ in central $\mathrm{A}+\mathrm{A}$ collisions. One sees that the large formation time assumption, always breaks down before enhancement, $\Delta E / E < 0$, is predicted.

One possibility for the increased rate and degree of large formation time breakdown once the short pathlength is included
---and the erroneously large correction at high $p_T$ for $\pi$ mesons---is the emphasis placed on short pathlengths by the exponential distribution of scattering centers. The exponential distribution was originally chosen to make the analytics simple, with the physical motivation that it captures the Bjorken expansion in the medium \cite{Djordjevic:2003zk}.

Bjorken expansion leads to a power law decay of the plasma density in time \cite{Bjorken:1982qr}, and so an exponential distribution likely overestimates the amount of expansion. This biases scatterings to occur at smaller $\Delta z$ than is physical, and likely overestimates the contribution from the short pathlength correction. Additionally it is not obvious how to model the time dependence of the collision geometry before thermalization $\tau \lesssim \tau_0$, as in principle the medium should be thermalizing during this time. Furthermore the treatment of scatters that occur for times $\tau \lesssim \tau_0$ is not obvious as it is possible that the well-separated scattering centers assumption $\lambda  \gg \mu^{-1}$ breaks down in this phase of the plasma. It was found numerically that DGLV energy loss results are insensitive to the exact distribution of scattering centers used \cite{Armesto:2011ht}.  Perhaps not surprisingly, the small pathlength correction has a large sensitivity to the exact distribution of scattering centers used \cite{Kolbe:2015rvk}.

We are thus motivated to consider an alternative distribution of scattering centers as we consider whether or not the various assumptions made in the energy loss derivations are consistent with our final energy loss numerics.  In this paper we will consider, in addition to the usual exponential distribution, the \textit{truncated step} distribution from \cite{Kolbe:2015rvk}.  The truncated step distribution is given by $\bar{\rho}_{\text{step}}(\Delta z) \equiv (L-a)^{-1} \Theta(\Delta z-a) \Theta(L-\Delta z)$, where $a$ is a small distance cut off.  The truncated step function attempts to capture the effect of a ``turn on'' of the QGP, before which no energy loss takes place, with subsequent equal probability for a scattering to occur until the end of the pathlength.  We think of the exponential distribution and truncated step distributions as limiting cases for what the real distribution of scattering centers may be. The exponential distribution maximally emphasizes the effect of early-time physics, while the truncated step distribution completely neglects early-time physics. A more realistic distribution is likely somewhere in between these two extremes. 

One choice for $a$ is $a = \mu^{-1}$, since for $\Delta z \lesssim \mu^{-1}$, the production point and scattering center are too close to be individually resolved. Another choice is $a = \tau_0$, the hydrodynamics turn-on time; $\tau_0$ is a particularly reasonable choice since we already only consider the medium density evolution for times $\tau>\tau_0$ in computing our pathlengths, \cref{eqn:mod_effective_length}.  For a typical value of $\mu = 0.5~\mathrm{GeV}$ in central $\mathrm{Pb} + \mathrm{Pb}$ collisions, $\tau_0 \simeq \mu^{-1} \simeq 0.4~\mathrm{fm}$; however in central $\mathrm{p} + \mathrm{Pb}$ collisions $\mu^{-1} \simeq 0.25 ~\mathrm{fm} < \tau_0 = 0.4~\mathrm{fm}$, and so the distinction between the two options for $a$ might be important. In this report we have chosen to use $a = \tau_0$ throughout for simplicity. 

\begin{figure}[H]
    \centering
    \begin{subfigure}[t]{0.49\textwidth}
        \centering
        \includegraphics[width=\linewidth]{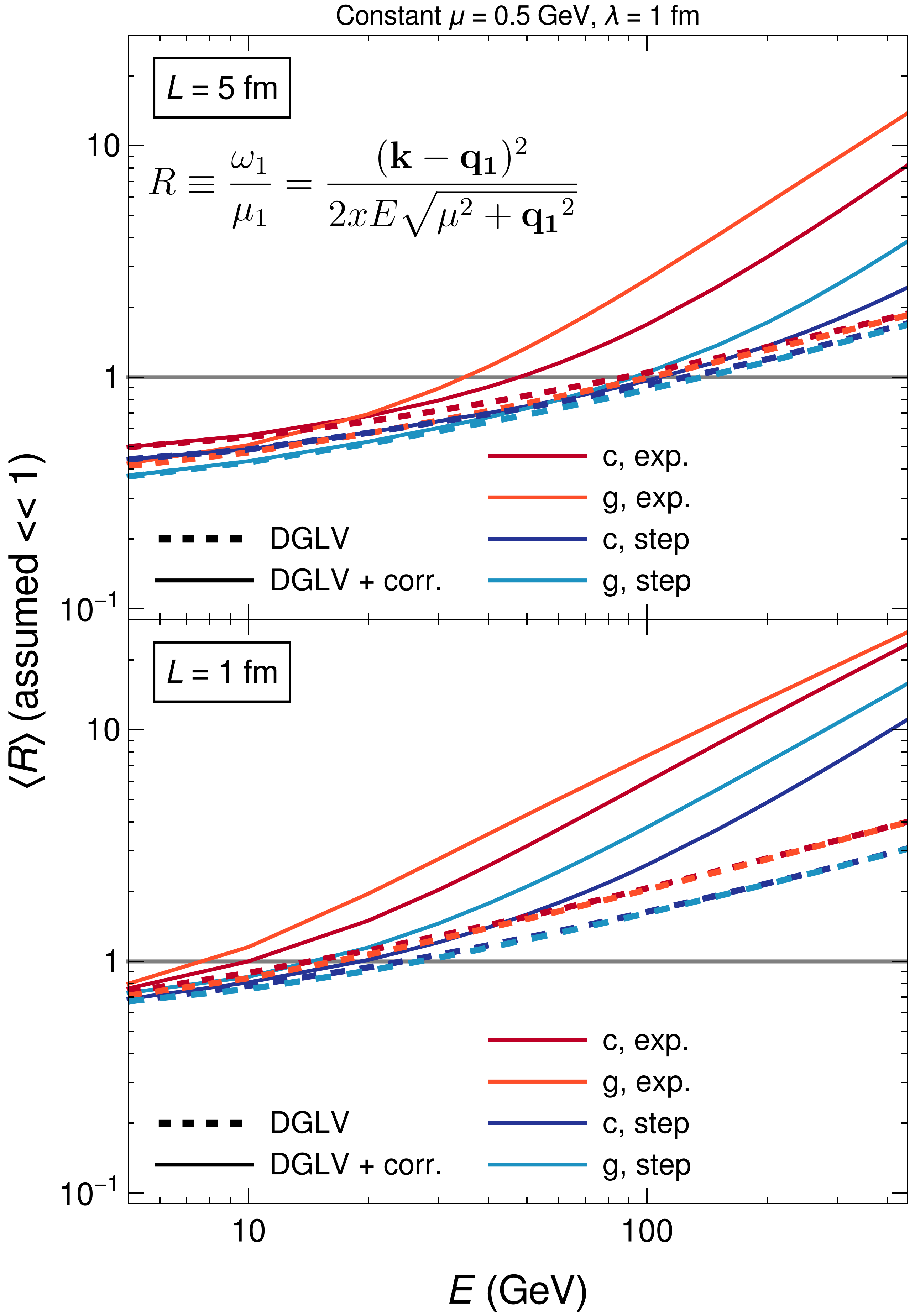}
        \caption{Plot of $\langle R\rangle \equiv \langle \omega_1/\mu_1 \rangle$ as a function of parent parton energy $E$.}
        \label{fig:SPL_largeformationtime2}
    \end{subfigure}\hfill
    \begin{subfigure}[t]{0.49\textwidth}
        \centering
        \includegraphics[width=\linewidth]{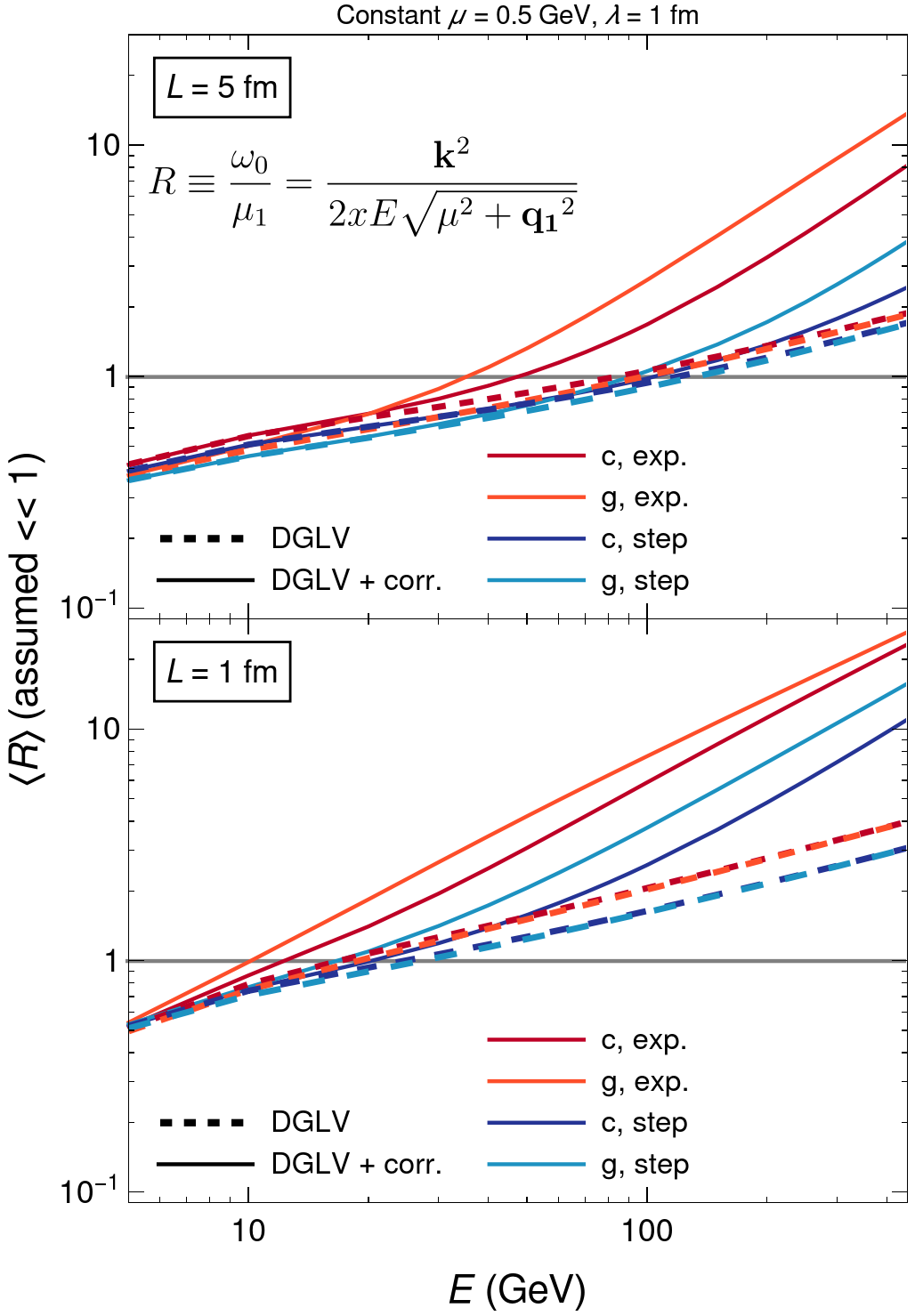}
        \caption{Plot of $\langle R\rangle \equiv \langle \omega_0/\mu_1 \rangle$ as a function of parent parton energy $E$.}
        \label{fig:SPL_largeformationtime3}
    \end{subfigure}
    \caption{Plots of $\langle R\rangle$ as a function of parent parton energy $E$, where $\langle R \rangle \ll 1$ implies consistency with the large formation time assumption. $\langle R \rangle$ is computed without (dashed) and with (solid) the short pathlength correction for charm quarks [gluons] with scattering centers distributed according to the exponential distribution (dark red [orange]) and truncated step function (dark blue [light blue]). $L=5$ fm in the top panes and $L=1$ fm in the bottom panes. All curves use constant $\lambda_g=1~\mathrm{fm}$ and $\mu=0.5~\mathrm{GeV}$.}
    \label{fig:SPL_largeformationtime_combined}
\end{figure}

\begin{figure}[H]
    \centering
    \begin{subfigure}[t]{0.49\textwidth}
        \centering
        \includegraphics[width=\linewidth]{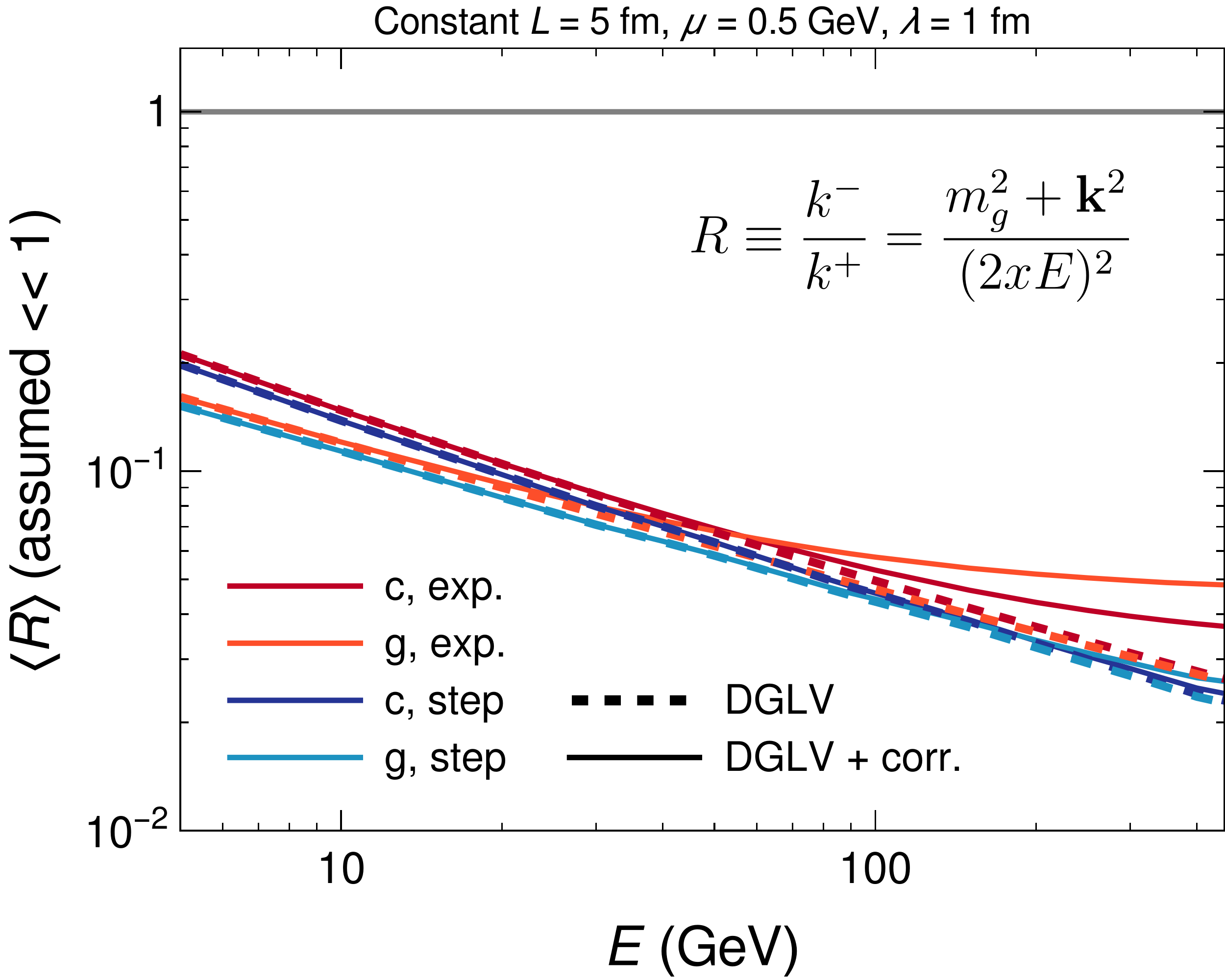}
        \caption{Plot of $\langle R\rangle \equiv \langle k^-/k^+ \rangle$ as a function of parent parton energy $E$.}
        \label{fig:SPL_collinear}
    \end{subfigure}\hfill
    \begin{subfigure}[t]{0.49\textwidth}
        \centering
        \includegraphics[width=\linewidth]{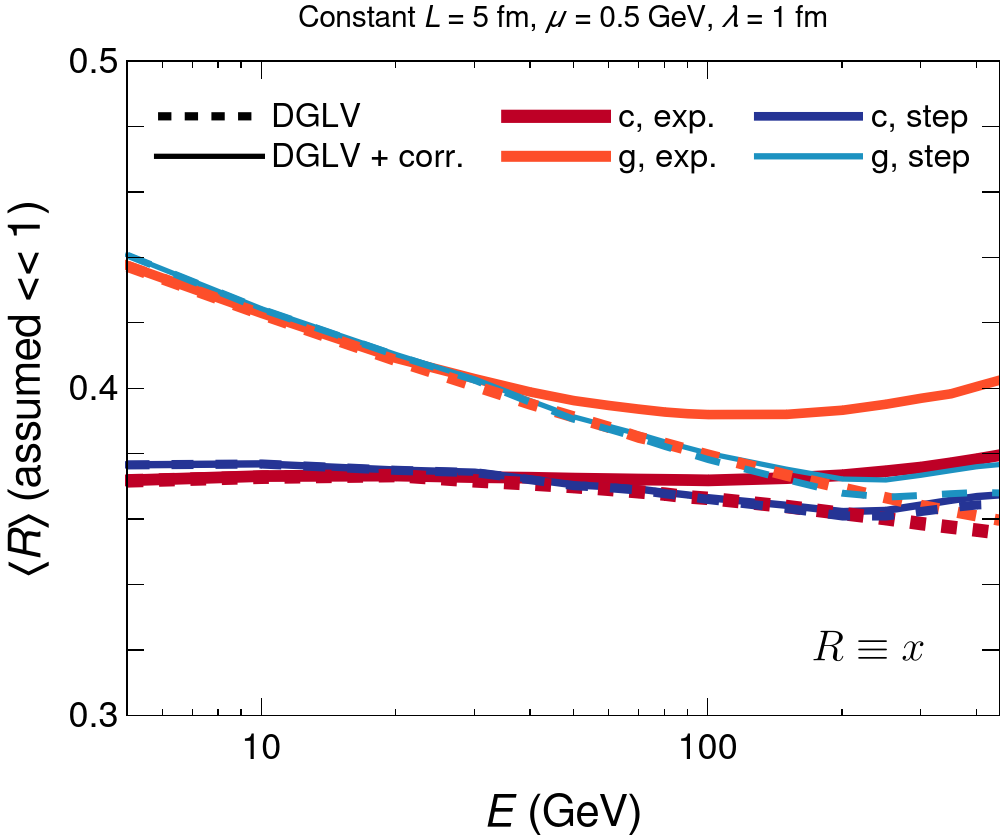}
        \caption{Plot of $\langle R\rangle \equiv \langle x \rangle$ as a function of parent parton energy $E$.}
        \label{fig:SPL_soft}
    \end{subfigure}
    \caption{Plots of $\langle R\rangle$ as a function of parent parton energy $E$. $\langle R \rangle \ll 1$ implies consistency with the respective approximations (collinear in \protect\subref{fig:SPL_collinear}, soft in \protect\subref{fig:SPL_soft}). $\langle R \rangle$ is computed without (dashed) and with (solid) the short pathlength correction for charm quarks [gluons] with scattering centers distributed according to the exponential distribution (dark red [orange]) and truncated step function (dark blue [light blue]). $L=5$ fm in the top panes and $L=1$ fm in the bottom panes. All curves use constant $L=5$ fm, $\lambda_g=1~\mathrm{fm}$, and $\mu=0.5~\mathrm{GeV}$.}
    \label{fig:SPL_collinear_soft}
\end{figure}

\begin{figure}[H]
    \centering
    \begin{subfigure}[t]{0.49\textwidth}
        \centering
        \includegraphics[width=\linewidth]{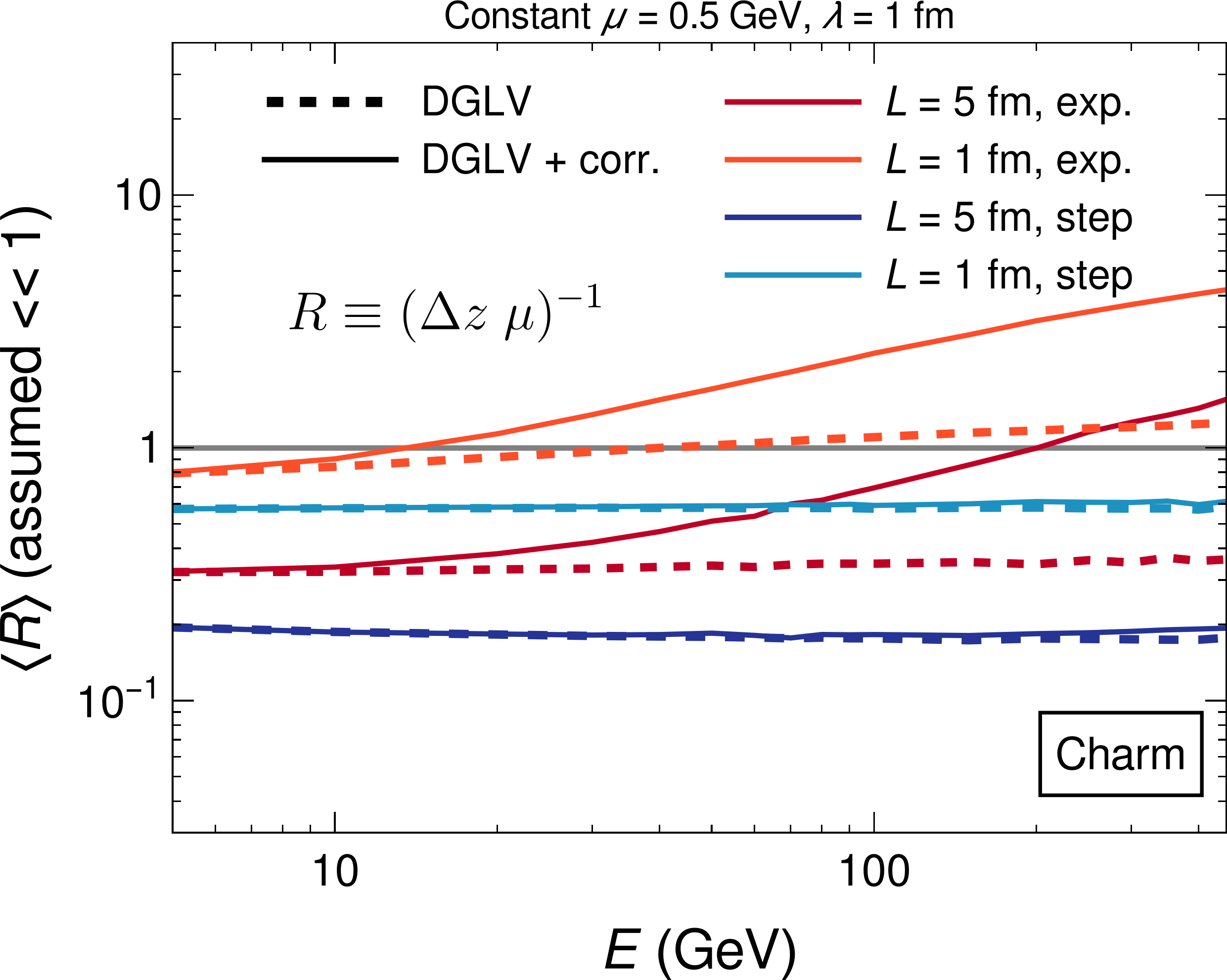}
        \caption{Plot of $\langle R\rangle \equiv \langle 1/\Delta z \, \mu \rangle$ as a function of parent parton energy $E$. A value of $\langle R \rangle \ll 1$ implies self-consistency of the assumption.}
        \label{fig:SPL_largepathlengthassumption}
    \end{subfigure}\hfill
    \begin{subfigure}[t]{0.49\textwidth}
        \centering
        \includegraphics[width=\linewidth]{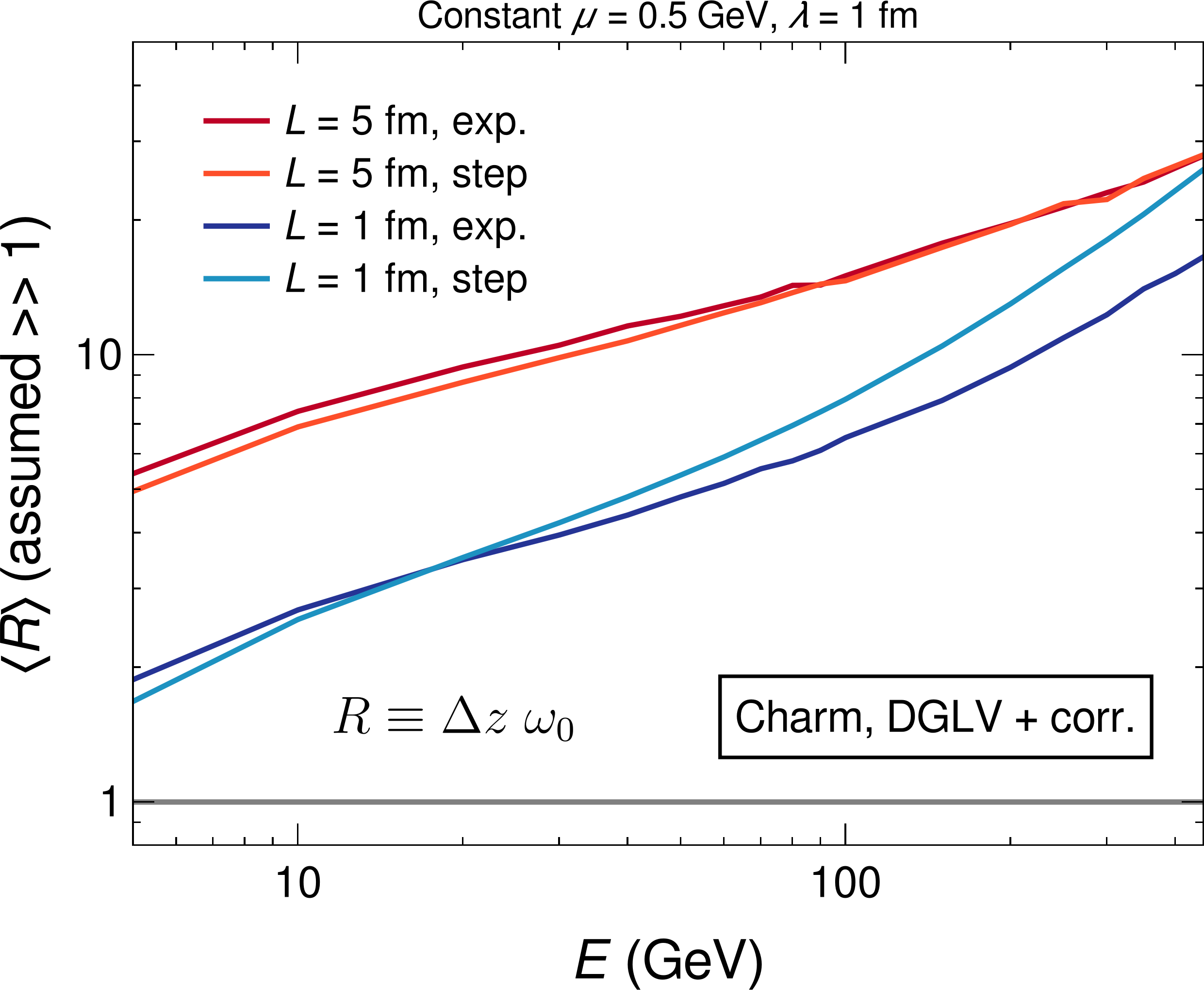}
        \caption{Plot of $\langle R\rangle \equiv \langle \Delta z \, \omega_0 \rangle$ as a function of parent parton energy $E$. A value of $\langle R \rangle \gg 1$ implies self-consistency of the assumption.}
        \label{fig:SPL_implicitlargepathlengthassumption}
    \end{subfigure}
    \caption{Plots of $\langle R\rangle$ as a function of parent parton energy $E$ to test the validity of large pathlength assumptions. $\langle R \rangle$ is computed with and without (solid and dashed, respectively) the short pathlength correction for charm quarks [gluons] with scattering centers distributed according to the exponential distribution (dark red [orange]) and truncated step function (dark blue [light blue]). $L=5$ fm in the top panes and $L=1$ fm in the bottom panes. All curves use constant $\lambda_g=1~\mathrm{fm}$ and $\mu=0.5~\mathrm{GeV}$.}
    \label{fig:SPL_largepathlengthassumptions}
\end{figure}

We now check all of the assumptions made in the computation of the DGLV radiative energy loss for: the corrected and uncorrected results; the exponential and truncated step scattering center distributions; charm quarks and gluons; and for large ($L = 5~\mathrm{fm}$) and small ($L = 1~\mathrm{fm}$) systems (\crefrange{fig:SPL_largeformationtime2}{fig:SPL_implicitlargepathlengthassumption}). All calculations use constant $\mu=0.5~\mathrm{GeV}$, and $\lambda_g = 1~\mathrm{fm}$ which are approximate averages in $\mathrm{A} + \mathrm{A}$ collisions, and standard benchmark choices \cite{Armesto:2011ht}. 

\sloppy\cref{fig:SPL_largeformationtime2,fig:SPL_largeformationtime3} show the expectation values of $\omega_1 / \mu_1 = (\mathbf{k} - \mathbf{q}_1)^2 / 2xE \sqrt{\mu^2 + \mathbf{q}_1^2}$ and $\omega_0 / \mu_1 = \mathbf{k}^2/2xE\sqrt{\mu^2+\mathbf{q}_1^2}$, respectively.  The large formation time assumption is equivalent to both ratios much less than one. 

For an exponential distribution of scattering centers with $L = 5~\mathrm{fm}$ (top panes of \cref{fig:SPL_largeformationtime2,fig:SPL_largeformationtime3}), the large formation time assumption: breaks down for $E \gtrsim 40~\mathrm{GeV}$ for the corrected result, and breaks down for $E \gtrsim 100$ GeV for the uncorrected result. For the truncated step distribution with $L = 5~\mathrm{fm}$, the large formation time assumption breaks down for both the corrected and uncorrected results for $E \gtrsim 100~\mathrm{GeV}$. We see in the plots the known numerical insensitivity to the distribution of scattering centers in the uncorrected DGLV radiative energy loss result.

For $L = 1~\mathrm{fm}$ (bottom panes of \cref{fig:SPL_largeformationtime2,fig:SPL_largeformationtime3}), the shape of all curves are approximately the same, but scaled by a factor $\sim1~\mathrm{fm} / 5~\mathrm{fm}$ in $E$. Thus the breakdown in the large formation time assumptions occurs roughly five times earlier in $E$ for the $L = 1$ fm pathlengths compared to the $L=5$ fm pathlengths. The reason for this simple approximate scaling is that all of the nontrivial dependence of $\Delta E / E$ on $E$ and $L$ in the distribution of emitted gluons \cref{eqn:mod_full_dndx} comes from terms $\sim \omega_\alpha \Delta z$ where $\alpha \in \{0, 1, m\}$. Once integrated these terms become $\omega_\alpha L \sim L / E$; assuming $k$ and $q$ have negligible dependence on $E$. If the kinematic cutoffs on the $\mathbf{k}$, and $\mathbf{q}$ integrals are important then this scaling breaks down; any deviation from this simple scaling must be due to the effects of the cutoff. 

We found that finite kinematic bounds do not significantly affect the consistency of the assumptions. The above scaling argument holds for all expectation values $\langle R \rangle$ so long as $R$ does not depend on $\Delta z$. Thus, in order to keep the number of plots shown to a manageable number, most assumption consistency plots from now on will be shown for only $L=5~\mathrm{fm}$.

The collinear and soft assumptions are tested for consistency in \cref{fig:SPL_collinear,fig:SPL_soft}, respectively. We find that both of these assumptions are consistently satisfied for both the short pathlength corrected and uncorrected DGLV results, for both the exponential and the truncated step distributions of scattering centers.

\Cref{fig:SPL_soft} shows $\langle x \rangle$ as a function of parent parton energy $E$, where $\langle x \rangle \ll 1$ is assumed under the soft approximation. The expectation value $\langle x \rangle$ decreases monotonically in energy for the uncorrected result with an exponential distribution of scattering centers; all other expectation values appear to converge numerically to a constant nonzero value. 
For the DGLV result with an exponential distribution, one can calculate $\langle x \rangle$ analytically for asymptotically high energies. For asymptotically high energies we take $m_g \to 0$, $M \to  0$, $k_{\text{max}} \to \infty$, and $q_{\text{max}}\to \infty$. These simplifications allow the angular, $\mathbf{k}$, and $\mathbf{q}$ integrals to be done analytically, as described in \cite{Gyulassy:2000er, Djordjevic:2003zk}. Proceeding in this way we obtain the following asymptotic expression for $\langle x \rangle $:
\begin{align}
  \langle x \rangle^{\text{DGLV}}_{\text{exp.}} &=  \frac{1}{\log (\frac{4 E}{L \mu^2})} + \mathcal{O}\left( \frac{L \mu^2}{4 E} \right)\label{eqn:SPL_asymptotic_x_DGLV_exp}\\
  \implies \langle x \rangle &\to 0 \text{ as } E \to \infty.\nonumber
\end{align}

In a similar way we can derive the same result for the short pathlength correction with an exponential distribution, using the asymptotic result from \cite{Kolbe:2015rvk} (see \cref{eqn:mod_asymptotics_correction_xint}) 
\begin{align}
  \langle x \rangle^{\text{corr.}}_{\text{exp.}} &= \frac{1}{2} \left[\frac{-\frac{1}{2}  + \log\left(\frac{2 E L}{2+L\mu}\right)}{-1+\log\left(\frac{2 E L}{2+L\mu}\right)}\right]\label{eqn:SPL_asymptotic_x_Short_exp}\\
  &\to \frac{1}{2} \text{ as } E\to \infty. \nonumber
\end{align}
Note that numerical investigation of $\langle x \rangle$ indicates that the convergence to the asymptotic values is slow for both the corrected and uncorrected energy loss.
For the truncated step distribution it is more difficult to perform asymptotic calculations, but numerical investigation shows that the uncorrected result with truncated step distribution converges to $\langle x \rangle \approx \frac{1}{3}$, and the corrected result with truncated step distribution converges to $\langle x \rangle \approx \frac{1}{2}$.

\Cref{fig:SPL_largepathlengthassumption} tests the consistency of the large pathlength assumption with the DGLV result as a function of parent parton energy $E$ for charm quarks. For $L=1~\mathrm{fm}$, we see that the large pathlength assumption is not a good approximation for either the uncorrected or small pathlength corrected DGLV calculation for the exponential scattering center distribution; with $\langle 1 / \Delta z \, \mu \rangle\sim 0.6$, the large pathlength assumption is not a particularly good approximation for the truncated step distribution, either. Even for $L=5~\mathrm{fm}$, the large pathlength assumption breaks down for the short pathlength corrected energy loss when the exponential distribution of scattering centers is used. While not shown, results for gluons are essentially identical.  We see that, as expected, one must quantitatively determine the importance of the short pathlength correction terms to the radiative energy loss derivation for short pathlengths, $L\sim1/\mu$ for all values of $E$ and for all parton types.

We note that there was one final assumption implicitly made in the derivation of the short pathlength correction to the DGLV radiated gluon distribution, the short formation time (with respect to scattering centers) assumption, $\Delta z \: \omega_0 \gg 1$ \cite{Kolbe:2015rvk, Kolbe:2015suq}. This in conjunction with the large formation time assumption furnishes a separation of scales $\Delta z^{-1} \ll \omega_0 \ll \mu_1$. Alternatively one can view this as a large pathlength assumption, wherein we have replaced $\Delta z \: \mu \gg 1$ with $\Delta z \: \omega_0 \gg 1$; which is guaranteed to be a weaker assumption, according to the large formation time assumption.  \cref{fig:SPL_implicitlargepathlengthassumption} shows $\langle\Delta z \: \omega_0\rangle$ as a function of parent parton energy $E$ for charm quarks with the short pathlength corrected energy loss.  We find that for both large and small systems, and exponential and truncated step distributions this assumption holds self-consistently. While not shown, results for gluons are essentially identical.  The short formation time (with respect to scattering centers) assumption is intimately tied to the large formation time assumption, and so if future work aims to remove the large formation time assumption, then this assumption should also be removed. Note that we computed $\langle \Delta z \: \omega_0 \rangle$ (assumed $\gg 1$) instead of $\langle (\Delta z \: \omega_0)^{-1} \rangle$ (assumed $\ll 1$), due to numerical convergence issues with the latter.

\section{Suppression with Exponential and Step Distributions}
\label{sec:SPL_final_results}

We saw in the previous section that for $p_T\gtrsim40$ GeV the large formation time approximation breaks down for energy loss calculations using the exponential distribution of scattering centers (which bias the scattering to shorter pathlengths) while the large formation time approximation breaks down only for $p_T\gtrsim100$ GeV for an energy loss calculation using the truncated step distribution of scattering centers.  Additionally, we claimed that the exponential distribution of scattering centers and the truncated step distribution of scattering centers represent two extreme possibilities for what a more realistic distribution of scattering centers likely will be.  We are thus motivated to explore the sensitivity of our suppression predictions to the choice of distribution of scattering centers.  For pure DGLV energy loss without the short pathlength correction, one saw an insensitivity to this choice of scattering center distributions \cite{Armesto:2011ht}.  We will see that when including the short pathlength correction to the radiative energy loss, the heavy flavor observables are still relatively insensitive to the distribution of scattering centers.  However, we find that hadron observables that include a contribution from gluons are very sensitive to the choice of scattering center distribution, with the use of the truncated step function dramatically reducing the effect of the short pathlength correction to pion $R_{AA}(p_T)$.  This dramatic reduction in the effect of the short pathlength correction to the phenomenologically accessible pion suppression observable is expected from the dramatic reduction in the effect of the short pathlength correction to the average radiative energy loss, as seen in \cite{Kolbe:2015rvk}.

\begin{figure}[H]
    \centering
    \begin{subfigure}[t]{0.49\textwidth}
        \centering
        \includegraphics[width=\linewidth]{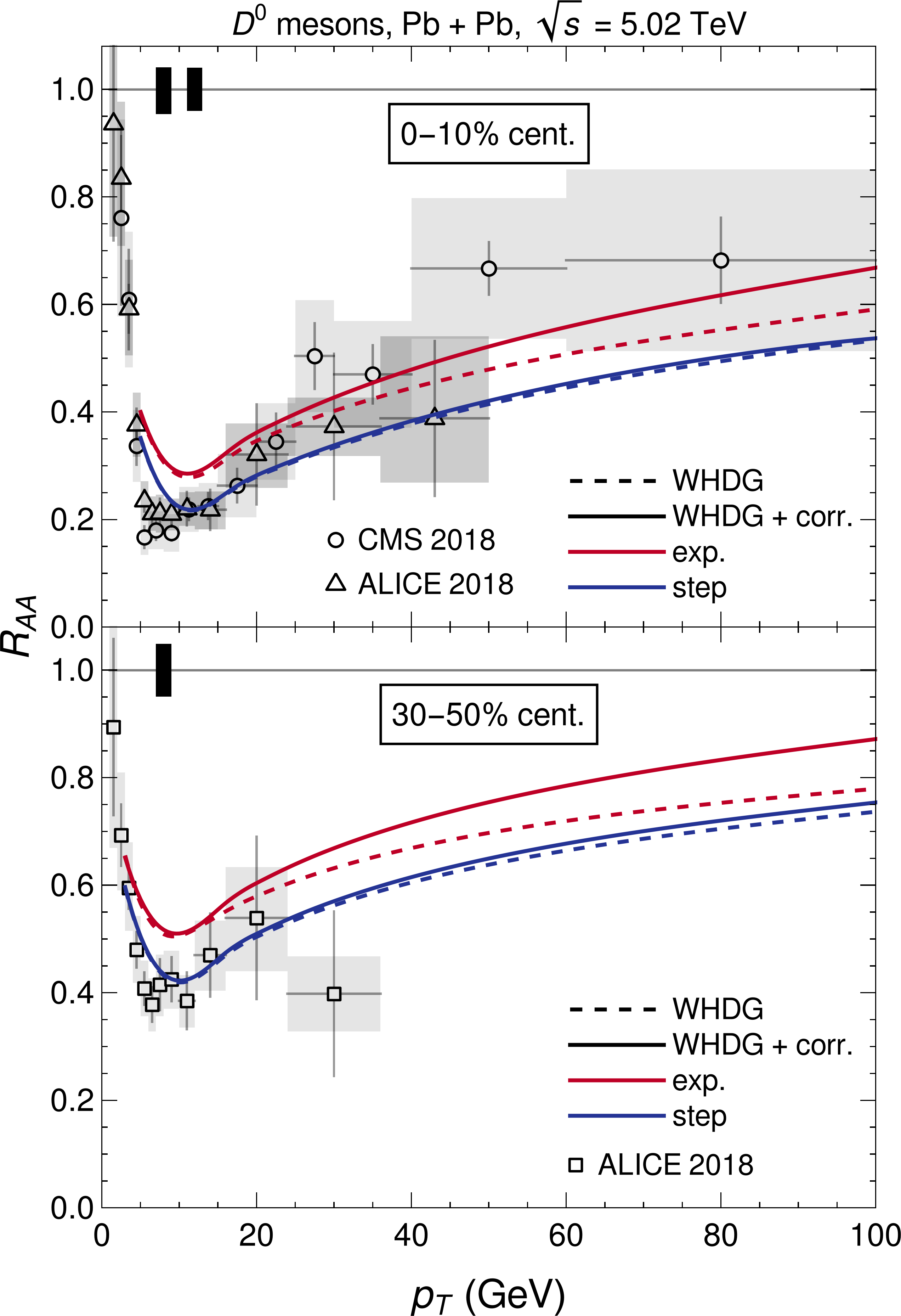}
        \caption{$R_{AA}$ for $D$ mesons in central 0--10\% and semi-central 30--50\% $\mathrm{Pb} + \mathrm{Pb}$ collisions.}
        \label{fig:SPL_raa_D_mesons_with_step}
    \end{subfigure}\hfill
    \begin{subfigure}[t]{0.49\textwidth}
        \centering
        \includegraphics[width=\linewidth]{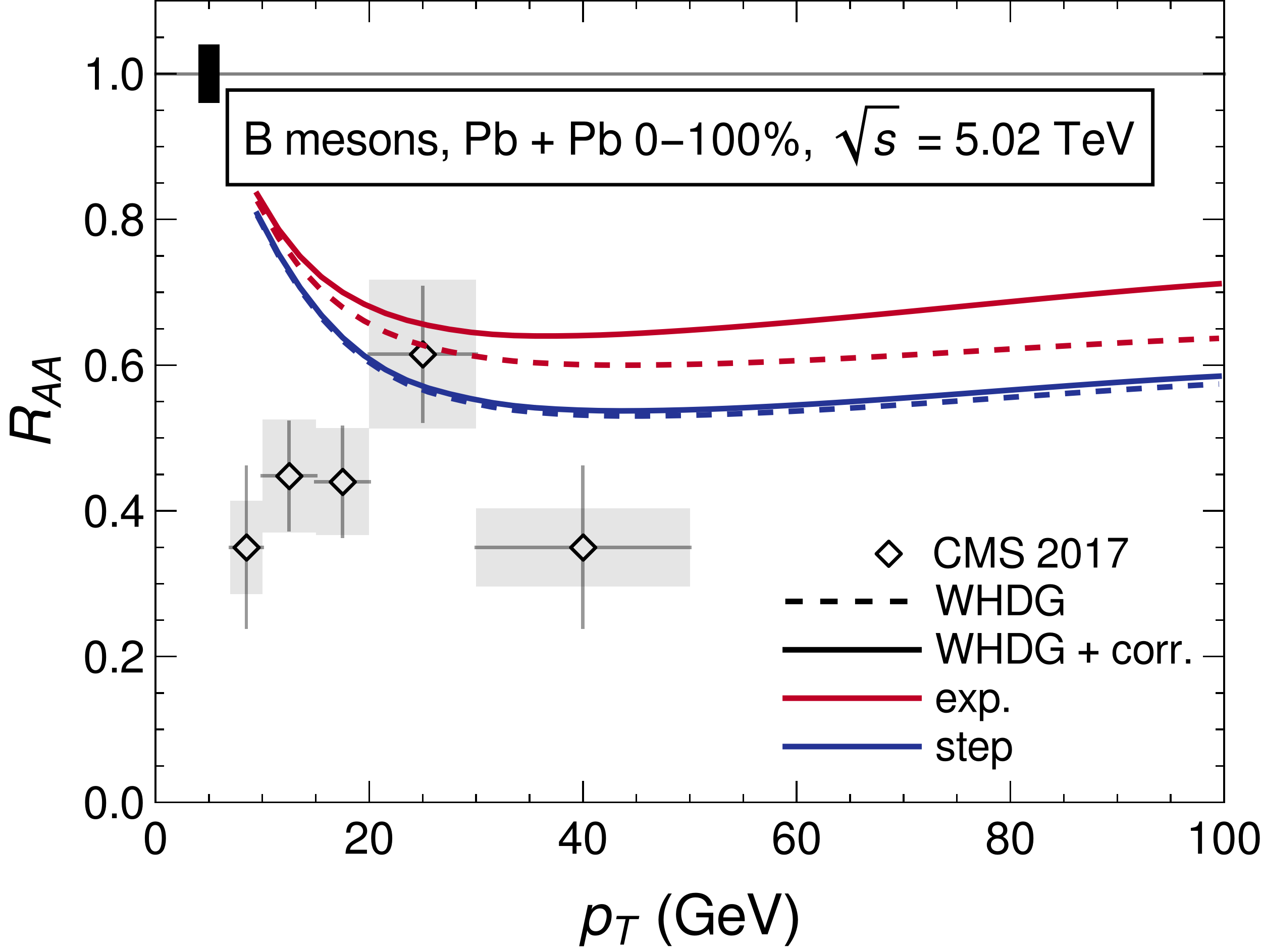}
        \caption{$R_{AA}$ for $B$ mesons in $\mathrm{Pb} + \mathrm{Pb}$ collisions.}
        \label{fig:SPL_raa_B_mesons_with_step}
    \end{subfigure}
    \caption{Plot of the nuclear modification factor $R_{AA}$ as a function of final transverse momentum $p_T$ in $\sqrt{s} = 5.02$ TeV $\mathrm{Pb} + \mathrm{Pb}$ collisions. Predictions with (solid) and without (dashed) the short pathlength correction to the radiative energy loss are shown using the exponential (red) and truncated step (blue) distributions for the scattering centers. Data are from ALICE \cite{ALICE:2018lyv} and CMS \cite{CMS:2017uoy}. The global normalization uncertainty on the number of binary collisions is indicated by solid boxes in the top left corner of the plot.}
    \label{fig:SPL_raa_mesons_with_step}
\end{figure}

\Cref{fig:SPL_raa_D_mesons_with_step} shows $R_{AA}(p_T)$ for $D$ mesons in central 0--10\% and semi-central 30--50\% $\mathrm{Pb}+\mathrm{Pb}$ collisions at $\sqrt{s}=5.02$ TeV for both the exponential and truncated step distributions of scattering centers. \Cref{fig:SPL_raa_B_mesons_with_step} shows $R_{AA}(p_T)$ for $B$ mesons in central 0--10\% $\mathrm{Pb}+\mathrm{Pb}$ collisions at $\sqrt{s}=5.02$ TeV for both the exponential and truncated step distributions of scattering centers.  For both heavy meson cases the truncated step distribution decreases the $R_{A A}$ by up to 10\% (20\%) for the original (corrected) WHDG results. For both $D$ and $B$ mesons with a truncated step distribution, the short pathlength correction is negligible for $p_T \lesssim 100$. As seen before, with an exponential distribution of scattering centers the effect of the short pathlength correction is $\lesssim 10\%$, which is small compared to other theoretical uncertainties in the model (e.g.\ higher orders in $\alpha_s$, treatment of the early times, etc.). Agreement with data is good for all predictions (corrected/uncorrected, and exp./trunc. step distributions), except for $p_T \lesssim 10$ where bulk effects may be important and the eikonal approximation is likely breaking down.

\begin{figure}[H]
  \centering
  \includegraphics[width=0.5\linewidth]{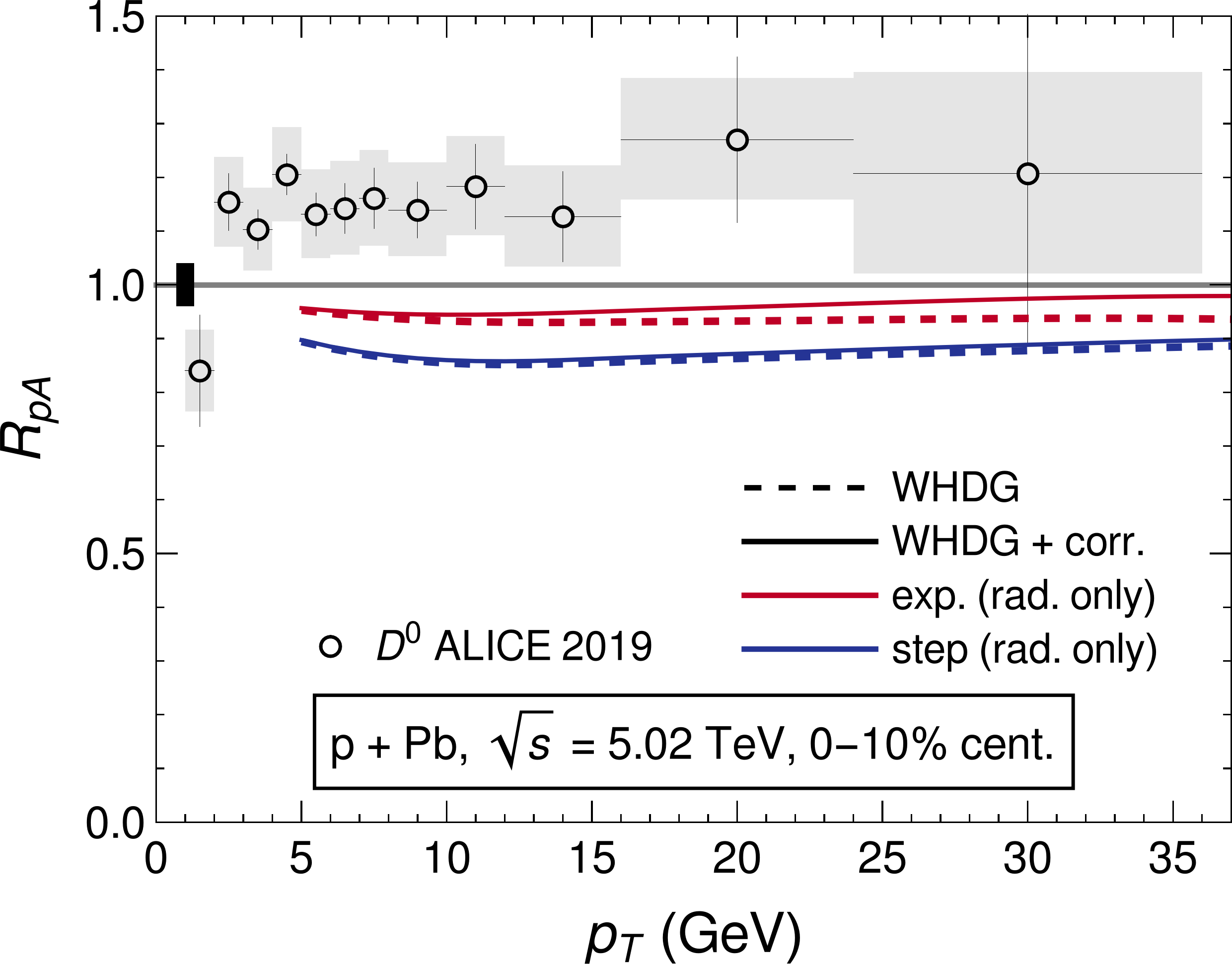}
    \caption{The nuclear modification factor $R_{pA}$ for $D$ mesons as a function of final transverse momentum $p_T$ in 0--10\% central $\mathrm{p} + \mathrm{Pb}$ collisions at $\sqrt{s}=5.02$ TeV.  Only radiative energy loss is included; predictions with (solid) and without (dashed) the short pathlength correction are shown using the exponential (red) and truncated step (blue) distributions for the scattering centers.  Data are from ALICE \cite{ALICE:2019fhe}. The experimental global normalization uncertainty on the number of binary collisions is indicated by the solid box in the top left corner of the plot.}
  \label{fig:SPL_raa_D_mesons_pA_with_step}
\end{figure}

The radiative-only nuclear modification factor $R_{pA}$ for $D^0$ mesons in 0--10\% most central $\mathrm{p}+\mathrm{A}$ collisions is shown in \cref{fig:SPL_raa_D_mesons_pA_with_step}, with and without the short pathlength correction, and with both an exponential and a truncated step distribution of scattering centers. We only show the radiative-only $R_{p A}$ since in \cref{sec:SPL_results} we determined that the elastic contribution was erroneously large in $\mathrm{p}+\mathrm{A}$ collisions due to the inapplicability of the average elastic energy loss in previous WHDG calculations in small collision systems. The $R_{pA}$ for the exponential distribution without the correction and the truncated step distribution with and without the correction all agree with each other to within $5\%$ and predict mild suppression of $R_{pA} \approx 0.9$ for all $p_T$. The corrected $R_{pA}$ with an exponential distribution predicts mild suppression at low $p_T \lesssim 20 ~\mathrm{GeV}$ and consistency with unity at moderate $p_T \gtrsim 20 ~\mathrm{GeV}$. Measured $R_{p A}$ is shown for $D^0$ mesons produced in 0--5\% most central $\mathrm{p} + \mathrm{Pb}$ collisions from ALICE \cite{ALICE:2019fhe}. Data predicts mild enhancement for all $p_T$, not inconsistent with our results shown here.

\begin{figure}[H]
    \centering
    \begin{subfigure}[t]{0.49\textwidth}
        \centering
        \includegraphics[width=\linewidth]{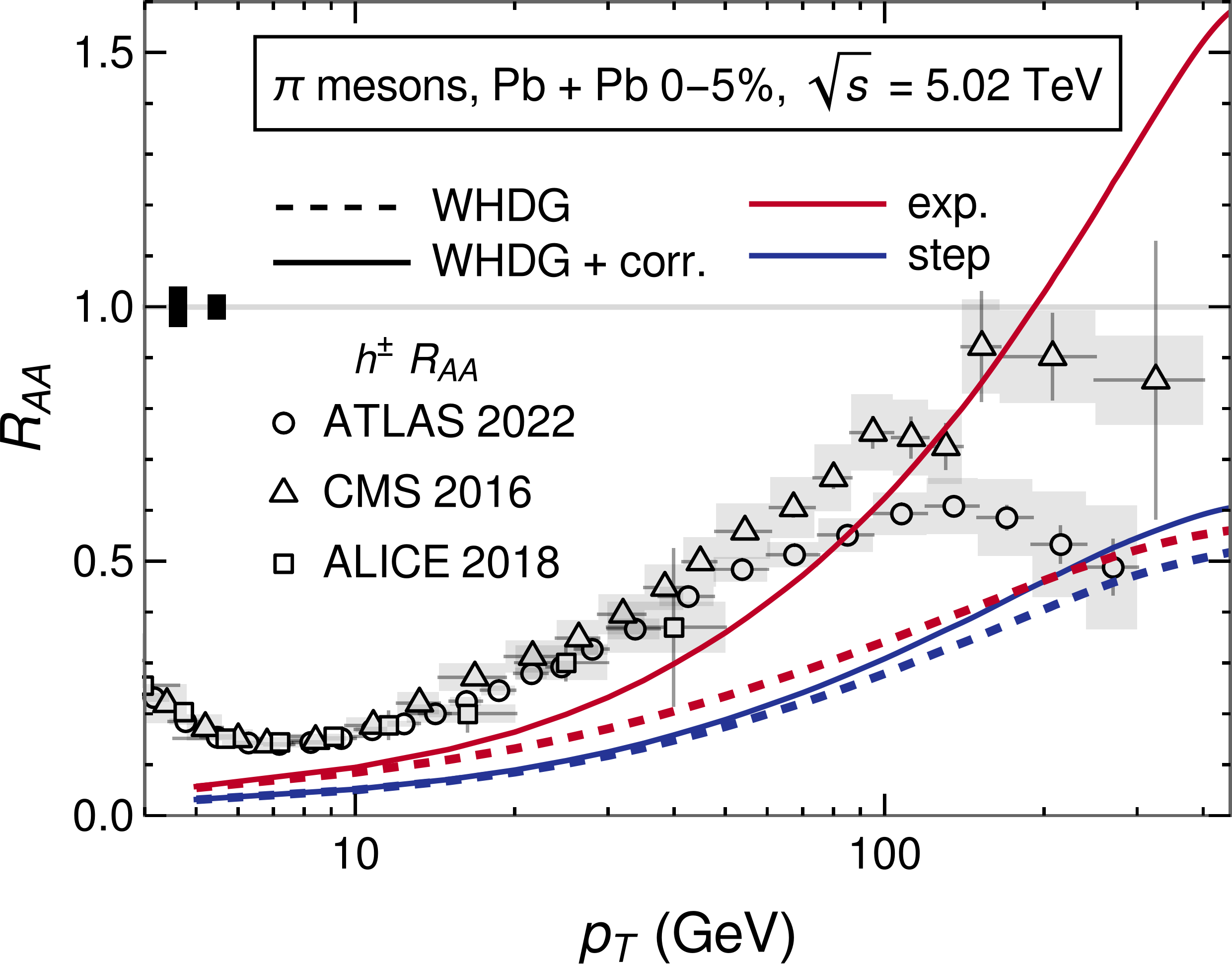}
        \caption{$R_{AA}$ for $\pi$ mesons in 0--5\% central $\mathrm{Pb} + \mathrm{Pb}$ collisions.}
        \label{fig:SPL_raa_pions_with_step}
    \end{subfigure}\hfill
    \begin{subfigure}[t]{0.49\textwidth}
        \centering
        \includegraphics[width=\linewidth]{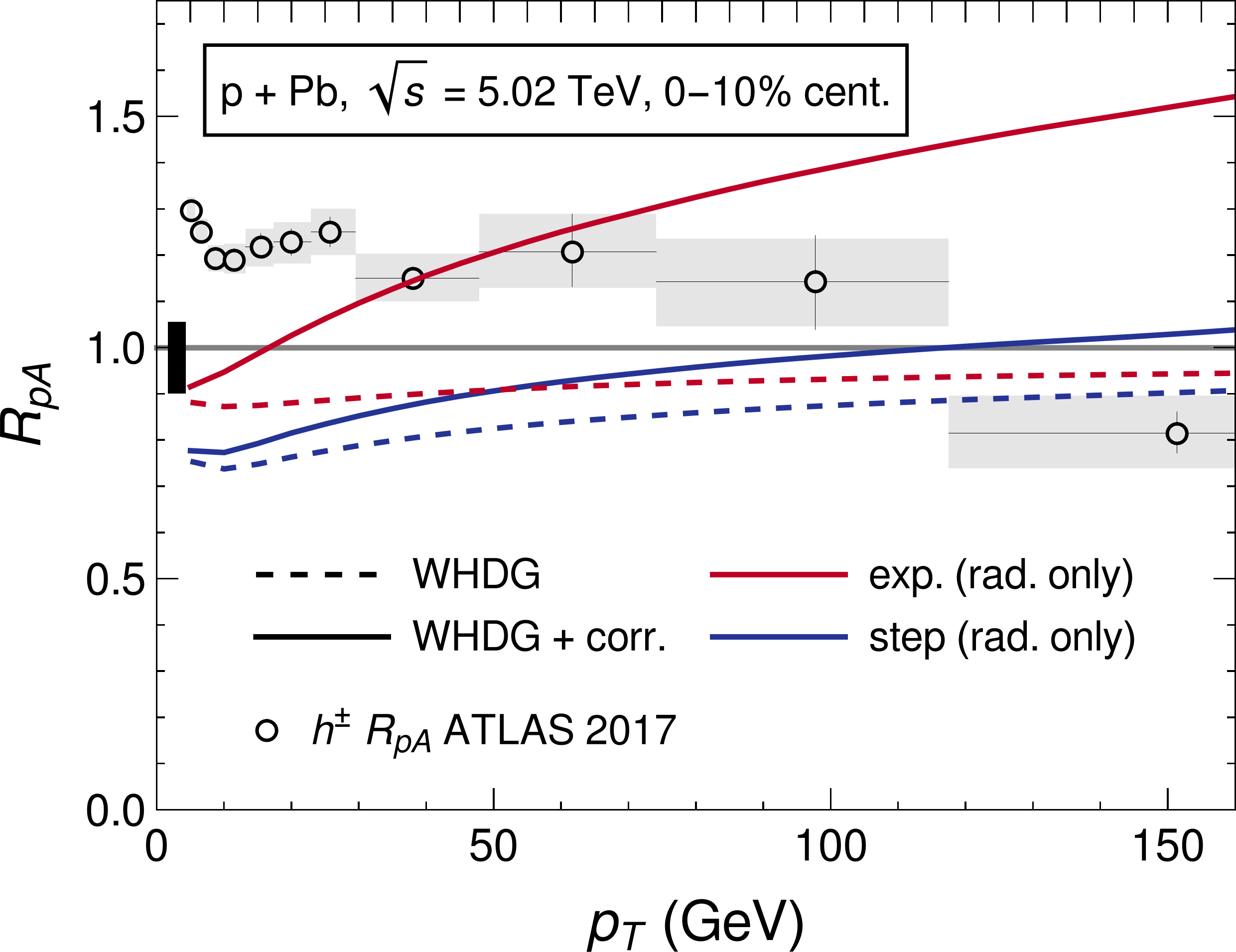}
        \caption{$R_{pA}$ for $\pi$ mesons in 0--10\% central $\mathrm{p} + \mathrm{Pb}$ collisions.}
        \label{fig:SPL_raa_pi_mesons_pA_with_step}
    \end{subfigure}
    \caption{Plot of the nuclear modification factor $R_{AA}$ and $R_{pA}$ for $\pi$ mesons as a function of final transverse momentum $p_T$ in $\mathrm{Pb} + \mathrm{Pb}$ and $\mathrm{p} + \mathrm{Pb}$ collisions at $\sqrt{s}=5.02$ TeV. Predictions with (solid) and without (dashed) the short pathlength correction to the radiative energy loss are shown using the exponential (red) and truncated step (blue) distributions for the scattering centers. Data are from ATLAS \cite{ATLAS:2022kqu, Balek:2017man}, CMS \cite{CMS:2016xef}, and ALICE \cite{Sekihata:2018lwz}. The experimental global normalization uncertainty on the number of binary collisions is indicated by solid boxes.}
    \label{fig:SPL_raa_pions_combined}
\end{figure}

\Cref{fig:SPL_raa_pions_with_step} shows $R_{AA}(p_T)$ for pions in 0--5\% central $\mathrm{Pb}+\mathrm{Pb}$ collisions at $\sqrt{s}=5.02$ TeV.  The convolved collisional and radiative energy loss model predictions both include and exclude the short pathlength correction to radiative energy loss, and use either the exponential or the truncated step distribution for the scattering centers.  The predictions are compared to data from ATLAS \cite{ATLAS:2022kqu}, CMS \cite{CMS:2016xef}, and ALICE \cite{Sekihata:2018lwz}.  While the predictions excluding the short pathlength correction are insensitive to the particular scattering center distribution chosen, one can see the tremendous sensitivity of the predictions to the scattering center distribution when the short pathlength correction to the radiative energy loss is included; when the truncated step distribution is used, the effect on $R_{AA}(p_T)$ of the short pathlength correction to the DGLV radiative energy loss is dramatically reduced.

\Cref{fig:SPL_raa_pi_mesons_pA_with_step} shows $R_{p A}(p_T)$ for $\pi$ mesons and charged hadrons in 0--10\% central $\mathrm{p}+\mathrm{Pb}$ collisions at $\sqrt{s}=5.02$ TeV.  For our theoretical predictions of pion suppression we only include radiative energy loss, as the average elastic energy loss of the WHDG model is inappropriate to use here.  We show predictions with (solid) and without (dashed) the short pathlength correction to the radiative energy loss, and we show predictions when using either the exponential (red) or truncated step (blue) distribution of scattering centers.  Charged hadron suppression data is from ATLAS \protect\cite{Balek:2017man}.

The difference between the $R^\pi_{pA}(p_T)$ predictions from the two scattering center distributions is small when excluding the short pathlength correction.  We see again that the effect on nuclear modification factor from the short pathlength correction to the radiative energy loss is large when using the exponential distribution of scattering centers and relatively small when using the truncated step distribution of scattering centers.  The prediction of enhancement by the corrected $R_{p A}$ with an exponential distribution is qualitatively similar to the observed enhancement for moderate momenta $p_T \lesssim 60~\mathrm{GeV}$. \rev{While a realistic distribution of scattering centers likely lies somewhere between the exponential and truncated step distributions, our results hint at the possibility that the experimentally measured $R_{pA}(p_T) > 1$ may be due---at least in part---to final state effects rather than initial state or normalization effects.}

\section{Discussion}
\label{sec:SPL_discussion}

The primary goal of this work, is to implement the WHDG convolved energy loss model \cite{Wicks:2005gt, Horowitz:2011gd} with the novel inclusion of the short pathlength correction to the radiative energy loss \cite{Kolbe:2015rvk}.

Due to the complexity of this model, there are many points at which one must decide on the level of approximation to proceed with. In line with the motivation of this work, we have always chosen to maintain consistency with previous work such as WHDG and DGLV. This has occasionally led us to overlook a more physically reasonable prescription (in our view) for a component of the energy loss model. One such instance concerns the treatment of the realistic collision geometry (see \cref{sec:mod_geometry}) and how it is mapped to the brick geometry. We will now present a derivation of a more realistic effective pathlength and effective temperature, which could be implemented in future work.

An integral part of radiative energy loss is the distribution of scattering centers, $\bar{\rho}(\Delta z)$, normalized to a single hard scatter at first order in opacity (see \cref{eqn:mod_density_scattering_centers}). The quantity $\bar{\rho}(\Delta z)$ is a model for the shape of the plasma, as the parton propagates through it. In theory, it is possible to improve the realism of the model by integrating through the plasma, in some sense replacing $\bar{\rho}(\Delta z)$ with a realistic plasma density $\rho(\vec{x}, \tau)$ and correspondingly $T$ with a realistic plasma temperature $T(\vec{x}, \tau)$. In this case the pathlength $L$ no longer needs to be specified \emph{a priori}. This approach would allow for a more accurate simulation of the hard partons propagation through the plasma, and is implemented similarly in the CUJET model \cite{Buzzatti:2011vt, Xu:2014ica}. Unfortunately implementing this approach is exceedingly computationally expensive, even for the original DGLV result, and would present significant computational challenges for the short pathlength corrected results. 

To account for a realistic collision geometry, we instead need \emph{effective} temperatures, densities, and lengths as inputs to the simple brick models for elastic and radiative energy loss. Essentially, we establish a brick with characteristic $\{L_{\text{eff}}, T_{\text{eff}}, \bar{\rho}\}$ for each parton that propagates through the plasma. The relationship between the plasma density $\rho(\mathbf{x}, \tau)$ and the distribution of scattering centers $\bar{\rho}(\Delta z)$, is a separation of the shape of the plasma density from its magnitude. Schematically, we can write this separation as
\begin{align}
  \frac{\mathrm{d}N^g}{\mathrm{d}x} =& \int \mathrm{d} \Delta z \; \rho(\mathbf{x}_i + \boldsymbol{\hat{\phi}}\Delta z, \Delta z) \left.\frac{\mathrm{d} N^{g}}{\mathrm{d}\Delta z \, \mathrm{d} x}\right|_{\mu = \mu(z)}
  \label{eqn:SPL_CUJET_dNdX_schematic}\\
  \approx& \int \mathrm{d} \Delta z' \; \rho(\mathbf{x}_i + \boldsymbol{\hat{\phi}}\Delta z', \Delta z') \nonumber\\
  &\times \int \mathrm{d} \Delta z \; \bar{\rho}(\Delta z) \left.\frac{\mathrm{d} N^{g}}{\mathrm{d}\Delta z \, \mathrm{d}x} \right|_{\mu = \mu_{\text{eff}}},
  \label{eqn:SPL_DGLV_dNdX_schematic}
\end{align}
where $\int \mathrm{d} \Delta z \; \bar{\rho}(\Delta z) = 1$. In the above $\mathbf{x}_i$ is the hard parton production point, $\boldsymbol{\hat{\phi}}$ is the direction of propagation, and $\rho$ is the density of the QGP. \Cref{eqn:SPL_CUJET_dNdX_schematic} is equivalent to what is done in CUJET \cite{Buzzatti:2011vt, Xu:2014ica} while \cref{eqn:SPL_DGLV_dNdX_schematic} is the approximation made in GLV \cite{Gyulassy:2000er}, DGLV \cite{Djordjevic:2003zk}, WHDG \cite{Wicks:2005gt, Horowitz:2011gd} and the short pathlength correction to DGLV \cite{Kolbe:2015rvk}. Note that the step from \cref{eqn:SPL_CUJET_dNdX_schematic} to \cref{eqn:SPL_DGLV_dNdX_schematic} is exact for a brick of plasma. 

The power of the approximation made in \cref{eqn:SPL_DGLV_dNdX_schematic} lies in separating the dependence of the path taken by the parton through the plasma from the rest of the energy loss calculation. This approach allows us to prescribe $\bar{\rho}(\Delta z)$, for instance exponential decay or truncated step, and perform the $\Delta z$ integral analytically. 
A more realistic approach to this, but less numerically intensive than integrating through the realistic plasma, would be to fit a trial $\bar{\rho}(\Delta z | \mathbf{x}_i, \phi)$ to the realistic plasma density $\rho(\mathbf{x}_i, \phi)$ for each path taken by a parton through the plasma.

The magnitude of the density is definitionally related to the opacity $\bar{n}$ via \cite{Gyulassy:2000er}
\begin{align}
  \bar{n} \equiv \frac{L_{\text{eff}}}{\lambda_{\text{eff}}} \equiv & \int d z \int d^2 \mathbf{q} \frac{d \sigma_{gg}(z)}{d^2 \mathbf{q}} \rho(\mathbf{x}_i + z \boldsymbol{\hat{\phi}}, \tau=z)\label{eqn:SPL_opacity_definition}\\
    \approx& ~ \sigma_{gg}^{\text{eff}} \int d z \;  \rho(\mathbf{x}_i + z \boldsymbol{\hat{\phi}}, \tau=z)
  \label{eqn:SPL_opacity_approximation}
\end{align}
where $\rho$ is the density from \cref{eqn:mod_effective_density}, and we have used \cref{eqn:SPL_opacity_approximation} to define the effective length $L_{\text{eff}}$ and effective gluon mean free path $\lambda_{\text{eff}}$. \Cref{eqn:mod_mean_free_path,eqn:SPL_opacity_approximation} yield
\begin{align}
    \frac{L_{\text{eff}}}{\lambda_{\text{eff}}} =& \left(\lambda_{\text{eff}} \; \rho_{\text{eff}}\right)^{-1} \int \mathrm{d} z \; \rho(\mathbf{x}_i + z \boldsymbol{\hat{\phi}}, \tau=z)\\
    \implies L_{\text{eff}} =& \frac{1}{\rho_{\text{eff}}} \int \mathrm{d} z \; \rho(\mathbf{x}_i + z \boldsymbol{\hat{\phi}}, \tau=z),
  \label{eqn:SPL_effective_length2}
\end{align}
where we have the freedom to prescribe $\rho_{\text{eff}}$. Note that \cref{eqn:SPL_effective_length2} differs from \cref{eqn:mod_effective_length} in the fact that the density is evaluated at $\tau = z$, which follows from the definition of the opacity in \cref{eqn:SPL_opacity_definition}.

Breaking apart the opacity $\bar{n}$ as $\bar{n} = L_{\text{eff}} / \lambda_{\text{eff}}$ serves to:
\begin{itemize}
  \item make contact with the effective pathlength prescription in WHDG \cite{Wicks:2005gt}, with a more rigorous derivation;
  \item obtain a prescription for the effective density $\rho_{\text{eff}}$ which can then be used to calculate other thermodynamic quantities in \cref{eqn:mod_thermodynamic_quantities}, most importantly the Debye mass $\mu$;
  \item and obtain a length scale $L_{\text{eff}}$ in the problem, which is important for prescribing the distribution of scattering centers $\bar{\rho}(\Delta z)$.
\end{itemize}

In this manuscript, we have followed WHDG in using \cref{eqn:mod_effective_density} as the effective density, which prescribes a single density to the entirety of the plasma. A more natural approach can be motivated by considering the step form \cref{eqn:SPL_CUJET_dNdX_schematic} to \cref{eqn:SPL_DGLV_dNdX_schematic}. In this step, we have approximated $\mu(z) \approx \mu_{\text{eff}}$, which leads to a natural definition for $\rho_{\text{eff}}$ as the average temperature along the path through the plasma, weighted by the plasma density.
\begin{equation}
  \rho_{\text{eff}} = \langle \rho \rangle(\mathbf{x}_i, \phi) = \frac{\int \mathrm{d} \Delta z \; \rho^2(\mathbf{x}_i + \boldsymbol{\hat{\phi}} \Delta z, \Delta z)}{\int \mathrm{d} \Delta z \; \rho(\mathbf{x}_i + \boldsymbol{\hat{\phi}} \Delta z, \Delta z)}.
  \label{eqn:SPL_effective_density2}
\end{equation}
This means both $L_{\text{eff}}$ and $\rho_{\text{eff}}$ depend on the specific path that the parton takes. This dramatically increases the numerical complexity as we must now evaluate the energy loss distribution for a distribution in $(L_{\text{eff}}, \rho_{\text{eff}})$. Note that Bjorken expansion is naturally taken into account with this prescription for the effective density $\rho_{\text{eff}}$ and so we do not need to use the approximation in \cref{eqn:mod_bjorken_expansion}.

This approach could be implemented in future work, offering the advantage of a more realistic collision geometry compared to the implementation in this work and WHDG \cite{Wicks:2005gt}; while still being significantly less computationally expensive than integrating through the realistic plasma as in CUJET \cite{Xu:2014ica}.

\section{Conclusions}
\label{sec:SPL_conclusions}

In this chapter we presented the first predictions for the suppression of leading high-$p_T$ hadrons from an energy loss model with explicit short pathlength corrections to the radiative energy loss.  We included collisional energy loss in the model, as well as averages over realistic production spectra for light and heavy flavor partons that propagate through a realistic QGP medium geometry generated by second order viscous hydrodynamics.  Thus our calculations here are, to a very good approximation, those of the WHDG energy loss model \cite{Wicks:2008zz}, but with short pathlength corrections \cite{Kolbe:2015rvk} to the DGLV opacity expansion \cite{Gyulassy:2000er,Djordjevic:2003zk}.  Predictions were presented for central and semi-central $\mathrm{Pb}+\mathrm{Pb}$ collisions and central $\mathrm{p}+\mathrm{Pb}$ collisions and compared to data from the LHC.  

We saw that the inclusion of the short pathlength correction to the radiative energy loss led to a \emph{reduction} of the suppression of leading hadrons.  This reduction is well understood as a result of the short pathlength correction enhancing the effect of the destructive LPM interference between the zeroth order in opacity DGLAP-like production radiation and the radiation induced by the subsequent collisions of the leading parton with the medium quanta.  The reduction in the suppression also increases as a function of $p_T$, which is a result of the different asymptotic energy scalings of DGLV energy loss ($\Delta E\sim\log E$) compared to the short pathlength correction ($\Delta E\sim E$).  

For heavy flavor observables, the inclusion of the short pathlength corrections leads to only a modest $\sim10$\% enhancement of $R_{AA}(p_T)$ in $\mathrm{Pb}+\mathrm{Pb}$ and $\mathrm{p}+\mathrm{Pb}$ collisions.  Even though the relative $R_{pA}$ enhancement of $\sim10\%$ is similar to that of $R_{AA}$, as one can see in \cref{fig:mod_deltaEoverE_vs_L_gluon} the influence of the short pathlength correction on the energy loss is significantly larger for shorter pathlengths and therefore also in the smaller collision system.  The reason that the $R_{pA}$ and $R_{AA}$ have similar relative enhancements is due to the scaling $R_{AA}\sim(1-\epsilon)^{n-1}$, where $\epsilon\equiv\Delta E/E$ is the fractional energy lost by the leading parton.  One can determine that the effective short pathlength correction, averaged over the Poisson convolution and geometry, is about 100\% stronger in $\mathrm{p}+\mathrm{Pb}$ compared to central $\mathrm{Pb}+\mathrm{Pb}$.  (Note that the Poisson convolution, with its large probability of no interaction or energy loss for short pathlengths, is crucial for in fact \emph{reducing} the enormous short pathlength correction influence seen in \cref{fig:mod_deltaEoverE_small_large,fig:mod_deltaEoverE_small_large_gluon} in the energy loss model.)  Thus, since $R_{pA}\sim1$ and $n\sim6$, even though the short pathlength correction is about 100\% stronger in $\mathrm{p}+\mathrm{Pb}$, the influence on $R_{pA}$ is similar to that in $R_{AA}$.

When we assume that the distribution of scattering centers that stimulate the emission of gluon radiation from high-$p_T$ partons is given by an exponential, which biases the leading partons to scatter at shorter distances, the short pathlength correction to $R_{AA}^\pi(p_T)$ grows dramatically with $p_T$.  This very fast growth in $p_T$ is due to the very large short pathlength correction to the gluonic radiative energy loss: the short pathlength correction to radiative energy loss breaks color triviality, and the correction to gluonic radiative energy loss is about ten times that of quark radiative energy loss (instead of the approximately factor of two that one would expect from color triviality) \cite{Kolbe:2015rvk}.  When a truncated step function is used as the distribution of scattering centers that stimulate the emission of gluon radiation, the short pathlength correction to $R^\pi_{AA}(p_T)$ becomes a much more modest $\sim10$\%.  It is interesting, but perhaps not totally surprising, that the short pathlength correction to the energy loss introduces an enhanced sensitivity to the precise distribution of scattering centers used in the energy loss model.  In either case of distributions of scattering centers, the more rapid growth in the short pathlength correction as a function of $p_T$ than that of the uncorrected DGLV energy loss \cite{Horowitz:2011gd} suggests that at least part of the faster-than-expected growth of the measured $R^\pi_{AA}(p_T)$ as a function of $p_T$ may be due to the influence of the short distance corrections to the energy loss of hard partons; cf.\ the reduction of suppression due to running coupling \cite{Buzzatti:2012dy}. 

We found that the average collisional energy loss, with fluctuations of this energy loss given by a Gaussian whose width is dictated by the fluctuation-dissipation theorem, of the WHDG energy loss model \cite{Wicks:2008zz} is inappropriate for small colliding systems.  Considering radiative energy loss only, $R_{pA}(p_T)$ for heavy flavor hadrons was again only modestly, $\lesssim10$\%, affected by the short pathlength correction to energy loss.  $R_{pA}^\pi(p_T)$ was significantly affected, $\sim 50$\%, by the short pathlength correction for an exponential distribution of scattering centers, but more modestly so, $\sim 10$\%, for a truncated step distribution of scattering centers.  For both distributions of scattering centers, the short pathlength corrected pion nuclear modification factor sees a tantalizing enhancement above 1, similar to data \cite{Balek:2017man,ALICE:2018lyv}.  One may then provocatively suggest that the experimentally measured enhancement of $R_{pA}(p_T)>1$ may be due---at least in part---to final state effects.  

We also investigated the self-consistency of the approximations used in the derivations of DGLV and short pathlength correction to DGLV single inclusive radiative gluon emission kernels for the phenomenologically relevant physical situations of RHIC and LHC.  We constructed dimensionless quantities that represented the approximations and checked whether, when averaged with a weight given by the strength of the energy loss kernel determined by DGLV or the short pathlength corrected DGLV, those quantities were small (or large) as required by the approximations that went into deriving those same DGLV and short pathlength corrected DGLV single inclusive radiated gluon spectrum kernels.  

We found that, when weighted by the energy loss kernel, the soft and collinear approximations were self-consistently satisfied when computed with phenomenologically relevant parameters.  
We further found that both the original DGLV derivation and the DGLV derivation with the inclusion of the short pathlength correction are \emph{not} consistent with the large formation time approximation for modest $\mathcal O(10\text{--}100 \, \mathrm{GeV})$ energies and pathlengths $\mathcal O(1\text{--}5 \, \mathrm{fm})$, independent of the choice of distribution of scattering centers.  The short pathlength corrected DGLV is similarly inconsistent with the large formation time approximation also for modest $\mathcal O(10\text{--}100 \, \mathrm{GeV})$ energies for pathlengths $\mathcal O(1\text{--}5 \, \mathrm{fm})$ for either the exponential or the truncated step function distribution of scattering centers. Finally, we see that the large pathlength approximation breaks down for small pathlengths $\sim 1$ fm and even for large pathlengths $\sim5$ fm for large enough $\gtrsim100$ GeV energies.  We noted in \cref{sec:mod_radiative_energy_loss} one more assumption, that of large transverse area.  This assumption is very difficult to assess using the methods of this article, especially as the utilization of the assumption occurs very early in the derivation.  Qualitatively, even in $\mathrm{p}+\mathrm{A}$ collisions, one has that the transverse size of the system will be $\sim(1\,\mathrm{fm})^2$ whereas the typical scale set by the scattering process itself is $1/\mu^2\sim(0.5\,\mathrm{fm})^2$.  It thus seems likely that this large transverse size assumption holds even for small collision systems.

Instead of thinking of the self-consistency of the numerics with the assumptions that went into the derivation of the energy loss, one may rather formulate the issue as whether or not one is integrating the matrix element (modulus squared) beyond the region under which its derivation is under control.  Thus one way of understanding that, e.g., $\langle \omega_0/\mu_1\rangle>1$ as shown in \cref{fig:SPL_largeformationtime3} is that the matrix element is integrated over regions of kinematics under which the derivation is not under control.  One may consider restricting the kinematics that are integrated over to those for which the derivation is under control.  We find, for example, that the expectation of the collinear approximation $\langle k^-/k^+\rangle$ is self-consistently less than 1, but note that the gluon kinematics are restricted in such a way as to enforce collinearity $|\mathbf{k}|^{\text{max}}=2x(1-x)E \Leftrightarrow k^- < k^+$.  As was shown in \cite{Horowitz:2009eb,Armesto:2011ht}, the DGLV inclusive gluon emission kernel is \emph{not} under good control near the kinematic bound $k^-\sim k^+$.  One may thus consider restricting the gluon kinematics such that the large formation time assumption is respected, for example, by taking $|\mathbf{k}|^{\text{max}} = \text{Min}(\sqrt{2xE\mu_1},2x(1-x)E) \Leftrightarrow \omega_0 < \mu_1$ and $k^- < k^+$.  Then presumably one would find that $\langle \omega_0/\mu_1\rangle\lesssim1$.  However, as was shown in \cite{Horowitz:2009eb}, there would then be a significant sensitivity in the energy loss model predictions to the exact kinematic bound chosen.

We thus conclude that in order to confidently compare leading hadron suppression predictions in $\mathrm{A}+\mathrm{A}$ collisions at $\gtrsim100$ GeV or in $\mathrm{p}+\mathrm{A}$ collisions at $\gtrsim 10$ GeV from an energy loss model based on an opacity expansion of the single inclusive gluon emission kernel, future work is needed to re-derive the opacity expansion single inclusive gluon emission kernel with both the large pathlength and the large formation time approximations relaxed.  Further work of numerically implementing finite pathlength effects in elastic energy loss \cite{Djordjevic:2006tw,Wicks:2008zz} will also play an important role in any quantitative comparison of an energy loss model and leading hadron suppression in small systems such as $\mathrm{p}+\mathrm{A}$ collisions.

 \chapter{Collisional and radiative energy loss in small systems}
 \label{sec:EL_paper}
 
\section*{Comment}

Chapter 5 of this thesis is an almost verbatim reproduction of the preprint manuscript \cite{Faraday:2024gzx} which will be submitted to a journal shortly. The only difference between the paper and this chapter, is that the model section of the paper was combined with the model section of a published paper \cite{Faraday:2023mmx} and was presented in \cref{sec:model}.

This paper was written in collaboration with Dr.\ W.\ A.\ Horowitz, and I have permission from my coauthor to reproduce this work in this thesis. I was the first author of the paper, and wrote the first draft of the paper. Comments were provided on multiple drafts of the manuscript by WAH, and I wrote all drafts of the manuscript. All numerics were performed by me with my own code, except for the codes for the fragmentation functions and production spectra which were provided to me by WAH. I produced all figures in the manuscript. It was the idea of both WAH and myself to incorporate the more realistic Poissonian elastic energy loss in the calculation. It was my idea to explain the surprising similarity of the Poisson and Gaussian elastic energy loss results with an expansion of the $R_{AA}$ in terms of the moments of the underlying energy loss distributions, as well as the proportion of elastic vs.\ radiative energy loss in \cref{sec:EL_surprising_similarity_gaussian_poisson}. I realized that the derivation of the power law approximation to the $R_{AA}$ was incorrect, and provided a new derivation of the power law $R_{AA}$ in \cref{sec:EL_validity_of_power_law_approximation_to_r_aa}. Conception of the project was joint between me and WAH.

\newpage

\section{Introduction}

Due to their sensitivity to final state in-medium effects, hard probes, including jets and leading hadrons, are a crucial component of the heavy-ion program at RHIC and LHC. One of the most useful hard probe observables is the nuclear modification factor $R_{AA}$, which captures the amount of in-medium energy loss suffered by high-$p_T$ partons. A measured $R_{AA} \sim 0.2$ for leading hadrons \cite{PHENIX:2001hpc, STAR:2002ggv} in central \coll{Au}{Au} collisions at RHIC---when compared to the photon $R_{AA} \simeq 1$ in the same collision \cite{PHENIX:2005yls}---is strong evidence of medium modifications to the hadron spectra. 
When one contrasts this large suppression with $R_{dA} \simeq 1$ of pions produced in minimum bias \coll{d}{Au} collisions \cite{PHENIX:2003qdw, STAR:2003pjh} and the independence of the $R_{AA}$ on the final state light hadron mass in \coll{Au}{Au} collisions \cite{PHENIX:2006mhb}, one is driven to conclude that the modification of the spectrum is due primarily to partonic energy loss. %
Various semi-classical perturbative Quantum Chromodynamics (pQCD) models, which make different sets of approximations, have been successful in making qualitative predictions for the energy loss of hard probes in \coll{A}{A} collisions \cite{Dainese:2004te, Schenke:2009gb, Wicks:2005gt, Horowitz:2012cf}.

The $p_T \sim 5\text{--}20$ GeV charged leading hadron $R_{AA}\sim 0.2$ in central \coll{Pb}{Pb} collisions at LHC \cite{ALICE:2010mlf} tells a similar story, when compared to null controls of the photon \cite{CMS:2012oiv} and $Z$ boson \cite{CMS:2011zfr} $R_{AA}\sim 1$, and charged leading hadron $R_{AA}\sim 1$ in minimum bias \coll{p}{Pb} collisions.
The LHC also allows for additional insight due to its high $\sqrt{s}$, which leads to a hotter medium and its higher integrated luminosity, which allows for better statistics.
The high integrated luminosity additionally allows for the measurement of hadrons which fragment from heavy quarks \cite{ALICE:2012ab, ALICE:2012ab, ALICE:2015vxz}, produced in the earliest stages of the collision, making them an invaluable tomographic tool in probing the QGP. Heavy-quarks also probe the mass and color charge dependence predicted by energy loss models (see \cite{Andronic:2015wma} for a review).

Another vital set of insights from RHIC and LHC was the near-perfect fluidity of the strongly-coupled low momentum modes of the QGP formed in semi-central \coll{Au}{Au} \cite{STAR:2000ekf, PHENIX:2003qra} and \coll{Pb}{Pb} \cite{ALICE:2010suc} collisions, as inferred by relativistic, viscous, hydrodynamic model predictions \cite{Romatschke:2007mq, Song:2007ux, Schenke:2010nt}.

More recently, the very same collective signatures of QGP formation have been experimentally observed at both RHIC and LHC in high multiplicity \coll{p}{p} \cite{ATLAS:2015hzw, ALICE:2023ulm}, \coll{p}{Pb} \cite{ATLAS:2013jmi, ALICE:2014dwt, CMS:2015yux} and \collFour{p}{d}{He3}{Au} \cite{PHENIX:2013ktj, PHENIX:2014fnc, PHENIX:2015idk, PHENIX:2016cfs, PHENIX:2017xrm}
collisions, which are qualitatively consistent with predictions from hydrodynamic models \cite{Weller:2017tsr, Schenke:2020mbo}.
In addition to these collective signatures, other QGP signatures have been detected in these small systems, including quarkonium suppression \cite{ALICE:2016sdt} and strangeness enhancement \cite{ALICE:2015mpp, ALICE:2013wgn}, which appear to depend only on the final multiplicity and not the collision system.
This evidence suggests that small droplets of QGP are formed at high multiplicity in even the smallest collision systems.

	Operating under the premise that QGP forms in central small collision systems, it follows that final state energy loss suffered by high-$p_T$ partons moving through the medium should result in a nuclear modification factor less than one in these collisions. Experimental measurements of the nuclear modification factor $R_{AB}$ in small systems from RHIC and LHC provide an inconclusive suppression pattern, particularly as a function of centrality. ALICE \cite{ALICE:2016yta}  and ATLAS \cite{ATLAS:2022kqu} report leading hadron nuclear modification factors in central \coll{p}{Pb} collisions consistent with no suppression, and even a $20\%$ \emph{enhancement}. PHENIX \cite{PHENIX:2021dod} measures the nuclear modification factor in central \collFour{p}{d}{He3}{Au} as $R_{AB} \simeq 0.75$, with a pronounced enhancement in peripheral collisions.

However, measuring the centrality-dependent $R_{AB}$ in small systems presents experimental challenges, notably centrality bias \cite{ALICE:2014xsp, PHENIX:2013jxf, Kordell:2016njg, PHENIX:2023dxl, Bzdak:2014rca}. In the determination of the nuclear modification factor, the Glauber model \cite{Glauber:1970jm,Miller:2007ri} is typically used to map between the measured event activity, related to the number of soft particles produced, and the number of binary collisions, related to the number of hard particles produced. Centrality bias is a non-trivial correlation between the hard and soft particles, which may lead to an incorrect normalization---the number of hard collisions---for the nuclear modification factor.
A recent PHENIX study \cite{PHENIX:2023dxl} experimentally measures the number of binary collisions using the prompt photon $R_{AB}$, which they then use to determine the $R_{AB}$ for pions produced in central \coll{d}{Au} collisions, independently of the standard Glauber model mapping. They find $R_{AB} \simeq 0.75$ for central \coll{d}{Au} collisions, that the enhancement \cite{PHENIX:2021dod} in peripheral \coll{d}{Au} collisions disappears, and that the $0\text{--}100\%$ centrality $R_{AB}$ is consistent with the minimum bias value. These results are qualitatively consistent with a picture of QGP formation in central \coll{d}{Au} collisions wherein final state energy loss leads to $R_{AB} < 1$. Other experimental work on this front includes measuring self-normalized observables which are not sensitive to this centrality bias \cite{ATLAS:2022iyq}, further understanding of high-$p_T$ $v_2$ measurements \cite{ATLAS:2019vcm}, and minimum bias measurements in intermediate-sized systems like \coll{O}{O} \cite{Huss:2020whe, Huss:2020dwe}. 

Given these experimental challenges, theoretical input becomes a vital and complementary approach to determining the consistency of suppression measurements in small systems with a final state energy loss scenario.

However, applying existing theoretical models---which have seen success in broadly describing data in large systems---to small systems presents its own set of challenges. Many of the assumptions that underlie canonical semi-classical pQCD energy loss models are likely to be inapplicable in small collision systems \cite{Faraday:2023mmx}, making quantitative predictions in small systems difficult. An instance of an explicit assumption that the system size is large is in the dropping of terms exponentially suppressed according to the system size, which all radiative energy loss approaches based on the opacity expansion \cite{Gyulassy:2000er, Djordjevic:2003zk, Zakharov:1997uu, Baier:1996kr} do.  
A correction that includes these previously neglected terms has been derived \cite{Kolbe:2015rvk,Kolbe:2015suq}, and we discussed the phenomenological implications of this correction for small and large systems in our previous work \cite{Faraday:2023mmx}. 

BDMPS-Z based models \cite{Baier:1996kr, Baier:1996sk, Baier:1996vi, Baier:1998kq, Zakharov:1996fv, Zakharov:1997uu} rely on applying the central limit theorem \cite{Armesto:2011ht}, by assuming that the high-$p_T$ parton undergoes a large number of collisions. The $\mathcal{O}(5)$ scatters \cite{Armesto:2011ht} that are present in large collision systems does not well motivate this assumption in large collision systems, let alone in small collision systems with $\mathcal{O}(0\text{--}1)$ collisions \cite{Faraday:2023mmx}. %
A similar assumption is commonly utilized for the elastic energy loss, where it is assumed that incident high-$p_T$ partons undergo a large number of collisions, which allows one to model the elastic energy loss probability distribution as Gaussian \cite{Wicks:2005gt, Moore:2004tg, Zigic:2021rku, Faraday:2023mmx, Djordjevic:2013xoa}. %

In our previous work \cite{Faraday:2023mmx}, we found that elastic energy loss comprises almost all energy loss in small systems, which we attributed to the breakdown of the assumption that the elastic energy loss distribution is Gaussian according to the central limit theorem. We reasoned that the small number of scatters present in small systems would result in a large probability weight associated with no energy loss, which cannot be captured by the Gaussian distribution, motivating a study of the validity of the Gaussian assumption.

In this work, we utilize an elastic energy loss kernel \cite{Wicks:2008zz}, which is derived in the HTL formalism but keeps the full kinematics of the hard exchanges (as opposed to a strict HTL calculation). With this approach, one may calculate the differential number of elastic scatters, allowing insight into the shape of the elastic energy loss distribution. This facilitates the treatment of the elastic and radiative energy loss in an identical manner by assuming that the number of scatters and number of radiated gluons emitted are separately independent and, therefore, that we may convolve the distributions according to a Poisson distribution \cite{Gyulassy:2001nm}.  
We also present results where the HTL elastic energy loss is modeled as a Gaussian distribution according to the central limit theorem \cite{Moore:2004tg}, which allows us to assess the validity of the Gaussian approximation. Finally, we compare the HTL-based and the Braaten and Thoma elastic energy loss models, which differ predominantly in the usage of vacuum and HTL propagators, to probe the theoretical uncertainty in the transition between HTL and vacuum propagators \cite{Romatschke:2004au,Gossiaux:2008jv,Wicks:2008zz}.	

	In this paper we present nuclear modification factor $R_{AB}$ results from our model, which receives small system size corrections to both the elastic and radiative energy loss. We present model results for final state pions, $D$ mesons, and $B$ mesons produced in central, semi-central, and peripheral \coll{Au}{Au} and \coll{Pb}{Pb} collisions as well as central \coll{p}{Pb} and \collFour{p}{d}{He3}{Au} collisions.
	We produce various model results by varying the elastic energy loss distribution between the Poisson HTL, Gaussian HTL \cite{Wicks:2008zz}, and Gaussian Braaten and Thoma \cite{Braaten:1991jj, Braaten:1991we} distributions, and by varying the radiative energy loss distributions between the Poisson DGLV radiative energy loss \cite{Djordjevic:2003zk, Gyulassy:2001nm} and the Poisson DGLV radiative energy loss which receives a short pathlength correction \cite{Kolbe:2015rvk, Kolbe:2015suq}. These various model results allow us to understand the phenomenological effect of the central limit theorem approximation used in the literature \cite{Wicks:2005gt, Horowitz:2011gd, Faraday:2023mmx,Zigic:2021rku}, the uncertainty in calculating the elastic energy loss with Hard Thermal Loops (HTL) vs.\ vacuum propagators, and the short pathlength correction to the radiative energy loss \cite{Faraday:2023mmx, Kolbe:2015suq, Kolbe:2015rvk}.

\section{Results}

In this section, we present nuclear modification factor results from our model for final-state pions, $D$ mesons, and $B$ mesons produced in \coll{p}{Pb} and \coll{Pb}{Pb} collisions at the LHC, as well as for final-state pions produced in \collFour{p}{d}{He3}{Au} collisions, and pions and $D$ mesons produced in \coll{Au}{Au} collisions at RHIC.

	In this work, our objective is to \emph{qualitatively} understand the influence of small system size corrections to the elastic and radiative energy loss on the nuclear modification factor in different collision systems. For this reason, we produce all theoretical curves with a constant coupling of $\alpha_s = 0.3$ and make no comparison with data; however, the choice of displayed collision systems \emph{is} motivated by available RHIC and LHC data. Future work in preparation \cite{Faraday:2024} will examine the extent to which both small and large system suppression data can be \emph{quantitatively and simultaneously} described by pQCD final state energy loss models, through a one-parameter fit to data of the strong coupling $\alpha_s$.

There are six sets of results presented for each collision system, which are constructed by varying both the radiative and elastic energy loss kernels. The radiative energy loss kernel is varied between the DGLV radiative energy loss kernel (\textit{DGLV}) \cite{Djordjevic:2003zk} and the DGLV radiative energy loss kernel which receives a short pathlength correction (\textit{DGLV + SPL}) \cite{Kolbe:2015rvk, Kolbe:2015suq}. The elastic energy loss kernel is varied between the Braaten and Thoma elastic energy loss with a Gaussian distribution (\textit{Gaussian BT}), the HTL result from Wicks \cite{Wicks:2008zz} with a Poisson distribution (\textit{Poisson HTL}), and the HTL result from Wicks \cite{Wicks:2008zz} with a Gaussian distribution (\textit{Gaussian HTL}).
We presented the model implementation details for each theoretical curve in \cref{sec:model}.

\subsection{Large system suppression at LHC}
\label{sec:EL_large_system_suppression_LHC}

\begin{figure}[!htpb]
	\centering
	\includegraphics[width=0.6\linewidth]{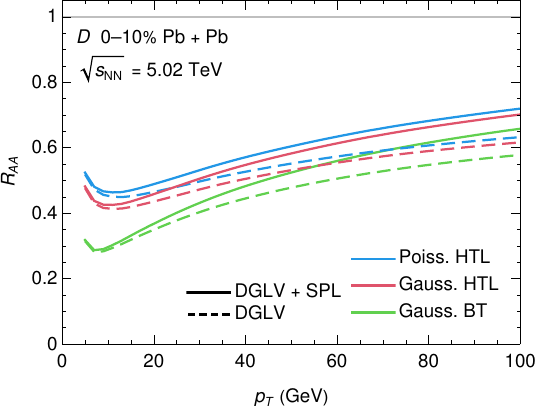}
	\caption{ Nuclear modification factor $R_{AA}$ as a function of $p_T$ for $D$ mesons produced in $0\text{--}10\%$ most central \coll{Pb}{Pb} collisions at $\sqrt{s_{NN}} = 5.02$ TeV. We produce theoretical results through our convolved radiative and elastic energy loss model by varying the elastic model between Gaussian BT \cite{Braaten:1991jj, Braaten:1991we}, Gaussian HTL, and Poisson HTL \cite{Wicks:2008zz}, and the radiative model between DGLV \cite{Djordjevic:2003zk} and DGLV + SPL \cite{Kolbe:2015rvk, Kolbe:2015suq}. }
	\label{fig:EL_raa_pbpb_D0010}
\end{figure}

\begin{figure}[!htpb]
\centering
\begin{subfigure}[t]{0.49\textwidth}
    \centering
    \includegraphics[width=\linewidth]{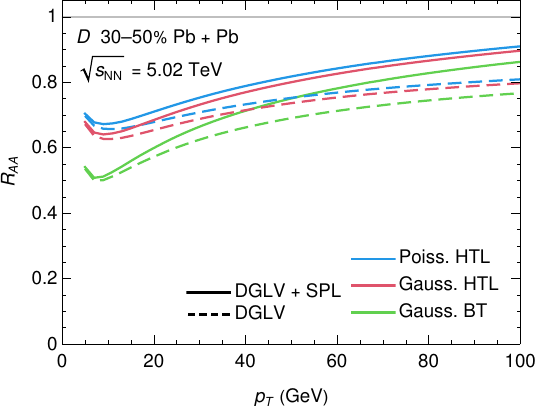}
    \caption{Nuclear modification factor $R_{AA}$ as a function of $p_T$ for $D$ mesons produced in $30\text{--}50\%$ most central \coll{Pb}{Pb} collisions at $\sqrt{s_{NN}} = 5.02$ TeV.}
    \label{fig:EL_raa_pbpb_D3050}
\end{subfigure}\hfill
\begin{subfigure}[t]{0.49\textwidth}
    \centering
    \includegraphics[width=\linewidth]{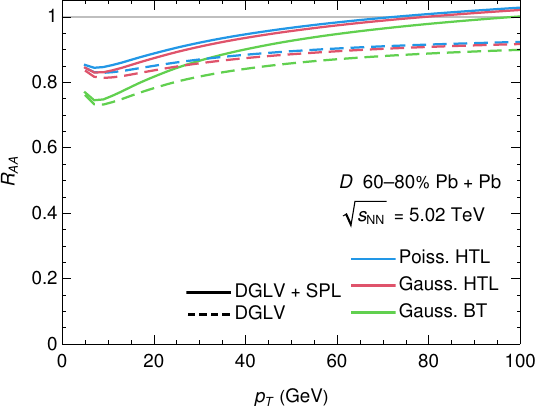}
    \caption{Nuclear modification factor $R_{AA}$ as a function of $p_T$ for $D$ mesons produced in $60\text{--}80\%$ most central \coll{Pb}{Pb} collisions at $\sqrt{s_{NN}} = 5.02$ TeV.}
    \label{fig:EL_raa_pbpb_D6080}
\end{subfigure}
\caption{Comparison of the nuclear modification factor $R_{AA}$ for $D$ mesons at different centrality ranges in \coll{Pb}{Pb} collisions at $\sqrt{s_{NN}} = 5.02$ TeV. Theoretical results are produced through our convolved radiative and elastic energy loss model, varying the elastic model between Gaussian BT \cite{Braaten:1991jj, Braaten:1991we}, Gaussian HTL, and Poisson HTL \cite{Wicks:2008zz}, and the radiative model between DGLV \cite{Djordjevic:2003zk} and DGLV + SPL \cite{Kolbe:2015rvk, Kolbe:2015suq}.}
\label{fig:EL_raa_pbpb_D_comparison}
\end{figure}

\begin{figure}[!htpb]
	\centering
	\includegraphics[width=0.6\linewidth]{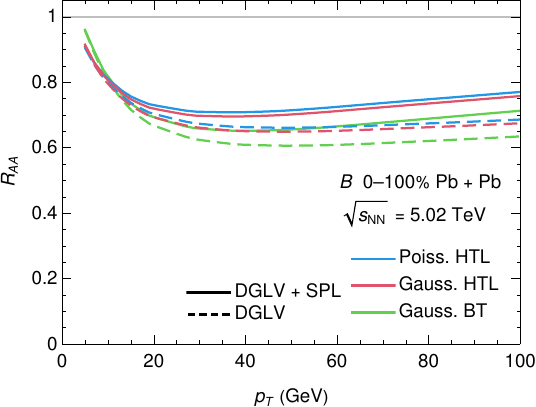}
	\caption{ Nuclear modification factor $R_{AA}$ as a function of $p_T$ for $B$ mesons produced in $0\text{--}100\%$ most central \coll{Pb}{Pb} collisions at $\sqrt{s_{NN}} = 5.02$ TeV. Theoretical results are produced through our convolved radiative and elastic energy loss model, by varying the elastic model between Gaussian BT \cite{Braaten:1991jj, Braaten:1991we}, Gaussian HTL, and Poisson HTL \cite{Wicks:2008zz}, and the radiative model between DGLV \cite{Djordjevic:2003zk} and DGLV + SPL \cite{Kolbe:2015rvk, Kolbe:2015suq}. }
	\label{fig:EL_raa_pbpb_B}
\end{figure}

\Cref{fig:EL_raa_pbpb_D0010,fig:EL_raa_pbpb_D3050,fig:EL_raa_pbpb_D6080} show the nuclear modification factor $R_{AA}$ of $D$ mesons produced in $0\text{--}10\%$, $30\text{--}50\%$, and $60\text{--}80\%$ most central \coll{Pb}{Pb} collisions at $\sqrt{s_{NN}} = 5.02$ TeV from our convolved elastic and radiative energy loss model. \Cref{fig:EL_raa_pbpb_B} shows the nuclear modification factor $R_{AA}$ of $B$ mesons produced in $0\text{--}100\%$ most central \coll{Pb}{Pb} collisions at $\sqrt{s_{NN}} = 5.02$ TeV from our convolved elastic and radiative energy loss model. We obtain theoretical model results by varying both the elastic and radiative energy loss kernels used in the calculation, as previously described.

Comparing the results calculated with the Poisson HTL elastic energy loss kernel to those calculated with the Gaussian HTL elastic energy loss kernel, shown in \cref{fig:EL_raa_pbpb_D0010,fig:EL_raa_pbpb_D3050,fig:EL_raa_pbpb_D6080,fig:EL_raa_pbpb_B}, we see that the heavy-quark $R_{AA}$ is largely insensitive to the shape of the elastic energy loss distribution. We conclude that approximating the elastic energy loss distribution as a Gaussian distribution is not important phenomenologically for heavy flavor $D$ and $B$ mesons.

At low momenta, a comparison between the Gaussian BT results and the Gaussian HTL results is sensitive to the uncertainty in the crossover region between HTL and vacuum propagators. 

We observe in \cref{fig:EL_raa_pbpb_D0010,fig:EL_raa_pbpb_D3050,fig:EL_raa_pbpb_D6080} that the $D$ meson $R_{AA}$ is acutely sensitive to this choice for all centrality classes. 
This sensitivity is significantly reduced in the $B$ meson $R_{AA}$ shown in \cref{fig:EL_raa_pbpb_B}, because the energy loss goes to zero as $p_T$ goes to zero. 

We conclude that quantitative suppression predictions that include heavy-flavor predictions, must consider the theoretical uncertainty due to the crossover between HTL and vacuum propagators in the elastic energy loss sector.

	In lieu of a theoretical framework which can resolve this uncertainty, 

future phenomenological models may use the different dependencies of elastic vs.\ radiative energy loss on system size and $p_T$ to phenomenologically extract the changeover between the HTL and vacuum propagators.

Comparing the DGLV and DGLV + SPL curves in \cref{fig:EL_raa_pbpb_D0010,fig:EL_raa_pbpb_D3050,fig:EL_raa_pbpb_D6080,fig:EL_raa_pbpb_B}, we see that the short pathlength correction has a negligible impact on the heavy flavor $R_{AA}$ in central collisions, but the size of the correction grows in $p_T$ in agreement with our previous results \cite{Faraday:2023mmx}. The relative size of the short pathlength correction grows as a function of centrality simply because of the smaller average pathlength in less central collisions.

\begin{figure}[!htpb]
    \centering
    \includegraphics[width=0.6\linewidth]{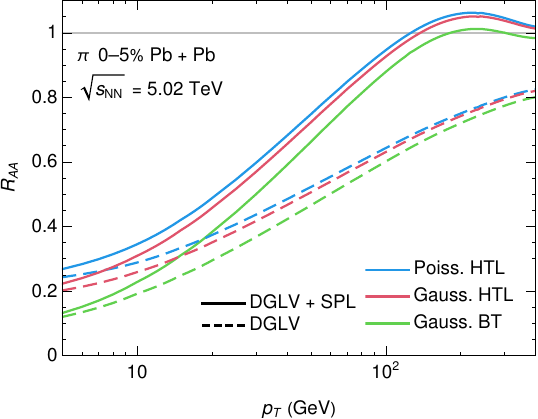}
    \caption{Nuclear modification factor $R_{AA}$ for pions in $0\text{--}5\%$ central \coll{Pb}{Pb} collisions. Theoretical results include variations in elastic models (Gaussian BT, Gaussian HTL, Poisson HTL) and radiative models (DGLV, DGLV + SPL).}
    \label{fig:EL_raa_pbpb_pions}
\end{figure}

\begin{figure}[!htpb]
    \centering
    \begin{subfigure}[t]{0.49\textwidth}
        \centering
        \includegraphics[width=\linewidth]{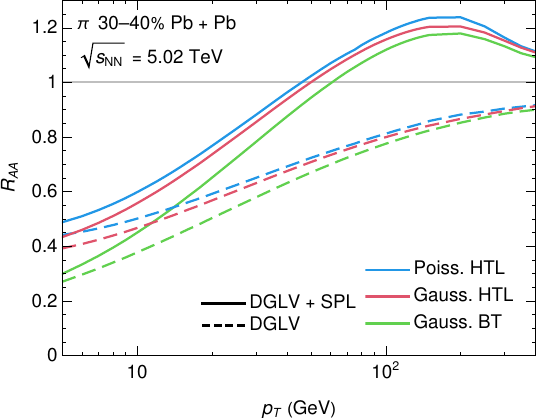}
        \caption{Nuclear modification factor $R_{AA}$ for pions in $30\text{--}40\%$ central \coll{Pb}{Pb} collisions. Theoretical results include variations in elastic models (Gaussian BT, Gaussian HTL, Poisson HTL) and radiative models (DGLV, DGLV + SPL).}
        \label{fig:EL_raa_pbpb_pions_3040}
    \end{subfigure}\hfill
    \begin{subfigure}[t]{0.49\textwidth}
        \centering
        \includegraphics[width=\linewidth]{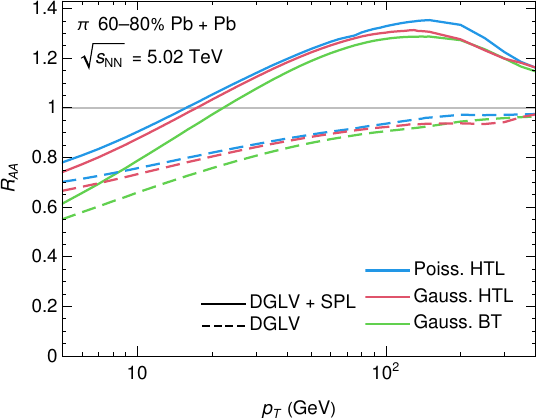}
        \caption{Nuclear modification factor $R_{AA}$ for pions in $60\text{--}80\%$ central \coll{Pb}{Pb} collisions. Theoretical results are produced for pions through our convolved radiative and elastic energy loss model, by varying the elastic model between Gaussian BT \cite{Braaten:1991jj, Braaten:1991we}, Gaussian HTL, and Poisson HTL \cite{Wicks:2008zz}, and the radiative model between DGLV \cite{Djordjevic:2003zk} and DGLV + SPL \cite{Kolbe:2015rvk, Kolbe:2015suq}.}
        \label{fig:EL_raa_pbpb_pions_6080}
    \end{subfigure}
    \caption{Comparison of the nuclear modification factor $R_{AA}$ for pions in $30\text{--}40\%$ and $60\text{--}80\%$ central \coll{Pb}{Pb} collisions at $\sqrt{s_{NN}} = 5.02$ TeV. Theoretical results are obtained using different combinations of elastic and radiative energy loss models.}
    \label{fig:EL_raa_pbpb_pions_combined}
\end{figure}

In \cref{fig:EL_raa_pbpb_pions,fig:EL_raa_pbpb_pions_3040,fig:EL_raa_pbpb_pions_6080} we plot the nuclear modification factor $R_{AA}$ as a function of $p_T$ for pions produced in $0\text{--}5\%$, $30\text{--}40\%$, and $60\text{--}80\%$ most central \coll{Pb}{Pb} collisions at $\sqrt{s_{NN}} = 5.02$ TeV for all aforementioned radiative and elastic energy loss kernels.

We observe in \cref{fig:EL_raa_pbpb_pions} that $R_{AA}$ results calculated with Poisson HTL and Gaussian HTL elastic energy loss kernels differ negligibly, and this difference is reduced as a function of $p_T$, similar to the heavy-flavor results. For small momenta $p_T \lesssim 20$ GeV, there is a large $\sim 50\text{--}100\%$ difference between the $R_{AA}$ calculated with the Gaussian BT elastic energy loss and that calculated with the Gaussian HTL elastic energy loss. The sensitivity to the choice of elastic energy loss kernel reduces as a function of $p_T$.

We note that at higher momenta the pion $R_{AA}$ results are not consistent with our previous work \cite{Faraday:2023mmx}. This difference is due to an error in our pion hadronization code which effectively led to our code neglecting light quark contributions to the pion energy loss. 
The corrected results presented in \cref{fig:EL_raa_pbpb_pions} show a faster rise of the $R_{AA}$ in $p_T$, now correctly capturing the changeover from pions fragmenting primarily from gluons to pions fragmenting primarily from light quarks \cite{Horowitz:2011gd}. We note the interesting phenomenological effect of a steep rise in the $R_{AA}$, caused by including the short pathlength correction to the radiative energy loss. Interestingly, the crossover from gluon-dominated energy loss to light quark-dominated energy loss leads to a reduction of the effect of the short pathlength correction, which is especially large for gluons, causing the $R_{AA}$ to flatten for $p_T \gtrsim 200$ GeV. 

Contrasting \cref{fig:EL_raa_pbpb_pions,fig:EL_raa_pbpb_pions_3040,fig:EL_raa_pbpb_pions_6080} we see that the short pathlength correction increases dramatically for more peripheral collisions. One may understand the increase in the short pathlength correction in peripheral systems as a function of the shorter pathlengths involved in these smaller collision systems.
The relaxation of the large formation time assumption---an assumption which we found was not satisfied self-consistently at large momenta within the DGLV formalism \cite{Faraday:2023mmx}---will likely significantly reduce the size of the short pathlength correction \cite{Faraday:2023uay}. 
Future phenomenological work \cite{Faraday:2024} will examine the effect of placing a cut on the radiated transverse momentum, which ensures that the large formation time approximation is never explicitly violated \cite{Faraday:2023mmx}.

\subsection{Large system suppression at RHIC}
\label{sec:EL_large_system_predictions_rhic}

\begin{figure}[!htbp]
    \centering
    \includegraphics[width=0.6\linewidth]{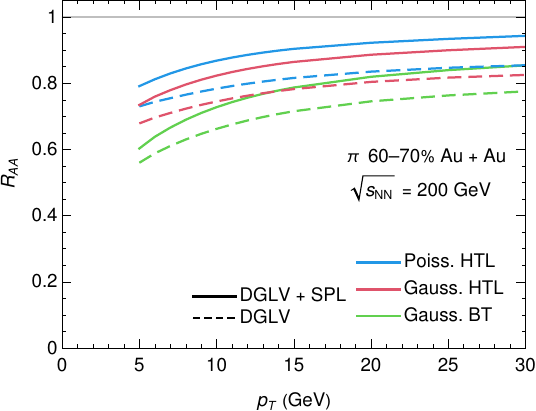}
    \caption{Nuclear modification factor $R_{AA}$ for pions produced in $60\text{--}70\%$ central $\mathrm{Au} + \mathrm{Au}$ collisions at $\sqrt{s_{NN}} = 200$ GeV, calculated using the convolved elastic and radiative energy loss model with various elastic and radiative energy loss kernels.}
    \label{fig:EL_raa-auau-pion-6070}
\end{figure}

\begin{figure}[!htbp]
    \centering
    \begin{subfigure}[t]{0.49\textwidth}
        \centering
        \includegraphics[width=\linewidth]{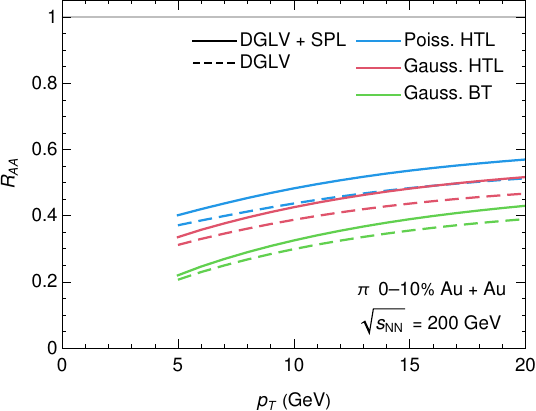}
        \caption{Nuclear modification factor $R_{AA}$ for pions in $0\text{--}10\%$ central $\mathrm{Au} + \mathrm{Au}$ collisions at $\sqrt{s_{NN}} = 200$ GeV. Theoretical results are obtained using different elastic (Gaussian BT, Gaussian HTL, Poisson HTL) and radiative models (DGLV, DGLV + SPL).}
        \label{fig:EL_raa_auau_pions}
    \end{subfigure}\hfill
    \begin{subfigure}[t]{0.49\textwidth}
        \centering
        \includegraphics[width=\linewidth]{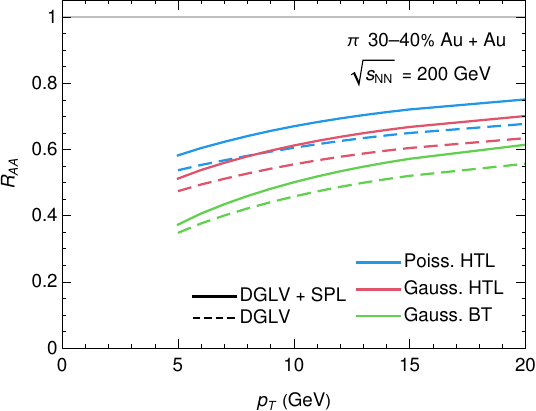}
        \caption{Nuclear modification factor $R_{AA}$ for pions in $30\text{--}40\%$ central $\mathrm{Au} + \mathrm{Au}$ collisions at $\sqrt{s_{NN}} = 200$ GeV. Theoretical results include variations in elastic (Gaussian BT, Gaussian HTL, Poisson HTL) and radiative models (DGLV, DGLV + SPL).}
        \label{fig:EL_raa-auau-pion-3040}
    \end{subfigure}
    \caption{Nuclear modification factor $R_{AA}$ as a function of transverse momentum $p_T$ for pions produced in $\mathrm{Au} + \mathrm{Au}$ collisions at $\sqrt{s_{NN}} = 200$ GeV. Theoretical results are obtained using different combinations of elastic and radiative energy loss models.}
    \label{fig:EL_raa_auau_pions_combined}
\end{figure}

\begin{figure}[!htpb]
	\centering
	\includegraphics[width=0.6\linewidth]{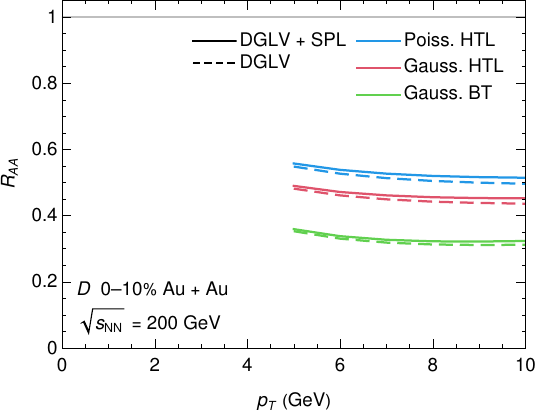}
	\caption{ Nuclear modification factor $R_{AA}$ as a function of $p_T$ for $D$ mesons produced in $0\text{--}10\%$ most central \coll{Au}{Au} collisions at $\sqrt{s_{NN}} = 200$ GeV. Theoretical results are produced for pions through our convolved radiative and elastic energy loss model, by varying the elastic model between Gaussian BT \cite{Braaten:1991jj, Braaten:1991we}, Gaussian HTL, and Poisson HTL \cite{Wicks:2008zz}, and the radiative model between DGLV \cite{Djordjevic:2003zk} and DGLV + SPL \cite{Kolbe:2015rvk, Kolbe:2015suq}. }
	\label{fig:EL_raa_auau_dmeson}
\end{figure}

\Cref{fig:EL_raa_auau_pions,fig:EL_raa-auau-pion-3040,fig:EL_raa-auau-pion-6070} show the nuclear modification factor $R_{AA}$ for pions produced in $0\text{--}10\%$, $30\text{--}40\%$, and $60\text{--}80\%$ most central \coll{Au}{Au} collisions at $\sqrt{s_{NN}} = 200$ GeV based on our theoretical model. \Cref{fig:EL_raa_auau_dmeson} shows the nuclear modification factor $R_{AA}$ for $D$ mesons produced in $0\text{--}10\%$ most central \coll{Au}{Au} collisions at $\sqrt{s_{NN}} = 200$ GeV based on our theoretical model.

Both the pion and $D$ meson $R_{AA}$ exhibit a similar dependence on the choice of elastic energy loss distribution as observed for the same final states in \coll{Pb}{Pb} collisions at LHC.
There is an $\mathcal{O}(10\text{--}25\%)$ relative difference between the $R_{AA}$ results using the Poisson HTL and Gaussian HTL elastic energy loss distributions, and this difference decreases as a function of $p_T$.
This relative difference is significantly larger than the equivalent relative difference of the $R_{AA}$ calculated with the same two elastic energy loss distributions in central \coll{Pb}{Pb} collisions at LHC, as shown in \cref{fig:EL_raa_pbpb_pions}.
The difference between the $R_{AA}$ calculated with the Gaussian HTL and Gaussian BT elastic energy loss is $\mathcal{O}(20\text{--}50\%)$, and it decreases as a function of $p_T$. This difference is similar to that in \coll{Pb}{Pb} at LHC, shown in \cref{fig:EL_raa_pbpb_D0010,fig:EL_raa_pbpb_pions,fig:EL_raa_pbpb_pions_3040,fig:EL_raa_pbpb_pions_6080,fig:EL_raa_pbpb_D3050,fig:EL_raa_pbpb_D6080}.

The impact of the short pathlength correction is small for both pions and $D$ mesons in central \coll{Au}{Au} collisions; however, it is smaller for $D$ mesons than for pions, as expected due to the breaking of color triviality in the short pathlength correction. The correction is significantly smaller for pions in central \coll{Au}{Au} collisions than for pions in central \coll{Pb}{Pb} collisions (compare \cref{fig:EL_raa_pbpb_pions,fig:EL_raa_auau_pions}) due to the much lower maximum momentum shown in the figure, as well as the smaller ratio of gluons to light quarks produced at RHIC compared to LHC \cite{Horowitz:2011gd}.
At low momenta, the short pathlength correction is small because it grows faster in $p_T$ than the uncorrected DGLV radiative energy loss \cite{Kolbe:2015suq, Kolbe:2015rvk}; see \cref{fig:mod_deltaEoverE_small_large,fig:mod_deltaEoverE_vs_L_gluon}. Additionally, because fewer gluons fragment to pions at RHIC compared to LHC, and because the short pathlength correction is significantly larger for gluons compared to quarks, the short pathlength correction is reduced at RHIC compared to LHC.

Comparing \cref{fig:EL_raa_auau_pions,fig:EL_raa-auau-pion-3040,fig:EL_raa-auau-pion-6070}, we see that the relative effect of the short pathlength correction grows for more peripheral collisions, consistent with the behavior observed in \coll{Pb}{Pb} collisions in \cref{fig:EL_raa_pbpb_pions,fig:EL_raa_pbpb_pions_3040,fig:EL_raa_pbpb_pions_6080}. The effect is not nearly as dramatic as it is for pions produced at LHC, simply because the short pathlength correction is significantly smaller at RHIC compared to LHC.

\subsection{Small system suppression at LHC}
\label{sec:EL_small_system_predictions_LHC}

\begin{figure}[!htbp]
    \centering
    \begin{subfigure}[t]{0.49\textwidth}
        \centering
        \includegraphics[width=\linewidth]{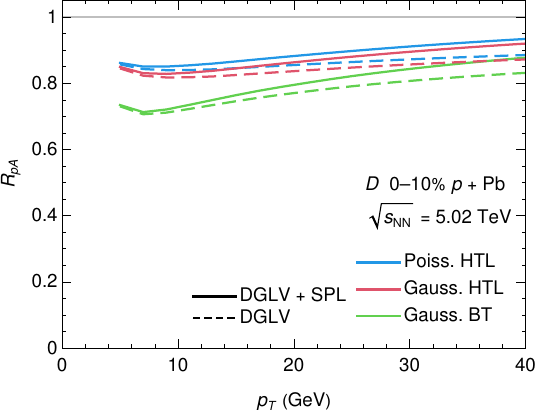}
        \caption{Nuclear modification factor $R_{AB}$ for $D$ mesons in $0\text{--}10\%$ central $\mathrm{p} + \mathrm{Pb}$ collisions at $\sqrt{s_{NN}} = 5.02$ TeV. Results include variations in elastic (Gaussian BT, Gaussian HTL, Poisson HTL) and radiative models (DGLV, DGLV + SPL).}
        \label{fig:EL_raa_ppb_D}
    \end{subfigure}\hfill
    \begin{subfigure}[t]{0.49\textwidth}
        \centering
        \includegraphics[width=\linewidth]{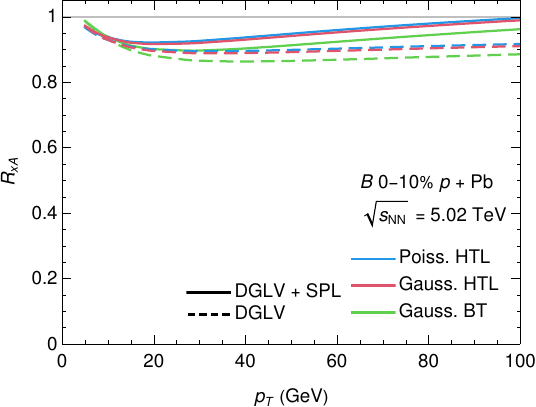}
        \caption{Nuclear modification factor $R_{AB}$ for $B$ mesons in $0\text{--}10\%$ central $\mathrm{p} + \mathrm{Pb}$ collisions at $\sqrt{s_{NN}} = 5.02$ TeV. Results include variations in elastic (Gaussian BT, Gaussian HTL, Poisson HTL) and radiative models (DGLV, DGLV + SPL).}
        \label{fig:EL_raa_ppb_B}
    \end{subfigure}
    \caption{Nuclear modification factor $R_{AB}$ as a function of transverse momentum $p_T$ for $D$ and $B$ mesons produced in $\mathrm{p} + \mathrm{Pb}$ collisions at $\sqrt{s_{NN}} = 5.02$ TeV. Theoretical results are obtained using different combinations of elastic and radiative energy loss models.}
    \label{fig:EL_raa_ppb_combined}
\end{figure}

\Cref{fig:EL_raa_ppb_D} shows the nuclear modification factor $R_{pA}$ as a function of $p_T$ for $D$ mesons produced in central \coll{p}{Pb} collisions at $\sqrt{s_{NN}} = 5.02$ TeV based on our theoretical energy loss model. \Cref{fig:EL_raa_ppb_B} shows the same results for $B$ mesons.

We see from \cref{fig:EL_raa_ppb_D} that for $D$ mesons produced in central \coll{p}{Pb} collisions, there is a negligible difference between the $R_{pA}$ calculated with the Poisson HTL elastic energy loss kernel, and the $R_{pA}$ calculated with the Gaussian HTL elastic energy loss kernel. This negligible difference is surprising because the $\mathcal{O}(0\text{--}1)$ elastic scatters in small collision systems is significantly fewer than the $\sim \!\! 80$ scatters required for the Poisson distribution to converge to a Gaussian distribution \cite{Wicks:2008zz}.
We discuss this point further in \cref{sec:EL_surprising_similarity_gaussian_poisson}. 

The difference between $R_{pA}$ results with and without the short pathlength correction to the radiative energy loss is negligible, even in this small collision system where we expect that the impact of the correction is largest (see \cref{fig:mod_deltaEoverE_small_large,fig:mod_deltaEoverE_small_large_gluon,fig:mod_deltaEoverE_vs_L_gluon}).
The small effect of the short pathlength correction arises because most energy loss in small collision systems is due to elastic energy loss and not radiative energy loss \cite{Faraday:2023mmx}, as evidenced by the plot of $\Delta E / E (L)$ shown in \cref{fig:mod_deltaEoverE_vs_L_gluon}. 
One can understand the dominance of the elastic over the radiative energy loss by noting that the elastic energy loss scales like $L$, while the radiative energy loss scales like $L^2$ due to LPM interference effects. \Cref{sec:EL_relative_elastic_radiative} provides a detailed discussion of the relative contribution of radiative vs.\ elastic energy loss as a function of system size.

We additionally see in \cref{fig:EL_raa_ppb_D} that the $D$ meson $R_{pA}$ at low $p_T$ shows a large $\mathcal{O}(20\%)$ difference between results calculated with the HTL elastic energy loss and those calculated with the BT elastic energy loss. This considerable sensitivity is even more pronounced when considering the degree of suppression: the quantity $1 - R_{pA}$ displays a $\mathcal{O}(70\%)$ relative difference between results computed with these two elastic energy loss kernels---much larger than the equivalent relative difference in \coll{Pb}{Pb} collisions, shown in \cref{fig:EL_raa_pbpb_D0010}. As in \coll{Pb}{Pb}, the relative difference between results calculated with these two elastic energy loss distributions decreases as a function of $p_T$ and displays reduced sensitivity to the choice of elastic energy loss kernel for $B$ mesons, shown in \cref{fig:EL_raa_ppb_B}, compared to $D$ mesons, shown in \cref{fig:EL_raa_ppb_D}.

\begin{figure}[!htpb]
	\centering
	\includegraphics[width=0.6\linewidth]{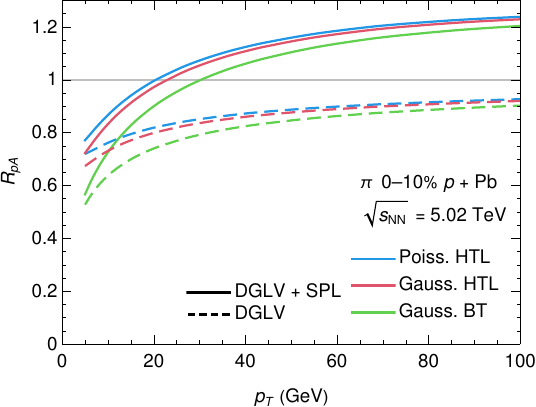}
	\caption{ Nuclear modification factor $R_{AB}$ as a function of $p_T$ for pions produced in $0\text{--}100\%$ most central \coll{Pb}{Pb} collisions at $\sqrt{s_{NN}} = 5.02$ TeV. Theoretical results for pions are produced through our convolved radiative and elastic energy loss model, by varying the elastic model between Gaussian BT \cite{Braaten:1991jj, Braaten:1991we}, Gaussian HTL, and Poisson HTL \cite{Wicks:2008zz}, and the radiative model between DGLV \cite{Djordjevic:2003zk} and DGLV + SPL \cite{Kolbe:2015rvk, Kolbe:2015suq}. }
	\label{fig:EL_raa_ppb_pions}
\end{figure}

\Cref{fig:EL_raa_ppb_pions} shows the nuclear modification factor $R_{pA}$ of pions produced in central \coll{p}{Pb} collisions at $\sqrt{s_{NN}} = 5.02$ TeV based on our model. The difference between results calculated with the Poisson HTL and the Gaussian HTL elastic energy loss distributions is negligible, as it was for $D$ mesons in \coll{p}{Pb} collisions. 
We observe a similarly large sensitivity to the choice between Gaussian HTL and Gaussian BT elastic energy loss distributions for pions in \coll{p}{Pb} as for $D$ mesons in \coll{p}{Pb}.

Including the short pathlength correction produces a nuclear modification factor $R_{pA} \sim 1.2$ for $p_T \gtrsim 40$ GeV with slow $p_T$ dependence. The flattening of the $R_{pA}$ can be understood as a crossover from gluon-dominated energy loss to light quark-dominated energy loss, similar to what was observed for pions produced in \coll{Pb}{Pb} collisions.

\subsection{Small system suppression at RHIC}
\label{sec:EL_small_system_predictions_rhic}

\begin{figure}[!htbp]
    \centering
    \begin{subfigure}[t]{0.49\textwidth}
        \centering
        \includegraphics[width=\linewidth]{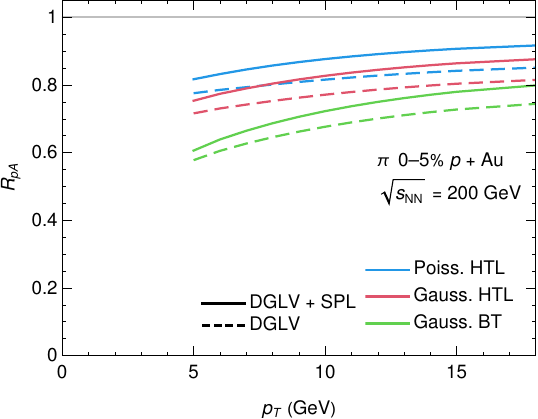}
        \caption{Nuclear modification factor $R_{pA}$ for pions in $0\text{--}5\%$ most central $\mathrm{p} + \mathrm{Au}$ collisions at $\sqrt{s_{NN}} = 200$ GeV. Results include variations in elastic (Gaussian BT, Gaussian HTL, Poisson HTL) and radiative models (DGLV, DGLV + SPL).}
        \label{fig:EL_raa-pion-pau-0005}
    \end{subfigure}\hfill
    \begin{subfigure}[t]{0.49\textwidth}
        \centering
        \includegraphics[width=\linewidth]{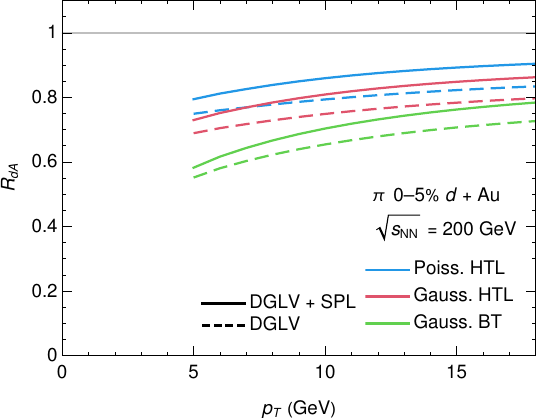}
        \caption{Nuclear modification factor $R_{dA}$ for pions in $0\text{--}5\%$ most central $\mathrm{d} + \mathrm{Au}$ collisions at $\sqrt{s_{NN}} = 200$ GeV. Results include variations in elastic (Gaussian BT, Gaussian HTL, Poisson HTL) and radiative models (DGLV, DGLV + SPL).}
        \label{fig:EL_raa_dau_pions}
    \end{subfigure}
    \caption{Nuclear modification factor for pions as a function of transverse momentum $p_T$ in $\mathrm{p} + \mathrm{Au}$ and $\mathrm{d} + \mathrm{Au}$ collisions at $\sqrt{s_{NN}} = 200$ GeV. Theoretical results are obtained using different combinations of elastic and radiative energy loss models.}
    \label{fig:EL_raa_pion_combined}
\end{figure}

\begin{figure}[!htpb]
	\centering
	\includegraphics[width=0.6\linewidth]{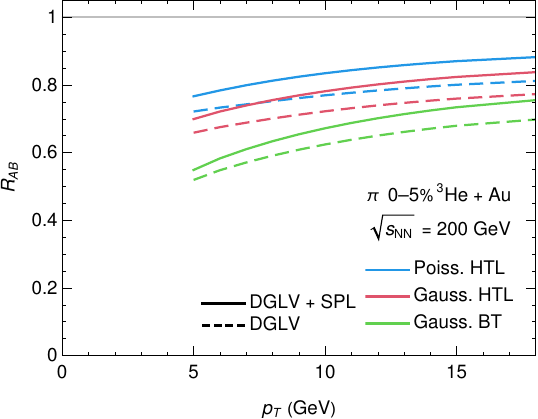}
	\caption{Nuclear modification factor $R_{{}^3\mathrm{He}A}$	as a function of $p_T$ for $0\text{--}5\%$ most central \coll{He3}{Au} collisions at $\sqrt{s_{NN}} = 200$ GeV. Theoretical results are produced through our convolved radiative and elastic energy loss model, by varying the elastic model between Gaussian BT \cite{Braaten:1991jj, Braaten:1991we}, Gaussian HTL, and Poisson HTL \cite{Wicks:2008zz}, and the radiative model between DGLV \cite{Djordjevic:2003zk} and DGLV + SPL \cite{Kolbe:2015rvk, Kolbe:2015suq}. }
	\label{fig:EL_raa_he3au_pions}
\end{figure}

\Cref{fig:EL_raa-pion-pau-0005,fig:EL_raa_dau_pions,fig:EL_raa_he3au_pions} show the nuclear modification factor $R_{AB}$ as a function of $p_T$ for pions produced in central \coll{p}{Au}, \coll{d}{Au}, and \coll{He3}{Au} collisions at $\sqrt{s_{NN}} = 200$ GeV based on our theoretical energy loss model.

We see that the nuclear modification factor $R_{AB}$ model results for \coll{p}{Au}, \coll{d}{Au}, and \coll{He3}{Au} collisions are almost identical. This can be understood by the similarity of the collisions, as evidenced by the length and temperature distributions shown in \cref{fig:mod_path_length_distribution,fig:mod_temperature_distribution}. 
The absolute difference between the Poisson HTL and Gaussian HTL pion $R_{AB}$ in \collFour{p}{d}{He3}{Au} collisions is similar to the same difference in the pion $R_{AB}$ in \coll{Au}{Au} collisions, as shown in \cref{fig:EL_raa_auau_pions}.
The largest difference between the theoretical curves is the choice of elastic energy loss kernel (HTL vs.\ BT), accounting for the $\sim \! 50\%$ difference in the predicted $R_{AB}$. The short pathlength correction to the radiative energy loss has a relatively small effect in \collFour{p}{d}{He3}{Au} collisions compared to its impact in \coll{p}{Pb} collisions, as shown in \cref{fig:EL_raa_ppb_pions}. The small impact of the short pathlength correction is due to the same reasons as the \coll{Pb}{Pb} vs.\ \coll{Au}{Au} case: the lower maximum $p_T$ and smaller fraction of gluons relative to light quarks at RHIC compared to LHC leads to a significantly smaller short pathlength correction.

\section{Surprising independence of the \texorpdfstring{$R_{AB}$}{R_AB} on the elastic energy loss distribution}
\label{sec:EL_surprising_similarity_gaussian_poisson}

\begin{figure}[!htpb]
	\centering
	\includegraphics[width=0.6\linewidth]{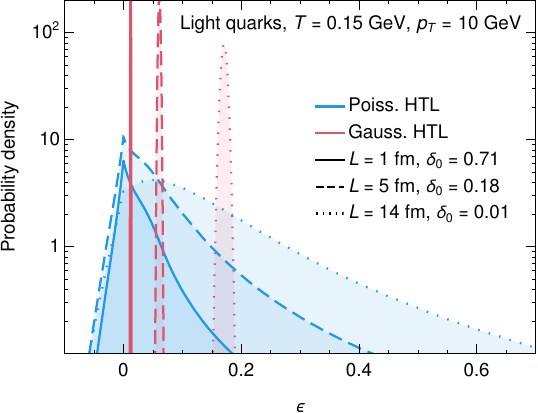}
	\caption{Comparison of realistic Poisson distribution to Gaussian approximation for various pathlengths $L$. The Gaussian is constrained to have the same expectation value $\langle \epsilon \rangle$, and its variance is given by the fluctuation dissipation theorem \cite{Moore:2004tg}. The probability weight attached to the delta function at $\epsilon = 0$ for the Poisson distribution is given by the value of $\delta_0$ in the legend.} 
	\label{fig:EL_comparison_gaussian_poisson}
\end{figure}

In this section, we aim to understand the surprisingly small difference between the $R_{AB}$ calculated with the Poisson HTL elastic energy loss kernel and the Gaussian HTL elastic energy loss kernel (see \cref{sec:mod_elastic_energy_loss} for a comparison). \Cref{fig:EL_comparison_gaussian_poisson} plots the Poisson HTL and Gaussian HTL elastic energy loss probability distributions for pathlengths of $L = 1$ fm, $L = 5$ fm, and $L = 14$ fm for incident light quarks at constant temperature $T = 0.15$ GeV and momentum $p_T = 10$ GeV. 
Note that we choose a relatively low temperature for illustrative purposes.

According to \cref{fig:mod_path_length_distribution} we may interpret the $L = 1$ fm results as typical of small system results in central \coll{p}{Pb} and \collFour{p}{d}{He3}{Au}, the $L = 5$ fm results as typical of large system results in central \coll{Pb}{Pb} and \coll{Au}{Au}, and the $L=14$ fm results as being from an unrealistically large system. \Cref{fig:EL_comparison_gaussian_poisson} shows that for all system sizes, the Gaussian approximation does not visually appear to be a reasonable approximation to the full Poisson distribution. One may conclude that the small difference observed between the Gaussian HTL and Poisson HTL results in all systems and final states shown in \crefrange{fig:EL_raa_pbpb_D0010}{fig:EL_raa_dau_pions} is not attributable to convergence of the Poisson distribution to the Gaussian distribution according to the central limit theorem.

\begin{figure}[!htpb]
	\centering
	\includegraphics[width=0.6\linewidth]{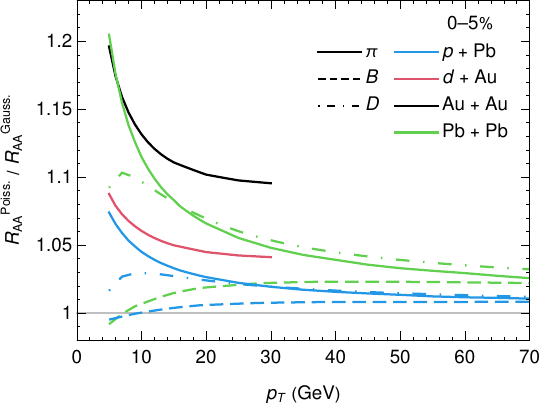}
	\caption{Plot of the double ratio $R_{\text{PG}} \equiv R_{AB}^{\text{Pois.}} / R_{AB}^{\text{Gauss.}}$, where the $R_{AB}$ is calculated with the DGLV radiative energy loss \cite{Djordjevic:2003zk} convolved with the HTL Poisson elastic energy loss for $R^{\text{Pois.}}_{\text{AB}}$, and HTL Gaussian elastic energy loss for $R^{\text{Gauss.}}_{AB}$. The ratio $R^{\text{PG}}_{AB}$ is plotted as a function of $p_T$ for pions, $D$ mesons, and $B$ mesons produced in $0\text{--}5\%$ most central \coll{d}{Au}, \coll{Au}{Au}, \coll{p}{Pb}, and \coll{Pb}{Pb} collisions. 
} 
	\label{fig:EL_raa_poisson_over_gaussian_vs_pt}
\end{figure}

As further evidence that the approximation of the Poisson distribution to a Gaussian distribution does not make sense on the grounds of convergence according to the central limit theorem, we compute the double ratio $R_{\text{PG}} \equiv R_{AB}^{\text{Pois.}} / R_{AB}^{\mathrm{Gauss.}}$, where $R_{AB}^{\text{Pois.}}$ is the $R_{AB}$ calculated with Poisson HTL elastic energy loss convolved with the DGLV radiative energy loss and $R_{AB}^{\text{Gauss.}}$ is the $R_{AB}$ calculated with the is the Gaussian HTL elastic energy loss convolved with the DGLV radiative energy loss. The ratio $R_{\text{PG}}$ is a measure of the relative difference between the $R_{AB}$ calculated with Poisson HTL and Gaussian HTL elastic energy loss kernels, respectively. \Cref{fig:EL_raa_poisson_over_gaussian_vs_pt} plots $R_{\text{PG}}$ as a function of $p_T$ for pions, $B$ mesons, and $D$ mesons produced in central \coll{p}{Pb}, \coll{d}{Au}, \coll{Au}{Au}, and \coll{Pb}{Pb} collisions.

	\Cref{fig:EL_raa_poisson_over_gaussian_vs_pt} displays several qualitative features that warrant explanation, especially in light of the inadequacy of the appeal to the central limit theorem according to \cref{fig:EL_comparison_gaussian_poisson}.
We explain these features in order in \cref{sec:EL_distribution_dependence,sec:EL_relative_elastic_radiative}. We see that the ratio $R_{\text{PG}}$
\begin{itemize}
	\item \emph{Is close to one} for all final states and collision systems of interest.
	\item \emph{Increases as a function of system size} (comparing systems at the same $\sqrt{s_{NN}}$). This system size dependence is opposite to that which the central limit theorem predicts.
	\item \emph{Decreases as a function of $p_T$}.
	\item Exhibits a clear \emph{mass ordering} at low $p_T$ of $R_{\text{PG}}^{B}< R_{\text{PG}}^{D}< R_{\text{PG}}^{\pi}$, based on the final state hadron.
	\item \emph{Decreases as a function of $\sqrt{s_{NN}}$}, when considering systems of similar geometrical size (refer to \cref{fig:mod_path_length_distribution,fig:mod_temperature_distribution})---\coll{d}{Au} vs \coll{p}{Pb} and \coll{Pb}{Pb} vs \coll{Au}{Au}---at RHIC and LHC energies.
\end{itemize}

Two primary mechanisms contribute to the observed effects listed above. First, the relative contributions of elastic and radiative energy loss helps to explain why $R_{\text{PG}} \sim 1$, and the system size and  $p_T$ dependence of $R_{\text{PG}}$. The effect of the relative contribution of radiative vs.\ elastic energy loss is discussed in \cref{sec:EL_relative_elastic_radiative}. Second, the dependence of the $R_{AB}$ on the various moments of the elastic energy loss distribution helps to explains why $R_{\text{PG}} \sim 1$ as well as the system size, $\sqrt{s_{NN}}$, and final state dependence. We discuss the effect of the various moments of the energy loss distributions in \cref{sec:EL_distribution_dependence}. These two mechanisms comprehensively account for all the aforementioned features of the $R_{\text{PG}}$ ratio.

\subsection{Relative contribution of elastic and radiative energy loss to the \texorpdfstring{$R_{AB}$}{RAB}}
\label{sec:EL_relative_elastic_radiative}

While asymptotically the $p_T$ dependence of the radiative and elastic energy loss is the same, $\Delta E \sim \log E$, %
we saw numerically in \cref{sec:mod_numerical_elastic_radiative} that at non-asymptotic energies the $p_T$ dependence differs dramatically. We saw explicitly in \cref{fig:mod_deltaEoverE_small_large,fig:mod_deltaEoverE_small_large_gluon} that while the final asymptotic ratio of elastic vs.\ radiative energy loss depends strongly on both the pathlength $L$ and the energy loss model used for both the elastic and radiative energy loss, the trend in $p_T$ is clear: as $p_T$ increases the ratio of elastic energy loss to radiative energy loss $\Delta E_{\text{el.}} / \Delta E_{\text{rad.}}$ decreases. This trend in $p_T$ accounts for the $p_T$ dependence of \cref{fig:EL_raa_poisson_over_gaussian_vs_pt}.

	To quantify this effect, we define

\begin{equation}
	\left(\frac{\Delta E^{\text{el.}}}{\Delta E^{\text{rad.}}}\right)_{\text{eff.}} \equiv \frac{1-R_{AB}^{\mathrm{el.}}}{1-R_{AB}^{\mathrm{rad.}}},
	\label{eqn:EL_effective_deltaE_elastic_over_radiative}
\end{equation}

where $R_{AB}^{\text{el.}}$ is the nuclear modification factor calculated with the radiative energy loss turned off, and similarly $R_{AB}^{\text{rad.}}$ has the elastic energy loss turned off.

We consider the ratio of $1-R_{AB}$ in the above because of the following expansion\footnote{For a detailed discussion of such an expansion in the lost fractional energy $\epsilon$ refer to \cref{sec:EL_validity_of_power_law_approximation_to_r_aa}} in the lost fractional energy $\epsilon$. From \cref{eqn:mod_full_raa_spectrum_ratio} we have
\begin{align}
	R_{AB}^{\text{tot.}} =& \frac{1}{f\left(p_T\right)} \int \frac{d \epsilon}{1-\epsilon} f\left(\frac{p_T}{1-\epsilon}\right) P_{\text {tot.}}\left(\epsilon \left | \frac{p_T}{1-\epsilon}\right. \right)\nonumber\\
R_{AB}^{\text{tot.}}=& \int d \epsilon \left[1 + (1 + p_T f'(p_T)) \epsilon + \mathcal{O}(\epsilon^2) \right] \times \nonumber\\
& \int dx \; P_{\text{rad.}}(x) P_{\text{el.}}(\epsilon - x) \nonumber\\
1 - R_{AB}^{\text{tot.}} =& \int d \epsilon' \left[\left(1 + p_T f'(p_T) \right) (\epsilon' + x) + \mathcal{O}(x + \epsilon')^2\right] \times  \nonumber\\
& \int dx \; P_{\text{rad.}}(x) P_{\text{el.}}(\epsilon') \nonumber\\
=& \left(1 + p_T \frac{f'(p_T)}{f(p_T)} \right) \left[ \frac{\Delta E^{\text{rad.}}}{E} + \frac{\Delta E ^{\text{el.}}}{E} \right]\nonumber\\
&+ \mathcal{O}(\langle \epsilon^2 \rangle_{\text{tot.}})\nonumber\\
	1-R_{AB}^{\text{tot.}} =& (1-R_{AB}^{\text{rad.}}) + (1-R_{AB}^{\text{el.}}) + \mathcal{O}(\langle \epsilon^2 \rangle_{\text{tot.}}),
	\label{eqn:EL_one_minus_RAB}
\end{align}
which facilitates the interpretation of \cref{eqn:EL_effective_deltaE_elastic_over_radiative} as the fraction of elastic energy loss vs.\ radiative energy loss which contributes to the $R_{AB}$.  In \cref{eqn:EL_one_minus_RAB} we see that to leading order in $\left\langle  \epsilon \right\rangle_{\text{tot.}}$ in both the numerator and denominator, $(\Delta E^{\text{el.}}/\Delta E^{\text{rad.}})_{\text{eff.}} = \Delta E^{\text{rad.}}/\Delta E^{\text{el.}}$. The ratio $(\Delta E^{\text{el.}}/\Delta E^{\text{rad.}})_{\text{eff.}}$ is normalized such that it is equal to one when radiative and elastic energy loss contribute equally to suppression, below one when radiative energy loss is the dominant suppression mechanism, and above one when elastic energy loss is the dominant suppression mechanism. Higher order corrections to \cref{eqn:EL_one_minus_RAB} indicate an interplay between the radiative and elastic energy loss distributions, and one cannot separate these corrections into purely radiative or purely elastic contributions.

\begin{figure}[!htpb]
	\centering
	\includegraphics[width=0.6\linewidth]{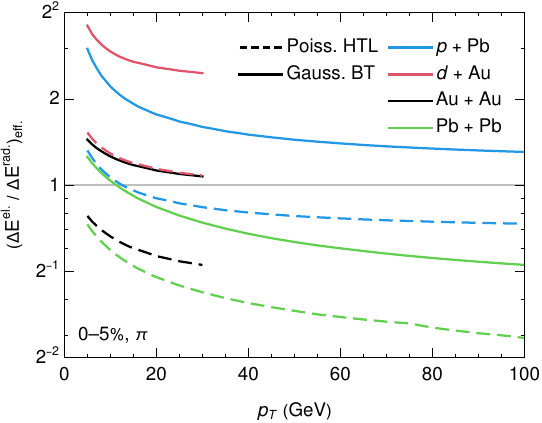}
	\caption{Plot of the ratio $\left(\Delta E^{\text{el.}} / \Delta E^{\text{rad.}}\right)_{\text{eff.}} \equiv (1-R_{AB}^{\mathrm{el.}})/(1-R_{AB}^{\mathrm{rad.}})$, where $R_{AB}^{\text{el.}}$ is the nuclear modification factor calculated with the radiative energy loss turned off, and similarly $R_{AB}^{\text{rad.}}$ has the elastic energy loss turned off. The ratio $(\Delta E^{\text{el.}}/\Delta E^{\text{rad.}})_{\text{eff.}}$ is plotted as a function of $p_T$ $0\text{--}5\%$ most central \coll{d}{Au}, \coll{Au}{Au}, \coll{p}{Pb}, and \coll{Pb}{Pb} collisions. We calculate the ratio for both Poisson HTL \cite{Wicks:2008zz} and Gaussian BT \cite{Braaten:1991jj,Braaten:1991we} elastic energy loss distributions convolved with the Poisson DGLV radiative energy loss distribution \cite{Djordjevic:2003zk}.
	}
	\label{fig:EL_raa_elastic_over_radiative_vs_pt}
\end{figure}

\Cref{fig:EL_raa_elastic_over_radiative_vs_pt} plots the ratio $(\Delta E^{\text{el.}}/\Delta E^{\text{rad.}})_{\text{eff.}}$ as a function of $p_T$ for $0\text{--}5\%$ most central \coll{d}{Au}, \coll{Au}{Au}, \coll{p}{Pb}, and \coll{Pb}{Pb} collisions. We use the DGLV radiative energy loss kernel and the Gaussian BT elastic energy loss kernel for all curves and vary the elastic energy loss kernel between the Poisson HTL and Gaussian BT elastic energy loss kernels.

From \cref{fig:EL_raa_elastic_over_radiative_vs_pt}, the elastic energy loss is $\sim 1\text{--}3$  times more important in small systems than large collision systems. Additionally, using the Poisson HTL vs.\  Gaussian BT elastic energy loss distributions dramatically changes the relative importance of the elastic energy loss. We conclude that further theoretical control is needed in the elastic energy loss sector to make quantitative predictions in small systems. Indeed, even \textit{qualitative} questions, such as whether elastic or radiative energy loss is of primary importance in small and large systems, are sensitive to this uncertainty.

The extensive range of relative contributions of the elastic vs radiative energy loss as a function of system size and $\sqrt{s_{NN}}$ leads us to propose that a system size scan will help to disentangle the relative contribution of radiative and elastic energy loss.

\subsection{Distribution dependence of the nuclear modification factor}
\label{sec:EL_distribution_dependence}

One approach to understanding how the shape of energy loss distributions affect the nuclear modification factor is by expanding the integrand of the $R_{AB}$ in powers of the fractional energy loss $\epsilon$. This approach is motivated by the fact that the Gaussian and Poisson distributions are constrained to have identical\footnote{In practice, they may differ slightly due to the kinematic cut at $\epsilon =1$. See \cref{fig:mod_deltaEoverE_small_large_gluon,fig:mod_deltaEoverE_small_large,fig:mod_deltaEoverE_vs_L_gluon} for a comparison of the first moment ($\Delta E / E$) between the Gaussian and Poisson distributions.} zeroth moments $\left\langle \epsilon^0 \right\rangle \equiv \int \mathrm{d} \epsilon P(\epsilon) = 1$ (probability conservation) and identical first moments $\left\langle \epsilon^1 \right\rangle \equiv \int \mathrm{d} \epsilon P(\epsilon) = \Delta E / E$. Starting with the expression for $R_{AB}$ from \cref{eqn:mod_full_raa_spectrum_ratio} one may schematically expand the integrand in powers of $\epsilon$ to obtain 
\begin{align}
	R_{AB}(p_T) =& \frac{1}{f\left(p_T\right)} \int \frac{d \varepsilon}{1-\varepsilon} f\left(\frac{p_T}{1-\varepsilon}\right) P_{\text {tot.}}\left(\varepsilon \left | \frac{p_T}{1-\varepsilon}\right. \right)\nonumber\\
	R_{AB}(p_T) =& \sum_n c_n(p_T) \int d \epsilon \; \epsilon^n P_{\text{tot.}}(\epsilon | p_T)\nonumber\\
	=& \sum_n c_n(p_T) \left\langle \epsilon^n(p_T) \right\rangle_{\text{tot.}}
	\label{eqn:EL_schematic_moment_expansion}
\end{align}
where $\left\langle \epsilon^n(p_T) \right\rangle_{\text{tot.}} \equiv \int d \epsilon \; \epsilon^n P_{\text{tot.}}(\epsilon | p_T)$ is the $n^{\text{th}}$ raw moment of the $P_{\text{tot.}}$ distribution and $c_n(p_T)$ are the coefficients of the expansion. Importantly, this form of the $R_{AB}$ partitions the effect of the production spectrum into the $c_n(p_T)$ and the effects of energy loss into the $\left\langle \epsilon^n(p_T) \right\rangle_{\text{tot.}}$. The $c_n$ do not have a simple analytic form; however we list the first few:

\begin{align}
	c_0(p_T) =& 1 \nonumber\\
	c_1(p_T) =& 1 + p_T \frac{f'(p_T)}{f(p_T)} \nonumber\\
	c_2(p_T) =& 1 + p_T \frac{f'(p_T)}{f(p_T)} + \frac{1}{2} p_T^2 \frac{f''(p_T)}{f(p_T)}\\
	c_3(p_T) =& 1 + 3 p_T \frac{f'(p_T)}{f(p_T)} + \frac{3}{2} p_T^2 \frac{f''(p_T)}{f(p_T)} + \frac{1}{6} p_T^3 \frac{f'''(p_T)}{f(p_T)}.\nonumber
	\label{eqn:EL_cn}
\end{align}

Since the total energy loss distribution is the convolution of the radiative and elastic energy loss distributions, the moments $\left\langle \epsilon^n(p_T) \right\rangle_{\text{tot.}}$ of the total probability distribution $P_{\text{tot.}}$ are related to the moments $\left\langle \epsilon^n(p_T) \right\rangle_{\text{rad.}}$ and $\left\langle \epsilon^n(p_T) \right\rangle_{\text{el.}}$ of the radiative $P_{\text{rad.}}$ and elastic $P_{\text{el.}}$ distributions respectively through a binomial expansion
\begin{align}
	\left\langle \epsilon^n \right\rangle_{\text{tot.}} \equiv& \int d \epsilon \; \epsilon^n P_{\text{tot.}}(\epsilon | p_T)\\
	=& \int d \epsilon \; \epsilon^n \int d x \; P_{\text{el.}}(\epsilon - x | p_T) P_{\text{rad.}}(x | p_T)\\
	=& \int d \epsilon' \; (\epsilon' + x)^n \int d x \; P_{\text{el.}}(\epsilon' | p_T) P_{\text{rad.}}(x | p_T)\\
  =&\sum_k \binom{n}{k} \left\langle \epsilon^k \right\rangle_{\text{rad.}} \left\langle \epsilon^{n-k} \right\rangle_{\text{el.}}.
	\label{eqn:EL_moment_binomial_radiative_elastic}
\end{align}

\begin{figure}[!htbp]
	\centering
	\includegraphics[width=0.6\linewidth]{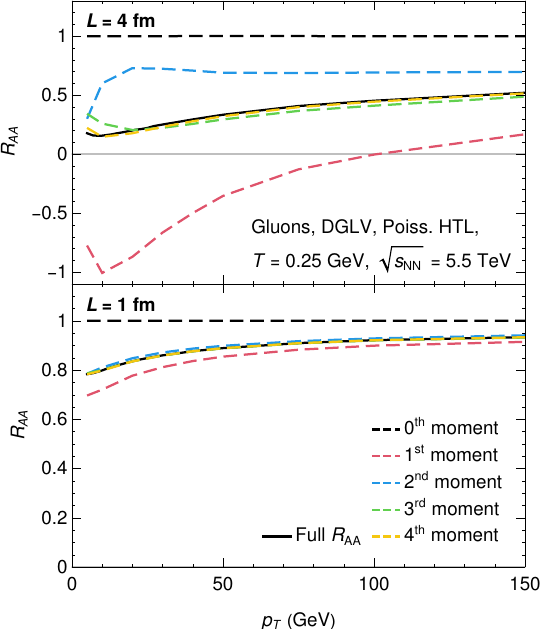}
	\caption{Figure showing the breakdown of the $R_{AB}$ in terms of the moments $\langle \epsilon^i \rangle \equiv \int d \epsilon \epsilon^i P_{\text{tot.}}(\epsilon)$ of the total energy loss probability distribution as a function of $p_T$ for a gluon at constant temperature $T=0.25$~GeV. The top pane is calculated at constant pathlength $L = 4$~fm and the bottom pane at constant pathlength $L=1$~fm. The solid curve shows the $R_{AB}$ calculated according to \cref{eqn:mod_full_raa_spectrum_ratio} while the dashed curves indicate the sum of the contributions up to the $i^{\text{th}}$ moment (indicated in the legend). All curves are computed with DGLV radiative energy loss \cite{Djordjevic:2003zk} and Poisson HTL elastic energy loss \cite{Wicks:2008zz} and at $\sqrt{s_{NN}} = 5.5$ TeV.}
	\label{fig:EL_raa_moment_breakdown}
\end{figure}

\Cref{fig:EL_raa_moment_breakdown} shows the order-by-order contributions to $R_{AA}$ as a function of $p_T$ for the expansion described by \cref{eqn:EL_schematic_moment_expansion}. This calculation is performed for a gluon moving through a constant brick of length $L = 4$~fm (top pane) and $L = 1$~fm (bottom pane), and constant temperature $T = 0.25$~GeV. From the figure, we see that when there is a large amount of suppression of $R_{AA} \sim 0.2$ (top pane), we need up to the fourth moment to correctly converge to the full $R_{AA}$. In contrast, for smaller suppression of $R_{AA} \sim 0.75$ (bottom pane), only the first one or two moments are needed to correctly reproduce the full $R_{AA}$.

From this illustrative example, one may already understand the system size dependence of $R_{\text{PG}}$ in \cref{fig:EL_raa_poisson_over_gaussian_vs_pt}. In small systems, there is a small amount of suppression, and so only the $0^{th}$ and $1^{st}$ moments are necessary to describe the $R_{AB}$, for which both the Gaussian and Poisson distributions produce identical values. In larger systems, more moments are required for a satisfactory fit; however, the elastic energy loss is only a small contribution to suppression compared to the radiative energy loss (see \cref{sec:EL_relative_elastic_radiative}).

In order to illustrate this mechanism for a more extensive range of collision systems and final states, we summarize this effect by introducing the \textit{average important moment}

\begin{equation}
\left\langle n \right\rangle \equiv \frac{\sum_n n \; \left| c_n  \left\langle \epsilon^n \right\rangle \right|}{\sum_n \left| c_n \left\langle \epsilon^n \right\rangle \right|},
	\label{eqn:EL_average_important_moment}
\end{equation}
as a measure of how much each moment of the expansion in \cref{eqn:EL_schematic_moment_expansion} contributes to the $R_{AB}$. Note that the reason we consider the absolute value of the summed terms is that the expansion in \cref{eqn:EL_schematic_moment_expansion} is generically oscillatory\footnote{The oscillation in the expansion is due to the power-law behavior of the spectrum meaning that successive derivatives of the spectrum $f(p_T)$ have alternating signs.}, as evidenced in \cref{fig:EL_raa_moment_breakdown}, and so to capture the ``impact" of a particular moment we consider the absolute value of that moment's contribution.
Since the zeroth and first moments of both the Poisson and Gaussian elastic energy loss distributions are identical, we expect that smaller values of $\left\langle n \right\rangle$ correspond to a more similar $R_{AB}$ for the two distributions. 

\begin{figure}[!htbp]
	\centering
	\includegraphics[width=0.6\linewidth]{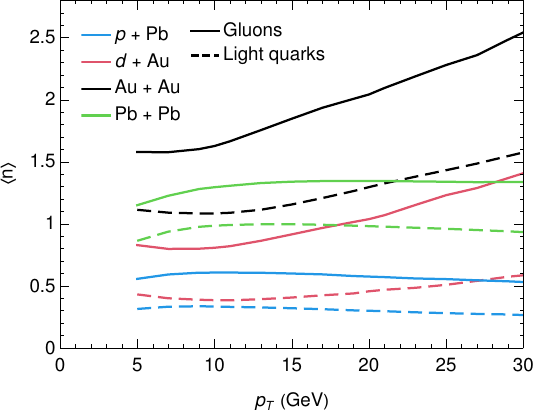}
	\caption{Plot of the average important moment $\left\langle n \right\rangle$ as a function of $p_T$ for gluons and light quarks produced in $0\text{--}5\%$ most central \coll{p}{Pb}, \coll{d}{Au}, \coll{Au}{Au}, and \coll{Pb}{Pb} collisions. All curves are computed with Poisson HTL elastic energy loss and DGLV radiative energy loss.}
	\label{fig:EL_average_important_moment}
\end{figure}

\Cref{fig:EL_average_important_moment} plots $\left\langle n \right\rangle$ as a function of $p_T$ for gluons and light quarks produced in central \coll{p}{Pb}, \coll{d}{Au}, \coll{Au}{Au}, and \coll{Pb}{Pb} collisions. In the figure, we observe $\left\langle n \right\rangle$ is larger in \coll{Au}{Au} and \coll{Pb}{Pb} collisions than it is in \coll{d}{Au} and \coll{p}{Pb} collisions. 
This system size dependence of $\left\langle n \right\rangle$ explains the system size dependence of the ratio $R_{\text{PG}}$ in \cref{fig:EL_raa_poisson_over_gaussian_vs_pt}---in large \coll{Au}{Au} and \coll{Pb}{Pb} collision systems there is more energy loss and hence the higher order moments are more important in comparison to the smaller collision systems of \coll{p}{Pb} and \coll{d}{Au}. Since the zeroth and first moment contributions to the $R_{AB}$ are constrained to be identical, the smaller $\left\langle n \right\rangle$ means that the $R_{AB}$ computed with Gaussian and Poisson elastic energy loss distributions respectively will be more similar.

The $p_T$ dependence of $\left\langle n \right\rangle$ shown in \cref{fig:EL_average_important_moment} is weak at LHC but moderate at RHIC. On may understand this $p_T$ dependence as a competition between the $p_T$ dependence of the production spectrum and the $p_T$ dependence of the energy loss. Less energy loss corresponds to a lower $\left\langle n \right\rangle$ because $\left\langle \epsilon^n \right\rangle \sim (\Delta E / E)^n \sim ( \log E / E)^n$, if one neglects the effects of the shape of the distribution. The $p_T$ dependence of the spectra affects the $p_T$ dependence of the $R_{AB}$ through its effect on the $c_n(p_T)$ coefficients of \cref{eqn:EL_schematic_moment_expansion}, listed in \cref{eqn:EL_cn}. From \cref{eqn:EL_cn}, we see that a faster-changing spectra leads to a larger $f'(p_T)$ and a larger magnitude for the various $c_n$. We conclude that at LHC, the $p_T$ dependence of the energy loss is more important to understand the $p_T$ dependence of $\left\langle n \right\rangle$, while at RHIC, the $p_T$ dependence of the spectra is more important.

\Cref{fig:EL_average_important_moment} shows a strong $\sqrt{s_{NN}}$ dependence in $\left\langle n \right\rangle$; the \coll{Au}{Au} and \coll{d}{Au} collisions at $\sqrt{s_{NN}} = 200$ GeV have a larger $\left\langle n \right\rangle$ than the corresponding \coll{Pb}{Pb} and \coll{p}{Pb} collisions at LHC at $\sqrt{s_{NN}} = 5.02$ TeV. One may understand this $\sqrt{s_{NN}}$ dependence from the steeper spectra at lower $\sqrt{s_{NN}}$ leading to larger values for the $c_n$ coefficients, thereby making the average important moment $\left\langle n \right\rangle$ larger. The $\sqrt{s_{NN}}$ dependence explains why in \cref{fig:EL_raa_poisson_over_gaussian_vs_pt} the $R_{AB}$ results calculated with Poisson and Gaussian elastic energy loss distributions respectively, are more similar in \coll{p}{Pb} and \coll{Pb}{Pb} collisions than in the corresponding similarly sized systems of \coll{d}{Au} and \coll{Au}{Au} collisions respectively.

The flavor dependence of the $R_{\text{PG}}$ ratio in \cref{fig:EL_raa_poisson_over_gaussian_vs_pt} can be understood simply as a result of the flavor dependence of the suppression. Since $\Delta E^{B} < \Delta E^{D} < \Delta E^{\pi}$ we have that $\left\langle n \right\rangle^{B} < \left\langle n \right\rangle^{D} < \left\langle n \right\rangle^{\pi}$ from which it follows that $R_{\text{PG}}^{B} < R_{\text{PG}}^{D} < R_{\text{PG}}^{\pi}$.

We conclude that the sensitivity of the $R_{AB}$ to the shape of the underlying energy loss probability distribution decreases as a function of $\sqrt{s_{NN}}$. Additionally, when our model predicts a small amount of energy loss, as in small systems, then the $R_{AB}$ is largely insensitive to the shape of the underlying elastic energy loss distribution.

\section{Power law approximation to nuclear modification factor}
\label{sec:EL_validity_of_power_law_approximation_to_r_aa}

In our previous work \cite{Faraday:2023mmx}, we calculated the $R_{AB}$ under the assumption that the production spectra followed a power law
\begin{equation}
d N^q_{pp} / d p_T \equiv f(p_T) \approx A p_T^{-n(p_T)},
\label{eqn:EL_eqn-power-law}
\end{equation}
where $A$ is a normalization constant. This power law approximation followed previous work \cite{Wicks:2005gt, Wicks:2008zz, Horowitz:2011gd} and is similar to other power law approximations in the literature \cite{Arleo:2017ntr, Baier:2002tc}.
This assumption allows one to simplify the expression for the $R_{AB}$ in \cref{eqn:mod_full_raa_spectrum_ratio} as follows \cite{Faraday:2023mmx, Horowitz:2010dm}
\begin{align}
	R_{A B}^q\left(p_T\right) =& \frac{1}{f(p_T)} \int d \epsilon  \frac{1}{1 - \epsilon} f\left( \frac{p_T}{1-\epsilon}\right) P_{\text{tot.}}\left(\epsilon \left| \frac{p_T}{1-\epsilon}\right) \right.\nonumber\\
  =& \frac{\int \frac{\mathrm{d} \epsilon}{1-\epsilon} \frac{A}{\left(p_T / 1-\epsilon\right)^{n\left(p_T / [1-\epsilon]\right)}} P\left(\epsilon \mid \frac{p_T}{1-\epsilon}\right)}{A \; p_T^{- n\left(p_T\right)}}\label{eqn:EL_eqn-raa-power-law-simplifications}\\
	=& \int d \epsilon (1-\epsilon)^{n(p_T / 1-\epsilon) - 1} \frac{p_T^{n(p_T / 1 - \epsilon)}}{p_T^{n(p_T)}} P\left(\epsilon | \frac{p_T}{1 - \epsilon}\right).\nonumber
\end{align}
If one then assumes that $n(p_T)$ is slowly varying and that $\epsilon \ll 1$, then $n(p_T / 1 - \epsilon) \simeq n(p_T)$ and \cref{eqn:EL_eqn-raa-power-law-simplifications} simplifies to \cite{Horowitz:2010dm}
\begin{equation}
	R_{AB} = \int d \epsilon (1 - \epsilon)^{n(p_T) - 1} P(\epsilon | p_T).
	\label{eqn:EL_power-law-raa}
\end{equation}

The practical method in which the function $n(p_T)$ was calculated in our previous work \cite{Faraday:2023mmx} and the literature \cite{Wicks:2005gt, Horowitz:2010dm} is to differentiate \cref{eqn:EL_eqn-power-law} resulting in
\begin{equation}
	\frac{f^{\prime}(p_T)}{f(p_T)} = -n^{\prime}(p_T) \log p_T -\frac{n(p_T)}{p_T}.
\end{equation}
If one further assumes that $n'(p_T) \log p_T$ is small according to the assumption that $n(p_T)$ varies slowly we obtain
\begin{equation}
	n(p_T) = - p_T \frac{f^{\prime}(p_T)}{f(p_T)}.
\label{eqn:EL_eqn-npt-derivative}
\end{equation}

\begin{figure}[!htpb]
	\centering
	\includegraphics[width=0.6\linewidth]{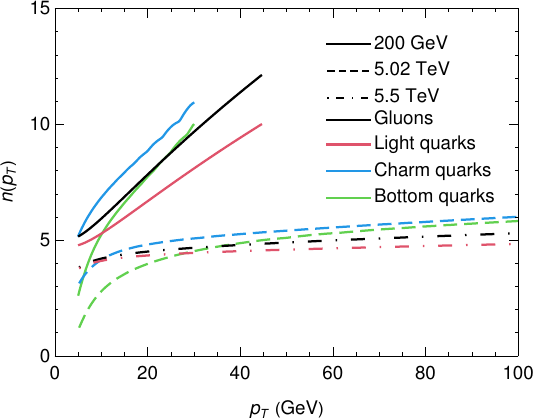}
	\caption{Production spectrum power $n(p_T)$ as a function of $p_T$ for gluons, light quarks, charm quarks, and bottom quarks produced at RHIC and LHC.}
	\label{fig:EL_npt}
\end{figure}

\Cref{fig:EL_npt} shows the $n(p_T)$ calculated according to \cref{eqn:EL_eqn-npt-derivative} for various partons at RHIC and LHC energies. We see in the figure that lower $\sqrt{s_{NN}}$ has a larger $n(p_T)$, which corresponds to a steeper spectrum. 

	Also evident in the figure is that $n(p_T)$ is not always slowly varying. At RHIC energies $\sqrt{s_{NN}} = 200$ GeV, $n(p_T)$ grows linearly from $\mathcal{O}(5)\text{--}\mathcal{O}(10)$ over the $p_T$ range $\mathcal{O}(5)\text{--}\mathcal{O}(40)$ GeV, and for bottom and charm quarks at LHC energies $\sqrt{s_{NN}} = 5.02$ TeV the charm and bottom quark $n(p_T)$ grows from $\mathcal{O}(1)\text{--}\mathcal{O}(5)$ over the $p_T$ range $\mathcal{O}(5)\text{--}\mathcal{O}(40)$ GeV. Based on these apparent inconsistencies, it is reasonable to question the validity of the above set of approximations.

The utility of the slowly-varying power law $R_{AB}$ expression in \cref{eqn:EL_power-law-raa} is largely in its explanatory power. The slowly-varying power law $R_{AB}$ cleanly separates the effect of the production spectra from the effect of energy loss, which is not as obvious in the original expression in \cref{eqn:mod_full_raa_spectrum_ratio}. This clean separation of the $p_T$ dependencies in the power law $R_{AB}$ allows for a good qualitative description of, for instance, the $p_T$ dependence of the $R_{AB}$. To first order in $\epsilon$ we have that \cref{eqn:EL_power-law-raa} is
\begin{equation}
	R^q_{AB} \sim 1 - (n(p_T) - 1) \frac{\Delta E}{E} (p_T)
\end{equation}
One may then understand the slow $p_T$ dependence of the $R^{\pi}_{AA}$ at RHIC in \cref{fig:EL_raa_auau_pions,fig:EL_raa-auau-pion-3040,fig:EL_raa-auau-pion-6070} as a partial cancellation of the growth in $p_T$ of $n(p_T)$ (see \cref{fig:EL_npt}) with the decrease in $p_T$ of $\Delta E / E (p_T)$ (see \cref{fig:mod_deltaEoverE_small_large,fig:mod_deltaEoverE_small_large_gluon}). 

\begin{figure}[!htpb]
	\centering
	\includegraphics[width=0.6\linewidth]{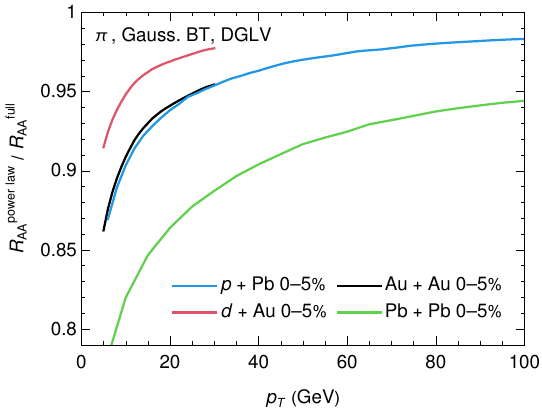}
	\caption{Plot of the ratio of the full $R_{AB}$ to the slowly-varying power law approximation to the full $R_{AB}$, $R_{AB}^{\text{power law}} / R_{AB}^{\text{full}}$ as a function of $p_T$. Curves are produced for $0\text{--}5\%$ most central \coll{p}{Pb}, \coll{d}{Au}, \coll{Au}{Au}, and \coll{Pb}{Pb} collisions.}
	\label{fig:EL_spectrum-ratio-vs-full-raa}
\end{figure}

To examine the validity of the slowly-varying power law approximation we plot in \cref{fig:EL_spectrum-ratio-vs-full-raa} the ratio of the slowly-varying power law nuclear modification factor $R_{AB}^{\text{power law}}$ (\cref{eqn:EL_power-law-raa}) to the full $R_{AB}$ (\cref{eqn:mod_full_raa_spectrum_ratio}) as a function of $p_T$ for pions produced in central \coll{p}{Pb}, \coll{d}{Au}, \coll{Au}{Au}, and \coll{Pb}{Pb} collisions. We see that the difference is maximal in \coll{Pb}{Pb} collisions at low momenta, reaching around $20\%$, while the difference is $\lesssim 10\%$ in all other colliding systems. 
We note that \cref{fig:EL_spectrum-ratio-vs-full-raa} is generated only for the elastic energy loss kernel set to Gaussian BT and the radiative energy loss kernel set to DGLV, as the dependence on the specific energy loss kernel used was minimal.

\begin{figure}[!htpb]
	\centering
	\includegraphics[width=0.6\linewidth]{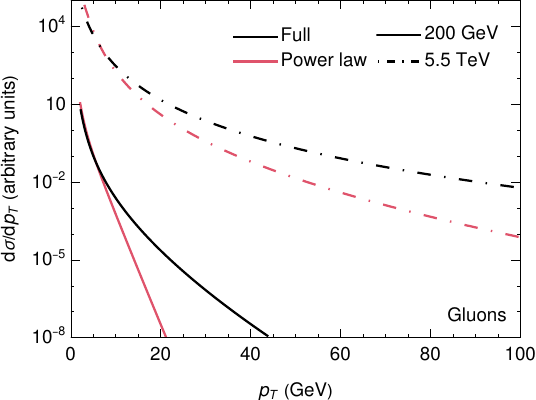}
	\caption{Comparison of the production spectra $d \sigma / d p_T$ to the slowly-varying power law approximation in \cref{eqn:EL_eqn-power-law} as a function of $p_T$ for gluons produced at $\sqrt{s_{NN}} = 200$ GeV and $\sqrt{s_{NN}} = 5.5$ TeV.}
	\label{fig:EL_spectra-comparison}
\end{figure}

The relatively good agreement between the full $R_{AB}$ and the slowly-varying power law approximation to the $R_{AB}$ apparently validates the set of approximations used to derive \cref{eqn:EL_power-law-raa} from \cref{eqn:mod_full_raa_spectrum_ratio}. \Cref{fig:EL_spectra-comparison} plots the production spectra $d \sigma / d p_T$ and the slowly-varying power law approximation to the production spectra calculated according to \cref{eqn:EL_eqn-power-law}, where we found the parameter $A$ by fitting the power law spectra to the original spectra. 

The figure demonstrates that the slowly varying power law spectra deviate from the original spectra by many orders of magnitude. This substantial disagreement makes the agreement at the level of the $R_{AB}$ shown in \cref{fig:EL_spectrum-ratio-vs-full-raa} astounding and calls into question the validity of the steps taken to arrive at the result in \cref{eqn:EL_power-law-raa}.

In order to understand why the slowly-varying power law $R_{AB}$ in \cref{eqn:EL_power-law-raa} is a good approximation to the full $R_{AB}$ for most phenomenologically relevant systems, we consider an expansion of the integrand of the full $R_{AB}$ in \cref{eqn:mod_full_raa_spectrum_ratio} in powers of the lost fractional energy $\epsilon$. We will perform the expansion presented in \cref{eqn:EL_schematic_moment_expansion} explicitly for the full spectrum $R_{AB}$ in \cref{eqn:mod_full_raa_spectrum_ratio} and the slowly-varying power law approximation to the $R_{AB}$ in \cref{eqn:EL_power-law-raa}. For simplicity, we neglect the Jacobian $1 / 1 - \epsilon$ since it is common to both expressions and simply expand the factor $f(p_T / 1-\epsilon) / f(p_T)$ in powers of $\epsilon$. We obtain
\begin{equation}
	\begin{aligned}[t]
		\frac{f(p_T/1-\varepsilon)}{f(p_T)} =& \sum_{m \geq 0} p_T^m \frac{f^{(m)} (p_T)}{f(p_T)} \frac{1}{m!} \left( \frac{\varepsilon}{1-\varepsilon}\right)^m\\
  =& 1 +p_T \frac{f^{\prime}\left(p_T\right)}{f\left(p_T\right)} \varepsilon 
	+ \left[p_T \frac{f^{\prime}\left(p_T\right)}{f\left(p_T\right)}+\frac{1}{2} p_T^2 \frac{f^{\prime \prime}\left(p_T\right)}{f\left(p_T\right)}\right] \varepsilon^2 \\
	&+ \left[p_T \frac{f^{\prime}\left(p_T\right)}{f\left(p_T\right)}+\frac{2}{2 !} p_T^2 \frac{f^{\prime \prime}\left(p_T\right)}{f\left(p_T\right)}+\frac{1}{3 !} p_T^3 \frac{f^{\prime \prime \prime}\left(p_T\right)}{f\left(p_T\right)}\right] \varepsilon^3 + \mathcal{O}(\epsilon^4) \\
	  =& 1 -n(p_T) \varepsilon + \left[-n(p_T)+\frac{1}{2} p_T^2 \frac{f^{\prime \prime}\left(p_T\right)}{f\left(p_T\right)}\right] \varepsilon^2 \\
	&+ \left[-n(p_T) + \frac{2}{2 !} p_T^2 \frac{f^{\prime \prime}\left(p_T\right)}{f\left(p_T\right)}+\frac{1}{3 !} p_T^3 \frac{f^{\prime \prime \prime}\left(p_T\right)}{f\left(p_T\right)}\right] \varepsilon^3 + \mathcal{O}(\varepsilon^4),
\end{aligned}
\label{eqn:EL_eqn-raa-taylor-expansion}
\end{equation}
where in the last line we have substituted $n(p_T) \equiv - p_T f'(p_T) / f(p_T)$ according to \cref{eqn:EL_eqn-npt-derivative}. The equivalent procedure can be carried out for \cref{eqn:EL_eqn-raa-power-law-simplifications} by expanding the factor $(1-\epsilon)^{n(p_T)}$,
\begin{align}
	(1-\varepsilon)^{n(p_T)}=& \sum_{m \geq 0} \frac{n(p_T)[n(p_T) - 1] \cdots [n(p_T)-m+1]}{m!} (-1)^m \epsilon^m \label{eqn:EL_eqn-npt-expansion}\\
 =&\; 1-n\left(p_T\right) \varepsilon+\frac{n(p t)\left[n\left(p_T\right)-1\right]}{2} \varepsilon^2 -\frac{n\left(p_T\right)\left[n\left(p_T\right)-1\right]\left[n\left(p_T\right)-2\right]}{6} \varepsilon^3 +\mathcal{O}(\varepsilon^4).\nonumber
\end{align}

\begin{figure}[!htbp]
	\centering
	\centering
	\includegraphics[width=0.6\linewidth]{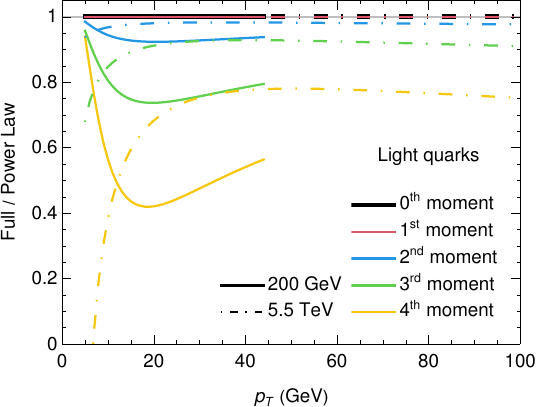}
	\caption{Order-by-order ratio of the $\mathcal{O}(\langle \epsilon^i \rangle)$ contribution to the $R_{AB}$ for the full result, to the same order contribution for the power law approximation. The order-by-order ratio is shown for the $0^{\text{th}}--4^{\text{th}}$ order contributions for light quarks produced at both $\sqrt{s_{NN}} = 200$ GeV and $\sqrt{s_{NN}} = 5.5$ TeV.
	}
	\label{fig:EL_order_by_order_full_vs_npt_raa}
\end{figure}

It is immediately evident that \cref{eqn:EL_eqn-raa-taylor-expansion} and \cref{eqn:EL_eqn-npt-expansion} agree at zeroth and first order in $\epsilon$; however, the second and third order terms are not immediately comparable. \Cref{fig:EL_order_by_order_full_vs_npt_raa} shows the contributions of various orders in $\epsilon$ to the $R_{AB}$ for the full $R_{AB}$ calculated with \cref{eqn:mod_full_raa_spectrum_ratio} and the slowly-varying power law approximation to the $R_{AB}$ calculated with \cref{eqn:EL_power-law-raa}. 

In the figure, we see that there are significant contributions from terms of order $\geq 2$ in $\epsilon$, for which the expansions in \cref{eqn:EL_eqn-npt-expansion,eqn:EL_eqn-raa-taylor-expansion} appear to be different.

We proceed by calculating the coefficients of the terms in the full $R_{AB}$ expansion in \cref{eqn:EL_eqn-raa-taylor-expansion} in terms of the $n(p_T)$ function. Starting from $n(p_T) \equiv - p_T f'(p_T) / f(p_T)$ which implies $f'(p_T) = - f(p_T) n(p_T) / p_T$, we obtain
\begin{align}
p_T^2 \frac{f''(p_T)}{f(p_T)}  =& n\left(p_T\right)\left[n\left(p_T\right)+1\right] + p_T n'(p_T) \\
p_T^3 \frac{f'''p_T)}{f(p_T)}  =& - n(p_T) \left[n(p_T) + 1\right] \left[n(p_T) + 2\right] \nonumber+ p_T n'(p_T) [3 n(p_T) + 2] - p_T^2 n''(p_T)
\end{align}
	Now, we can read off the coefficient of the second order in $\epsilon$ as
	\begin{multline}
		p_T \frac{f'(p_T)}{f(p_T)} + \frac{1}{2} p_T^2 \frac{f''(p_T)}{f(p_T)}  = \frac{1}{2} n(p_T) [n(p_T) - 1] + \frac{1}{2} p_T n'(p_T)\\
	\end{multline}
	The coefficient of the third order in $\epsilon$ is
	\begin{multline}
		p_T \frac{f'(p_T)}{f(p_T)} + p_T^2 \frac{f''(p_T)}{f(p_T)}  + \frac{1}{3!} \frac{f'''(p_T)}{f(p_T)} = - \frac{1}{6} n(p_T) \left[n(p_T) - 1\right] \left[n(p_T) - 2\right] \\
		+ \frac{1}{2} p_T n'(p_T) \left( \frac{4}{3} + n(p_T)\right) - \frac{p_T^2 n''(p_T)}{6}.
		\label{eqn:EL_third_order}
	\end{multline}

	The order by order difference between \cref{eqn:EL_eqn-raa-taylor-expansion,eqn:EL_eqn-npt-expansion} can then be written as
	\begin{multline}
		f(p_T/1-\varepsilon)/f(p_T) - (1-\varepsilon)^{n(p_T)} = \frac{1}{2} p_T n'(p_T) \epsilon^2 + \left[\frac{1}{2} p_T n'(p_T) \left( \frac{4}{3} + n(p_T)\right) - \frac{p_T^2n''(p_T)}{6}\right] \epsilon^3 \\+ \mathcal{O}(\epsilon^4).
		\label{eqn:EL_difference_taylor_npt_spectrum}
	\end{multline}

We note the surprising cancellation that occurs at each order in $\epsilon$, leaving only terms proportional to $n'(p_T)$ or higher order derivatives of $n$.

The expansion of the $R_{AB}$ in terms of the moments of the energy loss probability distribution for the slowly-varying power law $R_{AB}$ and the full spectrum $R_{AB}$ is identical up to $\mathcal{O}(\epsilon^2)$. At higher orders in $\epsilon$ the slowly-varying power law $R_{AB}$ from \cref{eqn:EL_power-law-raa} differs from the full $R_{AB}$ in \cref{eqn:mod_full_raa_spectrum_ratio}.

From this analysis, it is unsurprising that we see a large relative difference in the slowly-varying power law $R_{AB}$ and the full $R_{AB}$ at low $p_T$ in \coll{Pb}{Pb} and \coll{Au}{Au}, where there is a large amount of energy loss.

We conclude that the slowly-varying power law $R_{AB}$ expression in \cref{eqn:EL_eqn-power-law} still useful for its explanatory power, and provides a good approximation for most of the phenomenologically relevant phase space. 

	One should think of the slowly-varying power law $R_{AB}$ as an approximation in the limit of small fractional energy loss ($\epsilon \ll 1$) and slowly varying spectra ($p_T^m n^{(m)}(p_T) \ll 1 \forall m > 0$). Therefore, while the original calculation performed to derive \cref{eqn:EL_power-law-raa} is not valid, the expression is still a well-defined approximation to the full $R_{AB}$, as motivated by the expansion of the $R_{AB}$ in moments of the energy loss probability distribution performed in this work. While initially, the agreement at the level of the $R_{AB}$ of the slowly-varying power law approximation to the full $R_{AB}$ is astounding, this analysis in terms of the moments makes it clear that the mistake is thinking of the function $n(p_T)$ as the power law of the production spectrum. Instead it should simply be defined as $n(p_T) \equiv p_T f'(p_T) / f(p_T)$, where $f(p_T)$ is the production spectrum.

Since the full expression for the $R_{AB}$ in \cref{eqn:mod_full_raa_spectrum_ratio} is not significantly more numerically intensive, one should use this expression for all numerical results.

\section{Conclusions}
\label{sec:EL_conclusions}

In this chapter we presented results for the nuclear modification factor of leading high-$p_T$ hadrons from a pQCD based energy loss model which is, to a good approximation, the Wicks-Horowitz-Djordjevic-Gyulassy (WHDG) \cite{Wicks:2005gt} convolved radiative and elastic energy loss model which receives small system size corrections to \emph{both} the radiative \cite{Faraday:2023mmx, Kolbe:2015rvk, Kolbe:2015suq} \emph{and} elastic \cite{Wicks:2008zz} energy loss. 
Our model utilizes realistic production spectra, fragmentation functions, and a QGP medium geometry with fluctuating initial conditions generated by IP-Glasma \cite{Schenke:2020mbo,shen_private_communication,Schenke:2012hg, Schenke:2012wb}. In this work, we expanded our model \cite{Faraday:2023mmx} to include a more realistic elastic energy loss kernel \cite{Wicks:2008zz}, calculated within the HTL formalism \cite{Braaten:1989mz, Klimov:1982bv, Pisarski:1988vd, Weldon:1982aq, Weldon:1982bn}, and with reduced approximations in comparison to the elastic energy loss kernel \cite{Braaten:1991jj, Braaten:1991we} used previously \cite{Faraday:2023mmx}. This new elastic energy loss kernel allowed us to retain the distributional information of the elastic energy loss, thereby removing the need to apply the central limit theorem \cite{Moore:2004tg} in the calculation, a common assumption in the literature \cite{Wicks:2005gt, Horowitz:2011gd, Zigic:2021rku}.

We varied the radiative energy loss kernel between the DGLV radiative energy loss kernel \cite{Djordjevic:2003zk} and the DGLV radiative energy loss kernel which receives a short pathlength correction \cite{Kolbe:2015rvk, Kolbe:2015suq}. We varied the elastic energy loss kernel between the Braaten and Thoma elastic energy loss kernel \cite{Braaten:1991jj, Braaten:1991we} (used in our previous work \cite{Faraday:2023mmx}), and an HTL-based elastic energy loss kernel \cite{Wicks:2008zz}. Additionally, we considered the full Poissonian distribution \cite{Gyulassy:2001nm}, which arises naturally from the HTL-based approach, as well as the Gaussian approximation \cite{Moore:2004tg} to this elastic energy loss distribution. 

This set of elastic energy loss kernels allowed us to interrogate the theoretical uncertainty in the elastic energy loss due to the crossover from HTL propagators to vacuum propagators \cite{Wicks:2008zz, Romatschke:2004au}, %
as well as the effect of the Gaussian approximation of the elastic energy loss distribution \cite{Moore:2004tg} according to the central limit theorem.

Model results for the nuclear modification factor were presented for light and heavy-flavor hadrons produced in central \coll{p}{Pb} and central and peripheral \coll{Pb}{Pb} collisions at LHC, and in central \collFour{p}{d}{He3}{Au} and central and peripheral \coll{Au}{Au} collisions at RHIC.

We found that for both heavy and light flavor mesons produced in central \coll{Pb}{Pb} and \coll{Au}{Au} collisions, results calculated with the Poisson HTL elastic energy loss differed from results calculated with the Gaussian HTL elastic energy loss by only $\mathcal{O}(5\text{--}15\%)$ for mid--high $p_T$. For light and heavy-flavor mesons produced in central \coll{p}{Pb} and \collFour{p}{d}{He3}{Au} collisions, we found that the difference between $R_{AB}$ results calculated with the Poisson HTL and Gaussian HTL elastic energy loss distributions was extremely small, $\mathcal{O}(2\text{--}5\%)$, for all $p_T \gtrsim 5$ GeV. This negligible difference between the Gaussian and Poisson results is especially surprising given the $\mathcal{O}(0\text{--}1)$ scatters that occur in small systems, making the central limit theorem inapplicable. 
This trend of $R_{AB}$ as a function of system size contradicts the expectation that the Gaussian approximation should \emph{improve} with increasing system size, according to the central limit theorem.
Paradoxically, the approximation is most accurate for small systems, where the central limit theorem is inapplicable, and least accurate for large systems.

We demonstrated that the system size, $\sqrt{s_{NN}}$, and flavor dependence of the relative difference between $R_{AA}$ results calculated with Gaussian and Poisson elastic distributions can be explained by expanding the integrand of the $R_{AB}$ in the lost fractional energy $\epsilon$. Such an expansion leads to an expression for the $R_{AB}$ in terms of the moments of the underlying energy loss distributions.
We showed that when there is only a small degree of suppression, as there is in small systems, the $R_{AB}$ depends mostly on the first few moments, the zeroth and first of which are constrained to be identical for the Gaussian and Poisson elastic energy loss. Conversely, in large systems, the higher order moments are more important, exacerbating the difference between the $R_{AB}$ calculated with the Gaussian and the Poisson elastic energy loss distributions, respectively. The slope of the production spectra also impacts which moments are important in this expansion of the $R_{AA}$. At RHIC, the production spectra are steeper than for the equivalent partons at LHC, which leads to the higher-order moments being more important at RHIC vs.\ at LHC. This can be understood simply by approximating $R_{AA} \sim \int d \epsilon \; (1-\epsilon)^{n-1} P_{\text{tot.}}(\epsilon)$ where $d \sigma / d p_T \propto p_T^{-n}$ is the production spectrum and $n$ is a constant \cite{Horowitz:2010dm}. At RHIC $n \sim 6\text{--}8$ while at LHC $n \sim 4\text{--}5$, from which it is evident that at RHIC, the higher order moments of a distribution are more important than at LHC. The $p_T$ dependence of the relative difference between results calculated with the Poisson HTL and Gaussian HTL elastic energy loss distributions is a result of the decreasing importance of the elastic energy loss compared to the radiative energy loss as a function of $p_T$.

	We suggested previously \cite{Faraday:2023mmx} that if one assumes that medium-induced gluon emission and vacuum gluon emission are both independent, then a more realistic distribution for the \emph{additional} gluon emission in medium, may be a Skellam distribution. A further study of the distributional dependence of the $R_{AB}$, using the techniques introduced in this work, is necessary to understand how sensitive the $R_{AB}$ is to the modeling of the radiative energy loss distribution as Poissonian compared to, for instance, a Skellam distribution.

In our previous work \cite{Faraday:2023mmx}, we found that elastic energy loss accounted for $\sim 90\%$ of the suppression in \coll{p}{Pb} collisions, which we suggested may be due to a breakdown of the application of the central limit theorem in these systems with $\mathcal{O}(0\text{--}1)$ elastic scatterings. Our results in this work imply that the large contribution of the elastic energy loss, in comparison to the radiative energy loss, in small systems is \textbf{not} due to the inapplicability of the Gaussian approximation, but rather a physical effect. The different length dependencies of the elastic and radiative energy loss---due to LPM interference the radiative energy loss scales as $L^2$, while the elastic energy loss scales as $L$---explains the large contribution of the elastic energy loss compared to the radiative energy loss in small systems.  

We demonstrated that---over the range of phenomenologically interesting collision systems, final states, and momenta---the fractional contribution of the radiative vs.\ elastic energy loss varies from $\sim \! 0.25 \text{--} 4$, demonstrating the importance of including both elastic and radiative energy loss in theoretical energy loss models.

We also made comparisons for all aforementioned systems between $R_{AB}$ results calculated with the Gaussian HTL \cite{Wicks:2008zz} and the Gaussian Braaten and Thoma (BT) \cite{Braaten:1991jj, Braaten:1991we} elastic energy loss distributions. These calculations both utilize HTL propagators \cite{Braaten:1989mz, Klimov:1982bv, Pisarski:1988vd, Weldon:1982aq, Weldon:1982bn} to calculate the elastic energy loss suffered by a high-$p_T$ parton moving through the QGP. They differ in that the BT result uses vacuum propagators at high momentum transfer, and HTL at low momentum transfer, while the HTL result \cite{Wicks:2008zz} uses HTL propagators for all momentum transfers. One may, therefore, consider the two results as an approximate measure of the theoretical uncertainty in the crossover region from HTL to vacuum propagators \cite{Wicks:2008zz, Romatschke:2004au}.  %
We found that the difference between $R_{AA}$ results calculated with the Poisson HTL and BT elastic energy loss kernels was $\mathcal{O}(30\text{--}50\%)$ for $p_T$ of $\mathcal{O}(10\text{--}50)$ GeV for both light and heavy flavor hadrons produced in central \coll{Pb}{Pb} and \coll{Au}{Au} collisions. For light and heavy flavor mesons produced in central \coll{p}{Pb} collisions, the difference between $R_{pA}$ results calculated with Poisson HTL and BT elastic energy loss kernels was $\mathcal{O}(5\text{--}10\%)$ for $p_T$ in the range $\mathcal{O}(5\text{--}50)$ GeV. For pions produced in central \collFour{p}{d}{He3}{Au} collisions we found a $\mathcal{O}(10\text{--}25\%)$ relative difference between the two calculations. In all cases, the relative difference between the $R_{AB}$ results calculated with the two different elastic energy loss kernels decreases as a function of $p_T$, due to the decreasing contribution of the elastic energy loss in comparison to the radiative energy loss. While the sensitivity seems decreased in small systems, this is an artifact of the small amount of energy loss in small systems. If one instead considers the difference at the level of the $1 - R_{AA}$, which more closely corresponds to energy loss, the difference in results calculated with HTL and BT propagators becomes $\mathcal{O}(50\text{--}100)\%$. 

We conclude that the uncertainty in the choice of elastic energy loss kernel is essential for making rigorous and quantitative phenomenological predictions for suppression in \emph{both} small and large systems. 
For models applied to systems of similar sizes, one may absorb this fundamental uncertainty into one's choice of the effective coupling constant, for instance. However, this may not be possible for models applied to a wide range of system sizes because the elastic energy loss scales like $\alpha_s^2$, while the radiative energy loss scales like $\alpha_s^3$, and various system sizes receive a vastly different proportion of their total energy loss from the elastic and radiative energy loss kernels; meaning that a global change of $\alpha_s$ will not globally affect the $R_{AA}$ in the same way as a change of the elastic energy loss kernel will. We propose that a system size scan will help to elucidate which prescription for changing between HTL and vacuum propagators is favored by suppression data, in lieu of an analytical approach to this problem.

	In this work, we also investigated the impact of the short pathlength correction \cite{Kolbe:2015rvk, Kolbe:2015suq} to the DGLV radiative energy loss \cite{Djordjevic:2003zk}. As in our previous work \cite{Faraday:2023mmx}, we saw that the short pathlength correction is anomalously large for high-$p_T$ pions, which is likely due to the breakdown of the large formation time assumption, which we discussed in detail in \cite{Faraday:2023mmx}. We investigated the phenomenology of the short pathlength correction for the first time at RHIC energies,  where we found that the smaller momentum range, as well as the smaller fraction of gluons compared to light quarks, lead to the short pathlength correction being negligible in \emph{both} \coll{Au}{Au} and \collFour{p}{d}{He3}{Au} collisions. We also presented results for semi-central and peripheral \coll{Pb}{Pb} collisions for the first time, where we observed that the short pathlengths lead to anomalously large $R^{\pi}_{AA} \sim \mathcal{O}(1.1 \text{--}1.3)$ in the $p_T$ range $\mathcal{O}(50\text{--}400)$ GeV. We reiterate our conclusions from our previous work \cite{Faraday:2023mmx}: control over the large formation time approximation \cite{Faraday:2023mmx}---which breaks down dramatically for the short pathlength correction at high-$p_T$ and in small systems---will be crucial in making quantitative theoretical predictions. Future work \cite{Faraday:2024} will consider the effects of including a cut-off on the radiated transverse momentum, which ensures that no contributions from regions where the large formation time approximation is invalid are included in the calculation.

Finally, we investigated the validity of the slowly-varying power law approximation to the $R_{AA}$, which has previously been used \cite{Horowitz:2010dm, Wicks:2005gt, Horowitz:2011gd, Faraday:2023mmx}. The power law approximation to the $R_{AA}$ is given by $R_{AA} \simeq \int d \epsilon \; (1-\epsilon)^{n(p_T)-1}$, where $f(p_T) \equiv d \sigma / d p_T \propto p_T^{-n(p_T)}$ is the production spectrum and $n(p_T)$ is a slowly-varying function. We found that if one compares the spectra $f(p_T)$ and $ A p_T^{-n(p_T)}$, where $A$ is a fitted constant, the spectra differ by many orders of magnitude. However if one compares the power law $R_{AA}$ to the full $R_{AA}$, these quantities differ by $\mathcal{O}(5\text{--}15\%)$ only. We presented a new derivation of the power law $R_{AA} \simeq \int d \epsilon \; (1-\epsilon)^{n(p_T)}$, based on an expansion of the $R_{AA}$ in terms of the moments of the energy loss distribution $\langle \epsilon^i \rangle$. This expansion showed that $n(p_T)$ is \emph{not} the power of the spectrum but rather $n(p_T) \equiv p_T f'(p_T) / f(p_T)$, and that the power law $R_{AA}$ is valid in the limit of small fractional energy loss ($\langle \epsilon \rangle \ll 1$) and slowly varying $n(p_T)$ ($p_T^m n^{(m)}(p_T) \ll 1$ for all $m \geq 1$).

	Since this work aimed to understand various approximations and uncertainties in our energy loss model, we made no comparisons to experimental data. Future work in preparation \cite{Faraday:2024} will perform a global fit of the strong coupling $\alpha_s$ to available large system \coll{A}{A} data at LHC and RHIC. This large-system-constrained model may then be used to make predictions for small \collFour{p}{d}{He3}{A} systems, which will allow us to quantitatively infer the extent to which a single pQCD energy loss model can quantitatively describe both small and large system suppression data simultaneously.

 \chapter{The large formation time assumption and a one-parameter fit to experimental data}
 \label{sec:lft}
 
\section{Introduction}
\label{sec:lft_introduction}

In \cref{sec:EL_paper,sec:SPL_paper}, and the corresponding papers \cite{Faraday:2023mmx, Faraday:2024gzx}, we saw that the short pathlength correction \cite{Kolbe:2015suq, Kolbe:2015rvk} was extremely large at high-$p_T$ for pions produced in both \coll{Pb}{Pb} and \coll{p}{Pb} collisions. In \cref{sec:SPL_assumptions}, we concluded that this was likely due to the self-consistency  breakdown of the \emph{large formation time approximation}, a particular assumption in the DGLV approach. Ideally, one should remove this assumption, and calculate the resulting \emph{short formation time correction} in order to rigorously understand the implications of such a self-consistency breakdown. 

In this chapter, we will take a simpler, phenomenological approach to reducing the impact of the large formation time breakdown. We suggested previously \cite{Faraday:2023uay,Faraday:2023mmx} that one may limit contributions from regions where the large formation time approximation is invalid, through a bound on the transverse radiated gluon momentum $|\mathbf{k}|$. This approach is similar to the typical bound on the transverse radiated gluon momentum, which is motivated by the collinear approximation \cite{Djordjevic:2003zk, Wicks:2005gt, Horowitz:2009eb}. We will see that this prescription dramatically reduces the short pathlength correction and also impacts the uncorrected DGLV result, making the DGLV and DGLV + SPL results nearly identical. 
The impact of this upper bound is particularly large at high-$p_T$, due to the $\sqrt{E}$ scaling of the large formation time upper bound, in contrast to the $E$ scaling of the collinear upper bound.

In addition to the implementation of this upper bound, we explore the sensitivity of our result to the exact value chosen for the upper bound, through varying the upper bound by factors of two. We find that the sensitivity to the exact upper bound is $\mathcal{O}(30\text{--}70)\%$---even larger than the sensitivity to the collinear upper bound \cite{Horowitz:2009eb}. 

Moreover, in this chapter, we make a first quantitative comparison with experimental data. As discussed in \cref{sec:EL_paper}, we did not make any comparison with data in order to keep the discussion clear. Due to the different elastic energy loss models leading to different suppression patterns, it was difficult to draw conclusions from a comparison to data, as very different strong couplings are needed in order to compare favorably to data. In this section, we perform a one-parameter fit of the strong coupling $\alpha_s$ to available large system data from RHIC and LHC. 
Since this is a first comparison, we aim to keep the process simple, and broadly we follow the procedure outlined in \cite{PHENIX:2008ove, PHENIX:2008saf}. Future work may treat this issue with a Bayesian analysis \cite{JETSCAPE:2021ehl, Karmakar:2023ity}.

Many extractions of various model parameters have been conducted by various groups for various energy loss models \cite{JET:2013cls, JETSCAPE:2017eso, JETSCAPE:2022jer,JETSCAPE:2024cqe, Karmakar:2023ity,Karmakar:2024jak,Shi:2018izg, Stojku:2020tuk,Bass:2008rv,Armesto:2009zi,Chen:2010te,Qin:2009bk,Kang:2014xsa}, typically focused on extracting the \emph{jet quenching parameter} $\hat{q}$. We argue that because of the large theoretical uncertainties in our model due to the large formation time approximation, such an extraction is premature. In the conclusions, we will discuss the extent to which our work on the large formation time approximation will impact other energy loss models and the extractions of medium properties from such models. 
As such, we emphasize that this fit is \textbf{not} an attempt to extract a physical value of the strong coupling constant. We will see that the various uncertainties in the model lead to different best fit values for $\alpha_s$, in the range $\alpha_s = 0.31 \text{--} 0.49$, indicating that an attempt to extract a physical value is currently not possible (with any reasonable level of confidence). Furthermore, we are neglecting important effects including running coupling \cite{Peigne:2008nd, Buzzatti:2012pe}, higher orders in opacity \cite{Wicks:2008ta}, and a full integral through the evolving hydrodynamic medium \cite{Bert:2024}. Physical insight is still possible, however, even without such an extraction of properties of the plasma. The questions we will address are:
\begin{itemize}
	\item \emph{To what extent is the suppression pattern in small systems compatible with that of large systems?}
	\item \emph{Can the uncertainty in the choice of elastic energy loss kernel be absorbed into a different strong coupling constant $\alpha_s$?}
	\item \emph{Can the uncertainty in the large formation time cutoff be absorbed into a different strong coupling constant $\alpha_s$?}
	\item \emph{To what extent can a variety of suppression patterns for various collision systems, final state hadrons, centralities, and $\sqrt{s}$ be described by a model with a single tuned parameter?}
\end{itemize}

\section{The large formation time assumption}
\label{sec:the_large_formation_time_assumption}

The large formation time assumption assumes that $\omega_1 \ll \mu_1 \Leftrightarrow (\mathbf{k} - \mathbf{q_1})^2 / 2 x E \ll \sqrt{\mu^2 + q_1^2}$. This assumption, among many other assumptions discussed in \cref{sec:SPL_assumptions,sec:mod_radiative_energy_loss_correction}, is made in the GLV \cite{Gyulassy:2000er} and DGLV \cite{Djordjevic:2003zk} calculations of the radiative energy loss. In \cref{sec:SPL_assumptions} we discussed the extent to which the various approximations made in the radiative energy loss calculation were self-consistent, by calculating the expectation value of dimensionless ratios $\langle R \rangle$ which were assumed to be much smaller than one in the original DGLV \cite{Djordjevic:2003zk} calculation. We found that the large formation time assumption was explicitly violated at large energies for both the DGLV result and the DGLV result which receives a short pathlength correction (DGLV + SP); see \cref{fig:SPL_largeformationtime2,fig:SPL_largeformationtime3,fig:SPL_largeformationtime2NoStep}. The degree of breakdown was significantly larger for: DGLV + SPL result vs.\  DGLV result, gluons vs.\ charm quarks, and the exponential vs.\ step distribution of scattering centers.

We suggested in \cite{Faraday:2023mmx} and \cref{sec:SPL_paper} that this self-consistency breakdown was likely due to contributions from regions of the phase space where the large formation time was invalid. Further, we suggested that one could enforce that only contributions from regions where the large formation time is valid are included by limiting the \rev{transverse} radiated momentum \rev{magnitude} $|\mathbf{k}|$. In current works, the upper bound on $|\mathbf{k}|$, $|\mathbf{k}|_{\text{max}}$, is obtained by requiring both the radiated gluon and the parton post-radiation to both be traveling in the $+$ direction (collinearity). Explicitly, we require that $k^+ \gg k^-$ and $p^+ \gg p^-$ where $k$ is the four momentum of the radiated gluon, and $p$ the four momentum of the post-radiation parton. From \cref{eqn:mod_momenta_k}, it follows that
\begin{align}
	x P^+ \gg& \frac{m_g^2 + k^2}{x P^+} \nonumber\\
	\implies 2 x E \gg&  |\mathbf{k}|,
	\label{eqn:lft_condition_from_k}
\end{align}
where in the last line we have approximated $m_g \sim 0$. Similarly, from \cref{eqn:mod_momenta_p} we have
\begin{align}
	(1-x) P^+ \gg& \frac{M^2 + \mathbf{k}^2}{(1-x) P^+} \nonumber\\
	\implies (1-x) 2 x E \gg& |\mathbf{k}|,
	\label{eqn:lft_condition_from_p}
\end{align}
where in the last line we have approximated $M \sim 0$. This last approximation may be problematic for heavy quarks at low momenta, but discussing this is beyond the scope of this work. One possibility to limit the phase space is to then require both of these conditions to be true, i.e.\ set $|\mathbf{k}|_{\text{max}} = \operatorname{Min}\left[2 x E, (1 - x)E\right]$. The option we followed in this work, following the literature \cite{Djordjevic:2003zk, Wicks:2005gt}, was to set $|\mathbf{k}|_{\text{max}} = 2x (1-x) E \leq \operatorname{Min}\left[2 x E, (1 - x)E\right]$, which guarantees that the conditions in \cref{eqn:lft_condition_from_k,eqn:lft_condition_from_p} are both satisfied\footnote{Since both $0 \leq x \leq 1$ and $0 \leq (1-x) \leq 1$.}. 

We follow a similar approach for the large formation time assumption, $(\mathbf{k} - \mathbf{q_1})^2 / 2 x E \ll \sqrt{\mu^2 + q_1^2}$. Mandating that the large formation time assumption is followed leads to 
\begin{align}
	(\mathbf{k} - \mathbf{q_1})^2 / 2 x E \ll& \sqrt{\mu^2 + q_1^2} \nonumber\\
	\implies |\mathbf{k}| \ll& \sqrt{2 x E\sqrt{\mu^2 + q_1^2}}, \nonumber\\
	\label{eqn:lft_largeformationtime_bound_derivation}
\end{align}
where in the last line we have approximated that $|\mathbf{q_1}| \ll |\mathbf{k}|$. This motivates an upper bound $|\mathbf{k}|_{\text{max}} = \operatorname{Min}[2 x E (1-x), \sqrt{2 x E\sqrt{\mu^2 + \mathbf{q_1}^2}}]$. We note that cutoffs motivated in such a way, are motivated by scale, but the exact value chosen is not well defined. One may estimate the sensitivity of our results to the exact value chosen for the cutoff by varying the upper bound by a factor; typically a factor of two is used. Horowitz and Cole \cite{Horowitz:2009eb} performed such an estimate of the uncertainty associated with the collinear upper bound, framed in an ambiguity of the use of $x_E$ (euclidean) or $x^+$ (light cone) and the maximum angle of allowed radiation. There is not a simple analog of this procedure once the large formation time upper bound is included, and so we will estimate the sensitivity of our results to the cutoff by calculating $\Delta E / E$ with $|\mathbf{k}|_{\text{max}} = \kappa \operatorname{Min}[2 x E (1-x), \sqrt{2 x E\sqrt{\mu^2 + \mathbf{q_1}^2}}]$, where $\kappa$ will be varied in the range $[0.5, 2]$\footnote{In principle one could vary the $|\mathbf{k}|_{\text{max}}$ associated with the collinear and large formation time upper bound independently; i.e.\ set $|\mathbf{k}|_{\text{max}} = \operatorname{Min}[\kappa_1 2 x E (1-x), \kappa_2 \sqrt{2 x E\sqrt{\mu^2 + \mathbf{q_1}^2}}]$ with $(\kappa_1, \kappa_2) \in [0.5, 2] \otimes [0.5, 2]$. Presumably, each individual curve generated would be different in this prescription, however the band covered by the curves should be similar.}.

\begin{figure}[!b]
	\centering
	\includegraphics[width=0.9\linewidth]{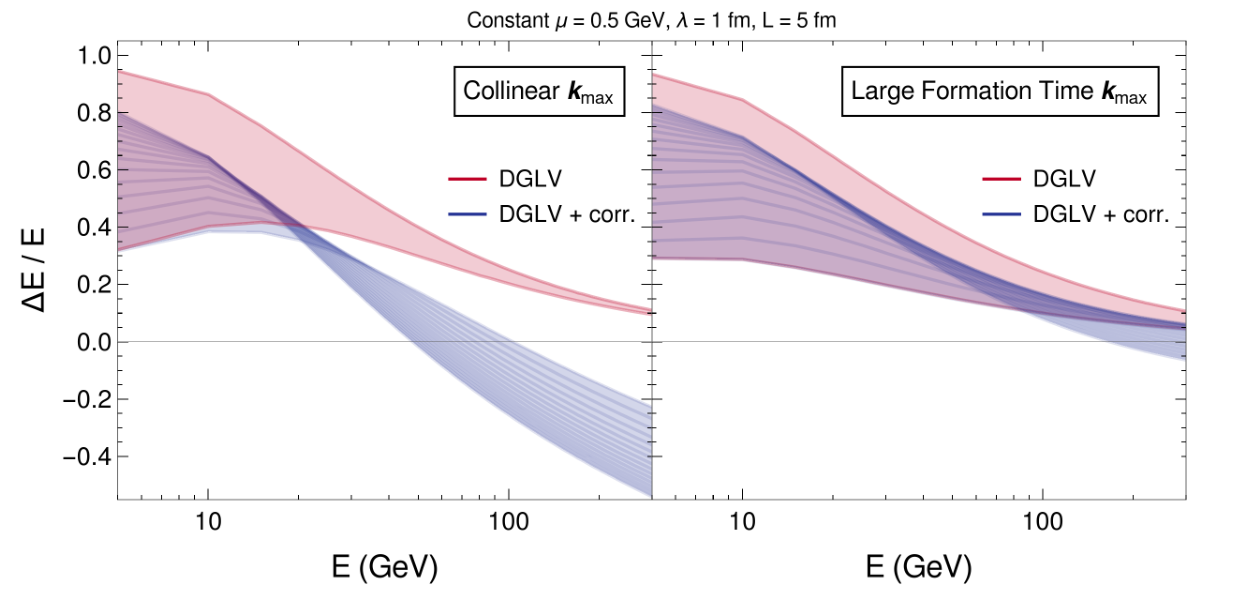}
	\caption{The fractional radiative energy loss calculated according to DGLV and short pathlength corrected DGLV \rev{for an incident gluon}. In the left pane the upper bound on the radiated gluon momentum $|\mathbf{k}|$ integral is set by the collinear assumption $| \mathbf{k}|_{\text{max}} = 2x E (1-x)$, while in the right pane the upper bound is taken as $|\mathbf{k}|_{\text{max}} = \text{Min}(\sqrt{2xE\mu_1},2x(1-x)E)$. Bands are calculated by varying the upper bound by factors of two. Calculations are performed with $\mu = 0.5~\mathrm{GeV}$, $\lambda = 1~\mathrm{fm}$, and $L = 5~\mathrm{fm}$. Figure originally produced for \cite{Faraday:2023uay}.}
	\label{fig:lft_vs_collinear_varying_deltaEoverE}
\end{figure}

\Cref{fig:lft_vs_collinear_varying_deltaEoverE} plots the $\Delta E  / E$ as a function of incident energy $E$ for the DGLV radiative energy loss and the DGLV radiative energy loss which receives a short pathlength correction. The left pane of the figure is plotted with the standard collinear bound $|\mathbf{k}|_{\text{max}} = \kappa 2 x E (1-x)$ and the right pane is plotted with our new collinear + large formation time bound $|\mathbf{k}|_{\text{max}} = \kappa \operatorname{Min}[2 x E (1-x), \sqrt{2 x E\sqrt{\mu^2 + \mathbf{q_1}^2}}]$. Bands are calculated by varying $\kappa$ in the range $[0.5,2]$ and calculations are performed with $\mu = 0.5~\mathrm{GeV}$, $\lambda = 1~\mathrm{fm}$, and $L = 5~\mathrm{fm}$. We note that while the DGLV result is completely monotonic as a function of $\kappa$, the DGLV result which receives a short pathlength correction is \emph{not} monotonic in $\kappa$. This nonmonotonicity is because at small energies the short pathlength correction is small, and so decreasing $\kappa$, decreases the energy loss, as it does for DGLV. At higher energies, however, the short pathlength correction is extremely large and negative, and so a decrease in $\kappa$ reduces the size of the short pathlength correction and thereby \emph{increases} the energy loss. One may observe this in \cref{fig:EL_raa_pbpb_D6080} through the crossing of the lines in the DGLV + corr.\ curve, where each line corresponds to a different value of $\kappa$.

Now that we have a sense of how the large formation time + collinear cutoff on the transverse radiated gluon momentum impacts the DGLV and DGLV + SPL results, it will be of value to see the sensitivity at the level of the $R_{AB}$. 
In \cref{sec:lft_results} we will see what the values of the $R_{AB}$ are for different collision systems, of particular interest will be pions at large momenta in \coll{Pb}{Pb} and \coll{p}{Pb} collisions. We saw in \cref{sec:SPL_paper,sec:EL_paper} that the short pathlength correction led to $R_{AB} > 1$ for pions at high $p_T$ produced in both \coll{Pb}{Pb} and \coll{p}{Pb} collisions at $\sqrt{s_{NN}} = 5.02 ~\mathrm{TeV}$. At this point, we want to understand further the uncertainty which arises from varying the multiplier $\kappa$ of the cutoff of the transverse radiated gluon momentum. 

\begin{figure}[!b]
	\centering
	\includegraphics[width=0.8\linewidth]{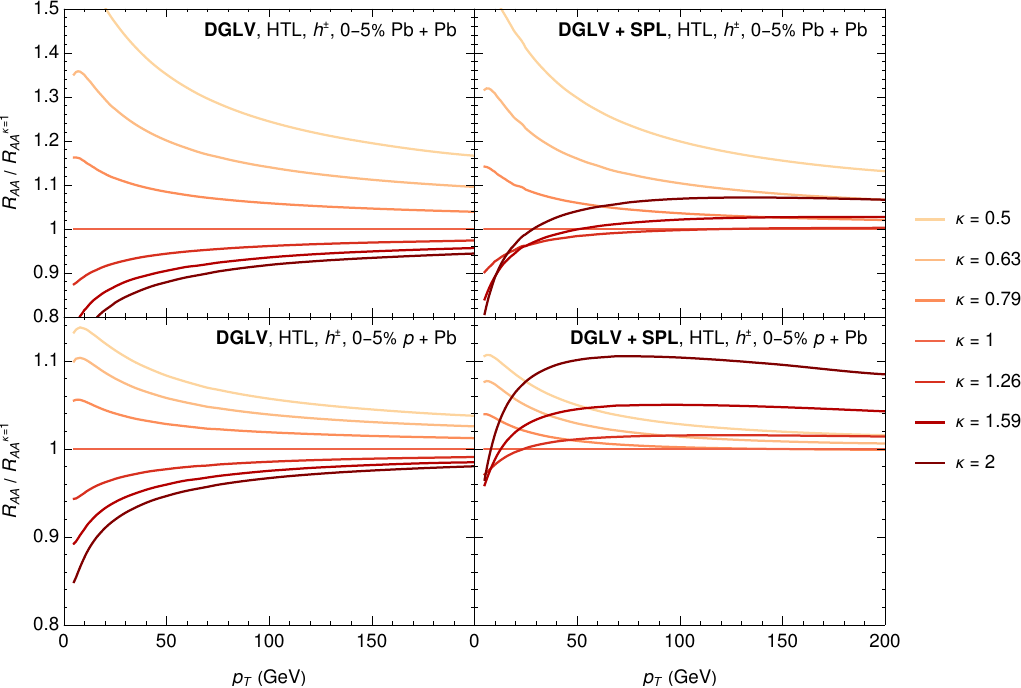}
	\caption{Plot of the ratio $R_{AA}/R_{AA}^{\kappa=1}$, as a function of transverse momentum $p_T$, where $R_{AA}$ is the nuclear modification factor for various values of the multiplier $\kappa$ of the cutoff on the transverse radiated gluon momentum and $R_{AA}^{\kappa = 1}$ is the same quantity with $\kappa = 1$. See text for details on the cutoff prescription. Theoretical curves are generated for pions produced in $0\text{--}5\%$ centrality \coll{Pb}{Pb} collisions (top) and $0\text{--}5\%$ centrality \coll{p}{Pb} collisions (bottom), comparing DGLV (left) and DGLV + SPL (right) radiative energy loss models. The elastic energy loss is HTL throughout. Different curves represent various $\kappa$ values ranging from 0.5 to 2}
	\label{fig:lft_sensitivity_at_raa_level_pbpb_and_ppb_pi}
\end{figure}

\Cref{fig:lft_sensitivity_at_raa_level_pbpb_and_ppb_pi} plots the ratio $R_{AB} / R_{AB}^{\kappa = 1}$, where $R_{AB}^{\kappa = 1}$ represents the $R_{AB}$ calculated with $\kappa = 1$. The left column shows the ratio plotted for the HTL elastic energy loss convolved with DGLV radiative energy loss, while the right column shows the ratio plotted for HTL elastic energy convolved with DGLV + SPL radiative energy loss. The top row shows the results for pions produced in $0\text{--}5\%$ \coll{Pb}{Pb} collisions at $\sqrt{s_{NN}} = 5.02 ~\mathrm{TeV}$, and the bottom row for pions produced in $0\text{--}5\%$ \coll{p}{Pb} collisions at $\sqrt{s_{NN}} = 5.02 ~\mathrm{TeV}$. All curves are produced for $\alpha = 0.4$, which we will see is the approximate global large system best fit value for this model in \cref{sec:lft_results}.

The most striking feature in \cref{fig:lft_sensitivity_at_raa_level_pbpb_and_ppb_pi} is the large sensitivity of the $R_{AA}$ to the chosen value of $\kappa$ for both the DGLV and DGLV + SPL results. 
We observe in the figure an $\mathcal{O}(35\text{--}70\%)$ variation in the charged hadron $R_{AA}$ in central \coll{Pb}{Pb} collisions and a $\mathcal{O}(10\text{--}20\%)$ variation in the charged hadron $R_{AA}$ in central \coll{p}{Pb} collisions for $10 ~\mathrm{GeV} \lesssim p_T \lesssim 100 ~\mathrm{GeV}$. While \cite{Horowitz:2009eb} found that the uncertainty stemming from varying the collinear upper bound was small at high-$p_T$, our results show that the sensitivity falls off slowly when the LFT constraint is included in the upper bound. This slow fall off of the sensitivity in $p_T$ may be understood by the different asymptotic scalings of the collinear and LFT upper bounds: the collinear upper bound scales as $E$ while the LFT upper bound scales as $\sqrt{E}$. This different scaling leads to the medium induced radiation distribution $dN / dx d |\mathbf{k}|$ being cutoff much earlier in $|\mathbf{k}|$ and where the integrand is larger, leading to a larger sensitivity.

\section{One-parameter fit to experimental data}
\label{sec:oneparameter_fit_to_experimental_data}

With our newfound control over the magnitude of the short pathlength correction, albeit with a large sensitivity to the choice of the exact value of $|\mathbf{k}|_{\text{max}}$ used, it will now be of value to make quantitative comparisons with experimental data. The goal of such a comparison is primarily to see whether the observed suppression pattern in small systems is compatible with that in large systems. In addition, we will comment on the uncertainty from the large formation time assumption, and the uncertainty in the elastic energy loss sector due to the crossover between HTL and vacuum propagators.

\subsection{Fitting procedure}
\label{sec:lft_fitting_procedure}

The model parameter which will be fitting to data is the strong coupling constant $\alpha_s$. We emphasize that the goal of such a fit is \emph{not} to extract a value of the strong coupling constant. Such an extraction would not be illuminating due to the large theoretical uncertainties present in the model; for instance the neglecting of running coupling effects, the uncertainty in the upper bound on the transverse radiated gluon energy, the neglecting of higher order terms in opacity, the approximation of the scattering centers as static (infinitely massive), the geometrical mapping to effective length and temperature functions, and the uncertainty in energy loss pre- and post-thermalization.

We treat each value of the $R_{AA}$ for a specific elastic energy loss kernel (HTL or BT), a specific radiative energy loss kernel (DGLV or DGLV + SPL), and a specific multiplier $\kappa$ of the cutoff on the radiated transverse gluon momentum $|\mathbf{k}|$. The value of $\kappa$ is varied between $0.5$ and $2$ in a logarithmic scale, i.e. $\kappa \in \{ 2^n \mid n \in [-1,1] \}$. For the DGLV radiative energy loss one could simply use $\kappa = 0.5$ and $\kappa = 2$, however, as we showed in \cref{sec:the_large_formation_time_assumption}, the DGLV + SPL result is nonmonotonic in $\kappa$. Due to the large computational cost of increasing the already large phase space of the model, we vary $\kappa$ over 7 different values, i.e. $\kappa \in \{2^{-1}, 2^{-2 / 3}, 2^{- 1 / 3}, 2^0, 2^{1 / 3}, 2^{2 / 3}, 2^1\} = \{ 0.5, 0.63, 0.79, 1, 1.26, 1.59, 2 \}$. We compute $\alpha_s$ over $\alpha_s \in \{0.3, 0.35, 0.4, 0.45, 0.5\}$, which we will see is a suitable fit for all energy loss models. The $R_{AA}$ is interpolated as a function of $\alpha_s$, in order to calculate the $\chi^2$ smoothly as a function of $\alpha_s$. We will see in \cref{sec:lft_results} that the $R_{AA}$ varies smoothly as a function of $\alpha_s$ and so this grid size is granular enough.

We will follow the fitting procedure used by PHENIX in \cite{PHENIX:2008ove, PHENIX:2008saf}, where they extracted one-parameter fits to their data for various energy loss models. Theory points $\mu_i(p)$ for the $R_{AB}$ are generated for each corresponding experimental data point $y_i$ as a function of a single parameter $p$. We then require that the experimental uncertainties can be divided into three categories \cite{PHENIX:2008saf}:
\begin{itemize}
	\item \textit{Type A ($\sigma_i$)}: $p_T$-uncorrelated uncertainties, including all statistical uncertainties and the $p_T$ uncorrelated contributions to the systematic uncertainties;
	\item \textit{Type B ($\sigma_{b_i}$)}: $p_T$-correlated contributions to the systematic uncertainties;
	\item \textit{Type C ($\sigma_{c_i}$)}: global uncertainties (normalization, uniform fractional shift for all points, etc.)
\end{itemize}
In general there will be many different sources of correlated systematic uncertainties contributing to each measured value of the observable $y_i$, which itself corresponding to the measured value at a specific $p_T$, $x_i$. One may think of the contribution of the $j^{th}$ correlated uncertainty (for instance 

We model each correlated systematic uncertainty $\Delta	y^b(\text{sys, uncertainty $j$})_i$, as it manifests on the experimental data point $y_i$ at the $p_T$ value $x_i$, as arising from a single underlying uncertainty $\Delta z^b_j$. The correlated type B variation on the measured value $y_i$ due to the $j^{th}$ correlated systematic uncertainty is then
\begin{equation}
	\Delta y^b(\text{sys, uncertainty $j$})_i  \equiv f(x_i) b_i^j \Delta z^b_j,
	\label{eqn:lft_systematic_uncertainty}
\end{equation}
where $\Delta z^b_j \equiv z^b_j - \langle z^b_j \rangle = z_b^j = (\sigma_b^j)^2$, and the random variable $z_b^j$ is the same for all measurements $y_i$. Note that we differ from \cite{PHENIX:2008ove} by introducing the function $f(x_i)$ which determines the manor in which the type B uncertainties are correlated in $p_T$. We define $\sigma_{b_i}^j \equiv b_i^j \sigma_{b_i}^j$, where $\sigma_{b_i}^j$ is the quoted experimental statistical correlated type B uncertainty associated with the data point at $(x_i, y_i)$ for the $j^{th}$ contribution to the type B uncertainty. We assume that the type B uncertainties follow Gaussian distributions. The type C uncertainty is modeled in a similar fashion, but where the reported uncertainty does not vary on a point-to-point basis. 

In principle, each of the $j$ correlated systematic uncertainty can be allowed to vary independently, however we will assume that the each of the uncertainties are completely independent which implies that $\sigma_b^2 \equiv \sum_j (\sigma_b^j)^2$. This is consistent with the procedure followed by PHENIX \cite{PHENIX:2008ove}. From \cite{PHENIX:2008ove}, we may then use the modified log likelihood
\begin{equation}
	-2 \log \mathcal{L}=\left[\left(\sum_{i=1}^n \frac{\left(y_i+\epsilon_b \sigma_{b_i}+\epsilon_c y_i \sigma_c-\mu_i(\vec{p})\right)^2}{\sigma_i^2}\right)+\epsilon_b^2+\epsilon_c^2\right] \equiv \chi^2\left(\epsilon_b, \epsilon_c, \vec{p}\right),
	\label{eqn:lft_chi_squared}
\end{equation}
where $\epsilon_b$ and $\epsilon_c$ are defined via $\Delta y_i^b \equiv f(x_i) \epsilon_b \sigma_{b_i}$ and $\Delta y_i^c \equiv \epsilon_c \sigma_c$. The above equation follows a $\chi^2$ distribution with $n+2$ degrees of freedom, if there are $n$ parameters $\vec{p}$. In our case $\vec{p} = ( \alpha_s )$ and so $n = 1$. One may find the best fit value of the parameter set $\vec{p}$, by minimizing \cref{eqn:lft_chi_squared}, where $\epsilon_b$ and $\epsilon_c$ are also varied to find a minimum for any particular value of $\vec{p}$. While the PHENIX paper was concerned with rejecting theory models, we will not take this approach, and so we will not calculate the $p$ value. Instead we use the $\chi^2$ procedure simply as a method to find the best fit, while taking into account experimental uncertainties in a reasonable manor.

As noted by the JETSCAPE collaboration \cite{JETSCAPE:2021ehl}, the $p_T$ dependence of the correlated systematic uncertainties is typically not reported by experiments. They suggest implementing a \emph{correlation length}, which allows uncertainties to be correlated for $p_T$ values that are close together, and uncorrelated for $p_T$ values which are further apart. Such an approach seems particularly important for ATLAS \cite{ATLAS:2022kqu} and CMS pion suppression data \cite{CMS:2016xef}, which have exceedingly small statistical uncertainties. We found that if one only allowed the correlated systematic uncertainties to follow the prescription outlined by PHENIX \cite{PHENIX:2008saf}, then obtaining a ``good" fit was not possible. 

For this first simple approach, we use three different prescriptions for the type B uncertainties when calculating the $\chi^2$, and use the minimum value of the produced $\chi^2$ value. The three procedures are (where the $f(x_k)$ is the function present in \cref{eqn:lft_systematic_uncertainty})
\begin{enumerate}
	\item \emph{Linear}: $f(x_i) = 1$, all points move by an equal fraction of their quoted type B uncertainty;
	\item \emph{Tilt}: $f(x_i)$ is a straight line passing through $(x_0, -1)$ and $(x_n, 1)$;
	\item \emph{Uncorrelated}: We assume that the systematic uncertainties are uncorrelated in $p_T$; i.e., they are absorbed into the type A uncertainty.
\end{enumerate}
This is an extreme case of the correlation length approach used by JETSCAPE \cite{JETSCAPE:2021ehl}, and will likely overestimate the goodness of fit. Since we are not interested in reporting the $p$ value or some similar measure of how well the model reproduces data, this will not be an issue at this stage of the work. We note that PHENIX \cite{PHENIX:2008ove} followed this same prescription with points 1.\ and 2.\ only.

Our final procedure is therefore to:
\begin{enumerate}
	\item Calculate the $\chi^2$, assuming that the type B uncertainties are linearly correlated in $p_T$;
	\item Calculate the $\chi^2$, assuming that the type B uncertainties are correlated in $p_T$ with a so called ``tilt", that is the points at low $p_T$ move in the opposite direction to the points at high $p_T$;
	\item Calculate the $\chi^2$, assuming that the type B uncertainties are uncorrelated in $p_T$;
	\item The minimum value of the $\chi^2$ calculated in these three ways is used;
	\item Perform the above steps for various models, including varying the elastic and radiative energy loss kernels and varying the multiplier of the $|\mathbf{k}|_{\text{max}}$ upper bound on the transverse radiated momentum;
	\item The $\alpha_s$ which corresponds to the minimum value of $\chi^2$ is referred to as the ``best fit value" of $\alpha_s$.
\end{enumerate}
One may perform this procedure on any subset of the data. We note that there are almost certainly correlations between data measured in different centrality classes, data measured by the same experiment, data of the same final states, data from the same collision systems, etc. None of these correlations are taken into account in this work, and every data set as a function of $p_T$ is treated as independent. Future work should consider these correlations in a more detailed manor. 

\subsection{Experimental data}
\label{sec:lft_experimental_data}

When choosing experimental data to fit on there are a number of considerations to take into account:
\begin{enumerate}
	\item \emph{Range of validity of our model}.
	\item \emph{Agreement between experimental data sets}.
\end{enumerate}
Because we neglect running coupling effects in our model, we expect that our model will not be valid over the full $p_T$ range for which experimental data exists. Of course, a rigorous calculation of the running of the strong coupling for radiative and elastic energy loss in a QGP has not been done, and so the scales at which the coupling runs is not known. It is expected that the hard scale enters into at least one of the couplings present in the calculation, common assumed scales being $Q^2 \sim ET$, $Q^2 \sim \mathbf{q}^2$, $Q^2  \sim \mathbf{k}^2$, and $Q^2 \sim T^2$ \cite{Peshier:2006ah, Xu:2014ica,Horowitz:2010yg,Zakharov:2008kt,Zakharov:2007pj}. The paradigm of applying perturbative QCD requires the hard scale to enter into at least one of the  strong couplings, and preferably all of them, in order for perturbation theory to be applicable. This also implies we should apply our model only when $E \gg \Lambda_{\text{QCD}}$ or perhaps $\sqrt{ET} \gg \Lambda_{\text{QCD}}$ or $\left\langle \mathbf{k}^2 \right\rangle \sim x^2 E^2 \gg \Lambda_{\text{QCD}}$ etc.  Finally, binary collision scaling is responsible for the production of high-$p_T$ particles, and so at lower momenta binary scaling does not hold. This argument implies we should not consider momenta which are too small, although it is perhaps not obvious what ``too small" means. The literature \cite{ATLAS:2022kqu, Wicks:2005gt, Zigic:2018smz, JETSCAPE:2021ehl} typically chooses a minimum value of momenta for pQCD based models in the range $5\text{--}10 ~\mathrm{GeV}$.

Based on these arguments, we choose to constrain our model based on the range $5\text{--}50 ~\mathrm{GeV}$. There is insight to be gained by performing fits on different subsets of the data, which we discuss in the conclusions of this chapter.

We choose to fit only on systems where the evidence for QGP formation is unequivocal. In this first analysis the data sets we consider in the fitting procedure are:
\begin{itemize}
	\item $h^\pm$ hadrons produced in $0\text{--}60\%$ centrality \coll{Pb}{Pb} collisions at $\sqrt{s_{NN}} = 5.02 ~\mathrm{TeV}$ as measured by ATLAS \cite{ATLAS:2022kqu}, CMS \cite{CMS:2016xef}, and ALICE \cite{ALICE:2018vuu};
	\item $D$ mesons produced in $0\text{--}50\%$ \coll{Pb}{Pb} collisions at $\sqrt{s_{NN}} = 5.02 ~\mathrm{TeV}$ as measured by ALICE \cite{ALICE:2018lyv};
	\item $\pi^0$ mesons produced in $0\text{--}60\%$ centrality \coll{Au}{Au} collisions at $\sqrt{s_{NN}} = 200 ~\mathrm{GeV}$ as measured by PHENIX \cite{PHENIX:2008saf, PHENIX:2012jha}.
\end{itemize}
The centrality classes used in the fit are the most granular provided by the experiment, which differs on an experiment-by-experiment basis. All post-fit $R_{AA}$ model results are plotted vs data in \cref{sec:all_fitted_results}.

We note that at this current stage we have not included data from \coll{Xe}{Xe} collisions at $\sqrt{s}_{NN} = 5.44 ~\mathrm{TeV}$ \cite{ALICE:2018hza, ATLAS:2022kqu} or \coll{Pb}{Pb} data at $\sqrt{s}_{\text{NN}} = 2.76 ~\mathrm{TeV}$ \cite{ALICE:2010yje}. Future work should include this in the global fit to data.

\subsection{Results}
\label{sec:lft_results}

In this section we present results from the global fit of our model to data. Due to the large number of experimental datasets compared to, we defer the full comparison of all model curves to all experimental datasets to \cref{sec:all_fitted_results}. Data for which model results were calculated were all centrality classes of charged hadrons produced in \coll{Pb}{Pb} collisions at $\sqrt{s_{NN}} = 5.02 ~\mathrm{TeV}$ \cite{ALICE:2018vuu, CMS:2016xef, ATLAS:2022kqu}; all centrality classes of $D^0$ mesons produced in \coll{Pb}{Pb} collisions at $\sqrt{s_{NN}} = 5.02 ~\mathrm{TeV}$ \cite{ALICE:2018lyv}; $B^{\pm}$ mesons produced in $0\text{--}100\%$ centrality \coll{Pb}{Pb} collisions at $\sqrt{s_{NN}} = 5.02 ~\mathrm{TeV}$ \cite{CMS:2017uoy}; all centrality classes of $\pi^0$ mesons produced in \coll{Au}{Au} collisions at $\sqrt{s_{NN}} = 200 ~\mathrm{GeV}$ \cite{PHENIX:2008saf, PHENIX:2012jha}; $\pi^0$ mesons produced in central \coll{p}{Pb} collisions at $\sqrt{s_{NN}} = 5.02 ~\mathrm{TeV}$ \cite{ATLAS:2022kqu}; $D^0$ mesons produced in central \coll{p}{Pb} collisions at $\sqrt{s_{NN}} = 5.02 ~\mathrm{TeV}$; $\pi^0$ mesons produced in central \collThree{p}{He3}{Au} collisions at $\sqrt{s_{NN}} = 200 ~\mathrm{GeV}$ \cite{PHENIX:2021dod}; and $\pi^0$ mesons produced in central \coll{d}{Au} collisions at $\sqrt{s_{NN}} = 5.02 ~\mathrm{TeV}$ \cite{PHENIX:2023dxl}. This list is not extensive, and future interesting comparisons will include minimum bias \collFour{p}{d}{He3}{Au} collisions, predictions for future \coll{O}{O} collisions at LHC, high-$p_T$ $v_2$ for all systems, and heavy-flavor data measured through more complicated decay patterns such as $b \to B \to D^0$ \cite{CMS:2018bwt, ALICE:2022tji}.

\begin{figure}[!htbp]
	\centering
	\includegraphics[width=0.8\linewidth]{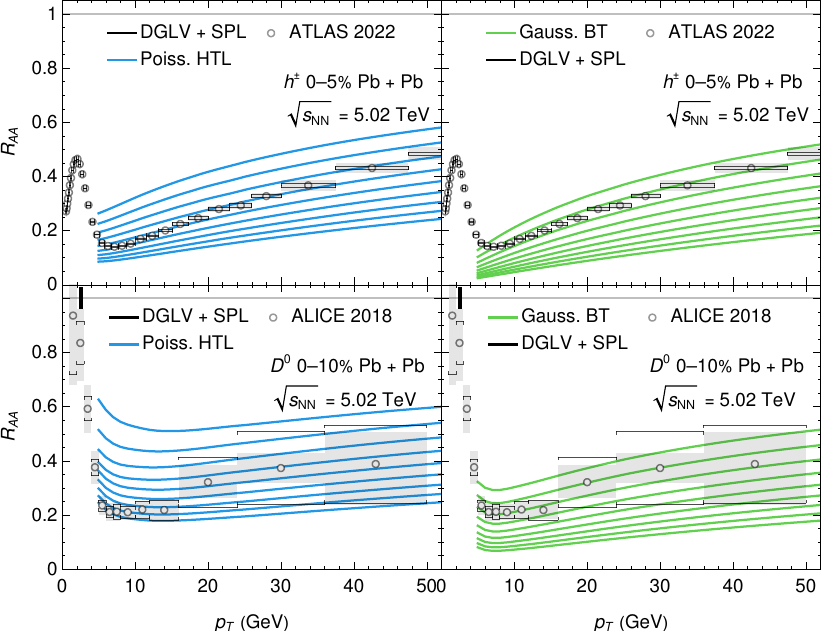}
	\caption{Plot of the nuclear modification factor $R_{AA}$ as a function of transverse momentum $p_T$ for strong coupling $\alpha_s$ in the range $[0.3, 0.5]$ (from large $R_{AA}$ to small large $R_{AA}$). Top panels show results for charged hadrons ($h^\pm$) produced in $0\text{--}5\%$ centrality \coll{Pb}{Pb} collisions and bottom panels show results for $D^0$ mesons produced in $0\text{--}10\%$ centrality \coll{Pb}{Pb} collisions. Left panels present model results for DGLV + SPL radiative energy loss convolved with HTL elastic energy loss, while right panels  present	model results for DGLV + SPL radiative energy loss convolved with BT elastic energy loss. Data from ALICE \cite{ALICE:2018lyv} and ATLAS \cite{ATLAS:2022kqu} is shown.}
	\label{fig:raa_varying_alpha_s}
\end{figure}

\Cref{fig:raa_varying_alpha_s} plots the $R_{AA}$ as a function of $p_T$ for charged hadrons produced in central \coll{Pb}{Pb} collisions at $\sqrt{s_{NN}} = 5.02 ~\mathrm{TeV}$ (top) and $D^0$ mesons produced in central \coll{Pb}{Pb} collisions at $\sqrt{s_{NN}} = 5.02 ~\mathrm{TeV}$ (bottom). Model results are generated for the HTL elastic energy loss (left) and the BT elastic energy loss (right), as well as by varying the coupling $\alpha_s$ between $0.3$ (largest $R_{AA}$) and $0.5$ (smallest $R_{AA}$) (largest $R_{AA}$ corresponds to small $\alpha_s$). All theory curves are generated with the large formation time + collinear upper bound with $\kappa = 1$ (see \cref{sec:the_large_formation_time_assumption}). Data from ATLAS \cite{ATLAS:2022kqu} and ALICE \cite{ALICE:2018vuu} is shown for reference.

In \cref{fig:raa_varying_alpha_s} we see that the range of coverage of the $R_{AA}$ is sufficient for the subset of data shown, which is representative of the full set of experimental data. We additionally observe that the $R_{AA}$ varies smoothly as a function of the strong coupling, implying that our grid size is sufficiently fine-grained for the current level of analysis. We see already that the HTL result will require a larger $\alpha_s$ than that required by the BT result to fit the data.

\begin{figure}[!htbp]
	\centering
	\includegraphics[width=0.9\linewidth]{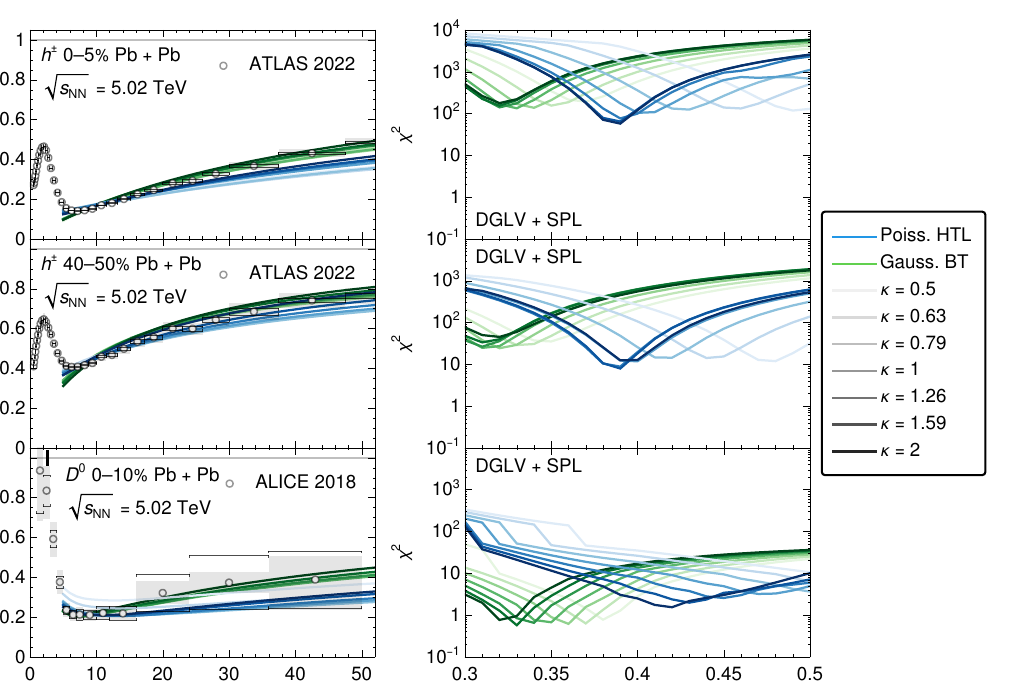}
	\caption{(left) Nuclear modification factor $R_{AA}$ as a function of transverse momentum $p_T$ for \coll{Pb}{Pb} collisions at $\sqrt{s_{NN}} = 5.02$ TeV. Theoretical curves are produced by varying the elastic energy loss between HTL and BT and the multiplier $\kappa$ of cutoff for transverse radiated gluon momentum between $0.5$ and $2$. The radiative energy loss is fixed as DGLV + SPL The top panels show results for charged hadrons in $0\text{--}5\%$ centrality collisions, the middle panels show results for charged hadrons in $40\text{--}50\%$ centrality collisions, and the bottom panels show results for $D$ mesons in $0\text{--}10\%$ collisions. Data which was used to generate the $\chi^2$ curves is also shown from ATLAS \cite{ATLAS:2022kqu} and ALICE \cite{ALICE:2018lyv}.(right) $\chi^2$ values for the corresponding $R_{AA}$ curves as a function of the strong coupling $\alpha_s$.}
	\label{fig:chisquare_raa_theory_comparison}
\end{figure}

\Cref{fig:chisquare_raa_theory_comparison} plots the $R_{AA}$ (left) of pions produced in central \coll{Pb}{Pb} collisions (top), pions produced in $40\text{--}50\%$ centrality \coll{Pb}{Pb} collisions (middle), and $D$ mesons produced in central \coll{Pb}{Pb} collisions (bottom) all at $\sqrt{s_{NN}} = 5.02 ~\mathrm{TeV}$. Data from ATLAS \cite{ATLAS:2022kqu} and ALICE \cite{ALICE:2018lyv} which was used to generate the $\chi^2(\alpha_s)$ is also shown. The corresponding $\chi^2$ as a function of $\alpha_s$ is shown in the right hand column, for each dataset. Multiple theory curves are generated from our model by varying the elastic energy loss between HTL (blue) \cite{Wicks:2008zz} and BT (green) \cite{Braaten:1991we, Braaten:1991jj}, as well as varying the multiplier of the large formation time cutoff on the transverse radiated gluon momentum between $0.5$ (lighter) and $2$ (darker). Each theory curve is independently fit to data, and the curves shown in the $R_{AA}$ are the best \emph{local} fits (only fit on this dataset).

We observe in the figure that a much larger $\alpha_s$ is needed to fit data when the HTL elastic energy loss is used. This was expected from the results presented in \cref{sec:EL_paper}, as the HTL elastic energy loss is smaller than the BT elastic energy loss (see \cref{fig:mod_deltaEoverE_small_large,fig:mod_deltaEoverE_small_large_gluon}). The different proportions of radiative and elastic energy loss in the HTL vs.\ BT case, lead to a flatter $p_T$ dependence of the HTL result than that of the BT result. This is because the BT result has a larger contribution from elastic energy loss compared to radiative energy loss, which falls off quickly as a function of $p_T$. The faster growth as a function of $p_T$ appears to be favored by both the heavy- and light-flavor data, however, this could also be due to neglecting running coupling effects in the model.

Multiple theory curves are also generated by varying $\kappa$, the multiplier of the upper bound on the transverse radiated gluon momentum, between $0.5$ and $2$. We see that our model is acutely sensitive to this choice, but most of this sensitivity can be absorbed into a different value of the strong coupling. This absorption leads to a thin band of theory curves for both the heavy- and light-flavor $R_{AA}$. The best-fit $\alpha_s$ values produced by the model varied in this way are $\alpha_s = 0.32\text{--}0.37$ for the BT case and $\alpha_s = 0.39\text{--}0.5$ for the HTL case, a $\sim 20\%$ variation. 

\begin{figure}[!htbp]
	\centering
	\includegraphics[width=0.85\linewidth]{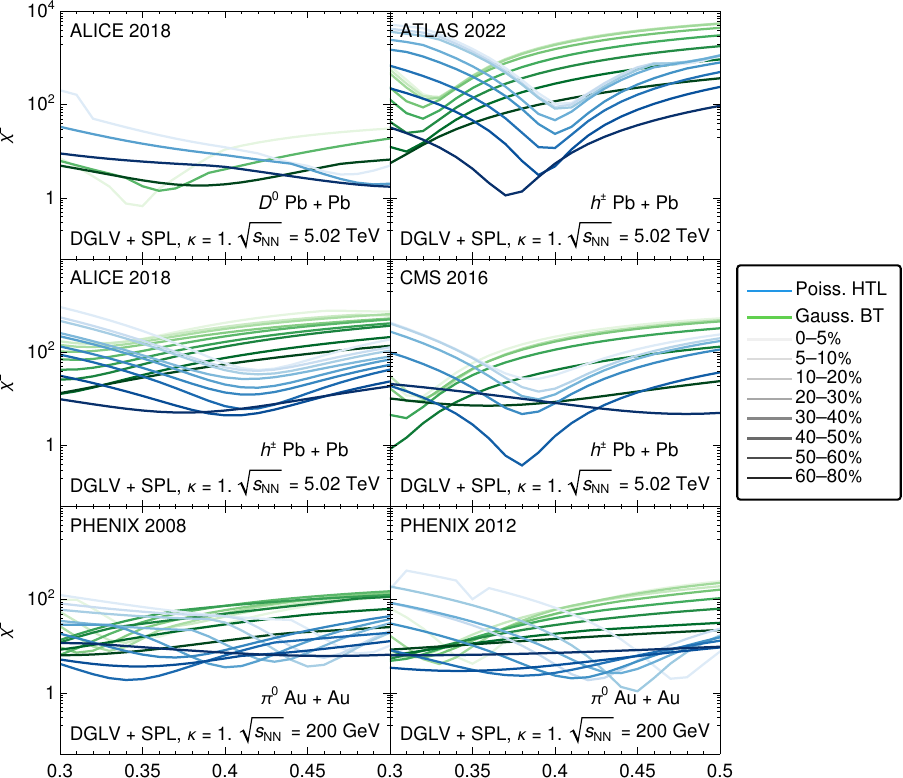}
	\caption{Plot of the $\chi^2$ as a function of $\alpha_s$. Theoretical curves are generated from the HTL (blue) \cite{Wicks:2008zz} as well as BT (green) \cite{Braaten:1991we, Braaten:1991jj} elastic energy loss convolved with DGLV + SPL \cite{Kolbe:2015suq, Kolbe:2015rvk} radiative energy loss with the large formation time + collinear cutoff and $\kappa = 1$. Different panels correspond to different collision systems. Left to right and top to bottom panels show: $D^0$ mesons produced in \coll{Pb}{Pb} collisions at $\sqrt{s_{NN}} = 5.02~\mathrm{TeV}$, charged hadrons produced in \coll{Pb}{Pb} collisions at $\sqrt{s_{NN}} = 5.02~\mathrm{TeV}$, charged hadrons produced in \coll{Pb}{Pb} collisions at $\sqrt{s_{NN}} = 5.02~\mathrm{TeV}$, charged hadrons produced in \coll{Pb}{Pb} collisions at $\sqrt{s_{NN}} = 5.02~\mathrm{TeV}$, neutral pions produced in \coll{Au}{Au} collisions at $\sqrt{s_{NN}} = 200~\mathrm{GeV}$, neutral pions produced in \coll{Au}{Au} collisions at $\sqrt{s_{NN}} = 200~\mathrm{GeV}$. Results for more central collisions are indicated with lighter curves while results for less central collisions are indicated with darker curves.}
	\label{fig:chi_square_function_of_centrality}
\end{figure}

\Cref{fig:chi_square_function_of_centrality} plots the $\chi^2$ as a function of $\alpha_s$ for different collision systems for various centrality classes. All curves are produced with DGLV + SPL radiative energy loss, and with a set $\kappa = 1$. We see in the figure that particularly for Pion data at LHC, all best fit values of the strong coupling constant in different centrality classes are extremely similar. For $D$ mesons the minimum of the $\chi^2$ is not well constrained due to large uncertainties present in the data, but the minima for various centrality classes are all compatible. At RHIC, again the minima is not well constrained due to large uncertainties, but the minima are still all compatible. We note in the figure that there is a minor false convergence in the bottom right pane around $\alpha_s = 0.35$. This false convergence is not important as it is far away from the minima of the $\chi^2$.

\begin{figure}[!htbp]
	\centering
	\includegraphics[width=0.49\linewidth]{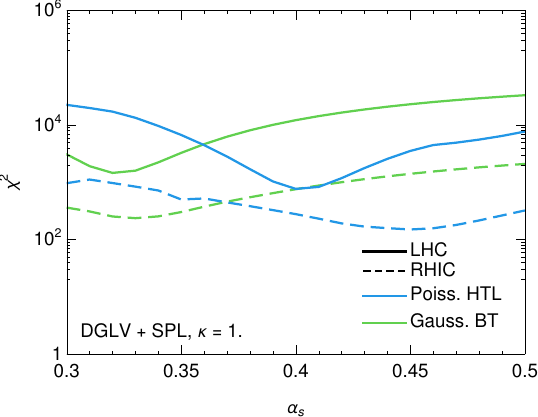}\hfill
	\includegraphics[width=0.49\linewidth]{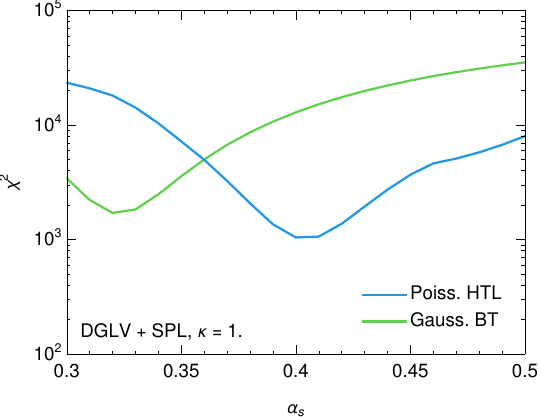}
	\caption{Plot of the $\chi^2$ for global fit to data with fixed $\kappa = 1$ (multiplier of $|\mathbf{k}|_{\text{max}}$) in central \coll{Pb}{Pb} and \coll{Au}{Au} collisions. All data for collision \coll{Pb}{Pb} and \coll{Au}{Au} collisions for centrality classes less than $60\%$ was included. The $p_T$ range considered was $5~\mathrm{GeV} \leq p_T \leq 50~\mathrm{GeV}$. Theoretical model fits are shown for elastic energy loss kernels of Poisson HTL and Gaussian BT. The left panel fits the data for LHC and RHIC separately, while the right panel shows the global $\chi^2$ for both RHIC and LHC data.}
	\label{fig:chiSquare_all_large_systems_rhic_lhc}
\end{figure}

The left pane of \cref{fig:chiSquare_all_large_systems_rhic_lhc} plots that $\chi^2$ as a function of $\alpha_s$ for HTL and BT elastic energy loss both convolved with DGLV + SPL radiative energy loss. The $\chi^2$ is produced from a global fit to large system data (see \cref{sec:lft_fitting_procedure}) at RHIC and the LHC separately. From the figure we see that a similar best value of $\alpha_s$ is obtained at both the LHC and RHIC for the BT elastic energy loss model, while two different minima are obtained for LHC and RHIC when using the HTL elastic energy loss model. As previously mentioned, the scale at which the various couplings run is not known, although one expects that the temperature enters into these scales in some way, for instance as $Q^2 = ET$ or $Q^2 = T^2$ \cite{Xu:2014ica}.  One may therefore expect that the larger temperature scale at the LHC leads to a smaller coupling coupling $\alpha_s(Q^2)$ with $Q^2 \sim ET \text{ or } \sim T^2 \text { etc.}$ than at RHIC. The fact that different elastic energy loss kernels can lead to different inferences about the value of $\alpha_s$ as a function of temperature, implies that an extraction of the scales at which the various couplings run from data may not be feasible. Future work should therefore aim to perform a rigorous calculation of the running of the strong coupling for both elastic and radiative energy loss in a QGP.

The right pane of \cref{fig:chiSquare_all_large_systems_rhic_lhc} plots the global $\chi^2$ for large system data (see \cref{sec:lft_fitting_procedure}), combined for both RHIC and LHC. We see that combining the LHC and RHIC data, leads to the best fit value of $\alpha_s$ being approximately the one at LHC from \cref{fig:chiSquare_all_large_systems_rhic_lhc}. This is because the data at LHC is more constraining due to the smaller statistical uncertainties, the larger number of data points, and the larger $p_T$ range. This is only impacts the HTL results, as the BT results have the same best fit value at RHIC and LHC, as previously mentioned.

\begin{figure}[!htbp]
	\includegraphics[width=\linewidth]{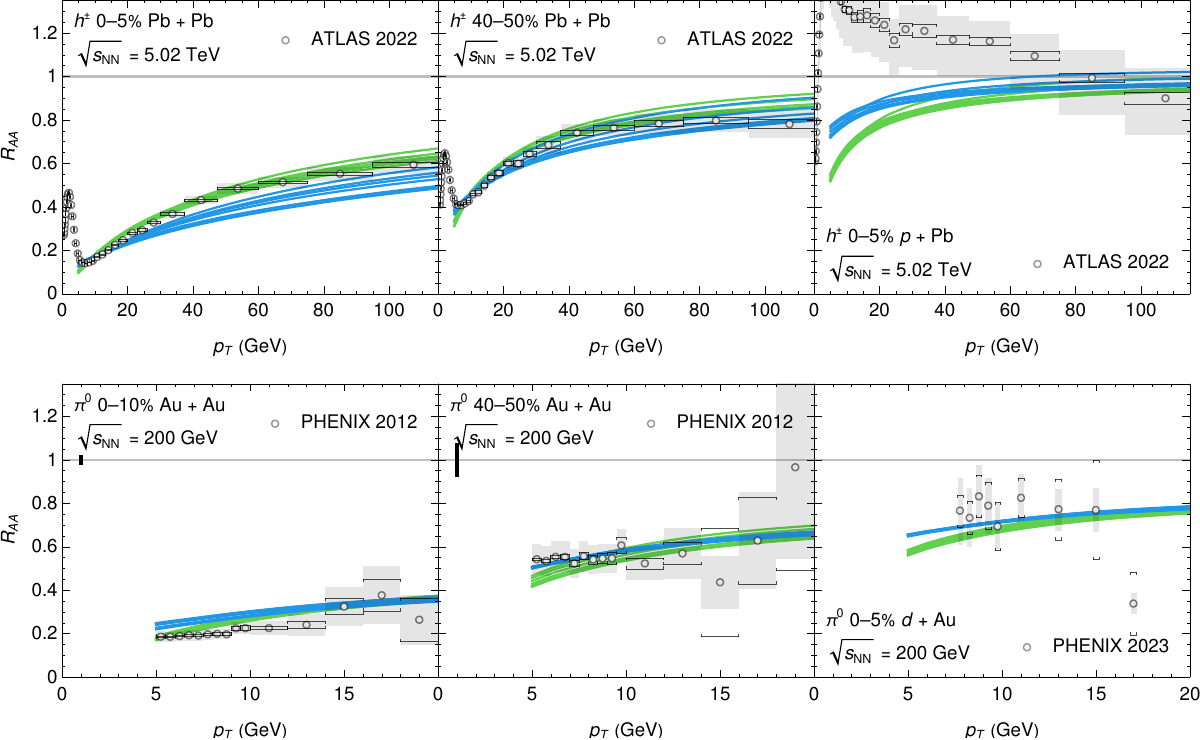}
	\caption{
Model results for the nuclear modification factor $R_{AA}$ as a function of $p_T$ for various collision systems at RHIC and LHC. Theoretical curves shown are calculated with the HTL elastic energy loss (blue) \cite{Wicks:2008zz} and the BT elastic energy loss (green) \cite{Braaten:1991we, Braaten:1991jj}, both convolved with the DGLV + SPL \cite{Kolbe:2015suq, Kolbe:2015rvk} radiative energy loss. Multiple theoretical curves are produced for each elastic energy loss model by varying the upper bound on the transverse radiated gluon momentum by factors of two up and down, where the upper bound is determined by ensuring neither the collinear nor the large formation time approximation are explicitly violated. Each theoretical curve shown has been constrained by a fit of the effective strong coupling $\alpha_s$ to experimental large system data, as described in detail in the text. Note that the value of the coupling differs between theory curves, but remains the same for different collision systems. Collision systems shown are (left to right and top to bottom) charged hadrons produced in $0\text{--}5\%$ \coll{Pb}{Pb} collisions at $\sqrt{s_{NN}} = 5.02 ~\mathrm{TeV}$; charged hadrons produced in $40\text{--}50\%$ \coll{Pb}{Pb} collisions at $\sqrt{s_{NN}} = 5.02 ~\mathrm{TeV}$; charged hadrons produced in $0\text{--}5\%$ \coll{p}{Pb} collisions at $\sqrt{s_{NN}} = 5.02 ~\mathrm{TeV}$; neutral pions produced in $0\text{--}10\%$ \coll{Au}{Au} collisions at $\sqrt{s_{NN}} = 200 ~\mathrm{GeV}$; neutral pions produced in $40\text{--}50\%$ \coll{Au}{Au} collisions at $\sqrt{s_{NN}} = 200 ~\mathrm{GeV}$; and neutral pions produced in $0\text{--}5\%$ \coll{d}{Au} collisions at $\sqrt{s_{NN}} = 200 ~\mathrm{GeV}$. Data is shown from \cite{ATLAS:2022kqu, PHENIX:2012jha, PHENIX:2023dxl}, with statistical and systematic uncertainties indicated by error bars and boxes, respectively. The global normalization uncertainty on the number of binary collisions is represented by solid boxes at unity (where supplied by the experiment).
}
	\label{fig:final_summary_plot_raa}
\end{figure}

The top  row of \cref{fig:final_summary_plot_raa} plots the $R_{AB}$ as a function of $p_T$ for charged hadrons produced in $0\text{--}5\%$ centrality \coll{Pb}{Pb} collisions (left), $40\text{--}50\%$ centrality \coll{Pb}{Pb} collisions (middle), and $0\text{--}5\%$ \coll{p}{Pb} collisions (right) all at $\sqrt{s_{NN}} = 5.02 ~\mathrm{TeV}$. Data from ATLAS \cite{ATLAS:2022kqu} is shown. The bottom row of \cref{fig:final_summary_plot_raa} plots the $R_{AB}$ as a function of $p_T$ for neutral pions produced in $0\text{--}10\%$ \coll{Au}{Au} collisions (left), $40\text{--}50\%$ \coll{Au}{Au} collisions, and $0\text{--}5\%$ \coll{d}{Au} collisions. Data from PHENIX \cite{PHENIX:2012jha, PHENIX:2023dxl} is also shown. Theory curves are produced by varying the elastic energy loss between HTL (blue) and BT (green) which is then convolved with the DGLV + SPL radiative energy loss. Bands are produced by varying the multiplier $\kappa$ of the cutoff on the radiated transverse gluon momentum by factors of two up and down. We note that one should not think of these bands as being a  one sigma confidence interval or similar, as there is not a notion of a probability distribution for this theoretical uncertainty. Each of the theory curves plotted are plotted with a coupling $\alpha_s$ which is globally fit to all large system data at RHIC and LHC, as described in \cref{sec:lft_fitting_procedure,sec:lft_experimental_data}. A sample $\chi^2$ of this fit is shown in the right pane of \cref{fig:chiSquare_all_large_systems_rhic_lhc} for varying elastic energy loss models, with set $\kappa = 1$. The plots shown in \cref{fig:final_summary_plot_raa} are a small, but representative, subset of all computed results. We present the equivalent results for a variety final states, collision systems, and centralities in \cref{sec:all_fitted_results}.

We see in \cref{fig:final_summary_plot_raa} that while the model predictions for the $R_{AB}$ are consistent with charged hadron \coll{Pb}{Pb} data in central and semi-central collisions \cite{ATLAS:2022kqu} for a large range in $p_T$, they are not consistent with the lack of suppression and even \emph{enhancement} in central \coll{p}{Pb} collisions at low-$p_T$ \cite{ATLAS:2022kqu}. For $p_T \gtrsim 40 ~\mathrm{GeV}$ the model results are consistent with data within the large systematic uncertainties. In the bottom row of \cref{fig:final_summary_plot_raa}, we see that the model results are consistent with both central and peripheral \coll{Pb}{Pb} data \emph{as well as} central \coll{d}{Au} suppression data. The smaller range of $p_T$ values at RHIC means that the ratio of elastic and radiative energy loss contributions to the total energy loss remains similar. In turn, this decreases the sensitivity of our final result to the elastic energy loss kernel used. We note that the RHIC \coll{d}{Au} result uses an $R_{AB}$ which is normalized according to the high-$p_T$ prompt photon yield. This means that the \coll{d}{Au} suppression results do not suffer from \emph{centrality bias}, which is a nontrivial correlation between the hard and soft particles produced in a heavy-ion collision \cite{ALICE:2014xsp, PHENIX:2013jxf, Kordell:2016njg, PHENIX:2023dxl, Bzdak:2014rca}. This centrality bias particularly effects small \collFour{p}{d}{He3}{A} collisions as well as peripheral \coll{A}{A} collisions.

\FloatBarrier

\section{Conclusion}
\label{sec:lft_conclusion}

In this chapter, we implemented a new upper bound on the transverse radiated gluon momentum $|\mathbf{k}|$ integral, which ensures that no contributions from the phase space where \emph{either} the large formation time or collinear approximation are explicitly violated are included. This procedure is similar to the one typically used to cut off the phase space by the Gyulassy-Levai-Vitev (GLV) \cite{Gyulassy:2000er,Djordjevic:2003zk} and Armesto-Salgado-Wiedemann-Single-Hard (ASW-SH) \cite{Wiedemann:2000za,Wiedemann:2000tf,Armesto:2003jh} models, wherein one ensures no contributions from regions of phase space where the collinear approximation is invalid are included. Under slightly different assumptions, this prescription leads to a maximum radiated gluon momentum of $|\mathbf{k}|_{\text{max}} \in \{x E, 2 x E, 2 x E( 1- x), \operatorname{Min}(2x E, (1-x)E)\}$ \cite{Armesto:2011ht}. The upper bound used in this work, which ensures neither the large formation time nor the collinear approximation are explicitly violated, is $|\mathbf{k}|_{\text{max}} = \kappa \operatorname{Min}[2 x E(1-x), \sqrt{2 x E \sqrt{\mu^2 + \mathbf{q}_1^2}}]$, where $\kappa$ is a dimensionless factor which is varied between 0.5 and 2. The variation of the factor $\kappa$ is used to capture the sensitivity of the model to the exact value chosen for the cutoff.

We showed that introducing this new upper bound on $|\mathbf{k}|$ dramatically reduced the magnitude of the short pathlength correction to the Djordjevic-Gyulassy-Levai-Vitev (DGLV) \cite{Djordjevic:2003zk} radiative energy loss (DGLV + SPL) \cite{Kolbe:2015suq, Kolbe:2015rvk}, leaving only a small difference between the DGLV and DGLV + SPL results for all phenomenologically relevant systems. However, we found that this upper bound significantly increases the sensitivity of our answer to the exact value chosen for the upper bound by varying the multiplier of this bound $\kappa$ by factors of two up and down. Typically one interprets such a sensitivity as a theoretical uncertainty, which is indicative of potential higher order effects in the expansion parameter \cite{Horowitz:2009eb}. The sensitivity to the upper bound leads to a $\mathcal{O}(35\text{--}70\%)$ variation in the charged hadron $R_{AA}$ in central \coll{Pb}{Pb} collisions and a $\mathcal{O}(10\text{--}20\%)$ variation in the charged hadron $R_{AA}$ in central \coll{p}{Pb} collisions for $10 ~\mathrm{GeV} \lesssim p_T \lesssim 100 ~\mathrm{GeV}$. This uncertainty is extremely large, even compared to other uncertainties in the problem, such as the $~30\%$ uncertainty due to NLO contributions in $\alpha_s$ not being included.
Our work here, therefore, suggests that corrections resulting from the removal of the large formation time approximation may be large. We note especially that the sensitivity to the upper bound is similar for both the DGLV and DGLV + SPL radiative energy loss models, and so such a large formation time correction will likely be important for, at least, all models based on the GLV framework. This includes canonical and contemporary phenomenological energy loss models such as the Wicks-Horowitz-Djordjevic-Gyulassy (WHDG) \cite{Wicks:2005gt, Horowitz:2011gd, Horowitz:2012cf} and CUJET \cite{Buzzatti:2011vt, Buzzatti:2012pe, Xu:2014ica, Xu:2015bbz} models. 

This work focuses on the GLV formalism, but we will briefly consider the implications of the large formation time breakdown for other models. The single gluon emission kernel $dN^g / dx$ is the same for both DGLV and ASW-SH \cite{Wiedemann:2000za, Wiedemann:2000tf} when the gluon mass is set to zero, suggesting that ASW-SH will be sensitive to the large formation time assumption. In the HT approach \cite{Guo:2000nz, Wang:2001ifa}, only the leading divergent term in $\mathbf{k}^2$ (denoted as $\ell_T^2$) is considered, and therefore terms proportional to $\mathbf{k}^2 / 2 x E \mu$ (which are ignored by the large formation time approximation) are presumably ignored.
This omission would affect the MATTER framework \cite{Majumder:2013re, Cao:2017qpx} and, by extension, analyses by the JETSCAPE collaboration \cite{JETSCAPE:2021ehl, JETSCAPE:2024cqe, JETSCAPE:2017eso, JETSCAPE:2022jer}. The AMY formalism \cite{Arnold:2001ms,Arnold:2002ja,Arnold:2001ba,Caron-Huot:2010qjx} receives contributions from topologically equivalent diagrams to the HT formalism \cite{Majumder:2010qh}, indicating that AMY shares a similar sensitivity to the large formation time approximation as HT. 
Since the GLV formalism is a limiting case of the BDMPS-Z path integral \cite{Baier:1996kr, Baier:1996sk, Zakharov:1997uu, Baier:1996vi, Zakharov:1996fv}, sensitivity to the large formation time approximation in GLV implies a similar sensitivity in the BDMPS-Z formalism, particularly in the dilute medium limit. 
Additionally, efforts to connect the BDMPS-Z and GLV results \cite{Mehtar-Tani:2019ygg, Barata:2021wuf, Mehtar-Tani:2019tvy, Andres:2020vxs, Andres:2023jao} suggest that the two approaches are strongly connected, implying that the BDMPS-Z result may also be sensitive to the large formation time approximation.
Djordjevic's dynamical scattering center model \cite{Djordjevic:2007at, Djordjevic:2008iz, Djordjevic:2009cr}, used in the DREENA framework \cite{Zigic:2018smz, Zigic:2021rku}, applies HTL propagators \cite{Braaten:1989mz, Klimov:1982bv, Pisarski:1988vd, Weldon:1982aq, Weldon:1982bn} for elastic and radiative energy loss. It assumes $\delta[q_z - q_0 - (\mathbf{k}^2)/(2 x E)] \approx \delta[q_z - q_0]$, which is only valid when $\mathbf{k}^2 / 2 x E q_0 \ll 1$. Given that the transferred momentum $q$ has a typical value of $\mu$, this approximation resembles the large formation time assumption, suggesting that the dynamical model shares the same sensitivity. 
In contrast, strongly coupled models \cite{Casalderrey-Solana:2011dxg, Gubser:2006bz, Herzog:2006gh, BitaghsirFadafan:2008adl, Mikhailov:2003er, Chesler:2008uy, Chesler:2008wd, Gubser:2008as, Morad:2014xla, Horowitz:2015dta, Casalderrey-Solana:2014bpa, Casalderrey-Solana:2016jvj, Hulcher:2017cpt} take a different approach to energy loss, where the formation time is effectively zero. Thus, the large formation time approximation is likely not important for strongly coupled energy loss models.

We performed a one-parameter fit of the strong coupling $\alpha_s$ to available large system data from RHIC and LHC. The fit was performed by minimizing the total $\chi^2$ for large system data as a function of $\alpha_s$. Within this fit we allowed the systematic uncertainties to be either completely correlated or completely uncorrelated in momentum, and chose the minimum value of the $\chi^2$. 

We saw that the HTL elastic energy loss model leads naturally to two different preferred $\alpha_s$ best fit values, where the coupling at RHIC is larger than that at LHC. We suggested that this may be an indication of the coupling running at scales which include the temperature. Common choices for the scale at which the coupling runs are $Q^2 \in \{ ET, \mu^2, \mathbf{q}^2, \mathbf{k}^2 / x (1-x), (2 \pi T)^2 \}$ \cite{Peshier:2006ah,Zigic:2018smz,Zigic:2021rku,Xu:2014ica}. For the Poisson HTL elastic energy loss, the extracted value of the strong coupling was\footnote{For the central value of the upper bound, $\kappa = 1$.} $\alpha_s = 0.4$ at LHC and $\alpha_s = 0.45$ at RHIC, with a relative difference of $\sim \! 12\%$. If all three couplings in the radiative energy loss and both couplings in the elastic energy loss ran\footnote{We use the running coupling equation from \cite{Peshier:2006ah}.} with the scale $Q^2 = ET$, we would expect a relative difference of $5\text{--}7\%$ for energies in the range $10\text{--}50 ~\mathrm{GeV}$. Alternatively, if the scale were set to $Q^2 = (2 \pi T)^2$, the expected fractional difference would be approximately $1 - \log(2 \pi T_{\text{RHIC}} / \Lambda_{\text{QCD}}) / \log(2 \pi T_{\text{LHC}} / \Lambda_{\text{QCD}}) \simeq 15\%$. Finally, if the couplings ran with the scale $Q^2 = \mu_D^2$, we would expect a larger relative difference of around 30\%\footnote{The consistency equation for $\mu^2(\alpha_s, T)$ was solved using the approach in \cite{Peshier:2006ah}.}. We found that the extracted coupling for the BT elastic energy loss was similar for RHIC and LHC, which favors a coupling which runs at the hard momentum scales, while the extracted coupling with the HTL elastic energy loss differed by $\sim 12\%$ which favors a coupling that runs, at least partially, with the softer momentum scales. Of course, different couplings likely run at different scales, and the scales are likely more complicated than the standard scales which are used. In addition, this simple analysis could not consider scales which are integrated over. The sensitivity of the extracted coupling at RHIC and LHC to the chosen elastic energy loss highlights the need for a rigorous understanding of the various scales at which the coupling runs. We note that the larger coupling at RHIC compared to LHC for the HTL result qualitatively agree with the results of the JET collaboration \cite{JET:2013cls}, who found that the extracted jet quenching parameter $\hat{q}$ was larger at RHIC compared to LHC.

Finally, we presented model results for a globally fitted value of $\alpha_s$ for each theoretical model. This global fit was performed for both HTL and BT elastic energy loss models---which captures the uncertainty in the crossover between HTL and vacuum propagators---each convolved with DGLV + SPL radiative energy loss. We showed that the value of $\alpha_s$ corresponding to the local minima of the $\chi^2$ for the various centrality classes of a particular collision were the same. We saw that the fit was visually satisfactory for all large system data at RHIC and LHC. Additionally, this large system constrained model successfully reproduced the measured suppression in central \coll{d}{Au} collisions at RHIC \cite{PHENIX:2023dxl}. However, the lack of suppression measured in \coll{p}{Pb} \cite{ATLAS:2022kqu} collisions is not reproduced by our model for $p_T \lesssim 30 ~\mathrm{GeV}$. We note that the centrality determination in small systems is difficult due to centrality bias \cite{ALICE:2014xsp, PHENIX:2013jxf, Kordell:2016njg, PHENIX:2023dxl, Bzdak:2014rca}, which may impact the \coll{p}{Pb} results from ATLAS. The PHENIX central \coll{d}{Au} $R_{AB}$ is insensitive to such a bias, since the $R_{AB}$ is normalized by the hard photon $R_{AB}$, which allows one to experimentally determine the number of binary collisions instead of using the Glauber model. 

Future work should consider a more detailed accounting of the correlations in the systematic experimental uncertainties. One may also infer empirically how the coupling runs with the hard momentum scale by performing a fit to data of $\alpha_s$ on different $p_T$ ranges, or otherwise include an \emph{ad hoc} prescription for the running coupling. The most theoretically satisfactory approach would be a rigorous calculation of the scales at which the coupling runs. A study of more collision systems including \coll{Xe}{Xe}, \coll{Pb}{Pb} at $\sqrt{s_{NN}} = 2.76 ~\mathrm{TeV}$, \coll{O}{O} will be valuable, as well as more observables such as the high-$p_T$ $v_2$, dihadron correlations $I_{AA}$ and $I_{pA}$, and dijet imbalance $A_J$ \cite{ALICE:2022wpn}.

 \chapter{Conclusions and Outlook}
 \label{sec:conclusion}

 The success of the heavy-ion programs at RHIC and LHC cannot be overstated. In the last two decades, overwhelming evidence has emerged that a Quark-Gluon Plasma (QGP) is formed in heavy-ion collisions, and reaches temperatures exceeding a trillion degrees \cite{WA97:1999uwz, PHENIX:2004vcz, PHOBOS:2004zne, STAR:2005gfr,BRAHMS:2004adc, ALICE:2022wpn}. This evidence generally comes in two forms: soft physics, where almost all experimental observables can be quantitatively described by relativistic, viscous hydrodynamic models \cite{Nijs:2020roc, Gale:2013da, Schenke:2020mbo, Weller:2017tsr}, and hard physics, for which a plethora of energy loss models exist that are at least qualitatively compatible with data \cite{Armesto:2011ht, JETSCAPE:2021ehl, Qin:2015srf, Wicks:2005gt, Horowitz:2011gd,Schenke:2009gb,Dainese:2004te, Cao:2020wlm,Casalderrey-Solana:2014bpa}. The promise of hard probes has been, and continues to be, that their creation from binary collisions at the beginning of the collision makes them sensitive to the entire evolution of the medium \cite{Connors:2017ptx,Busza:2018rrf,NSLongRangePlan2023}. This sensitivity may allow us to infer details of the initial state and subsequent evolution of the plasma from the modification of the spectra of high momentum particles and jets. Of course, such an inference about the thermodynamic properties of the plasma will require agreement between, or falsification of, various theoretical energy loss models, as well as a detailed analysis of the theoretical uncertainties in such energy loss models \cite{Horowitz:2009eb, Connors:2017ptx, Armesto:2011ht}.
This thesis argues that such theoretical uncertainties---which are often extremely large---need to be recognized and treated in any extraction of medium properties.

While it is often implied that we are in the precision era of extraction for the QGP formed in heavy-ion collisions \cite{Karmakar:2024jak,JETSCAPE:2021ehl,JETSCAPE:2017eso,JETSCAPE:2024cqe,Karmakar:2023ity,JET:2013cls}, the same argument cannot be made for potential QGP formation in small \collFour{p}{d}{He3}{A} collisions.
Although hydrodynamics can \emph{qualitatively} describe bulk data from small systems \cite{Zhao:2022ugy, Schenke:2020mbo, Weller:2017tsr, Bernhard:2019bmu}, including strangeness enhancement and elliptic flow, suppression models fail to offer even such a qualitative description of experimental data. Indeed, it seems difficult to reconcile the absence of suppression in central \coll{p}{Pb} collisions at LHC \cite{ATLAS:2022kqu, ALICE:2016yta, ATLAS:2022iyq, ALICE:2017svf} with the observed suppression in central \coll{d}{Au} collisions at RHIC \cite{PHENIX:2021dod, PHENIX:2023dxl}. Experimental issues in small systems include centrality bias, which likely leads to nontrivial correlations between the hard and soft modes of the plasma \cite{ALICE:2014xsp, PHENIX:2013jxf, Kordell:2016njg, PHENIX:2023dxl, Bzdak:2014rca}. This thesis makes canonical energy loss models more theoretically compatible with small system collisions, as well as progresses our understanding of the extent to which such a qualitative description of high-$p_T$ data in small and large systems is simultaneously possible. We will discuss the specific ways in which this thesis has aided such an understanding in the subsequent paragraphs of this conclusion.

In this thesis, we presented an energy loss model based on perturbative Quantum Chromodynamics (pQCD), which includes both radiative \cite{Djordjevic:2003zk, Kolbe:2015rvk, Kolbe:2015suq} and elastic energy loss \cite{Braaten:1991we, Braaten:1991jj, Wicks:2008zz}, to make phenomenological suppression predictions. The main focus was to understand the effects of including small system size corrections to the elastic \cite{Wicks:2008zz} and radiative \cite{Kolbe:2015suq, Kolbe:2015rvk} energy loss. The implementation details of the energy loss model were described in \cref{sec:model} and accompanying papers \cite{Faraday:2023mmx, Faraday:2024gzx}. Our energy loss model includes a realistic collision geometry generated from fluctuating IP-Glasma initial conditions \cite{Schenke:2012wb, Schenke:2020mbo}; realistic parton production spectra; elastic and radiative energy loss; Poisson energy loss distributions \cite{Gyulassy:2001nm}; and fragmentation functions \cite{deFlorian:2007aj, Cacciari:2005uk}. In \cref{sec:SPL_paper}, we presented our model results for the nuclear modification factor as a function of $p_T$ for $D$, $B$, and $\pi$ mesons produced in \coll{p}{Pb} and \coll{Pb}{Pb} collisions at $\sqrt{s_{NN}} = 5.02 ~\mathrm{TeV}$ for various centrality classes. We extended our model results in \cref{sec:EL_paper} to include the nuclear modification factor as a function of $p_T$ for $\pi$ and $D$ mesons produced in \collFour{p}{d}{He3}{Au} and \coll{Au}{Au} collisions at $\sqrt{s_{NN}} = 200 ~\mathrm{GeV}$ at RHIC for various centrality classes. We made quantitative comparisons with data, including a best fit of the effective strong coupling constant $\alpha_s$ to a large portion of available experimental large system data in \cref{sec:lft}. 

In \cref{sec:SPL_paper}, we expanded on the Wicks-Horowitz-Djordjevic-Gyulassy (WHDG) energy loss model \cite{Wicks:2005gt} by including a short pathlength (SPL) correction \cite{Kolbe:2015suq, Kolbe:2015rvk} to the Djordjevic-Gyulassy-Levai-Vitev (DGLV) radiative energy loss \cite{Djordjevic:2003zk}. This correction included contributions from previously neglected terms proportional to $e^{- \mu_D L}$, where $L$ is the pathlength and $\mu_D$ is the Debye mass. For small systems, $\mu_D L$ is not particularly large, and one should, in principle, include these contributions. We observed that the inclusion of the short pathlength correction to the radiative energy loss led to a \emph{reduction} of the suppression of leading hadrons. This reduction is well understood as a result of the short pathlength correction enhancing the effect of the destructive LPM interference between the zeroth order in opacity DGLAP-like production radiation and the radiation induced by the subsequent collisions of the leading parton with the medium quanta \cite{Kolbe:2015suq, Kolbe:2015rvk}. The reduction in suppression also increases as a function of $p_T$, which is a result of the different asymptotic energy scalings of DGLV energy loss ($\Delta E\sim\log E$) \cite{Gyulassy:2000er} compared to the short pathlength correction ($\Delta E\sim E$) \cite{Kolbe:2015suq, Kolbe:2015rvk}.

We additionally observed that the SPL correction was negligible for the phenomenologically relevant phase space for heavy-flavor mesons but grew extremely large for light-flavor mesons at high-$p_T$, resulting in $R_{AA} > 1$ in both \coll{Pb}{Pb} and \coll{p}{Pb} collisions. The anomalous size of the correction led us to perform a thorough examination of the consistency of the various assumptions used in the DGLV energy loss model. We found that while the collinear and soft approximations were satisfied self-consistently for both DGLV and DGLV + SPL, the large formation time assumption was \emph{not} satisfied self-consistently in our model. The breakdown became worse as a function of $p_T$. We attributed the anomalously large SPL correction to the breakdown of the large formation time assumption, as this approximation is used extensively in the derivation of the SPL correction. 

We addressed the breakdown of the large formation time approximation again in \cref{sec:lft} by noting that one may ensure only contributions for which the large formation time assumption holds are included in the result by restricting the transverse radiated gluon momentum. An identical procedure is typically used to ensure that only regions of phase space where the collinear approximation is valid are included, which results in a large sensitivity to the exact cutoff chosen \cite{Horowitz:2009eb}. We found that by requiring both the large formation time approximation and the collinear approximation hold, the short pathlength correction was dramatically reduced. Additionally, the DGLV radiative energy loss was also reduced, especially at large momenta. We understand the reduction of the DGLV and DGLV + SPL at large momenta as a difference in the asymptotic scalings of the two upper bounds on the transverse radiated gluon momentum: while the collinear upper bound scales as $E$, the large formation time upper bound scales as $\sqrt{E \mu_D}$. The different scalings of the cutoffs leads to a significantly larger region of the phase space being restricted at higher energies once the large formation time upper bound is included. We also examined the sensitivity to the exact value chosen for the upper bound on the transverse radiated gluon momentum by varying the upper bound by factors of two. 
For low momenta, the sensitivity of $\Delta E / E$ due to the exact value of the large formation time + collinear (LFT + collinear) upper bound is comparable to that of only the collinear upper bound \cite{Horowitz:2009eb}. However, at higher momenta, sensitivity to the precise value of the upper bound is significantly larger for the LFT + collinear upper bound in comparison to the collinear upper bound. The variation of the upper bound leads to a $\mathcal{O}(35\text{--}70\%)$ variation in the charged hadron $R_{AA}$ in central \coll{Pb}{Pb} collisions and a $\mathcal{O}(10\text{--}20\%)$ variation in the charged hadron $R_{AA}$ in central \coll{p}{Pb} collisions for $10 ~\mathrm{GeV} \lesssim p_T \lesssim 100 ~\mathrm{GeV}$. This is an extremely large uncertainty, even in comparison to other large uncertainties in the calculation such as the effect of NLO contributions in the coupling which are expected to be $\mathcal{O}(30 \%)$.

The $\mathcal{O}(35\text{--}70\%)$ sensitivity to the exact value of the upper bound, chosen so that the large formation time and collinear approximations are both satisfied, motivates the theoretical need for a \emph{short formation time correction}, i.e., a rederivation of DGLV and DGLV + SPL which removes the large formation time approximation. All models based on the GLV opacity expansion may potentially see large contributions from such a correction, since the results calculated without such a correction contain large contributions from regions of the phase space where the large formation time approximation is explicitly violated. The breakdown of the large formation time approximation will affect canonical and contemporary phenomenological energy loss models such as the Wicks-Horowitz-Djordjevic-Gyulassy (WHDG) \cite{Wicks:2005gt, Horowitz:2011gd, Horowitz:2012cf} and CUJET \cite{Buzzatti:2011vt, Buzzatti:2012pe, Xu:2014ica, Xu:2015bbz} models. 

While the focus of this work has been within the GLV formalism, we will take some time to discuss the extent to which the large formation time approximation may be violated in other works. 
The single emission kernel $dN^g / dx$ is identical for DGLV as it is for ASW-SH \cite{Wiedemann:2000za,Wiedemann:2000tf} if we set the gluon mass to zero, although there are various different details in the implementation of this kernel \cite{Armesto:2011ht}. It stands to reason that the ASW-SH formalism must be sensitive to the large formation time approximation. 
In the HT calculation \cite{Guo:2000nz,Wang:2001ifa}, only the leading order, divergent term in $\mathbf{k}^2$ (in their notation $\ell_T^2$) is considered, which implies terms $\sim \mathbf{k}^2 / 2 x E \mu_D$ (terms usually neglected according to the large formation time approximation) are neglected. Such an omission of these short formation time terms would impact the MATTER framework \cite{Majumder:2013re,Cao:2017qpx}, which would in turn have implications for various calculations and extractions performed by the JETSCAPE collaboration \cite{JETSCAPE:2021ehl,JETSCAPE:2024cqe,JETSCAPE:2017eso,JETSCAPE:2022jer}. 
Since the GLV opacity expansion formalism is a limiting solution of the BDMPS-Z \cite{Baier:1996kr, Baier:1996sk, Zakharov:1997uu, Baier:1996vi, Zakharov:1996fv} path integral \cite{Armesto:2011ht, Wiedemann:2000za}, it follows that a large sensitivity to the large formation time approximation in the GLV formalism would lead to a similarly large sensitivity in the BDMPS-Z formalism, at least in the dilute medium limit. It seems likely that the BDMSP-Z result is sensitive to the large formation time approximation as, at least in the original BDMPS formulation \cite{Baier:1996kr, Baier:1996sk, Baier:1996vi}, it receives contributions from the same diagrams as the GLV result \cite{Gyulassy:2000er}. Additionally, there is work that has been to link the BDMPS-Z and GLV results \cite{Mehtar-Tani:2019ygg, Barata:2021wuf, Mehtar-Tani:2019tvy, Andres:2020vxs, Andres:2023jao}, and so the physics underlying the two approaches should be the same. Of course, the exact limit in which one performs a calculation may change the impact of any approximations; however, the underlying physical situation---a high $p_T$ parton moving through a plasma formed in a heavy-ion collisions---is unchanged. 
The AMY \cite{Arnold:2001ms,Arnold:2002ja,Arnold:2001ba,Caron-Huot:2010qjx} energy loss model receives contributions from diagrams which are topologically equivalent to the diagrams in the HT energy loss model \cite{Majumder:2010qh}. While the limits in which these diagrams are evaluated may be different, it seems reasonable to presume a similar degree of sensitivity to the large formation time approximation between the HT and AMY approaches.
Djordjevic's dynamical scattering center framework \cite{Djordjevic:2007at, Djordjevic:2008iz, Djordjevic:2009cr}, and its implementation in the DREENA framework \cite{Zigic:2018smz, Zigic:2021rku}, uses HTL propagators \cite{Braaten:1989mz, Klimov:1982bv, Pisarski:1988vd, Weldon:1982aq, Weldon:1982bn} to calculate both the elastic and radiative energy loss. The approximation $\delta[q_z - q_0 - (\mathbf{k}^2)/(2 x E)] = \delta [q_z - q_0]$ is made in the derivation of the radiative energy loss \cite{Djordjevic:2007at}, which is true for $\mathbf{k}^2 / 2 x E q_0 \ll 1$. Since $q$ is a medium scale, it likely scales like $\mu_D$, and so this is a similar approximation to the large formation time approximation in this work. Due to the similarity of the dynamical and static scattering center results \cite{Djordjevic:2007at}, one might expect a similar degree of sensitivity in the dynamical result \cite{Djordjevic:2007at} as for the static result (this work). Such a sensitivity would impact the DREENA model results \cite{Zigic:2018smz, Zigic:2021rku}, and the corresponding extractions of medium properties \cite{Karmakar:2024jak, Karmakar:2023ity}. Strongly coupled approaches \cite{Casalderrey-Solana:2011dxg, Gubser:2006bz, Herzog:2006gh,BitaghsirFadafan:2008adl, Mikhailov:2003er,Chesler:2008uy, Chesler:2008wd, Gubser:2008as,Morad:2014xla,Horowitz:2015dta, Casalderrey-Solana:2014bpa, Casalderrey-Solana:2016jvj, Hulcher:2017cpt} have a very different physics picture of energy loss in the QGP and it would be inappropriate to make inferences from our work to these approaches. In purely strongly coupled approaches the formation time is zero and would not be impacted by this assumption.

The validity of the large formation time approximation in the context of calculating the medium-induced radiation from a highly offshell, high momentum particle propagating through a phenomenologically relevant plasma has \textbf{not} been explored in any context other than this work and our associated papers \cite{Faraday:2023mmx, Faraday:2023uay}. This thesis has shown that the standard and widely used DGLV calculation \cite{Djordjevic:2003zk} receives large contributions from regions of phase space where the large formation time approximation is invalid. One may artificially restrict the phase space to prevent such contributions from being included; however, this results in a large sensitivity to the exact point at which the restriction occurs. The large sensitivity to the large formation time cutoff offers circumstantial evidence that a short formation time correction, which removes the large formation time assumption, may be exceedingly large. Since such an examination of the large formation time approximation has not been performed in any other formalism, one may only speculate on the extent to which other formalisms are equally sensitive to the large formation time approximation. We motivate that an accounting of the large formation time approximation is required in order to place any energy loss formalism on a rigorous footing.

In \cref{sec:EL_paper}, we provided a detailed discussion of the elastic sector of our energy loss model. We presented results with three different elastic energy loss models. We included the Braaten and Thoma (BT) elastic energy loss \cite{Braaten:1991we, Braaten:1991jj}, which uses Hard Thermal Loop (HTL) propagators at small momentum transfers and vacuum propagators at large momentum transfers, and the HTL elastic energy loss \cite{Wicks:2008zz}, which used HTL propagators for all momentum transfers but kept the full kinematics of the exchange. Additionally, while the BT result only has finite $\Delta E / E$ and therefore no distributional shape, the HTL result has a finite number of scatters and one may therefore treat it with a full Poisson convolution, as in the radiative case. Therefore, we presented Gaussian BT, Gaussian HTL, and Poisson HTL elastic energy loss model results. These three sets of results allow us to (a) understand the uncertainty in the transition between HTL and vacuum propagators, through a comparison of the results calculated with the Gaussian BT and Gaussian HTL elastic energy loss kernels and (b) understand the applicability of the central limit theorem approximation in the elastic energy loss, through a comparison of the results computed with the Gaussian HTL and Poisson HTL elastic energy loss.

Comparing the Gaussian BT and Gaussian HTL results gave an estimate of the theoretical uncertainty associated with the transition between HTL and vacuum propagators. We found that the HTL and BT results for the $R_{AA}$ of light and heavy flavor hadrons produced in central \coll{Pb}{Pb} and \coll{Au}{Au} collisions differed by $\mathcal{O}(30\text{--}50\%)$ for $p_T$ of $\mathcal{O}(10\text{--}50)$ GeV. 
One is particularly sensitive to this uncertainty at lower momenta and in smaller collision systems, where elastic energy loss is relatively more important.

By comparing results calculated with the Gaussian HTL and Poisson HTL elastic energy loss, we assessed the validity of the application of the central limit theorem to the probability distribution of elastic energy loss. We were originally motivated to perform such a calculation as we found that almost all energy loss in small systems was from the elastic sector as opposed to the radiative sector, which we thought might be due to the breakdown of the central limit theorem assumption in small systems with $\mathcal{O}(0\text{--}1)$ scatters. Surprisingly, we found that the difference between the Gaussian and Poisson results was extremely small---particularly so in small systems. We showed that one may understand this small difference between the Poisson and Gaussian results through a decomposition of the $R_{AA}$ in terms of the moments of the underlying energy loss distribution. Small systems have small amounts of energy loss and therefore require fewer moments to accurately describe the $R_{AA}$. The Poisson and Gaussian distributions are constrained to have identical zeroth and first moments, which leads to almost identical $R_{AA}$ results.

Finally, we performed a detailed comparison to experimental data in \cref{sec:lft}. We conducted a $\chi^2$ goodness-of-fit analysis on available large system experimental data from RHIC and LHC for various final state hadrons to extract a best-fit value of the effective coupling constant. We found that a large amount of the sensitivity to the HTL vs.\ BT elastic energy loss could be absorbed into a different value of the strong coupling $\alpha_s$. However, even after independently finding the best global $\alpha_s$ value from a fit to large system data, fairly different $p_T$ and system size dependencies remained. The HTL elastic energy loss had a flatter $p_T$ dependence than the BT result, due to the higher proportion of radiative energy loss, which does not fall off as quickly as the elastic energy loss. We also saw that the uncertainty in the large formation time upper bound is largely absorbed by a change of the strong coupling. However, this uncertainty leads to extremely different best-fit values of the strong coupling. Such an uncertainty has not been included in \emph{any} extractions of the properties of the medium \cite{JET:2013cls, JETSCAPE:2017eso, JETSCAPE:2022jer,JETSCAPE:2024cqe, Karmakar:2023ity,Karmakar:2024jak,Shi:2018izg, Stojku:2020tuk}, for instance $\hat{q}$, and as such a short formation time correction may result in a dramatic change for such extracted parameters (or at least the associated confidence interval on the extracted parameters). We also note that in this work we assumed that any short formation time correction would have the same scale dependence as the leading order result. It is an open question whether such a short formation time correction would alter the $p_T$ or other scale dependence, which could mean that the full result cannot be captured by simply changing the effective coupling $\alpha_s$.

We first performed the fit of $\alpha_s$ separately for LHC \cite{ATLAS:2022kqu, CMS:2016xef, ALICE:2018lyv, ALICE:2018vuu} and RHIC \cite{PHENIX:2008saf, PHENIX:2012jha} large system data. The best-fit value of $\alpha_s$ for the Braaten and Thoma (BT) elastic energy loss \cite{Braaten:1991we, Braaten:1991jj} was the same for RHIC and LHC, while the HTL \cite{Wicks:2008zz} best-fit value was larger at RHIC compared to LHC---explicitly for $\kappa = 1$ we found $\alpha_s^{\text{RHIC}} / \alpha_s^{\text{LHC}} \sim 1.12$. The different couplings at RHIC and LHC may be explained by the coupling running, at least partially, with the temperature scale. Running coupling \cite{Peshier:2006ah, Xu:2014ica} effects are expected to be important, particularly at the large range of energies available at LHC. We argued in \cref{sec:lft_conclusion} that if the coupling ran only at the hard momentum scale $p_T$, then one expects $\alpha_s^{\text{RHIC}} / \alpha_s^{\text{LHC}} \sim 1$, while if the coupling runs only with an intermediary scale $\sqrt{ET}$ one expects $\alpha_s^{\text{RHIC}} / \alpha_s^{\text{LHC}} \sim 1.05 \text{--}1.07$, and if the coupling runs only with the soft momentum scale $2 \pi T$ or $\mu_D \sim g T$, one expects $\alpha_s^{\text{RHIC}} / \alpha_s^{\text{LHC}} \sim  1.15 \text{--}1.35$. Therefore our results for the BT elastic energy loss are qualitatively consistent with the coupling running at only the hard momentum scale, whereas the HTL elastic energy loss results are qualitatively consistent with the coupling running at the intermediary--soft momentum scales. These qualitatively different conclusions motivate a rigorous derivation of the scales at which the strong coupling runs. Such a running coupling calculation may help to understand whether energy loss in the QGP is strongly coupled, weakly coupled or involves a mixture of strongly- and weakly-coupled scales.

We additionally performed a global fit of our model simultaneously to data at both RHIC and LHC and presented the model results from this fit. We found a visually satisfactory description of all large system data at RHIC and LHC from a single parameter fit. We note that we did not compute the $\chi^2$ per degree of freedom nor the $p$ value of the fit, as the purpose of the work was not to determine whether the model was compatible with data. Such an analysis would not be appropriate at this time, due to the simple treatment of uncertainties in this work. Future work may more accurately model the systematic uncertainties, similar to the procedure followed in \cite{JETSCAPE:2021ehl}, to determine whether the model is compatible with all available data. Our large system fitted results were visually compatible with the measured suppression pattern of pions in $0\text{--}5\%$ centrality \coll{d}{Au} collisions at RHIC \cite{PHENIX:2023dxl}, but our model did not reproduce the lack of suppression of charged hadrons in $0\text{--}10\%$ centrality \coll{p}{Pb} collisions at LHC \cite{ATLAS:2022kqu}. The extent to which the lack of suppression is due to centrality bias, which the LHC results may be sensitive to \cite{ATLAS:2022kqu} but the RHIC results are insensitive to \cite{PHENIX:2023dxl}, is not currently obvious. Future work on this front may include theoretical calculations in our model of self-normalized observables at the LHC \cite{ATLAS:2022kqu}, minimum bias results in \coll{p}{Pb} and \coll{O}{O} collisions which are insensitive to centrality bias \cite{Huss:2020whe, Huss:2020dwe}, high momentum $v_2$ in both small and large systems \cite{ATLAS:2019vcm}, or jet substructure predictions \cite{Kolbe:2023rsq}.

Hard probes continue to be extremely promising tools for understanding the quark-gluon plasma (QGP) formed in heavy-ion collisions at the Large Hadron Collider (LHC) and Relativistic Heavy Ion Collider (RHIC). 
While quarkonia suppression \cite{ALICE:2016sdt}, elliptic flow \cite{CMS:2015yux,ATLAS:2015hzw}, and strangeness enhancement \cite{ALICE:2013wgn,ALICE:2015mpp} have all been observed in high-multiplicity small systems, the small-system suppression pattern---as evidenced by leading hadron $R_{AA}$ \cite{ATLAS:2022kqu, PHENIX:2023dxl} and high-$p_T$ $v_2$ \cite{ATLAS:2019vcm}---remains inconclusive. 
This puts hard probes at the forefront of the search for a consistent description of the physics of small collision systems.
However, the theoretical picture of energy loss in both small and large systems is plagued by uncertainties and, furthermore, all energy loss models employ some approximations that are inconsistent with the characteristics of a small plasma. 
Despite these significant challenges, this does not suggest that the situation is intractable; in fact there are clear avenues for theoretical progress. 
In this work we have improved the compatibility of theoretical energy loss models with the quark-gluon plasma that is formed in small systems, and future work should build upon this program.
Theoretical guidance on small-system suppression will be invaluable for answering one of the most tantalizing questions in the field: does a quark-gluon plasma form in even the smallest of collision systems?

	\clearpage
	\phantomsection

	\addcontentsline{toc}{chapter}{References}

 \bibliographystyle{apsrev4-2} %
 \bibliography{manual,msc} 

 \appendix
\renewcommand{\fullthechapter}{Appendix \thechapter}
 \chapter{All fitted results for \texorpdfstring{$R_{AB}$}{RAB} vs.\ experimental data}
 \label{sec:all_fitted_results}
 In \cref{sec:lft} we performed a global fit of the strong coupling constant $\alpha_s$ in our model to experimental data for pions and $D$ mesons in \coll{Pb}{Pb} and \coll{Au}{Au} collisions at RHIC and LHC. We additionally interpreted the predictions of these large-system-fitted results, in small systems. The plots shown were a representative sample of the actual data used for the global fit, and we reproduce the remaining figures here. There are an extremely large number of experimental datasets, particularly due to the large range of results across centrality reported by the different experiments, and the numerous sets of measurements for the same collision system from the same and different experiments.

\comm{Way to distinguish the overlapping DGLV and DGLV + SPL curves}
\comm{Bands in the plots}
\comm{}

\FloatBarrier
\section{Large System LHC Data}
\label{sec:large_system_lhc_data}

\begin{figure}[!htpb]
\centering
\begin{subfigure}[t]{0.49\textwidth}
    \centering
    \includegraphics[width=\linewidth]{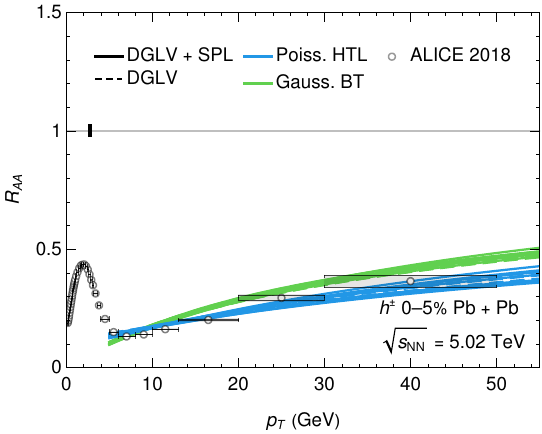}
    \caption{Nuclear modification factor $R_{AA}$ for charged hadrons in $0\text{--}5\%$ central \coll{Pb}{Pb} collisions.}
    \label{fig:ALICE_RAA_charged_0-5}
\end{subfigure}\hfill
\begin{subfigure}[t]{0.49\textwidth}
    \centering
    \includegraphics[width=\linewidth]{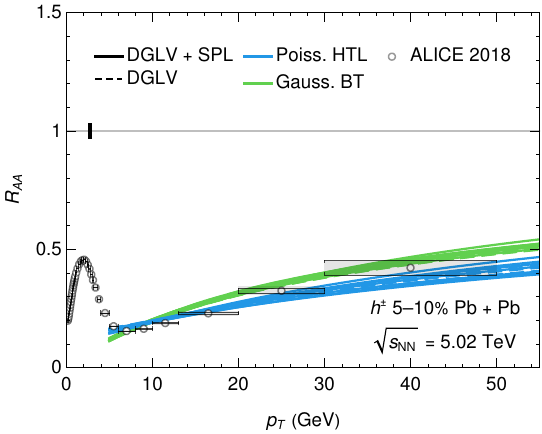}
    \caption{Nuclear modification factor $R_{AA}$ for charged hadrons in $5\text{--}10\%$ central \coll{Pb}{Pb} collisions.}
    \label{fig:ALICE_RAA_charged_5-10}
\end{subfigure}
\caption{Comparison of the nuclear modification factor $R_{AA}$ for charged hadrons in $0\text{--}10\%$ central \coll{Pb}{Pb} collisions at $\sqrt{s_{NN}} = 5.02$ TeV. Data is from \cite{ALICE:2018vuu}, with statistical and systematic uncertainties indicated by error bars and boxes, respectively. The global normalization uncertainty on the number of binary collisions is represented by solid boxes at unity.}
\label{fig:ALICE_RAA_charged_0-10}
\end{figure}

\begin{figure}[!htpb]
\centering
\begin{subfigure}[t]{0.49\textwidth}
    \centering
    \includegraphics[width=\linewidth]{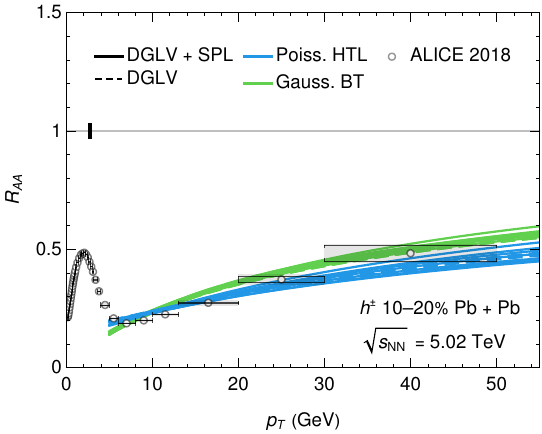}
    \caption{Nuclear modification factor $R_{AA}$ for charged hadrons in $10\text{--}20\%$ central \coll{Pb}{Pb} collisions.}
    \label{fig:ALICE_RAA_charged_10-20}
\end{subfigure}\hfill
\begin{subfigure}[t]{0.49\textwidth}
    \centering
    \includegraphics[width=\linewidth]{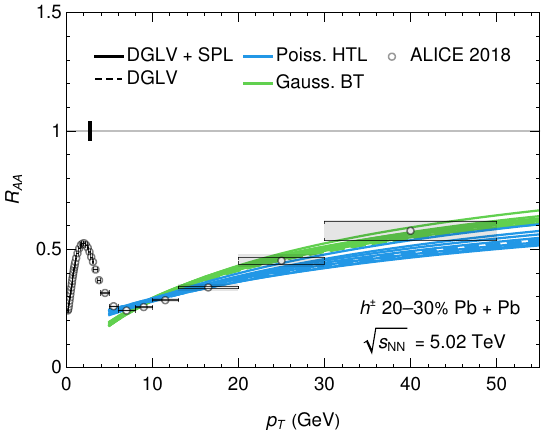}
    \caption{Nuclear modification factor $R_{AA}$ for charged hadrons in $20\text{--}30\%$ central \coll{Pb}{Pb} collisions.}
    \label{fig:ALICE_RAA_charged_20-30}
\end{subfigure}
\caption{Comparison of the nuclear modification factor $R_{AA}$ for charged hadrons in $10\text{--}30\%$ central \coll{Pb}{Pb} collisions at $\sqrt{s_{NN}} = 5.02$ TeV. Data is from \cite{ALICE:2018vuu}, with statistical and systematic uncertainties indicated by error bars and boxes, respectively. The global normalization uncertainty on the number of binary collisions is represented by solid boxes at unity.}
\label{fig:ALICE_RAA_charged_10-30}
\end{figure}

\begin{figure}[!htpb]
\centering
\begin{subfigure}[t]{0.49\textwidth}
    \centering
    \includegraphics[width=\linewidth]{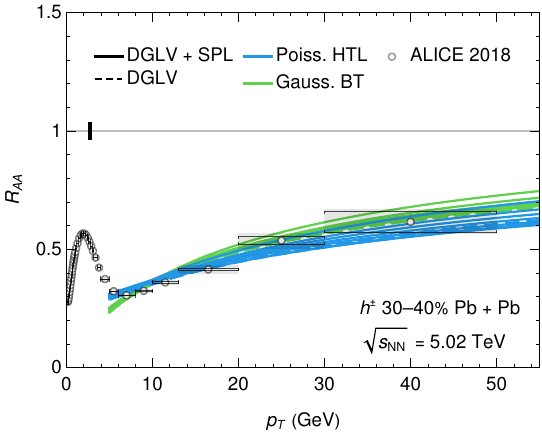}
    \caption{Nuclear modification factor $R_{AA}$ for charged hadrons in $30\text{--}40\%$ central \coll{Pb}{Pb} collisions.}
    \label{fig:ALICE_RAA_charged_30-40}
\end{subfigure}\hfill
\begin{subfigure}[t]{0.49\textwidth}
    \centering
    \includegraphics[width=\linewidth]{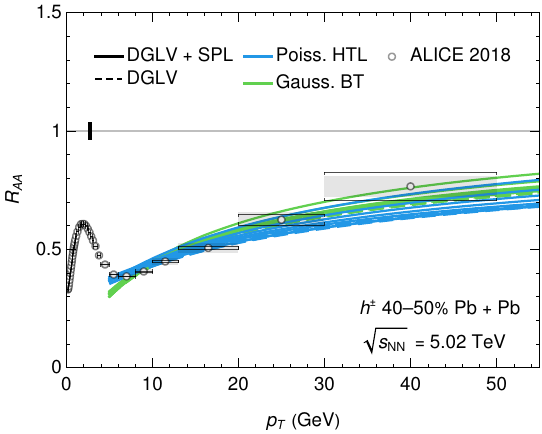}
    \caption{Nuclear modification factor $R_{AA}$ for charged hadrons in $40\text{--}50\%$ central \coll{Pb}{Pb} collisions.}
    \label{fig:ALICE_RAA_charged_40-50}
\end{subfigure}
\caption{Comparison of the nuclear modification factor $R_{AA}$ for charged hadrons in $30\text{--}50\%$ central \coll{Pb}{Pb} collisions at $\sqrt{s_{NN}} = 5.02$ TeV. Data is from \cite{ALICE:2018vuu}, with statistical and systematic uncertainties indicated by error bars and boxes, respectively. The global normalization uncertainty on the number of binary collisions is represented by solid boxes at unity.}
\label{fig:ALICE_RAA_charged_30-50}
\end{figure}

\begin{figure}[!htpb]
\centering
\begin{subfigure}[t]{0.49\textwidth}
    \centering
    \includegraphics[width=\linewidth]{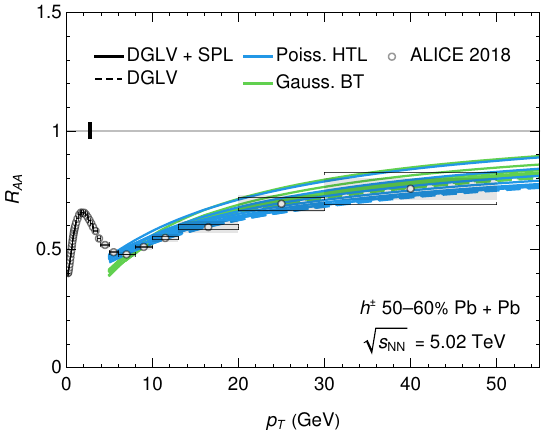}
    \caption{Nuclear modification factor $R_{AA}$ for charged hadrons in $50\text{--}60\%$ central \coll{Pb}{Pb} collisions.}
    \label{fig:ALICE_RAA_charged_50-60}
\end{subfigure}\hfill
\begin{subfigure}[t]{0.49\textwidth}
    \centering
    \includegraphics[width=\linewidth]{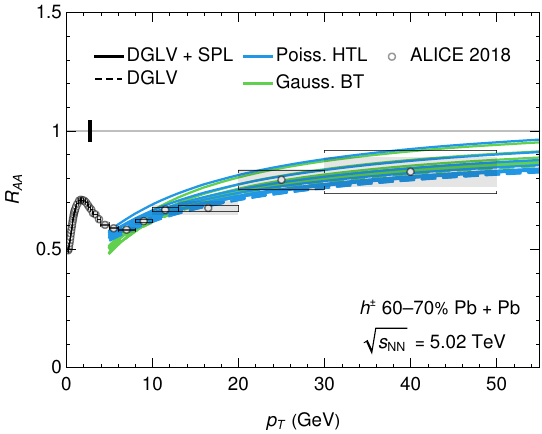}
    \caption{Nuclear modification factor $R_{AA}$ for charged hadrons in $60\text{--}70\%$ central \coll{Pb}{Pb} collisions.}
    \label{fig:ALICE_RAA_charged_60-70}
\end{subfigure}
\caption{Comparison of the nuclear modification factor $R_{AA}$ for charged hadrons in $50\text{--}70\%$ central \coll{Pb}{Pb} collisions at $\sqrt{s_{NN}} = 5.02$ TeV. Data is from \cite{ALICE:2018vuu}, with statistical and systematic uncertainties indicated by error bars and boxes, respectively. The global normalization uncertainty on the number of binary collisions is represented by solid boxes at unity.}
\label{fig:ALICE_RAA_charged_50-70}
\end{figure}

\begin{figure}[!htpb]
\centering
\includegraphics[width=0.49\textwidth]{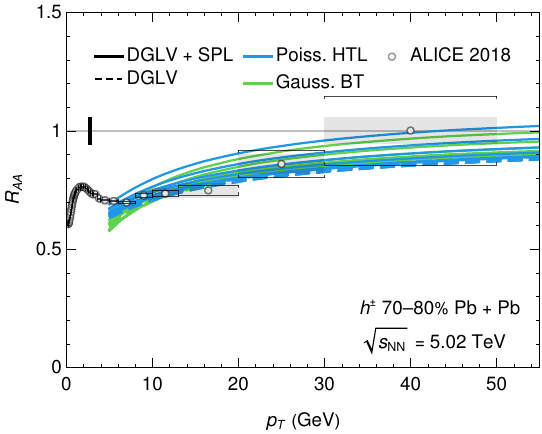}
\caption{Nuclear modification factor $R_{AA}$ for charged hadrons in $70\text{--}80\%$ central \coll{Pb}{Pb} collisions at $\sqrt{s_{NN}} = 5.02$ TeV. Data is from \cite{ALICE:2018vuu}, with statistical and systematic uncertainties indicated by error bars and boxes, respectively. The global normalization uncertainty on the number of binary collisions is represented by solid boxes at unity.}
\label{fig:ALICE_RAA_charged_70-80}
\end{figure}

\begin{figure}[!htpb]
\centering
\begin{subfigure}[t]{0.49\textwidth}
    \centering
    \includegraphics[width=\linewidth]{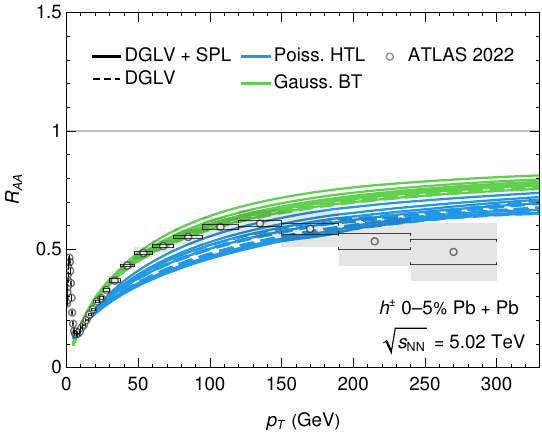}
    \caption{Nuclear modification factor $R_{AA}$ for charged hadrons in $0\text{--}5\%$ central \coll{Pb}{Pb} collisions.}
    \label{fig:ATLAS_RAA_charged_0-5}
\end{subfigure}\hfill
\begin{subfigure}[t]{0.49\textwidth}
    \centering
    \includegraphics[width=\linewidth]{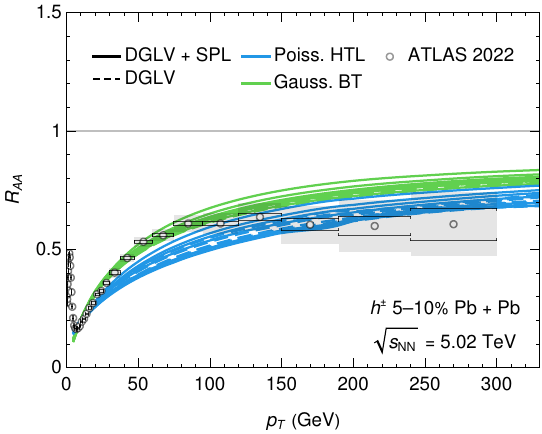}
    \caption{Nuclear modification factor $R_{AA}$ for charged hadrons in $5\text{--}10\%$ central \coll{Pb}{Pb} collisions.}
    \label{fig:ATLAS_RAA_charged_5-10}
\end{subfigure}
\caption{Comparison of the nuclear modification factor $R_{AA}$ for charged hadrons in $0\text{--}10\%$ central \coll{Pb}{Pb} collisions at $\sqrt{s_{NN}} = 5.02$ TeV. Data is from \cite{ATLAS:2022kqu}, with statistical and systematic uncertainties indicated by error bars and boxes, respectively. The global normalization uncertainty on the number of binary collisions is represented by solid boxes at unity.}
\label{fig:ATLAS_RAA_charged_0-10}
\end{figure}

\begin{figure}[!htpb]
\centering
\begin{subfigure}[t]{0.49\textwidth}
    \centering
    \includegraphics[width=\linewidth]{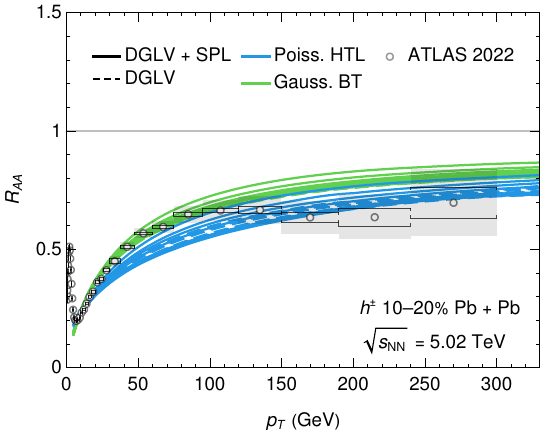}
    \caption{Nuclear modification factor $R_{AA}$ for charged hadrons in $10\text{--}20\%$ central \coll{Pb}{Pb} collisions.}
    \label{fig:ATLAS_RAA_charged_10-20}
\end{subfigure}\hfill
\begin{subfigure}[t]{0.49\textwidth}
    \centering
    \includegraphics[width=\linewidth]{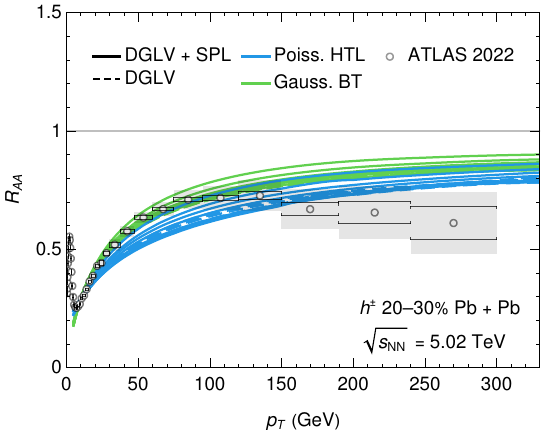}
    \caption{Nuclear modification factor $R_{AA}$ for charged hadrons in $20\text{--}30\%$ central \coll{Pb}{Pb} collisions.}
    \label{fig:ATLAS_RAA_charged_20-30}
\end{subfigure}
\caption{Comparison of the nuclear modification factor $R_{AA}$ for charged hadrons in $10\text{--}30\%$ central \coll{Pb}{Pb} collisions at $\sqrt{s_{NN}} = 5.02$ TeV. Data is from \cite{ATLAS:2022kqu}, with statistical and systematic uncertainties indicated by error bars and boxes, respectively. The global normalization uncertainty on the number of binary collisions is represented by solid boxes at unity.}
\label{fig:ATLAS_RAA_charged_10-30}
\end{figure}

\begin{figure}[!htpb]
\centering
\begin{subfigure}[t]{0.49\textwidth}
    \centering
    \includegraphics[width=\linewidth]{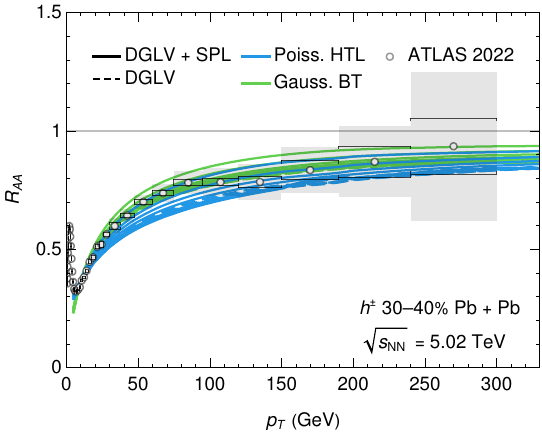}
    \caption{Nuclear modification factor $R_{AA}$ for charged hadrons in $30\text{--}40\%$ central \coll{Pb}{Pb} collisions.}
    \label{fig:ATLAS_RAA_charged_30-40}
\end{subfigure}\hfill
\begin{subfigure}[t]{0.49\textwidth}
    \centering
    \includegraphics[width=\linewidth]{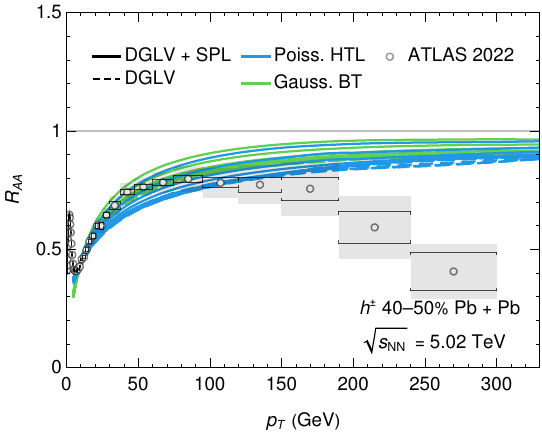}
    \caption{Nuclear modification factor $R_{AA}$ for charged hadrons in $40\text{--}50\%$ central \coll{Pb}{Pb} collisions.}
    \label{fig:ATLAS_RAA_charged_40-50}
\end{subfigure}
\caption{Comparison of the nuclear modification factor $R_{AA}$ for charged hadrons in $30\text{--}50\%$ central \coll{Pb}{Pb} collisions at $\sqrt{s_{NN}} = 5.02$ TeV. Data is from \cite{ATLAS:2022kqu}, with statistical and systematic uncertainties indicated by error bars and boxes, respectively. The global normalization uncertainty on the number of binary collisions is represented by solid boxes at unity.}
\label{fig:ATLAS_RAA_charged_30-50}
\end{figure}

\begin{figure}[!htpb]
\centering
\begin{subfigure}[t]{0.49\textwidth}
    \centering
    \includegraphics[width=\linewidth]{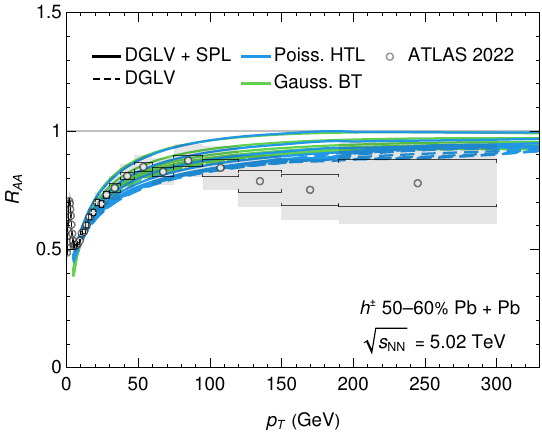}
    \caption{Nuclear modification factor $R_{AA}$ for charged hadrons in $50\text{--}60\%$ central \coll{Pb}{Pb} collisions.}
    \label{fig:ATLAS_RAA_charged_50-60}
\end{subfigure}\hfill
\begin{subfigure}[t]{0.49\textwidth}
    \centering
    \includegraphics[width=\linewidth]{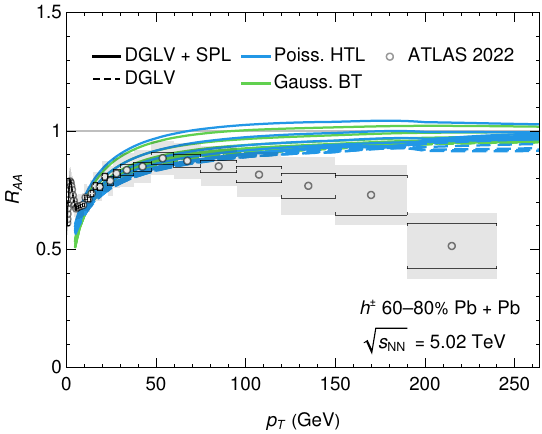}
    \caption{Nuclear modification factor $R_{AA}$ for charged hadrons in $60\text{--}80\%$ central \coll{Pb}{Pb} collisions.}
    \label{fig:ATLAS_RAA_charged_60-80}
\end{subfigure}
\caption{Comparison of the nuclear modification factor $R_{AA}$ for charged hadrons in $50\text{--}80\%$ central \coll{Pb}{Pb} collisions at $\sqrt{s_{NN}} = 5.02$ TeV. Data is from \cite{ATLAS:2022kqu}, with statistical and systematic uncertainties indicated by error bars and boxes, respectively. The global normalization uncertainty on the number of binary collisions is represented by solid boxes at unity.}
\label{fig:ATLAS_RAA_charged_50-80}
\end{figure}

\begin{figure}[!htpb]
\centering
\begin{subfigure}[t]{0.49\textwidth}
    \centering
    \includegraphics[width=\linewidth]{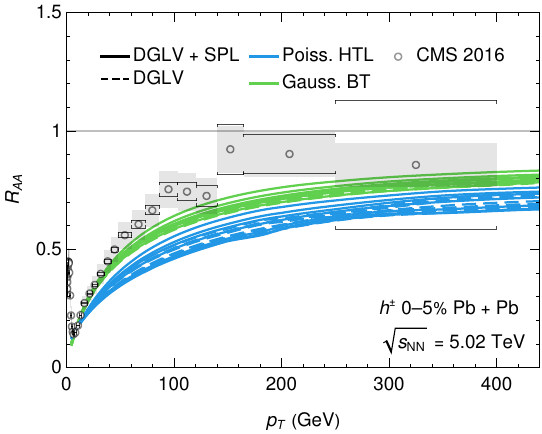}
    \caption{Nuclear modification factor $R_{AA}$ for charged hadrons in $0\text{--}5\%$ central \coll{Pb}{Pb} collisions.}
    \label{fig:CMS_RAA_charged_0-5_2016}
\end{subfigure}\hfill
\begin{subfigure}[t]{0.49\textwidth}
    \centering
    \includegraphics[width=\linewidth]{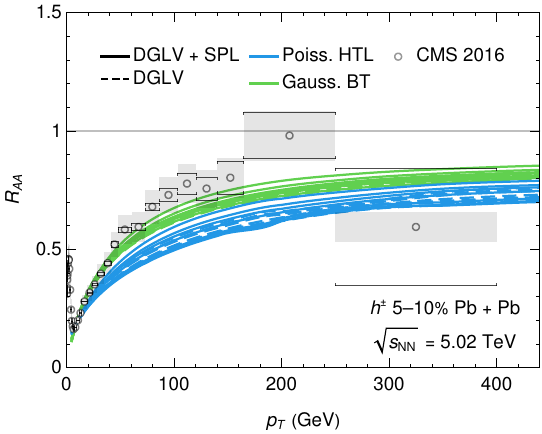}
    \caption{Nuclear modification factor $R_{AA}$ for charged hadrons in $5\text{--}10\%$ central \coll{Pb}{Pb} collisions.}
    \label{fig:CMS_RAA_charged_5-10_2016}
\end{subfigure}
\caption{Comparison of nuclear modification factors $R_{AA}$ for charged hadrons in $0\text{--}5\%$ and $5\text{--}10\%$ central \coll{Pb}{Pb} collisions at $\sqrt{s_{NN}} = 5.02$ TeV. Data is from \cite{CMS:2016xef}, with statistical and systematic uncertainties indicated by error bars and boxes, respectively.}
\label{fig:CMS_RAA_charged_0-10_2016}
\end{figure}

\begin{figure}[!htpb]
\centering
\begin{subfigure}[t]{0.49\textwidth}
    \centering
    \includegraphics[width=\linewidth]{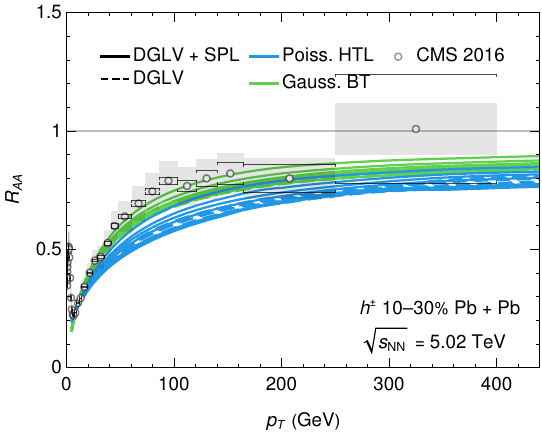}
    \caption{Nuclear modification factor $R_{AA}$ for charged hadrons in $10\text{--}30\%$ central \coll{Pb}{Pb} collisions.}
    \label{fig:CMS_RAA_charged_10-30_2016}
\end{subfigure}\hfill
\begin{subfigure}[t]{0.49\textwidth}
    \centering
    \includegraphics[width=\linewidth]{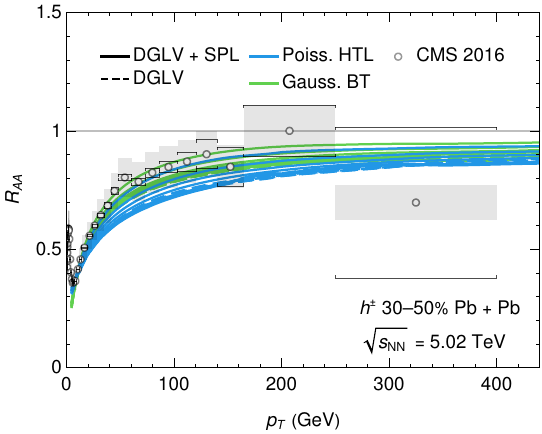}
    \caption{Nuclear modification factor $R_{AA}$ for charged hadrons in $30\text{--}50\%$ central \coll{Pb}{Pb} collisions.}
    \label{fig:CMS_RAA_charged_30-50_2016}
\end{subfigure}
\caption{Comparison of nuclear modification factors $R_{AA}$ for charged hadrons in $10\text{--}30\%$ and $30\text{--}50\%$ central \coll{Pb}{Pb} collisions at $\sqrt{s_{NN}} = 5.02$ TeV. Data is from \cite{CMS:2016xef}, with statistical and systematic uncertainties indicated by error bars and boxes, respectively.}
\label{fig:CMS_RAA_charged_10-50_2016}
\end{figure}

\begin{figure}[!htpb]
\centering
\begin{subfigure}[t]{0.49\textwidth}
    \centering
    \includegraphics[width=\linewidth]{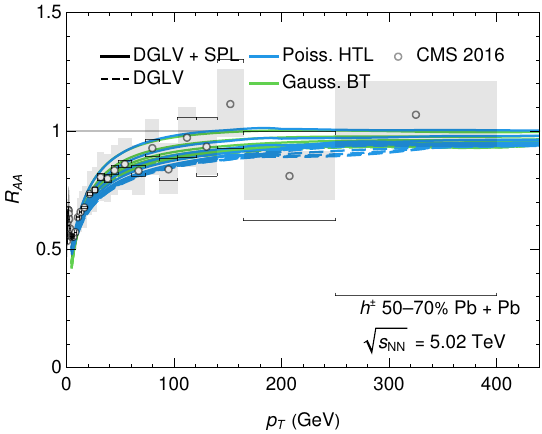}
    \caption{Nuclear modification factor $R_{AA}$ for charged hadrons in $50\text{--}70\%$ central \coll{Pb}{Pb} collisions.}
    \label{fig:CMS_RAA_charged_50-70_2016}
\end{subfigure}\hfill
\begin{subfigure}[t]{0.49\textwidth}
    \centering
    \includegraphics[width=\linewidth]{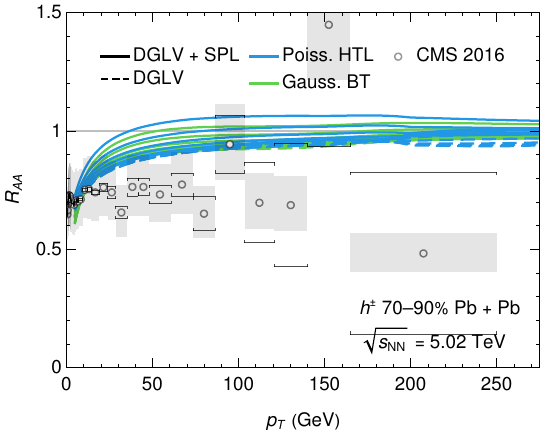}
    \caption{Nuclear modification factor $R_{AA}$ for charged hadrons in $70\text{--}90\%$ central \coll{Pb}{Pb} collisions.}
    \label{fig:CMS_RAA_charged_70-90_2016}
\end{subfigure}
\caption{Comparison of nuclear modification factors $R_{AA}$ for charged hadrons in $50\text{--}70\%$ and $70\text{--}90\%$ central \coll{Pb}{Pb} collisions at $\sqrt{s_{NN}} = 5.02$ TeV. Data is from \cite{CMS:2016xef}, with statistical and systematic uncertainties indicated by error bars and boxes, respectively.}
\label{fig:CMS_RAA_charged_50-90_2016}
\end{figure}

\begin{figure}[!htpb]
\centering
\begin{subfigure}[t]{0.49\textwidth}
    \centering
    \includegraphics[width=\linewidth]{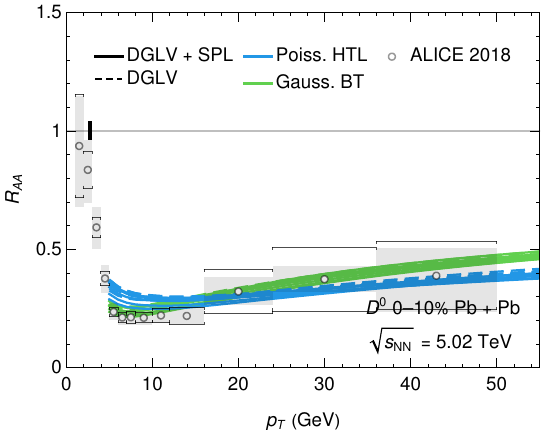}
    \caption{Nuclear modification factor $R_{AA}$ for $D^0$ mesons in $0\text{--}10\%$ central \coll{Pb}{Pb} collisions.}
    \label{fig:ALICE_RAA_D0_0-10_2018}
\end{subfigure}\hfill
\begin{subfigure}[t]{0.49\textwidth}
    \centering
    \includegraphics[width=\linewidth]{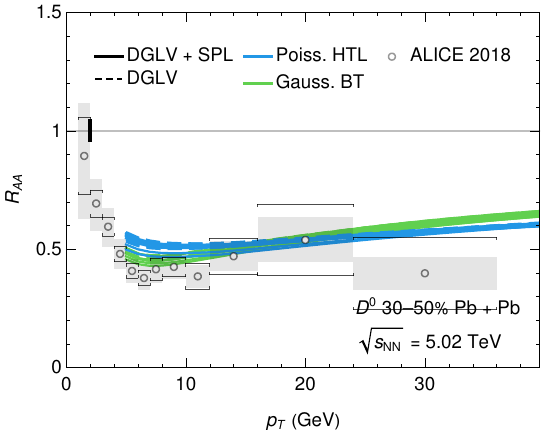}
    \caption{Nuclear modification factor $R_{AA}$ for $D^0$ mesons in $30\text{--}50\%$ central \coll{Pb}{Pb} collisions.}
    \label{fig:ALICE_RAA_D0_30-50_2018}
\end{subfigure}
\caption{Comparison of nuclear modification factors $R_{AA}$ for $D^0$ mesons in $0\text{--}10\%$ and $30\text{--}50\%$ central \coll{Pb}{Pb} collisions at $\sqrt{s_{NN}} = 5.02$ TeV. Data is from \cite{ALICE:2018lyv}, with statistical and systematic uncertainties indicated by error bars and boxes, respectively.}
\label{fig:ALICE_RAA_D0_0-30_2018}
\end{figure}

\begin{figure}[!htpb]
\centering
\includegraphics[width=0.49\textwidth]{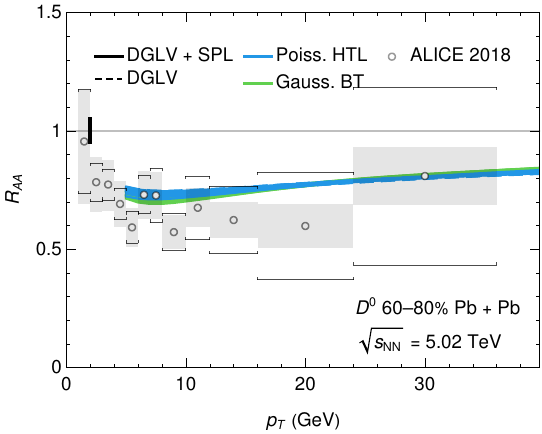}
\caption{Nuclear modification factor $R_{AA}$ for $D^0$ mesons in $60\text{--}80\%$ central \coll{Pb}{Pb} collisions.}
\label{fig:ALICE_RAA_D0_60-80_2018}
\end{figure}

\begin{figure}[!htpb]
\centering
\includegraphics[width=0.49\textwidth]{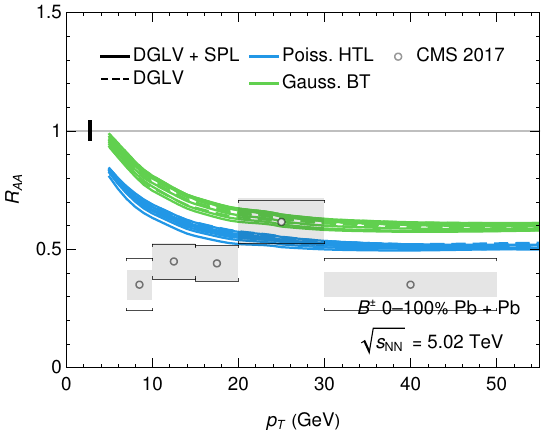}
\caption{Nuclear modification factor $R_{AA}$ for $B^\pm$ mesons in $0\text{--}100\%$ central \coll{Pb}{Pb} collisions.}
\label{fig:CMS_RAA_B+-_0-100_2017}
\end{figure}

\FloatBarrier
\section{Large system RHIC data}
\label{sec:large_system_rhic_data}

\begin{figure}[!htpb]
\centering
\begin{subfigure}[t]{0.49\textwidth}
    \centering
    \includegraphics[width=\linewidth]{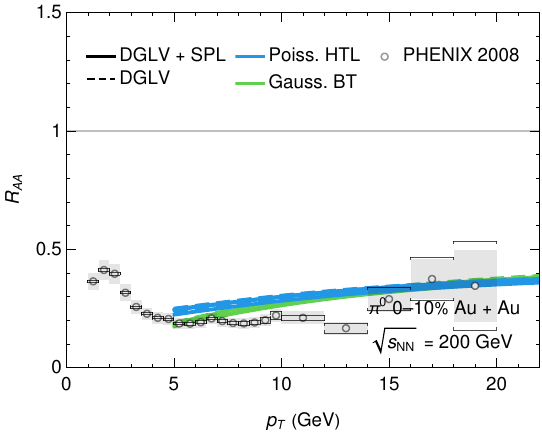}
    \caption{Nuclear modification factor $R_{AA}$ for $\pi^0$ in $0\text{--}10\%$ central \coll{Au}{Au} collisions.}
    \label{fig:PHENIX_RAA_Pi0_0-10_2008}
\end{subfigure}\hfill
\begin{subfigure}[t]{0.49\textwidth}
    \centering
    \includegraphics[width=\linewidth]{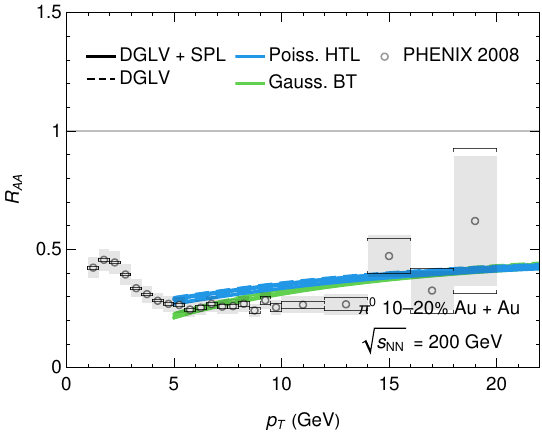}
    \caption{Nuclear modification factor $R_{AA}$ for $\pi^0$ in $10\text{--}20\%$ central \coll{Au}{Au} collisions.}
    \label{fig:PHENIX_RAA_Pi0_10-20_2008}
\end{subfigure}
\caption{Comparison of nuclear modification factors $R_{AA}$ for $\pi^0$ in $0\text{--}10\%$ and $10\text{--}20\%$ central \coll{Au}{Au} collisions at $\sqrt{s_{NN}} = 200$ GeV. Data is from \cite{PHENIX:2008saf}, with statistical and systematic uncertainties indicated by error bars and boxes, respectively.}
\label{fig:PHENIX_RAA_Pi0_0-20_2008}
\end{figure}

\begin{figure}[!htpb]
\centering
\begin{subfigure}[t]{0.49\textwidth}
    \centering
    \includegraphics[width=\linewidth]{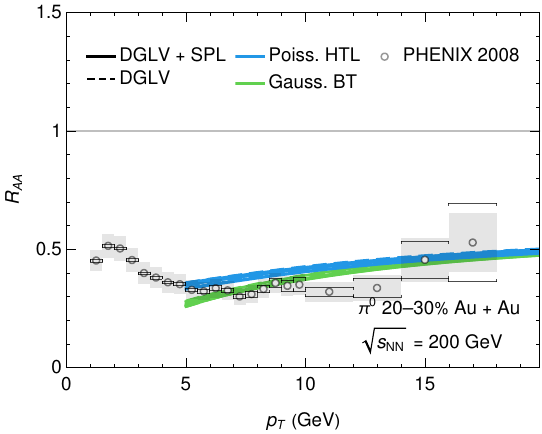}
    \caption{Nuclear modification factor $R_{AA}$ for $\pi^0$ in $20\text{--}30\%$ central \coll{Au}{Au} collisions.}
    \label{fig:PHENIX_RAA_Pi0_20-30_2008}
\end{subfigure}\hfill
\begin{subfigure}[t]{0.49\textwidth}
    \centering
    \includegraphics[width=\linewidth]{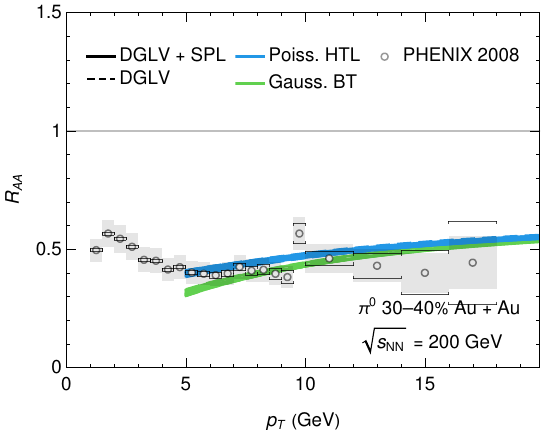}
    \caption{Nuclear modification factor $R_{AA}$ for $\pi^0$ in $30\text{--}40\%$ central \coll{Au}{Au} collisions.}
    \label{fig:PHENIX_RAA_Pi0_30-40_2008}
\end{subfigure}
\caption{Comparison of nuclear modification factors $R_{AA}$ for $\pi^0$ in $20\text{--}30\%$ and $30\text{--}40\%$ central \coll{Au}{Au} collisions at $\sqrt{s_{NN}} = 200$ GeV. Data is from \cite{PHENIX:2008saf}, with statistical and systematic uncertainties indicated by error bars and boxes, respectively.}
\label{fig:PHENIX_RAA_Pi0_20-40_2008}
\end{figure}

\begin{figure}[!htpb]
\centering
\begin{subfigure}[t]{0.49\textwidth}
    \centering
    \includegraphics[width=\linewidth]{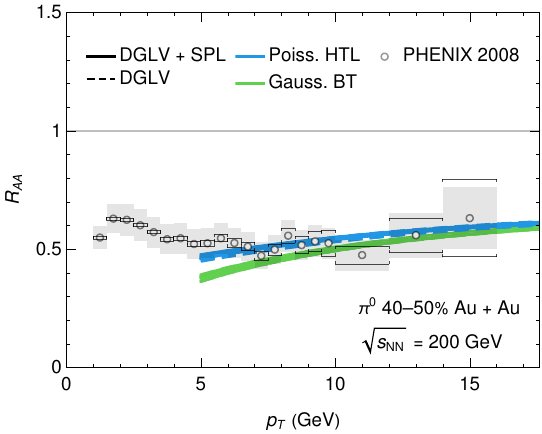}
    \caption{Nuclear modification factor $R_{AA}$ for $\pi^0$ in $40\text{--}50\%$ central \coll{Au}{Au} collisions.}
    \label{fig:PHENIX_RAA_Pi0_40-50_2008}
\end{subfigure}\hfill
\begin{subfigure}[t]{0.49\textwidth}
    \centering
    \includegraphics[width=\linewidth]{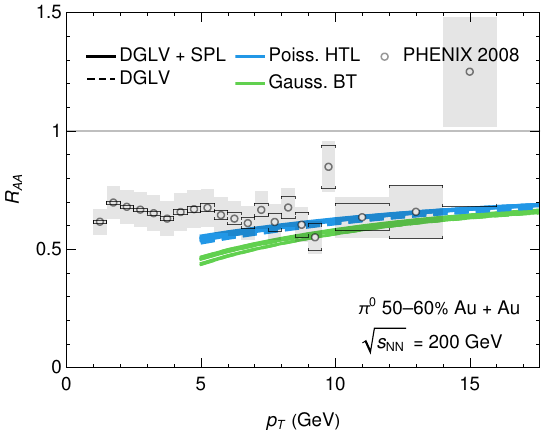}
    \caption{Nuclear modification factor $R_{AA}$ for $\pi^0$ in $50\text{--}60\%$ central \coll{Au}{Au} collisions.}
    \label{fig:PHENIX_RAA_Pi0_50-60_2008}
\end{subfigure}
\caption{Comparison of nuclear modification factors $R_{AA}$ for $\pi^0$ in $40\text{--}50\%$ and $50\text{--}60\%$ central \coll{Au}{Au} collisions at $\sqrt{s_{NN}} = 200$ GeV. Data is from \cite{PHENIX:2008saf}, with statistical and systematic uncertainties indicated by error bars and boxes, respectively.}
\label{fig:PHENIX_RAA_Pi0_40-60_2008}
\end{figure}

\begin{figure}[!htpb]
\centering
\begin{subfigure}[t]{0.49\textwidth}
    \centering
    \includegraphics[width=\linewidth]{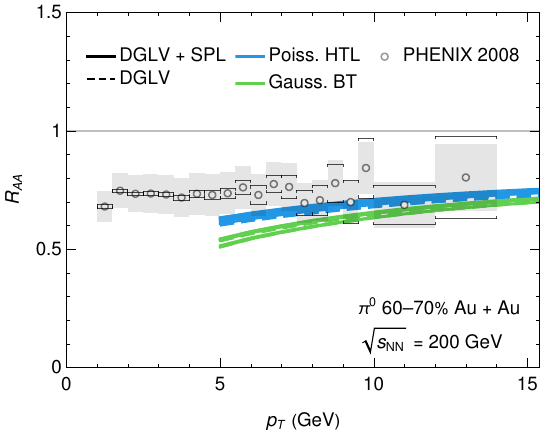}
    \caption{Nuclear modification factor $R_{AA}$ for $\pi^0$ in $60\text{--}70\%$ central \coll{Au}{Au} collisions.}
    \label{fig:PHENIX_RAA_Pi0_60-70_2008}
\end{subfigure}\hfill
\begin{subfigure}[t]{0.49\textwidth}
    \centering
    \includegraphics[width=\linewidth]{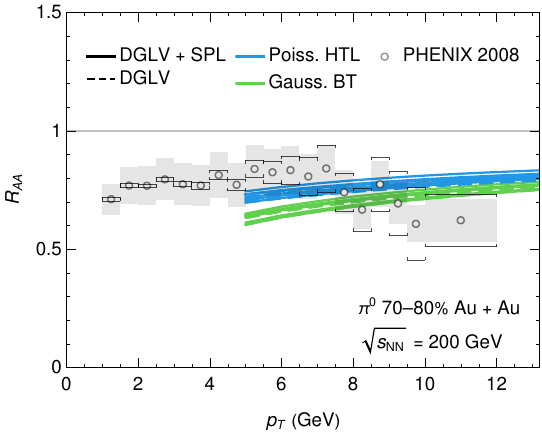}
    \caption{Nuclear modification factor $R_{AA}$ for $\pi^0$ in $70\text{--}80\%$ central \coll{Au}{Au} collisions.}
    \label{fig:PHENIX_RAA_Pi0_70-80_2008}
\end{subfigure}
\caption{Comparison of nuclear modification factors $R_{AA}$ for $\pi^0$ in $60\text{--}70\%$ and $70\text{--}80\%$ central \coll{Au}{Au} collisions at $\sqrt{s_{NN}} = 200$ GeV. Data is from \cite{PHENIX:2008saf}, with statistical and systematic uncertainties indicated by error bars and boxes, respectively.}
\label{fig:PHENIX_RAA_Pi0_60-80_2008}
\end{figure}

\begin{figure}[!htpb]
\centering
\includegraphics[width=0.49\textwidth]{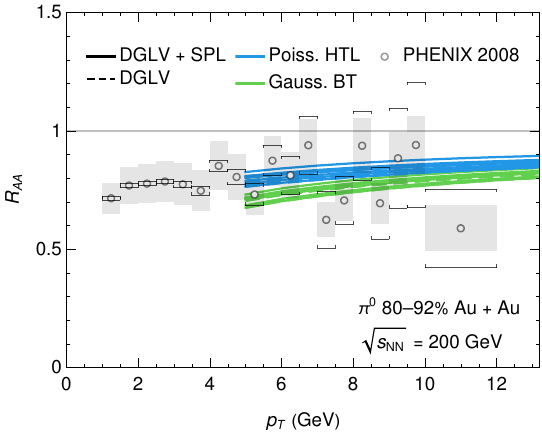}
\caption{Nuclear modification factor $R_{AA}$ for $\pi^0$ in $80\text{--}92\%$ central \coll{Au}{Au} collisions.}
\label{fig:PHENIX_RAA_Pi0_80-92_2008}
\end{figure}

\begin{figure}[!htpb]
\centering
\begin{subfigure}[t]{0.49\textwidth}
    \centering
    \includegraphics[width=\linewidth]{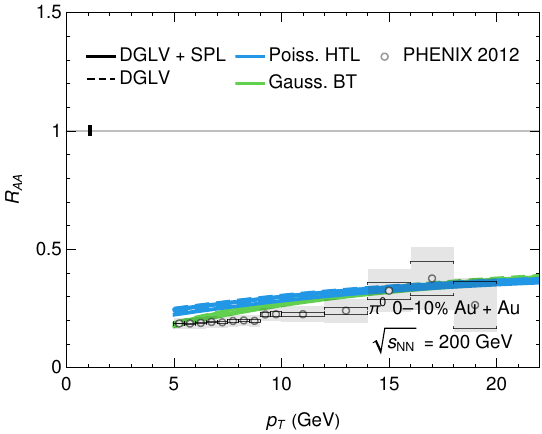}
    \caption{Nuclear modification factor $R_{AA}$ for $\pi^0$ in $0\text{--}10\%$ central \coll{Au}{Au} collisions.}
    \label{fig:PHENIX_RAA_Pi0_0-10_2012}
\end{subfigure}\hfill
\begin{subfigure}[t]{0.49\textwidth}
    \centering
    \includegraphics[width=\linewidth]{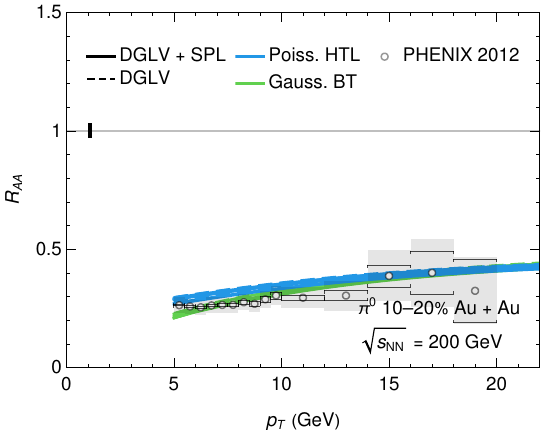}
    \caption{Nuclear modification factor $R_{AA}$ for $\pi^0$ in $10\text{--}20\%$ central \coll{Au}{Au} collisions.}
    \label{fig:PHENIX_RAA_Pi0_10-20_2012}
\end{subfigure}
\caption{Comparison of nuclear modification factors $R_{AA}$ for $\pi^0$ in $0\text{--}10\%$ and $10\text{--}20\%$ central \coll{Au}{Au} collisions at $\sqrt{s_{NN}} = 200$ GeV. Data is from \cite{PHENIX:2012jha}, with statistical and systematic uncertainties indicated by error bars and boxes, respectively.}
\label{fig:PHENIX_RAA_Pi0_0-20_2012}
\end{figure}

\begin{figure}[!htpb]
\centering
\begin{subfigure}[t]{0.49\textwidth}
    \centering
    \includegraphics[width=\linewidth]{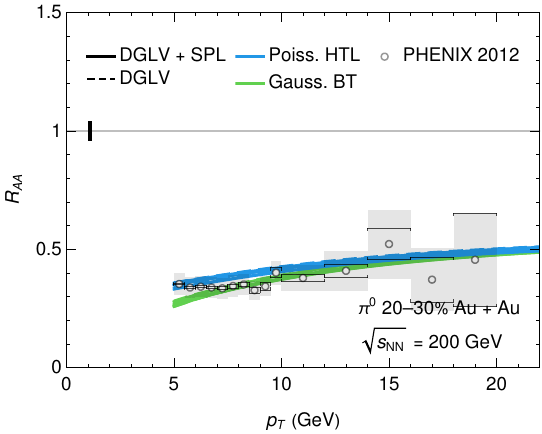}
    \caption{Nuclear modification factor $R_{AA}$ for $\pi^0$ in $20\text{--}30\%$ central \coll{Au}{Au} collisions.}
    \label{fig:PHENIX_RAA_Pi0_20-30_2012}
\end{subfigure}\hfill
\begin{subfigure}[t]{0.49\textwidth}
    \centering
    \includegraphics[width=\linewidth]{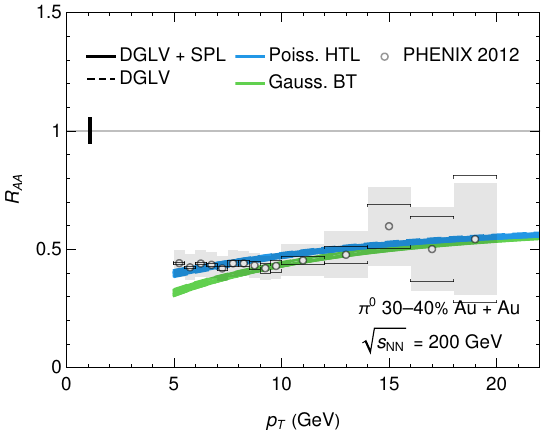}
    \caption{Nuclear modification factor $R_{AA}$ for $\pi^0$ in $30\text{--}40\%$ central \coll{Au}{Au} collisions.}
    \label{fig:PHENIX_RAA_Pi0_30-40_2012}
\end{subfigure}
\caption{Comparison of nuclear modification factors $R_{AA}$ for $\pi^0$ in $20\text{--}30\%$ and $30\text{--}40\%$ central \coll{Au}{Au} collisions at $\sqrt{s_{NN}} = 200$ GeV. Data is from \cite{PHENIX:2012jha}, with statistical and systematic uncertainties indicated by error bars and boxes, respectively.}
\label{fig:PHENIX_RAA_Pi0_20-40_2012}
\end{figure}

\begin{figure}[!htpb]
\centering
\begin{subfigure}[t]{0.49\textwidth}
    \centering
    \includegraphics[width=\linewidth]{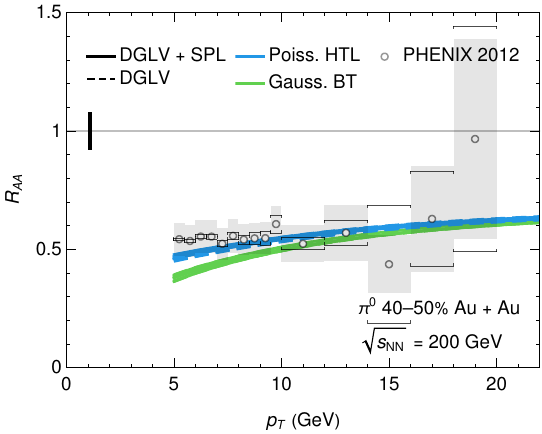}
    \caption{Nuclear modification factor $R_{AA}$ for $\pi^0$ in $40\text{--}50\%$ central \coll{Au}{Au} collisions.}
    \label{fig:PHENIX_RAA_Pi0_40-50_2012}
\end{subfigure}\hfill
\begin{subfigure}[t]{0.49\textwidth}
    \centering
    \includegraphics[width=\linewidth]{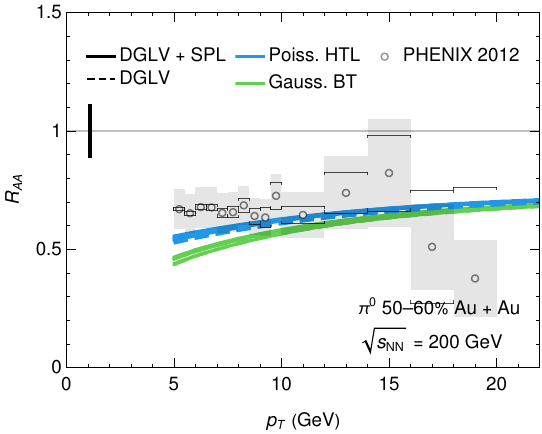}
    \caption{Nuclear modification factor $R_{AA}$ for $\pi^0$ in $50\text{--}60\%$ central \coll{Au}{Au} collisions.}
    \label{fig:PHENIX_RAA_Pi0_50-60_2012}
\end{subfigure}
\caption{Comparison of nuclear modification factors $R_{AA}$ for $\pi^0$ in $40\text{--}50\%$ and $50\text{--}60\%$ central \coll{Au}{Au} collisions at $\sqrt{s_{NN}} = 200$ GeV. Data is from \cite{PHENIX:2012jha}, with statistical and systematic uncertainties indicated by error bars and boxes, respectively.}
\label{fig:PHENIX_RAA_Pi0_40-60_2012}
\end{figure}

\begin{figure}[!htpb]
\centering
\begin{subfigure}[t]{0.49\textwidth}
    \centering
    \includegraphics[width=\linewidth]{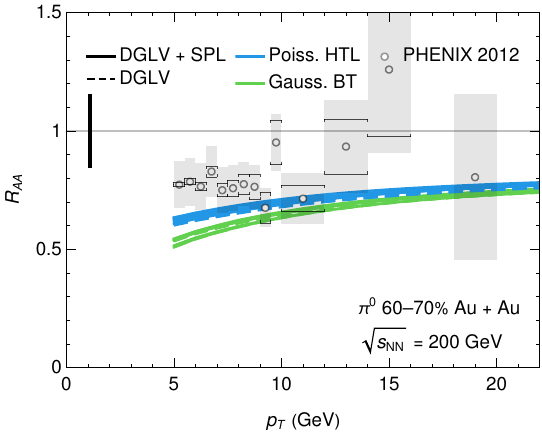}
    \caption{Nuclear modification factor $R_{AA}$ for $\pi^0$ in $60\text{--}70\%$ central \coll{Au}{Au} collisions.}
    \label{fig:PHENIX_RAA_Pi0_60-70_2012}
\end{subfigure}\hfill
\begin{subfigure}[t]{0.49\textwidth}
    \centering
    \includegraphics[width=\linewidth]{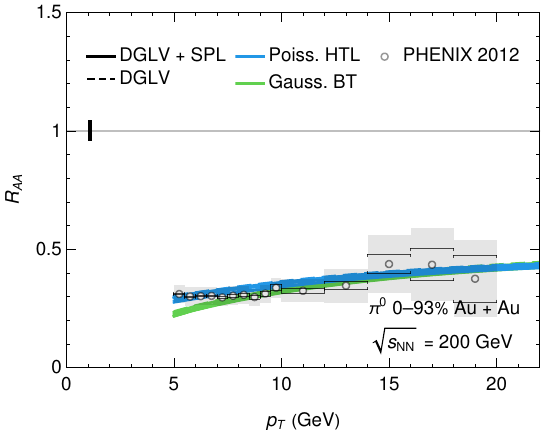}
    \caption{Nuclear modification factor $R_{AA}$ for $\pi^0$ in $0\text{--}93\%$ central \coll{Au}{Au} collisions.}
    \label{fig:PHENIX_RAA_Pi0_0-93_2012}
\end{subfigure}
\caption{Comparison of nuclear modification factors $R_{AA}$ for $\pi^0$ in $60\text{--}70\%$ and $0\text{--}93\%$ central \coll{Au}{Au} collisions at $\sqrt{s_{NN}} = 200$ GeV. Data is from \cite{PHENIX:2012jha}, with statistical and systematic uncertainties indicated by error bars and boxes, respectively.}
\label{fig:PHENIX_RAA_Pi0_60-93_2012}
\end{figure}

\begin{figure}[!htpb]
\centering
\begin{subfigure}[t]{0.49\textwidth}
    \centering
    \includegraphics[width=\linewidth]{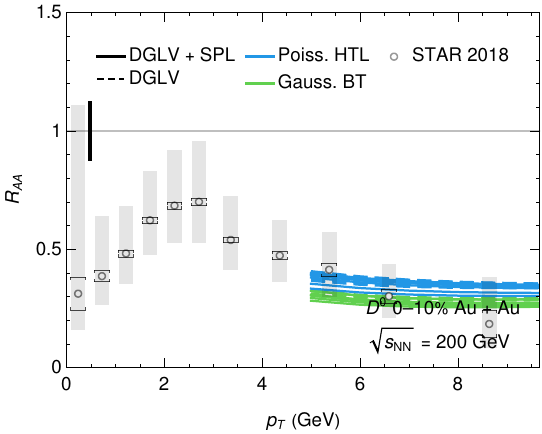}
    \caption{Nuclear modification factor $R_{AA}$ for $D^0$ in $0\text{--}10\%$ central \coll{Au}{Au} collisions.}
    \label{fig:STAR_RAA_D0_0-10_2018}
\end{subfigure}\hfill
\begin{subfigure}[t]{0.49\textwidth}
    \centering
    \includegraphics[width=\linewidth]{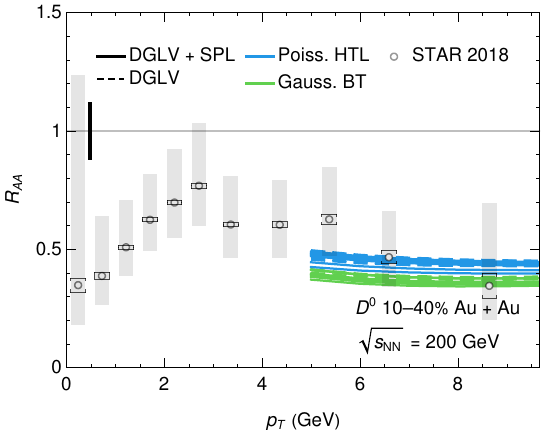}
    \caption{Nuclear modification factor $R_{AA}$ for $D^0$ in $10\text{--}40\%$ central \coll{Au}{Au} collisions.}
    \label{fig:STAR_RAA_D0_10-40_2018}
\end{subfigure}
\caption{Comparison of nuclear modification factors $R_{AA}$ for $D^0$ in $0\text{--}10\%$ and $10\text{--}40\%$ central \coll{Au}{Au} collisions at $\sqrt{s_{NN}} = 200$ GeV. Data is from \cite{STAR:2018zdy}, with statistical and systematic uncertainties indicated by error bars and boxes, respectively.}
\label{fig:STAR_RAA_D0_0-40_2018}
\end{figure}

\begin{figure}[!htpb]
\centering
\includegraphics[width=0.49\textwidth]{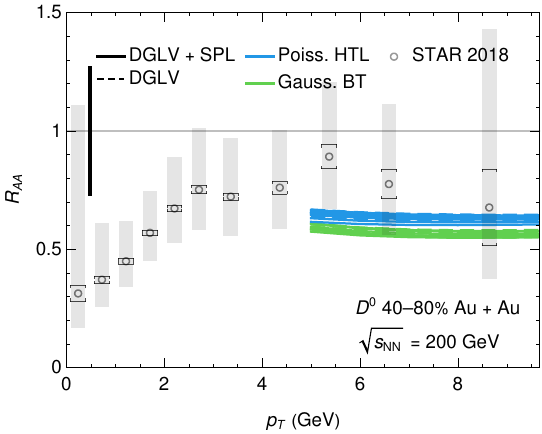}
\caption{Nuclear modification factor $R_{AA}$ for $D^0$ in $40\text{--}80\%$ central \coll{Au}{Au} collisions.}
\label{fig:STAR_RAA_D0_40-80_2018}
\end{figure}

\FloatBarrier
\section{Small system LHC data}
\label{sec:small_systems}

\begin{figure}[!htpb]
\centering
\begin{subfigure}[t]{0.49\textwidth}
    \centering
    \includegraphics[width=\linewidth]{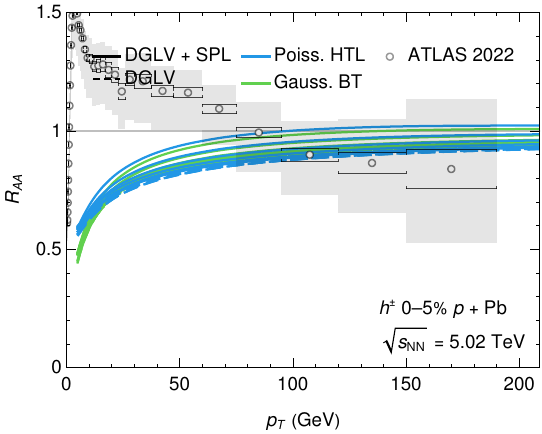}
    \caption{Nuclear modification factor $R_{AA}$ for charged hadrons in $0\text{--}5\%$ central \coll{p}{Pb} collisions.}
    \label{fig:ATLAS_RAA_Charged_0-5_2022}
\end{subfigure}\hfill
\begin{subfigure}[t]{0.49\textwidth}
    \centering
    \includegraphics[width=\linewidth]{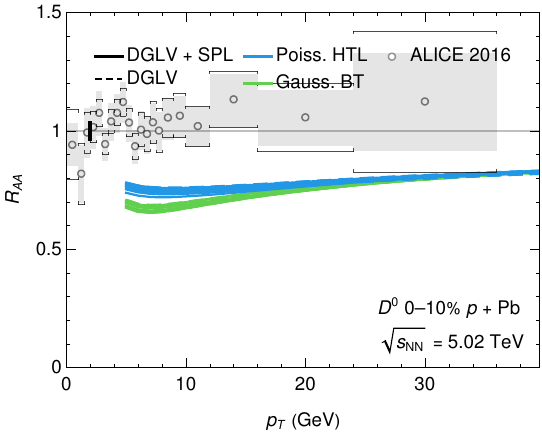}
    \caption{Nuclear modification factor $R_{AA}$ for $D^0$ in $0\text{--}10\%$ central \coll{p}{Pb} collisions.}
    \label{fig:ALICE_RAA_D0_0-10_2016}
\end{subfigure}
\caption{Comparison of nuclear modification factors $R_{AA}$ for charged hadrons and $D^0$ in central \coll{p}{Pb} collisions. Data is from \cite{ATLAS:2022kqu} and \cite{ALICE:2016yta}, with statistical and systematic uncertainties indicated by error bars and boxes, respectively.}
\label{fig:pPb_comparison}
\end{figure}

\FloatBarrier
\section{Small system RHIC data}
\label{sec:small_system_rhic_data}

\begin{figure}[!htpb]
\centering
\begin{subfigure}[t]{0.49\textwidth}
    \centering
    \includegraphics[width=\linewidth]{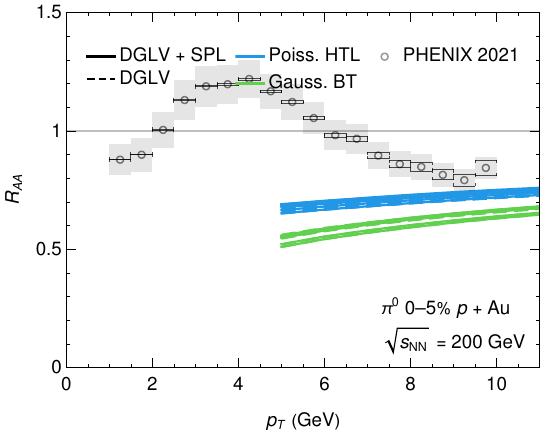}
    \caption{Nuclear modification factor $R_{AA}$ for $\pi^0$ in $0\text{--}5\%$ central \coll{p}{Au} collisions.}
    \label{fig:PHENIX_RAA_Pi0_pAu_0-5_2021}
\end{subfigure}\hfill
\begin{subfigure}[t]{0.49\textwidth}
    \centering
    \includegraphics[width=\linewidth]{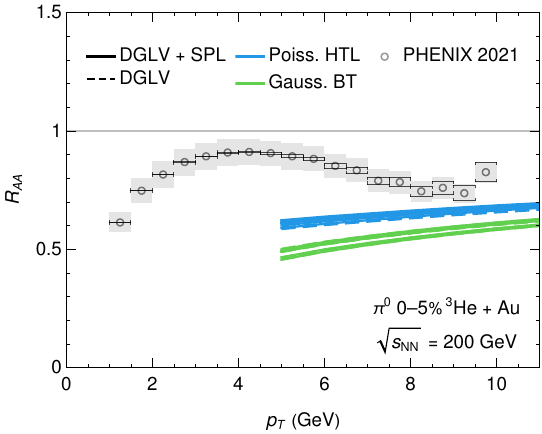}
    \caption{Nuclear modification factor $R_{AA}$ for $\pi^0$ in $0\text{--}5\%$ central \coll{He3}{Au} collisions.}
    \label{fig:PHENIX_RAA_Pi0_He3Au_0-5_2021}
\end{subfigure}
\caption{Comparison of $R_{AA}$ for $\pi^0$ in central \coll{p}{Au} and \coll{He3}{Au} collisions.}
\label{fig:PHENIX_RAA_Pi0_Comparison_1}
\end{figure}

\begin{figure}[!htpb]
    \centering
    \includegraphics[width=0.49\linewidth]{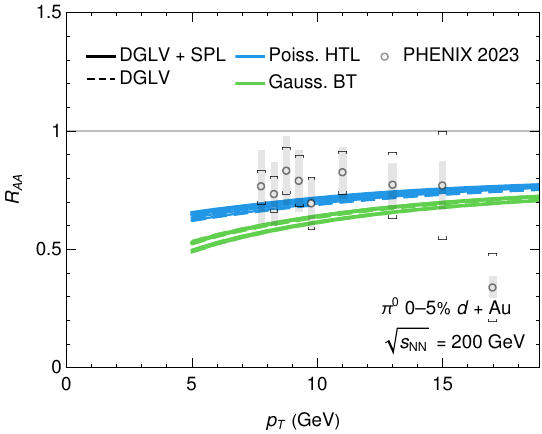}
    \caption{Nuclear modification factor $R_{AA}$ for $\pi^0$ in $0\text{--}5\%$ central \coll{d}{Au} collisions.}
	\label{fig:PHENIX_RAA_Pi0_dAu_0-5_2023}
\end{figure}

\end{document}